\documentclass[12pt,preprint]{aastex}

\usepackage{graphicx}
\usepackage{subfigure}

\shorttitle{2MASS Analysis of Bok Globules}
\shortauthors{Racca et al.}

\newcommand{\tco}{$^{13}$CO}
\newcommand{\ceio}{C$^{18}$O}

\begin{document}

\title{A 2MASS Analysis of the Stability of Southern Bok Globules\altaffilmark{1}}

\author{Germ\'an A. Racca and Jos\'e W. S. Vilas-Boas}
\affil{INPE-DAS, Av. dos Astronautas 1758, Jardim da Granja,
12227-010, S\~{a}o Jos\'e dos Campos-SP, Brazil.}
\email{german@das.inpe.br, jboas@das.inpe.br}
 \and
\author{Ramiro de la Reza}
\affil{Observat\'orio Nacional, Rua General Jos\'e Cristino 77, S\~{a}o Crist\'ov\~{a}o,
20921-400, Rio de Janeiro, Brazil.}
\email{delareza@on.br}

\altaffiltext{1}{Based on a Ph.D. Thesis made at Observat\'orio Nacional,
Rio de Janeiro, Brazil.}

\begin{abstract}

We used near-infrared 2MASS data to construct visual extinction maps of a sample of Southern
Bok globules utilizing the NICE method. We derived radial extinction profiles of dense cores identified
in the globules and analyzed their stability against gravitational collapse with isothermal Bonnor-Ebert
spheres. The frequency distribution of the stability parameter ($\xi_{max}$) of these cores shows that
a large number of them are located in stable states, followed by an abrupt decrease of cores in unstable
states. This decrease is steeper for globules with associated IRAS point sources than for starless globules.
Moreover, globules in stable states have a Bonnor-Ebert temperature of $T = 15 \pm 6$~K, while the
group of critical plus unstable globules has a different temperature of $T = 10 \pm 3$~K. Distances were
estimated to all the globules studied in this work and the spectral class of the IRAS sources was calculated.
No variations were found in the stability parameters of the cores and the spectral class of their associated
IRAS sources.
On the basis of \tco\ J = 1--0 molecular line observations, we identified and modeled a blue-assymetric line
profile toward a globule of the sample, obtaining an upper limit infall speed of 0.25 km s$^{-1}$.

\end{abstract}


\section{Introduction}
\label{sec:intro}

\citet{bok47} were the first to call attention to small dark regions, dense
and rounded, which called \emph{globules}, and suggested the possibility
that the star formation could occur within them. These objects were called
later Bok globules and in general they are isolated, cold ($T \sim$ 10~K),
dense ($n \sim 10^{4-5}$~cm$^{-3}$), and seen as potential sites for the
formation of low-mass stars ($<$ 10~M$_{\odot}$).
Following \citet{beichman86}, the globules that do not contain
protostars or young stellar objects (YSOs) associated, are called starless
cores. When they show some evidence of star formation, such as
Class 0 or I protostars, are called protostellar-cores \citep{andre93}.

As the majority of the stars form in molecular cloud complexes and in
clusters embedded in giant molecular clouds and not in isolation, it is very
important to know in detail the physical processes that give rise to the
formation of a single star in an isolated environment. For this purpose, Bok
globules constitute an ideal laboratory to study the isolated low-mass star
formation \citep[][hereafter BHR]{bourke95a}.

Visual extinction was used by \citet{clemens88} to study 248 Bok globules in
the Northern Hemisphere, and by BHR to study 169 globules in the Southern
Hemisphere.
Sub-millimeter and millimeter wavelength studies of starless cores
\citep{ward94}, indicate that the 1D profiles (mean of azimuthal
distributions) are flat for radii $\sim$ 3000-7000~AU, and steeper for
larger radii. These profiles deviate from the distribution $n(r) \propto r^{-2}$
of the singular isothermal sphere, proposed by \citet{shu77} as the
initial state of isolated dense cores before gravitational collapse. Similar
profiles are obtained by mapping the dust distribution in a dense core or
Bok globule. These configurations are consistent with ``Bonnor-Ebert
spheres'' \citep{ebert55,bonnor56}, which are non-singular solutions to the
equations of hydrostatic equilibrium confined by an external pressure. In
the 1D isothermal approximation, there is a family of Bonnor-Ebert solutions
parametrized by the central density $n_{c}$, where the density profile is
characterized by two regimes: a slow decrease in the density for small
radii, and a faster decrease, of power-law type ($ \propto r^{-2}$), for
larger radii.

Detailed modelling of profiles in the infrared \citep{kandori05,teixeira05}
and continuum emission \citep[e.g.,][]{ward99} confirm that Bonnor-Ebert
spheres are a good approach to the internal structure of starless cores.
However, it is not currently possibly to distinguish between a static
Bonnor-Ebert sphere and a collapsing one for profiles based in dust
absorption or emission, since the structure of the profile, for a given $n_c$,
does not change significantly until the last stages of collapse \citep{myers05}.

Using near-infrared observations, \citet{alves01} showed that a Bonnor-Ebert
sphere with stability parameter $\xi_{max}$ = 6.9 adjust perfectly the
observed column density profile of the starless globule Barnard 68.
\citet{harvey01} modeled the profile of the protostellar globule B335 with
$\xi_{max}$ = 12.5, and \citet{harvey03} estimated $\xi_{max}$ = 25 for the
starless core L694-2. The Coalsack globule 2 was modeled by \citet{racca02}
that derived $\xi_{max}$ = 7.3, while \citet{lada04} obtained $\xi_{max}$ = 5.8
using a more sensitive telescope. In recent works, \citet{huard06} found
$\xi_{max}$ = 35.8 for the protostellar core L1014, and \citet{kainu07}
obtained $\xi_{max}$ = 23 for the protostellar globule DCld303.8-14.2 and
$\xi_{max} \gtrsim$ 8 for the starless globule Thumbprint Nebula.

The largest sample of Bok globules studied in the near-infrared was done by
\citet{kandori05} where 14 globules were selected. They found that more than
half of the starless globules (7 out of 11) are located close to the
critical state, with $\xi _{max}$ = 6.5 $\pm$ 2. This led \citet{kandori05}
to suggest that a Bonnor-Ebert sphere in the critical state characterizes
the structure of typical starless cores. The remaining starless globules and
those that present evidence of star formation show clearly unstable states,
with $\xi _{max}>$ 10.

In this work, we present a study of 21 Southern Bok globules, where 11 are
starless and 10 have associated IRAS\footnote{Infrared Astronomical
Satellite \citep{neugebauer84}} point sources. In \S \ref{sec:observ}
we present the source list and the observations. Using 2MASS data, we
constructed visual extinction maps in \S \ref{sec:extmaps}, which enabled us
to detect dense cores embedded in the globules. In \S \ref{sec:distance} we
estimated new distances to the globules, and by means of the Bonnor-Ebert
model, in \S \ref{sec:structure} we studied their internal structure and
stability against gravitational collapse. In \S \ref{sec:starfrom} we
analized the IRAS point sources associated to the globules and in \S
\ref{sec:collapsing} we found evidence of infall motions in a globule using
millimetric observations of the \tco\ and \ceio\ molecules. Our
main conclusions are summarized in \S \ref{sec:summary}.

\section{Observations}
\label{sec:observ}

\subsection{Sample List}
\label{sec:sample}

In this work, we searched for a list of Bok globules
satisfying the following conditions:

\begin{enumerate}
\item that they were completely isolated, i.e., neither associated to bright
nebula nor to a molecular complex,

\item that they had a distance determination,
\end{enumerate}

\noindent
Using these criteria, we got 21 Southern Bok globules, selected
among 169 globules of the BHR sample, where 11 do not present any evidence
of star formation, and 10 have associated IRAS point sources. Following
previous definitions (see \S \ref{sec:intro}), in the first group there are
``starless globules'', and in the second there are ``IRAS globules''.

Table \ref{tab:sample} presents the selected objects. Column 1 and 2 show
the names of the globules as they appear in BHR and \citet{hartley86},
respectively; columns 3 and 4 give their equatorial coordinates; column 5
shows the optical sizes in arcmins, and column 6 gives the distances
determined by BHR.

%==============================
\begin{deluxetable}{crcccc}
\tablecaption{\label{tab:sample}Sample List}
\tabletypesize{\footnotesize}
\tablewidth{0pt}
\tablecolumns{6}
\tablehead{
\colhead{Name} & \colhead{DC No.} & \colhead{R.A.} & \colhead{Dec.} & \colhead{Size} & \colhead{Distance} \\
 & & \colhead{(J2000)} & \colhead{(J2000)} & \colhead{($a' \times b'$)} & \colhead{(pc)}
}

\startdata
\cutinhead{Starless Globules}
016	&	255.4-3.9		&	08 05 26	&	$-$39 08 54	&	6	$\times$	2	&	300	\\
044	&	269.5+4.0 	&	09 26 19	&	$-$45 11 00	&	2	$\times$	2	&	300	\\
053	&	274.2-0.4 	&	09 28 47	&	$-$51 36 42	&	8	$\times$	4	&	500	\\
059	&	291.1-1.7 	&	11 07 07	&	$-$62 05 48	&	5	$\times$	3	&	250	\\
074	&	300.0-3.7 	&	12 22 09	&	$-$66 27 06	&	3	$\times$	3	&	175	\\
075	&	300.2-3.5 	&	12 24 13	&	$-$66 10 42	&	6	$\times$	3	&	175	\\
111	&	327.2+1.8 	&	15 42 20	&	$-$52 49 06	&	3	$\times$	2	&	250	\\
113	&	331.0-0.7 	&	16 12 43	&	$-$52 15 36	&	8	$\times$	4	&	200	\\
133	&	340.5+0.5 	&	16 46 45	&	$-$44 30 48	&	10	$\times$	4 $\:$	&	700	\\
144	&	346.4+7.9 	&	16 37 28	&	$-$35 13 54	&	7	$\times$	1	&	170	\\
145	&	347.5-8.0 	&	17 48 01	&	$-$43 43 12	&	8	$\times$	5	&	450	\\
\cutinhead{IRAS Globules}
034	&	267.2-7.2 	&	08 26 34	&	$-$50 39 54	&	8	$\times$	2	&	400	\\
058	&	289.3-2.8 	&	10 49 00	&	$-$62 23 06	&	5	$\times$	3	&	250	\\
117	&	334.6+4.6 	&	16 06 18	&	$-$45 55 18	&	7	$\times$	2	&	250	\\
121	&	337.1-4.9 	&	16 58 42	&	$-$50 35 48	&	2	$\times$	2	&	300	\\
126	&	338.6+9.5 	&	16 04 29	&	$-$39 37 48	&	6	$\times$	5	&	170	\\
138	&	345.0-3.5 	&	17 19 36	&	$-$43 27 06	&	10	$\times$	6 $\:$	&	400	\\
139	&	345.2-3.6 	&	17 20 45	&	$-$43 20 30	&	4	$\times$	2	&	400	\\
140	&	345.4-4.0 	&	17 22 55	&	$-$43 22 36	&	6	$\times$	3	&	400	\\
148	&	349.0+3.0 	&	17 04 26	&	$-$36 18 48	&	2.5	$\times$	2.5	&	200	\\
149	&	349.2+3.1 	&	17 04 27	&	$-$36 08 24	&	3	$\times$	1	&	200	\\
\enddata

\tablecomments{Units of right ascension are hours, minutes, and seconds, and
units of declination are degrees, arcminutes, and arcseconds.}
\end{deluxetable}
%==============================

\subsection{Near-Infrared Data}
\label{sec:infrared}

The near-infrared data were obtained from the Two-Micron All Sky Survey
(2MASS) catalog \citep{skrutskie06}. The 2MASS Point Source Catalog (PSC)
contains 471 million stars, covering 99.998 \% of the sky in the J(1.24 $\mu$m),
H(1.66 $\mu$m) and K$_{s}$(2.16 $\mu$m)\footnote{From hereafter,
filter ``K$_{s}$'' will be called simply ``K''.} bands,
observed with two telescopes of 1.3 meters diameter, located in Mount
Hopkins (Arizona, USA) and Cerro Tololo (Chile). The 2MASS PSC is more than
99\% complete for J $<$ 15.8, H $<$ 15.1 and K $<$ 14.3.

For each of the globules in Table \ref{tab:sample}, we retrieved JHK
magnitudes and their corresponding uncertainties in the 2MASS PSC in regions
of 15$'$ $\times$ 15$'$ centered on the globules. Only those
stars whose photometric uncertainties were $\leq$ 0.1 mag in all three bands
were extracted, ensuring a signal-to-noise ratio $S/N \geq$ 10. The search
for stars in the 2MASS PSC was made through the web interface Gator\footnote{
\url{http://irsa.ipac.caltech.edu/applications/Gator/}}.

\subsection{Molecular Line Data}
\label{sec:molecular}

We searched for molecular lines observed toward these globules
having line profiles data available. We found only \tco\ and \ceio\ (J = 1--0)
line observations toward the globules BHR 138 and BHR 149 made
in 1992 October and 1993 May \citep[see][]{vilasb94}. These observations
were made with the 15~m Swedish-ESO Submillimeter Telescope (SEST) at La
Silla, Chile. The receiver front end was based on a Schottky diode waveguide
mixer followed by an intermediate-frequency amplifier. The SSB system
temperature was typically 390~K.

The back end used an acousto-optical spectrometer with resolution 43~kHz
(0.11 km s$^{-1}$) and total bandwidth of 100~MHz. The spectra were taken by
using overlap frequency switching mode (with 7~MHz frequency shift) and
integrating, on average, for periods of 2 minutes. The observations were
chopped against a cold load to obtain the correction for atmospheric
attenuation. The halfpower beam width was 48$''$ and the beam
efficiency 0.9. The rms pointing accuracy was better than 10$''$.

\section{Visual Extinction Maps}
\label{sec:extmaps}

We used the NICE method of \citet{lada94} to construct visual extinction
maps for the globules. This technique uses a control field region free of
reddening to refer the extinctions measured in the cloud region, and assumes
that the stellar population is the same in the control field and the cloud
region. If we know the intrinsic color of a star, then the color excess is:

\begin{equation}
E(H-K) = (H-K)_{observed} - (H-K)_{intrinsic}.
\end{equation}

\noindent
Intrinsic $(H-K)$ colors for main-sequence and giant stars are in
a small interval, 0.0 to 0.3 for stars with spectral types between A0 and
M5-7 \citep{koor83,bessell88}. Therefore, we can approximate the intrinsic
color of each on-cloud star by the average color of the control field, and
the color excess is:

\begin{equation}
\label{eq:excess}
E(H-K) = (H-K) - \overline{(H-K)}_{control}.
\end{equation}

Using equation~\ref{eq:excess}, we calculate the color excess for each star
in the globule region. Then, we can express this color excess in terms of
visual extinction with the knowledge of the reddening law in the 2MASS
system. Folowing \citet{nielbock05}, this relation is:

\begin{equation}
\label{eq:Av}
A_V = 19.4 \times E(H-K),
\end{equation}

\noindent
whose associated uncertainty is:

\begin{equation}
\label{eq:sigmaAv}
\sigma_{A_V} = 19.4 \times \sqrt{\sigma_i^2 + \sigma_{control}^2},
\end{equation}

\noindent
where $\sigma _{i}$ is the uncertainty in the observed $(H-K)$
color and $\sigma _{control}$ is the standard deviation of the colors in the
control field. Nearly 90\% of the regions have $\sigma _{control} \sim 0.1$
mag, which sets the minimum mensurable value for the extinction in $ \sim $ 2
mag. Only two regions in our sample have $\sigma_{control}\sim $ 0.2 mag,
giving a minimum of $\sim$ 4 mag of visual extinction.

In this way, we obtained the distribution of extinction in the fields
containing the Bok globules. However, the spatial distribution of stars is
non-uniform. In order to
produce a uniformly sampled extinction map, we smoothed the data by
spatially convolving the extinction measurements (equation~\ref{eq:Av}) with
a Gaussian filter (kernel) with a given smoothing
parameter (resolution), and finally sampled the map at the Nyquist frequency
\citep[e.g.,][]{lada99}. The form of the Gaussian kernel is given by:

\begin{equation}
\label{eq:kernel}
K(\alpha,\alpha_i,\delta,\delta_i) = \frac{1}{2\pi} \: \mathrm{exp} \left( -\frac{r_i^2}{2h^2}\right),
\end{equation}

\noindent
where

\begin{equation}
r_i^2 = (\delta - \delta_i)^2 + (\alpha - \alpha_i)^2 \mathrm{cos}^2(\delta),
\end{equation}

\noindent
and $h$ is the map resolution. Hence, the visual extinction $A_V$
at each point $(\alpha,\delta)$ in the map is calculated using:

\begin{equation}
\label{eq:Av2}
A_V (\alpha,\delta) = \frac{\sum_{i=1}^n K(\alpha,\alpha_i,\delta,\delta_i)
\times A_V (\alpha_i,\delta_i)}{\sum_{i=1}^n
K(\alpha,\alpha_i,\delta,\delta_i)},
\end{equation}

\noindent
where $n$ is the total number of stars observed in the cloud
region and $A_V (\alpha_i,\delta_i)$ is given by equation~\ref{eq:Av}, and
the associated uncertainty for the extinction at the point $(\alpha,\delta)$
in the map is \citep{lombardi01}:

\begin{equation}
\label{eq:sigmaAv2}
\sigma_{A_V} (\alpha,\delta) = \sqrt{\frac{\sum_{i=1}^n
K^2(\alpha,\alpha_i,\delta,\delta_i) \times \sigma_{A_V}^2
(\alpha_i,\delta_i)}{\sum_{i=1}^n K^2(\alpha,\alpha_i,\delta,\delta_i)}},
\end{equation}

\noindent
where $\sigma_{A_V} (\alpha_i,\delta_i)$ is given by equation~\ref{eq:sigmaAv}.

\subsection{Construction of the Extinction Maps}
\label{sec:construct}

To test the 2MASS data using the NICE method, we applied it to Barnard 68
(B68), a well known and studied Bok globule. \citet{alves01} observed B68
with the ESO-NTT telescope in the JHK bands to produce an extinction map of
this globule with a resolution of 10$''$.

\begin{figure}
\centering
\includegraphics[scale=0.85]{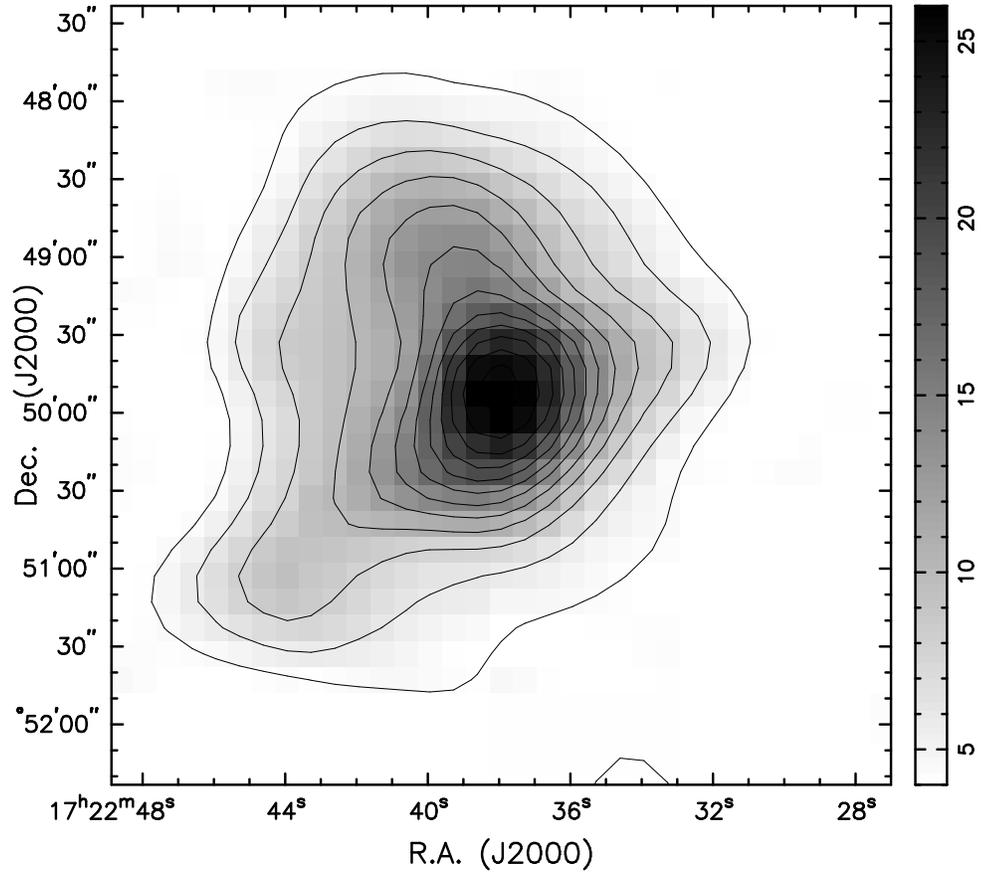}
\caption{Extinction map for B68 constructed using 2MASS data with resolution 20$''$. Contours
               start at $A_V$ = 4 mag and increases in steps of 2 to 26 mag. Compare with the map
               of \citet{alves01}.}
\label{fig:b68}
\end{figure}

The choice of the smoothing parameter $h$ is a compromise between a final
map with high signal-to-noise ratio and low spatial resolution, or a noisy
map with high resolution \citep{lombardi01}. To construct the extinction map
of B68 (and all the globules in this work) we choose $h$ = 20$''$.
Nonetheless, other values were tested: with $h$ = 10$''$ the map is very noisy,
and with $h$ = 30$''$ the map loses details. Therefore, a choice of $h$ = 20$''$
provides the best spatial resolution for a signal-to-noise ratio between 8
and 12 in the central region of B68. Figure \ref{fig:b68} shows the visual
extinction map for B68 constructed using data from the 2MASS catalog.
Comparing with the map of \citet{alves01}, we see a good correspondence
between them, taking into account that the 2MASS used a 1.3 meters
telescope, while \citeauthor{alves01} used a 3.5 meters telescope and longer
integration times. The sensitivity difference between these telescopes is
certainly responsible for the different values of extinction found in the center of
B68. From the 2MASS data, we obtain a maximum extinction of 24 magnitudes,
while \citet{alves01} and \citet{hotzel02b}, using the same data set, derived
30 magnitudes. This result shows that the extinctions derived toward the center
of the globules in our sample, as well as the density contrast, have to be seen
as lower limits. In despite of getting a lower extinction peak, the extinction distribution
and the peak position are coincident.

Figure \ref{fig:maps} shows the visual extinction maps for all Bok globules
in our sample. They were constructed with the NICE method as described in \S
\ref{sec:extmaps}. Each map has $\sim$~$15' \times 15'$,
centered on the position listed in Table \ref{tab:sample}. In the right side
of each map, a bar indicates the extinction in magnitudes. White crosses
superimposed denote the IRAS sources associated to the globules
(see \S \ref{sec:starfrom}).

%% This subsection was added to answer referee question number 1 %%
\subsection{Comparison with the Star Counts Method}
\label{sec:scounts}

In this item we compare star counts and the NICE methods in order to verify
which one is more apropriate to study extinction toward the Bok globules
of our sample, derived from 2MASS data.

The 2MASS maps constructed in this work were performed only with
those stars whose color excess $E(H-K) > 0$, thus removing almost all
foreground stars that can be present in each globule region. This
allowed us to minimize the effects that foreground stars can produce
in the construction of the extinction maps \citep[see Figure 6 of][]{cambresy02}.

In order to test if the 2MASS data together with the NICE method
can produce reliable dust extinction maps of Bok globules, we also
constructed maps using the traditional star counts method. The
number of stars per unit area brigher than $m_{\lambda}$ in the control
field, $N_{off}$, is:

\begin{equation}
 \label{eq:counts}
 \mathrm{log}(N_{off}) = a + b m_{\lambda}.
\end{equation}

\noindent
On the cloud region, the slope of equation \ref{eq:counts} remains the
same, but the number of stars per unit area, $N_{on}$, decreases as
extinction increases. The extinction is then:

\begin{equation}
 \label{eq:extcounts}
 A_{\lambda} = \frac{1}{b} \: \mathrm{log}\left(\frac{N_{off}}{N_{on}}\right),
\end{equation}

\noindent
where $b$ is the slope of the cumulative luminosity function
\citep[see, e.g.,][]{jane88}. We converted the $A_{\lambda}$ extinction
(with $\lambda$ = J,H,K) to optical $A_V$ extinction using the reddening
law in the 2MASS system \citep{nielbock05}.

We constructed extinction maps for B68, our test globule, and for BHR 059,
one of the smallest globules in our sample, using the star counts technique
with the condition that, at least, one star must be present in each counting
box. We obtained three Nyquist sampled maps for each globule, that is, one
map for counts in J, H and K bands, respectively. Table \ref{tab:counts}
shows the parameters used in equation \ref{eq:extcounts} to build the maps.

%==============================
\begin{deluxetable}{cccccccc}
\tablecaption{\label{tab:counts}Parameters of the Star Counts Maps}
%\tabletypesize{\scriptsize}
%\rotate
\centering
\tablewidth{0pt}
\tablecolumns{8}
\tablehead{
\colhead{} & \multicolumn{3}{c}{Barnard 68} & \colhead{} & \multicolumn{3}{c}{BHR 059} \\
\cline{2-4} \cline{6-8} \\
\colhead{Wavelenght} & \colhead{Bin} & \colhead{$N_{off}$} & \colhead{b} & \colhead{} & \colhead{Bin} & \colhead{$N_{off}$} & \colhead{b} \\
}

\startdata
J & 80$''$ &      21 & 0.36 $\pm$ 0.03 & & 70$''$ & 16 & 0.35 $\pm$ 0.02 \\
H & 50$''$ & $\phn$8 & 0.35 $\pm$ 0.03 & & 70$''$ & 15 & 0.34 $\pm$ 0.01 \\
K & 50$''$ & $\phn$7 & 0.37 $\pm$ 0.02 & & 70$''$ & 13 & 0.32 $\pm$ 0.01 \\
\enddata

\end{deluxetable}
%==============================

The extinction maps for B68 and BHR 059 are shown in Figure \ref{fig:countsmaps}.
We can see that B68 is not detected in the map based on K band star counts.
This is probably because the distribution of the stars in the K band is almost
uniform in the most opaque region of B68. The H and J band maps show B68 as a
region of extent $\sim$ 3-4 pixels without any details on the distribution of
dust extinction in the core. It is important to note that a non uniforme
distribution of field stars causes unrealistic peaks of high visual extinction,
which are not seen in the DSS red plates. The same situation occurs for BHR 059.

We estimated the erros in the derived visual extinctions \citep{dickman78}
and obtained $\delta A_V \sim$ 4, 8 and 12 magnitudes for J, H and K band maps,
respectively, in regions where $A_V$ = 0 (where $N_{on} = N_{off}$) and
$\delta A_V \sim$ 10, 20 and 30 magnitudes for J, H and K band maps, respectively,
in the most opaque regions (where $N_{on}$ = 1 in a single counting box).
On the other hand, the signal-to-noise ratio for the maps constructed through the
NICE method is always greater than 3 in the most extincted regions where we constructed
the radial extinction profiles. Even though the signal-to-noise ratio is not
as good for the more diffuse, low extinction regions, it is not crucial for
our purposes, because we are interested in the profile of the core embedded
in the globule to finally model it with a Bonnor-Ebert sphere. This discussion
demonstrate that, to our purposes, the NICE method applied to the 2MASS data
is the best way to construct visual extinction maps and trace
Bok globules to finally apply the stability analysis of the radial profile.

\section{Distance Determination}
\label{sec:distance}

The knowledge of the distances to the globules is extremely important, as
they are necessary to determine several fundamental parameters, such as
mass, size, density and also to calculate the luminosities of the young
objects embedded in the globules. Several methods have been used to
determine distances, being the more typical the star count method
\citep{bok41} and the Wolf diagram \citep{wolf23}. A newer technique,
proposed by \citet{maheswar04}, uses optical and near-infrared broadband
photometry of stars in the field of the cloud.

In this paper, we employ the technique of \citet{dickman83}, which uses the
plot of color excess $E(B - V)$ versus the distance to stars in the vicinity
of the cloud. In this graph, a jump in the reddening indicates the presence
of a cloud or a group of clouds. The same method was used by BHR to estimate
the distance to some globules. However, for the reasons described in \S \ref{sec:applic},
we decided to calculate ourselves these distances again.

\subsection{Application of the Method}
\label{sec:applic}

To estimate the distances to the Bok globules, we need to know the reddening 
$E(B-V)$ of the stars in a given area around each globule. \citet{dickman83}
used a region of 11$^{\circ}$ $\times $ 16$^{\circ}$ in their work. BHR
used circular regions of 5$^{\circ}$ radius. If the region had not a
sufficient amount of stars, they increased the radius to 7.5$^{\circ}$ or 10$^{\circ }$.
In this work, we used the SKY2000 catalog of \citet{sky2000},
which has information for $\sim$ 300,000 stars brighter than 8 magnitudes,
and searched for stars in circles of 3$^{\circ}$ radius centered on each
Bok globule. Figure \ref{fig:galactic} shows the galactic distribution of
the globules, and the circles are their neighborhood used to determine the
distance. As we see, some regions coincide, given the proximity between the
globules. In these cases, both are assumed to be at the same distance.

To calculate the color excess $E(B - V)$ for each star, we searched the
SKY2000 catalog for all those stars with known spectral types, apparent
visual magnitude $V$, and observed color index $(B - V)$. Having the
spectral type of a star, we know the absolute magnitude $M_V$ and the
intrinsic color index $(B - V)_0$ using the calibration of \citet{kaler82}.
Assuming a value of 3.1 for the total-to-selective extinction ratio
\citep{rieke85}, we calculate the extinction along the line of sight of a
star in the neighborhood of a globule as:

\begin{equation}
A_V = 3.1 \times E(B-V) = 3.1 \times \left[ (B-V) - (B-V)_0 \right],
\end{equation}

\noindent
and finally the distance in parsecs, corrected for extinction:

\begin{equation}
\label{eq:dist}
\mathrm{log}(r) = \frac{1}{5} (V - M_V + 5 - A_V).
\end{equation}

Figure \ref{fig:dist} shows the graphics of $E(B - V)$ versus distance for
the globules, where the dashed lines indicate the distances adopted here.
In Table \ref{tab:distance}, column 1 indicates the BHR globule name,
column 2 the distance obtained in this work, and column 3 the distance
obtained by BHR.

%==============================
\begin{deluxetable}{ccc}
\tablecaption{\label{tab:distance}Distances to the Bok Globules}
%% \tabletypesize{\small}
\tablewidth{0pt}
\tablecolumns{3}
\tablehead{
\colhead{$\: \:$ Name $\: \:$} & \colhead{ $\: \:$ Distance\tablenotemark{a} $\: \:$} & \colhead{ $\: \:$ Distance\tablenotemark{b} $\: \:$} \\
 & \colhead{(pc)} & \colhead{(pc)}
}

\startdata
016 & 250 & 300 \\
034 & 200 & 400 \\
044 & 200 & 300 \\
053 & 200 & 500 \\
058 & 200 & 250 \\
059 & 200 & 250 \\
074 & 175 & 175 \\
075 & 175 & 175 \\
111 & 250 & 250 \\
113 & 200 & 200 \\
117 & 175 & 250 \\
121 & 125 & 300 \\
126 & 250 & 170 \\
133 & 225 & 700 \\
138 & 225 & 400 \\
139 & 225 & 400 \\
140 & 225 & 400 \\
144 & 225 & 170 \\
145 & 150 & 450 \\
148 & 175 & 200 \\
149 & 175 & 200 \\
\enddata

\tablenotetext{a}{Distance obtained in this work.}
\tablenotetext{b}{Distance obtained by \citet{bourke95a}.}
\end{deluxetable}
%==============================

\section{Internal Structure: Bonnor-Ebert Models}
\label{sec:structure}

To study the internal structure of the Bok globules, they were fitted
with Bonnor-Ebert isothermal spheres. From the extinction maps
generated in \S \ref{sec:extmaps}, radial extinction profiles were built
for all the Bok globules considered here.

A sphere of self-graviting gas in hydrostatic equilibrium, where the
pressure and density at each point are related through the isothermal
equation of state, is called a Bonnor-Ebert sphere \citep{ebert55,bonnor56}.
\citet{bonnor56} studied the gravitational stability of isothermal spheres
and deducted a critical parameter that allows classify them as stable or
unstable against gravitational collapse.

The shape of a Bonnor-Ebert density profile consists of a flat central
region and a steeper external region, of the form $\rho(r) \propto r^{-2}$.
For a Bonnor-Ebert sphere with radius $R$ and central density $\rho _{c}$,
its normalized profile, i.e., $\rho(r)/\rho _{c}$ vs. $r/R$, is
characterized by a dimensionless radius $\xi _{max}$, whose critical value
is 6.5. This value of the critical parameter corresponds to a contrast
between the central density and the density at the edge of the sphere of 14.
Increasing values of this parameter denote spheres more centrally condensed.

\subsection{Radial Extinction Profiles, Masses and Densities}
\label{sec:profmassden}

In order to construct radial profiles, we define the position of the center
of the globule as the position of the extinction peak, which does not
necessarily coincide with the position listed in Table \ref{tab:sample}. As
some extinction maps have more than one core for the same globule, we will
assign a letter to the new cores identified within a single globule. Then,
we set concentric annuli at each core center, 20$''$ wide, and
averaged the extinction at the Nyquist frequency, obtaining a mean
extinction value every 10$''$ (impact parameters). The
resulting uncertainty for an impact parameter corresponds to the error
propagation of each pixel within the annulus, calculated with equation
\ref{eq:sigmaAv2}. The radial profile for B68 is shown in Figure \ref{fig:b68prof},
and the resulting radial extinction profiles for the globules
of our sample are shown in Figure \ref{fig:perfs}. The solid curve in each
plot corresponds to the Bonnor-Ebert fitting (see \S \ref{sec:bemodel}).

As shown in \S \ref{sec:bemodel} and \S \ref{sec:befitting} below, to fit these
profiles we modeled a radial density profile for each impact parameter, and
the radius $R$ of the core is the impact parameter for which we obtained the
best fit for that profile. \citet{teixeira05} define the radius of a core as
the distance for which the profile reaches a constant level of extinction or 
\emph{plateau}. The radius obtained in this article coincide with the
definition of \citeauthor{teixeira05} because the cores are embedded in a
more diffuse region in the globules. Table \ref{tab:coordrad} lists in
column 1 the BHR globules, where a letter attached represents a
substructure identified in the globule, columns 2 and 3 give the coordinates
of the cores as defined above, and column 4 lists their sizes.

%==============================
\begin{deluxetable}{lccc}
\tablecaption{\label{tab:coordrad}Coordinates and Sizes of Dense Cores in Bok Globules}
\tabletypesize{\scriptsize}
\tablewidth{0pt}
\tablecolumns{4}
\tablehead{
\colhead{Name} & \colhead{R.A.} & \colhead{Dec.} & \colhead{$R$} \\
 & \colhead{(J2000)} & \colhead{(J2000)} & \colhead{(pc)}
}

\startdata
\cutinhead{Starless Globules}
016 A & 08 05 18.8 & $-$39 08 53 & 0.07 \\
016 B & 08 05 27.2 & $-$39 08 20 & 0.06 \\
044 A & 09 26 09.3 & $-$45 11 10 & 0.07 \\
044 B & 09 26 20.4 & $-$45 10 56 & 0.06 \\
053 & 09 28 46.5 & $-$51 36 20 & 0.07 \\
059 & 11 07 10.1 & $-$62 05 36 & 0.06 \\
074 & 12 22 10.2 & $-$66 27 39 & 0.07 \\
075 & 12 24 14.2 & $-$66 10 58 & 0.09 \\
111 & 15 42 19.8 & $-$52 48 26 & 0.11 \\
113 & 16 12 51.6 & $-$52 16 23 & 0.07 \\
133 & 16 46 42.6 & $-$44 31 10 & 0.13 \\
144 A & 16 37 29.3 & $-$35 13 43 & 0.07 \\
144 B & 16 37 35.5 & $-$35 14 43 & 0.07 \\
145 & 17 47 51.9 & $-$43 42 15 & 0.07 \\
\cutinhead{IRAS Globules}
034 & 08 26 27.7 & $-$50 39 30 & 0.06 \\
058 & 10 49 02.4 & $-$62 22 18 & 0.05 \\
117 A & 16 06 25.2 & $-$45 54 17 & 0.06 \\
117 B & 16 06 11.8 & $-$45 56 29 & 0.06 \\
121 & 16 58 47.0 & $-$50 36 34 & 0.06 \\
126 & 16 04 29.2 & $-$39 37 47 & 0.11 \\
138 & 17 19 32.9 & $-$43 26 55 & 0.10 \\
139 & 17 20 51.7 & $-$43 19 44 & 0.07 \\
140 A & 17 22 56.2 & $-$43 22 26 & 0.07 \\
140 B & 17 22 52.9 & $-$43 21 40 & 0.06 \\
140 C & 17 23 29.6 & $-$43 25 10 & 0.06 \\
148 & 17 04 26.4 & $-$36 18 34 & 0.07 \\
149 A & 17 04 31.2 & $-$36 07 52 & 0.06 \\
149 B & 17 04 53.1 & $-$36 03 15 & 0.06 \\
\enddata

\tablecomments{Units of right ascension are hours, minutes, and seconds, and
units of declination are degrees, arcminutes, and arcseconds.}
\end{deluxetable}
%==============================

Once constructed the radial profiles of the cores, we calculate their
masses integrating the profile until the radius of the core:

\begin{equation}
\label{eq:mass1}
M = \mu m_H \int_{\Omega} N_H \: d \Omega,
\end{equation}

\noindent
where $\mu$ (= 1.36) is the mean atomic weight of the gas, $m_H$
is the mass of the H atom, $N_H$ is the column density and $\Omega$ is the
cloud area projected in the plane of the sky. To express the column density
in terms of visual extinction, we used the gas-to-dust ratio of \citet{bohlin78}:

\begin{equation}
\label{eq:g2d}
N_H = 2 \times 10^{21} \: A_V \: \mathrm{cm^{-2} \: mag^{-1}},
\end{equation}

\noindent
and substracting the contribution of the diffuse structure or
\emph{plateau} extinction, we have that the mass of an embedded
dense core is:

\begin{equation}
\label{eq:mass2}
M = 2\pi \left( \frac{N_H}{A_V} \right) \mu m_H D^2 \int_0^R \left( A_V - A_V^{plateau} \right) rdr,
\end{equation}

\noindent
where $D$ is the distance to the globule and $R$ is the core
radius. Finally, to calculate the mean volumetric density of each core, we used:

\begin{equation}
\label{eq:nmean}
\overline{n} = \frac{3M}{4\pi \mu m_H R^3}.
\end{equation}

The above results are listed in Table \ref{tab:massden}. Column 1 indicates
the core name, column 2 the \emph{plateau} extinction, column 3 and 4 are
the masses and mean volumetric densities of the cores without \emph{plateau}
substraction, and in columns 5 and 6 are given the masses and densities
substacting the \emph{plateau} contribution.

%==============================
\begin{deluxetable}{lccccc}
\tablecaption{\label{tab:massden}Masses and Densities of Dense Cores in Bok Globules}
\tabletypesize{\scriptsize}
\tablewidth{0pt}
\tablecolumns{6}
\tablehead{
\colhead{Name} & \colhead{$A_V^{plateau}$} & \colhead{$M$\tablenotemark{a}} & \colhead{$\overline{n}$\tablenotemark{a}} & \colhead{$M$\tablenotemark{b}} & \colhead{$\overline{n}$\tablenotemark{b}} \\
 & \colhead{(mag)} & \colhead{($M_{\odot}$)} & \colhead{(10$^4$ cm$^{-3}$)} & \colhead{($M_{\odot}$)} & \colhead{(10$^4$ cm$^{-3}$)}
}

\startdata
\cutinhead{Starless Globules}
016 A & 8.7 & 9.5 & 19.1 & 3.6 & 7.2 \\
016 B & 6.2 & 4.7 & 15.1 & 1.2 & 3.8 \\
044 A & 5.1 & 5.4 & 10.9 & 1.4 & 2.7 \\
044 B & 5.8 & 4.0 & 12.9 & 1.1 & 3.4 \\
053  & 7.1 & 8.0 & 16.1 & 2.4 & 4.9 \\
059  & 8.1 & 6.8 & 21.7 & 2.6 & 8.4 \\
074  & 2.8 & 3.6 & 7.4 & 1.5 & 3.0 \\
075  & 3.5 & 6.3 & 5.9 & 1.8 & 1.7 \\
111  & 9.8 & 22.7 & 11.8 & 5.5 & 2.9 \\
113  & 7.9 & 7.7 & 15.5 & 1.4 & 2.9 \\
133  & 7.5 & 20.4 & 6.4 & 3.2 & 1.0 \\
144 A & 11.7 & 11.5 & 23.2 & 2.6 & 5.3 \\
144 B & 11.1 & 9.4 & 18.9 & 2.2 & 4.4 \\
145  & 3.6 & 3.9 & 7.8 & 1.4 & 2.8 \\
\cutinhead{IRAS Globules}
034  & 2.5 & 2.2 & 7.0 & 0.9 & 2.8 \\
058  & 5.4 & 4.4 & 24.3 & 2.1 & 11.4 \\
117 A & 8.9 & 4.9 & 15.8 & 0.9 & 2.8 \\
117 B & 8.3 & 5.9 & 18.8 & 0.8 & 2.7 \\
121  & 3.1 & 2.4 & 7.8 & 0.7 & 2.4 \\
126  & 6.0 & 13.0 & 6.8 & 2.6 & 1.3 \\
138  & 1.9 & 5.0 & 3.5 & 2.3 & 1.6 \\
139  & 4.8 & 6.3 & 12.8 & 2.7 & 5.4 \\
140 A & 7.9 & 6.8 & 13.7 & 1.6 & 3.3 \\
140 B & 7.7 & 5.2 & 16.6 & 0.9 & 3.0 \\
140 C & 4.5 & 3.8 & 12.3 & 1.4 & 4.4 \\
148  & 4.5 & 4.8 & 9.8 & 1.4 & 2.8 \\
149 A & 4.6 & 4.0 & 12.8 & 1.5 & 4.9 \\
149 B & 3.6 & 2.8 & 8.9 & 0.9 & 2.8 \\
\enddata

\tablenotetext{a}{Value without {\it plateau} substraction.}
\tablenotetext{b}{Value with {\it plateau} substraction.}
\end{deluxetable}
%==============================

\subsection{Bonnor-Ebert Model}
\label{sec:bemodel}

A Bonnor-Ebert sphere is a pressure-truncated, self-graviting isothermal
sphere of gas in hydrostatic equilibrium. The equation that describes such a
gaseous sphere is the modified Lane-Emden equation:

\begin{equation}
\label{eq:lane}
{\frac{1 }{\xi^2}} {\frac{d }{d\xi}} (\xi^2 {\frac{d\phi }{d\xi}}) = e^{-\phi},
\end{equation}

\noindent
where $\rho = \rho_c e^{-\phi(\xi)}$. Here, $\xi = (r/a) \sqrt{ 4\pi G \rho_c}$
is a dimensionless radial parameter and $a$ is the isothermal
sound speed. Introducing standard boundary conditions
($\phi$(0) $=$ 0 and d$\phi$(0)/d$\xi$ $=$ 0), we solved equation~\ref{eq:lane}
numerically using a fourth order Runge-Kutta method. Assuming an external radius
$R$, solutions of equation~\ref{eq:lane} can be parametrized by $\rho_c$.
\citet{bonnor56} demonstrated that the value of $\xi$ evaluated at the external radius $R$:

\begin{equation}
\xi_{max} = \frac{R}{a} \sqrt{4 \pi G \rho_c}
\end{equation}

\noindent
provides a stability measure. Systems with $\xi_{max} >$ 6.5 are
unstable to gravitational collapse. Equivalently, the density contrast
between the center and the border of the cloud is a function of $\xi_{max}$:

\begin{equation}
\label{eq:contrast}
\frac{\rho_c}{\rho_R} = e^{\phi(\xi_{max})}.
\end{equation}

\noindent
If $\rho_R = \rho(R)$, and in the critical state $\xi_{max}$ = 6.5,
then $\rho_c / \rho_R$ = 14. Equation \ref{eq:contrast} shows that, for
stable states where $\xi_{max} <$ 6.5, the center-to-edge density contrast
is lesser than 14.

Finally, knowing the core radius $R$ and the stability parameter $\xi_{max}$,
we can calculate the physical parameters of the sphere, as the central density:

\begin{equation}
\label{eq:rhoc}
\rho_c = \frac{1}{4\pi G} \left( \frac{a\xi_{max}}{R} \right)^2,
\end{equation}

\noindent
or the volumetric central density:

\begin{equation}
\label{eq:nc}
n_c = \frac{\rho_c}{\mu m_H},
\end{equation}

\noindent
where $\mu$ (= 2.33) is the molecular weight of the H$_2$
molecule. The mass of the Bonnor-Ebert sphere is given by:

\begin{equation}
\label{eq:massbe}
M_{BE} = \frac{1}{\sqrt{4\pi \rho_c}} \left( \frac{a^2}{G} \right)^{3/2}
\xi_{max}^2 \left( \frac{d\phi}{d\xi} \right)_{\xi = \xi_{max}},
\end{equation}

\noindent
and the external pressure at the edge:

\begin{equation}
\label{eq:pext}
P_{ext} = a^2 \rho_c e^{-\phi(\xi_{max})}.
\end{equation}

\subsection{Bonnor-Ebert Fitting}
\label{sec:befitting}

In order to compare theoretical profiles with observed extinction profiles
for each core, we constructed a series of theoretical Bonnor-Ebert profiles
in the following way: for various values of the stability parameter
$\xi_{max}$ and temperature $T$, we obtained the volumetric density $n(r)$
solving equation~\ref{eq:lane} as explained above. Then, we integrated the
volumetric density profile along the line-of-sight to obtain the hydrogen
column density profile:

\begin{equation}
\label{eq:nbe}
N_{BE}(r) = 2 \times \int_r^{R} n(r') \frac{r' dr'}{\sqrt{r'^2 - r^2}},
\end{equation}

\noindent
where $r$ is the projected distance from the center of the core or
impact parameter. The stability parameter was varied from 3 to 15 in steps
of $\Delta \xi$ = 0.1, and the temperature was varied from 5 K to 30 K in
steps of $\Delta T$ = 0.1 K. To express the column density in terms of
visual extinction, we used again the gas-to-dust ratio of \citet{bohlin78}
given by equation \ref{eq:g2d}. The fitting was evaluated using a reduced
$\chi^2$ with 2 degrees of freedom:

\begin{equation}
\chi^2_r = \frac{1}{n-2} \sum_{i=1}^n \left[ \frac{A_V^{BE}(i) - A_V^{obs}(i)}{\sigma_i} \right]^2,
\end{equation}

\noindent
where $n$ is the number of points considered in the profile and
$\sigma_i$ is the uncertainty in the observed $A_V$. To estimate the
uncertainty of the best-fit parameters, we took the second derivative of
$\chi^2_r$ with respect to each parameter $\xi_{max}$ and $T$ in the region
where $\chi^2_r$ is minimum:

\begin{mathletters}

\begin{equation}
\sigma^2_{\xi_{max}} = 2 \left( \frac{\partial^2 \chi^2_r}{\partial \: \xi_{max}^2} \right)^{-1},
\end{equation}

\begin{equation}
\sigma^2_T = 2 \left( \frac{\partial^2 \chi^2_r}{\partial \: T^2} \right)^{-1},
\end{equation}

\end{mathletters}

\noindent
according to \citet{bevington92}.

The Bonnor-Ebert model applied to B68 using the 2MASS data gave the
following results:

\begin{displaymath}
 \xi_{max} = 6.9 \pm 0.3,
\end{displaymath}
\begin{displaymath}
 T = 16.8 \pm 0.9 \: {\mathrm K}.
\end{displaymath}

\noindent
The theoretical profile for B68 is shown in Figure
\ref{fig:b68prof} as a continuous line superimposed on the observed profile. The
stability parameter is in good concordance with that obtainded by
\citet{alves01}, who calculated $\xi _{max} = 6.9 \pm 0.2$ using a temperature
of 16 K measured by \citet{bourke95b}, which coincides with the value
determined in our fit.
However, more recent observations indicate a lower
value for the temperature of B68. \citet{hotzel02a} observed B68 in NH$_3$ and
obtained a gas kinetic temperature of 10 $\pm$ 1.2 K, while \citet{lai03} found
a value of 11 K using the same molecule.
\citet{hotzel02b}, using the same data, modeled B68
with a Bonnor-Ebert sphere and found central extinctions of $A_{V}^{c}$ = 30.3 mag,
which is different from the value derived in this work, $A_{V}^{c}$ = 19.5 mag.
This difference arises probabely because they observed B68 with a greater
resolution and sensitivity than the 2MASS, making it possible to detect a
larger number of stars toward B68 than 2MASS.

Table \ref{tab:befitting} summarizes the results of the Bonnor-Ebert
modelling for the Bok globules of our sample. Column 1 gives the names of
the cores, column 2 the stability parameters, column 3 the Bonnor-Ebert
temperatures, column 4 and 5 the central volumetric densities and masses,
column 6 the external pressures in units of the Boltzmann constant, column 7
the central extinctions and column 8 the minimum value of $\chi^2_r$. The
resulting theoretical profiles for the globules are shown in Figure \ref{fig:perfs}.

%==============================
\begin{deluxetable}{lccccccc}
\tablecaption{\label{tab:befitting}Physical Parameters of Dense Cores in Bok Globules from
                      Bonnor-Ebert Fitting}
\tabletypesize{\scriptsize}
\centering
\tablewidth{0pt}
\tablecolumns{8}
\tablehead{
\colhead{Name} & \colhead{$\xi_{max}$} & \colhead{$T$} & \colhead{$n_c$} & \colhead{$M$} & \colhead{$P_{ext}/k_B$} & \colhead{$A_V^c$} & \colhead{$\chi^2_r$} \\
 & & \colhead{(K)} & \colhead{(10$^4$ cm$^{-3}$)} & \colhead{($M_{\odot}$)} & \colhead{(10$^4$ cm$^{-3}$ K)} & \colhead{(mag)}
}

\startdata
\cutinhead{Starless Globules}
016 A & 5.6 $\pm$ 0.1 & 24.4 $\pm$ 1.4 & 18.8 & 3.1 & 46.0 & 18.2 & 0.90 \\
016 B & 8.2 $\pm$ 0.2 & $\phn$8.5 $\pm$ 1.2 & 16.9 & 1.1 & $\phn$5.6 & 10.8 & 0.62 \\
044 A & 8.1 $\pm$ 0.5 & $\phn$8.3 $\pm$ 1.0 & 11.2 & 1.2 & $\phn$3.7 & $\phn$8.7 & 0.45 \\
044 B & 3.9 $\pm$ 0.7 & 10.4 $\pm$ 1.7 & $\phn$5.1 & 1.0 & 11.6 & $\phn$5.6 & 0.95 \\
053 & 4.7 $\pm$ 0.5 & 16.6 $\pm$ 1.5 & $\phn$7.5 & 2.1 & 18.7 & $\phn$9.1 & 0.61 \\
059 & 5.3 $\pm$ 0.4 & 21.7 $\pm$ 1.2 & 19.6 & 2.4 & 48.6 & 17.3 & 1.84 \\
074 & 6.7 $\pm$ 0.5 & $\phn$9.3 $\pm$ 1.2 & $\phn$8.7 & 1.3 & $\phn$5.3 & $\phn$7.9 & 0.30 \\
075 & 6.5 $\pm$ 1.6 & $\phn$8.5 $\pm$ 1.3 & $\phn$4.5 & 1.6 & $\phn$2.6 & $\phn$5.4 & 0.08 \\
111 & 4.9 $\pm$ 0.2 & 25.2 $\pm$ 1.8 & $\phn$5.5 & 4.9 & 19.0 & $\phn$9.7 & 0.86 \\
113 & 6.1 $\pm$ 2.0 & $\phn$9.3 $\pm$ 2.0 & $\phn$7.1 & 1.3 & $\phn$5.4 & $\phn$7.0 & 0.28 \\
133 & 8.2 $\pm$ 0.8 & 11.0 $\pm$ 3.0 & $\phn$5.1 & 2.8 & $\phn$2.2 & $\phn$6.8 & 0.06 \\
144 A & 4.3 $\pm$ 0.1 & 19.4 $\pm$ 1.6 & $\phn$7.7 & 2.3 & 27.2 & $\phn$9.7 & 0.53 \\
144 B & 3.4 $\pm$ 0.3 & 20.7 $\pm$ 1.7 & $\phn$6.0 & 1.9 & 35.6 & $\phn$8.1 & 0.58 \\
145 & 5.1 $\pm$ 0.3 & $\phn$9.7 $\pm$ 1.1 & $\phn$5.7 & 1.2 & $\phn$6.9 & $\phn$6.2 & 0.24 \\
\cutinhead{IRAS Globules}
034 & 4.5 $\pm$ 0.7 & $\phn$7.9 $\pm$ 1.0 & $\phn$5.1 & 0.8 & $\phn$6.6 & $\phn$5.1 & 0.72 \\
058 & 3.8 $\pm$ 0.2 & 22.7 $\pm$ 1.3 & 12.5 & 1.9 & 65.8 & 12.8 & 3.02 \\
117 A & 3.9 $\pm$ 0.4 & $\phn$8.9 $\pm$ 1.3 & $\phn$4.8 & 0.8 & $\phn$9.4 & $\phn$5.0 & 0.27 \\
117 B & 3.8 $\pm$ 0.3 & $\phn$7.1 $\pm$ 1.6 & $\phn$2.7 & 0.7 & $\phn$4.5 & $\phn$3.4 & 0.21 \\
121 & 9.4 $\pm$ 0.6 & $\phn$5.1 $\pm$ 0.8 & 13.3 & 0.6 & $\phn$1.9 & $\phn$7.6 & 0.46 \\
126 & 6.8 $\pm$ 0.7 & 10.1 $\pm$ 1.9 & $\phn$4.2 & 2.2 & $\phn$2.7 & $\phn$5.8 & 0.08 \\
138 & 6.4 $\pm$ 0.4 & 10.7 $\pm$ 1.5 & $\phn$4.9 & 2.1 & $\phn$3.8 & $\phn$6.3 & 0.33 \\
139 & 6.4 $\pm$ 0.5 & 17.6 $\pm$ 1.3 & 15.6 & 2.5 & 20.0 & 14.4 & 1.95 \\
140 A & 4.8 $\pm$ 0.7 & 12.5 $\pm$ 1.4 & $\phn$7.3 & 1.5 & 13.0 & $\phn$7.8 & 0.49 \\
140 B & 7.0 $\pm$ 1.7 & $\phn$7.2 $\pm$ 1.1 & 10.7 & 0.9 & $\phn$4.5 & $\phn$7.7 & 0.44 \\
140 C & 6.1 $\pm$ 0.9 & 10.7 $\pm$ 1.1 & 12.0 & 1.3 & 10.5 & $\phn$9.8 & 0.97 \\
148 & 4.0 $\pm$ 0.4 & 10.6 $\pm$ 1.8 & $\phn$3.5 & 1.2 & $\phn$7.8 & $\phn$4.7 & 0.16 \\
149 A & 5.1 $\pm$ 0.3 & 12.7 $\pm$ 1.2 & 10.2 & 1.4 & 16.1 & $\phn$9.5 & 0.75 \\
149 B & 6.3 $\pm$ 0.6 & $\phn$6.7 $\pm$ 1.0 & $\phn$8.2 & 0.8 & $\phn$4.1 & $\phn$6.4 & 0.24 \\
\enddata

\end{deluxetable}
%==============================

For the physical parameters obtained in the Bonnor-Ebert fitting, we
calculated the mean, standard deviation (SD), median, lower quartile (LQ),
and upper quartile (UQ) for the whole sample, for the starless globules and
for the IRAS globules, respectively. These quantities are shown in Table
\ref{tab:statist}. Comparing starless with IRAS globules, we see from this table
that the mean value of the physical parameters are nearly identical for
both kind of globules. However, for the temperature and external pressure, starless globules
have larger values than IRAS globules. In order to better explore these trends,
a large sample of carefully selected globules has to be studied.

It is important to mention that we also used a slightly different fitting procedure,
in which the theoretical profile was convolved with a Gaussian kernel of FWHM = 20$''$,
the resolution of the extinction maps. We obtained that, while the temperature remains
nearly the same, the stability parameter $\xi_{max}$ seems to be greater in all cases.
Nevertheless, their associated uncertainty is also greater for all fittings carried out
with kernel convolution, and in more than 60$\%$ of the cases the error bars overlap.
Also, in all the fittings the reduced $\chi^2$ test is always bigger for convolved profiles,
suggesting that there is no need to convolve a theoretical Bonnor-Ebert profile with a Gaussian
kernel for fitting the dense cores of our sample using 2MASS data.

%==============================
\begin{deluxetable}{lccccccccccccccccc}
\tablecaption{\label{tab:statist}Statistics for Dense Cores in Bok Globules}
\tabletypesize{\scriptsize}
\rotate
\centering
\tablewidth{0pt}
\tablecolumns{18}
\tablehead{
\colhead{} & \multicolumn{5}{c}{All Globules} & \colhead{} & \multicolumn{5}{c}{Starless Globules} & \colhead{} & \multicolumn{5}{c}{IRAS Globules} \\
\cline{2-6} \cline{8-12} \cline{14-18} \\
\colhead{Physical Parameter} & \colhead{Mean} & \colhead{SD} & \colhead{Median} & \colhead{LQ} & \colhead{UQ} & \colhead{} & \colhead{Mean} & \colhead{SD} & \colhead{Median} & \colhead{LQ} & \colhead{UQ} & \colhead{} & \colhead{Mean} & \colhead{SD} & \colhead{Median} & \colhead{LQ} & \colhead{UQ} \\
}

\startdata
$R$ (pc) & 0.07 & 0.02 & 0.07 & 0.06 & 0.07 & & 0.08 & 0.02 & 0.07 & 0.07 & 0.07 & & 0.07 & 0.02 & 0.06 & 0.06 & 0.07 \\
$\xi_{max}$ & 5.7$\phn$ & 1.6$\phn$ & 5.5$\phn$ & 4.5$\phn$ & 6.6$\phn$ & & 5.8$\phn$ & 1.6$\phn$ & 5.5$\phn$ & 4.8$\phn$ & 6.7$\phn$ & & 5.6$\phn$ & 1.6$\phn$ & 5.6$\phn$ & 4.1$\phn$ & 6.4$\phn$ \\
$T$ (K) & 12.6$\phm{07}$ & 5.9$\phn$ & 10.5$\phm{07}$ & 8.5$\phn$ & 16.9$\phm{07}$ & & 14.5$\phm{08}$ & 6.5$\phn$ & 10.7$\phm{07}$ & 9.3$\phn$ & 20.4$\phm{07}$ & & 10.8$\phm{07}$ & 4.7$\phn$ & 10.4$\phm{06}$ & 7.4$\phn$ & 12.1$\phm{07}$ \\
$n_c$ (10$^4$ cm$^{-3}$) & 8.7$\phn$ & 4.7$\phn$ & 7.4$\phn$ & 5.1$\phn$ & 11.4$\phm{07}$ & & 9.2$\phn$ & 5.3$\phn$ & 7.3$\phn$ & 5.6$\phn$ & 10.6$\phm{07}$ & & 8.2$\phn$ & 4.2$\phn$ & 7.8$\phn$ & 4.8$\phn$ & 11.7$\phm{07}$ \\
$M$ (M$_\odot$) & 1.7$\phn$ & 0.9$\phn$ & 1.4$\phn$ & 1.0$\phn$ & 2.1$\phn$ & & 2.0$\phn$ & 1.1$\phn$ & 1.7$\phn$ & 1.3$\phn$ & 2.3$\phn$ & & 1.3$\phn$ & 0.6$\phn$ & 1.2$\phn$ & 0.8$\phn$ & 1.8$\phn$ \\
$P_{ext}/k_B$ (10$^4$ cm$^{-3}$ K) & 14.6$\phm{07}$ & 16.2$\phm{02}$ & 7.4$\phn$ & 4.4$\phn$ & 18.8$\phm{07}$ & & 17.0$\phm{08}$ & 16.2$\phm{02}$ & 9.3$\phn$ & 5.3$\phn$ & 25.2$\phm{07}$ & & 12.2$\phm{07}$ & 16.3$\phm{02}$ & 7.2$\phn$ & 4.2$\phn$ & 12.4$\phm{07}$ \\
$A_V^c$ (mag) & 8.5$\phn$ & 3.6$\phn$ & 7.8$\phn$ & 6.1$\phn$ & 9.7$\phn$ & & 9.3$\phn$ & 3.9$\phn$ & 8.4$\phn$ & 6.8$\phn$ & 9.7$\phn$ & & 7.6$\phn$ & 3.1$\phn$ & 7.0$\phn$ & 5.3$\phn$ & 9.0$\phn$ \\
\enddata

\end{deluxetable}
%==============================

\subsection{Modelling Results and Discussion}
\label{sec:modresbe}

Figure \ref{fig:evol} shows the stability parameter $\xi_{max}$ versus the
center-to-edge density contrast $\rho_c / \rho_R$ for the globules of our
sample. Gray data correspond to starless globules and black data to IRAS
globules. This figure shows that there is no difference between starless and
IRAS globules with respect to the stability parameter. Taking into account
the uncertainties in $\xi_{max}$, we can see three groups in this
plot: a first group showing clearly stable states, with $\xi_{max} = 4.5 \pm 0.7$
(8 starless + 7 IRAS), a second group distributed around the critical
state with $\xi_{max} = 6.5 \pm 0.2$ (3 starless + 6 IRAS), and a third
group showing clearly unstable states, with $\xi_{max} = 8.5 \pm 0.6$
(3 starless + 1 IRAS). Remarkably, the first group of stable globules has a
temperature of $T = 15 \pm 6$~K, and the critical plus unstable groups have
$T = 10 \pm 3$~K. This tendency of decreasing temperature with increasing
stability parameter can also be seen in Figure \ref{fig:xmax_temp}, where
both parameters are plotted with their respective uncertainties. The same
behavior is shown by the external pressure.

These results show a difference with respect to the sample of
\citet{kandori05}: they found that more than half of their starless globules
are located near the critical state, which is not our case. Secondly, their Bonnor-Ebert
temperature distribution (not shown by these authors) against stability parameter
did not show any tendency of decreasing temperature with increasing $\xi _{max}$,
as in our sample.

Figure \ref{fig:histog} shows histograms of the logarithmic density
contrast, with the vertical dashed line corresponding to the critical
Bonnor-Ebert sphere. The left histogram corresponds to the whole sample,
showing that there is a large number of cores in the stable regime, followed
by an abrupt decrease of cores in the unstable regime. The same behavior is
seen for starless globules (upper-right) and for IRAS globules
(lower-right), but the decrease in the number of cores for IRAS globules is
steeper than for starless globules.
This seems to be in agreement with the fact that globules with density contrasts
larger than 14 must have shorter lifetimes. The opposite happens when the density contrast
is smaller than the critical value. In order to state whether this is a real
trend or not a large sample of globules has to be considered.
The peak of the starless globules occurs
just below the critical value, while for IRAS globules the peak is around
the critical value. This is in contrast to the globule distribution showed
by \citet[][see their Figure 6]{kandori05}, where the majority of starless
globules are in the unstable regime and the star-forming globules are well
in the unstable regime.

Similarly to \citet{kandori05}, we plot in Figure \ref{fig:parameters} the
correlations between density contrast and the different parameters of the
globules derived from the Bonnor-Ebert fitting in a logarithmic scale. The
dotted lines correspond to the variation of each parameter for Bonnor-Ebert
spheres with minimum and maximum values of $T$ and $P_{ext}$ ($T_{min}$ =
5.1~K and $P_{ext,min} = 1.9 \times 10^{4}$~cm$^{-3}$~K; $T_{max}$ = 25.2~K and
$P_{ext,max} = 65.8 \times 10^{4}$~cm$^{-3}$~K), as indicated in Figures
\ref{fig:parameters} (a) and (b) for the temperature and external pressure,
respectively. With these two extreme Bonnor-Ebert spheres, we see in Figures
\ref{fig:parameters} (c)-(f) that we can cover nearly all the points in the
plots, which demonstrate that the dispersions in the distributions of the
parameters can be accounted by two Bonnor-Ebert spheres, one with fixed
minimum values of $T$ and $P_{ext}$, and other with fixed maximum values.

Finally, if the IRAS globules are star-forming globules, then our sample of
Bok globules do not show an evolutionary sequence, as is the case of
\citet{kandori05}, where their sample of globules suggest an evolutionary
sequence in the sense that the starless globules are distributed around a
critical Bonnor-Ebert sphere and the star-forming globules are highly
unstable.

\section{The Infrared Sources}
\label{sec:starfrom}

To study the properties the of IRAS sources associated with the globules of
our sample, we looked for flux measurements in the following catalogs:

\begin{itemize}
\item DENIS\footnote{Deep Near-Infrared Survey \citep{epchtein99}} for band I (0.79 $\mu$m)

\item 2MASS for bands J (1.24 $\mu$m), H (1.66 $\mu$m) and K (2.16 $\mu$m)

\item MSX\footnote{Midcourse Space Experiment \citep{price01}} for bands A (8.28 $\mu$m), C
(12.13 $\mu$m), D (14.65 $\mu$m) and E (21.34 $\mu$m)

\item IRAS for bands 12 $\mu$m, 25 $\mu$m, 60 $\mu$m and 100 $\mu$m
\end{itemize}

To determine the nature of the IRAS sources embedded in the globules, we
used two criteria to segregate the spectral classes: in \S \ref{sec:sed} we
build the spectral energy distribution (SED) for the IRAS sources and
calculated the spectral index, and in \S \ref{sec:blt} we calculated the
bolometric luminosity and temperature for the sources.

\subsection{Spectral Energy Distribution}
\label{sec:sed}

The shape of the broadband spectrum of a YSO depends on the nature and
distribution of the circumstellar material. This implies a correlation
between the SED and the evolutionary state of the YSO, where the initial
protostellar stages, during which a stellar embryo is surrounded by large
amounts of collapsing circumstellar material, have very different infrared
signatures than the more advances stages of pre-main sequence, where most of
the original material has been incorporated to the young star \citep[see][]{lada99s}.

Class 0 \citep{andre93} and I \citep{adams87,lada87} sources are
characterized by SEDs with peaks in the far-infrared and sub-milimeter, as a
result of the emission of a cold dust collapsing envelope. These objects are
the most deeply embedded and obscured. Once the collapse of the envelope ceases,
the dusty disk can still produce substantial emission in the infrared.
This part of the evolution gives raise to Class II and III sources
\citep{adams87,lada87}, whose SEDs are characterized by peaks in the opitcal
and near-infrared, where stellar fotospheric emission dominates. These
objects correspond to T Tauri stars (CTTS and WTTS).

Based on the shapes of SEDs of observed infrared sources, \citet{adams87}
defined the spectral index $\alpha$ between 2-25 $\mu$m as:

\begin{equation}
\alpha = \frac{d \: \mathrm{log}(\lambda F_{\lambda})}{d \: \mathrm{log} (\lambda)}
\end{equation}

\noindent
where $F_{\lambda}$ is the flux corresponding to wavelength $\lambda$.
The different spectral indices computed in regions of star
formation are interpreted as an evolutionary sequence, obtaining the
following classification: $\alpha >$ 0 for Class I sources, 0 $> \alpha >$ $-$2
for Class II sources, and $\alpha <$ $-$2 for Class III sources.

The SEDs of the IRAS globules are shown in Figure \ref{fig:seds}. To
calculate the spectral index we fitted a straight line using the last
squares method to the points between 2.2 and 25 $\mu$m, with the
requirement that these points have a determined uncertainty. Fluxes that are
upper limits (shown as arrows in Figure \ref{fig:seds}) where not taken into
account. Table \ref{tab:classif} shows the name of the globule in column 1,
the associated IRAS sources in column 2, the index $\alpha$ in column 3 and
the corresponding spectral classification according to that index in column
4. It was possible to determine the spectral index for 8 of 13 IRAS sources,
of which 6 correspond to Class I, one to Class II and one to Class III.

%==============================
\begin{deluxetable}{cccccrc}
\tablecaption{\label{tab:classif}Spectral Classification of YSOs in Bok Globules}
\tabletypesize{\small}
\tablewidth{0pt}
\tablecolumns{7}
\tablehead{
\colhead{Name} & \colhead{IRAS} & \colhead{$\alpha$} & \colhead{Class\tablenotemark{a}} & \colhead{$L_{bol}$} & \colhead{$T_{bol}$} & \colhead{Class\tablenotemark{b}} \\
 & & & & \colhead{($L_{\odot}$)} & \colhead{(K)} &
}

\startdata
034 & 08250-5030 & \nodata & \nodata & 1.0 & 29 & 0 \\
058 & 10471-6206 & 1.3 & I & 2.2 & 32 & 0 \\
117 & 16029-4548 & $-$1.0 $\:\:$ & II & 4.2 & 742 & II \\
121 & 16549-5030 & \nodata & \nodata & 0.6 & 227 & I \\
       & 16554-5031 & 1.7 & I & 1.3 & 98 & I \\
126 & 16009-3927 & $-$2.5 $\:\:$ & III & 37.6 $\:$ & 2478 & II \\
138 & 17159-4324 & \nodata & \nodata & 3.3 & 28 & 0 \\
139 & 17172-4316 & 0.7 & I & 2.8 & 123 & I \\
       & 17169-4314 & \nodata & \nodata & 1.9 & 90 & I \\
140 & 17193-4319 & 1.5 & I & 1.9 & 35 & 0 \\
       & 17195-4320 & \nodata & \nodata & 15.4 $\:$ & 1676 & II \\
148 & 17011-3613 & 1.6 & I & 1.6 & 31 & 0 \\
149 & 17012-3603 & 1.1 & I & 1.5 & 39 & 0 \\
\enddata

\tablenotetext{a}{Spectral classification according to index $\alpha$}
\tablenotetext{b}{Spectral classification according to $T_{bol}$}
\end{deluxetable}
%==============================

\subsection{The BLT Diagram}
\label{sec:blt}

A useful tool for studying the evolutionary state of a young object is the
graph of the bolometric luminosity ($L_{bol}$) vs. bolometric temperature
($T_{bol}$) (BLT diagram).
\citet{myers93} defined and used the BLT diagram to study the evolutionary
state of YSOs in the Taurus-Auriga complex. $T_{bol}$ is defined as the
temperature of a blackbody which has the same mean frequency of the observed
continnum spectrum. A main-sequence star whose spectrum is a blackbody has
$T_{bol} = T_{eff}$. A pre-main sequence star with infrared excess has a wider
and redder spectrum than a blackbody, and therefore $T_{bol} < T_{eff}$. An
embedded infrared source is very reddened to have an optical measure of
$T_{eff}$, its spectrum is much wider than a blackbody, and its $T_{bol}$ is
very small \citep[$<$ 500 K,][]{chen95}.

To derive the bolometric luminosity for the IRAS sources, we integrated
their SEDs across the whole range in wavelengths:

\begin{equation}
\label{eq:lum1}
L_{bol} = 4\pi D^2 \int_{0}^{\infty} F_{\lambda} \: d\lambda = 9.2\pi D^2
\int_{0}^{\infty} \lambda F_{\lambda} \: d\mathrm{log} \lambda.
\end{equation}

\noindent
In most cases, the lowest available wavelength to make the
integration is the J band. As the SEDs for all the sources grow up as the
wavelength increases (see Figure \ref{fig:seds}), we ignored the term
between $\lambda$ = 0 and the J band, which is very small compared to the
total luminosity \citep[see][]{baba06}. Therefore, the integration was
divided into two parts:

\begin{equation} 
\label{eq:lum2}
L_{bol} = 9.2\pi D^2 \left( \int_{\lambda_1}^{\lambda_2} \lambda F_{\lambda}
\: d\mathrm{log} \lambda + \int_{\lambda_2}^{\infty} \lambda F_{\lambda} \: d
\mathrm{log} \lambda \right),
\end{equation}

\noindent
where $\lambda_1 =$ J and $\lambda_2 =$ 100 $\mu$m, the longest
available wavelength. To compute the second part of the integral, as we do
not know the shape of the spectrum between $\lambda_2$ and infinity, we
assumed that the SED is proportional to $\lambda^{-1}$ \citep{wilking89}.

The bolometric temperature is defined by \citet{myers93} as:

\begin{equation}
\label{eq:tbol}
T_{bol} \equiv \frac{\zeta (4)}{4\zeta (5)} \frac{h \overline{\nu}}{k_B} =
1.25 \times 10^{-11} \: \overline{\nu} \: \mathrm{(K \: Hz^{-1})},
\end{equation}

\noindent
where $\zeta(n)$ is the Riemann zeta function of argument $n$ and
the mean frequency $\overline{\nu}$ is the ratio of the first and zeroth
moments of the source spectrum:

\begin{equation}
\label{eq:meanf}
\overline{\nu} = \frac{\int_{0}^{\infty} \nu F_{\nu} \: d\nu}{\int_{0}^{\infty} F_{\nu} \: d\nu}.
\end{equation}

Once we calculated these parameters, we constructed the BLT diagram.
\citet{chen95} established a correlation between the temperature ranges in
the BLT diagram and the evolutionary class of a YSO, getting as a result the
following classification: $T_{bol} <$ 70 K for Class 0, 70 K $< T_{bol} <$
650 K for Class I, 650 K $< T_{bol} <$ 2800 K for Class II, and $T_{bol} >$
2800 K for Class III sources. The values of $T_{bol}$, $L_{bol}$ and
spectral classification according to this criterion are shown in columns 5,
6 and 7 of Table \ref{tab:classif}, respectively. Out of 13 IRAS sources, 6
were classified as Class 0, 4 as Class I, and 3 as Class II. The BLT diagram
is shown in Figure \ref{fig:blt}, where the solid line is the ZAMS from
\citet{siess00} and the vertical dashed lines represent the intervals
corresponding to different spectral classes.

The mean value of the Bonnor-Ebert stability parameter is $\sim$ 5 for
each spectral class, suggesting that there is no difference between the average
value of the stability parameters of the cores and the spectral class
of their associated IRAS sources.

\section{BHR 138: A Collapsing Globule?}
\label{sec:collapsing}

In this section, we will use the millimetric observations
data of the \tco\ and \ceio\ molecules, in the rotational
transition J = 1 $\rightarrow $ 0, to study BHR 138 and BHR 149 (core A).
Both globules are part of a survey of more than 30 condensations identified
in Scorpius \citep{vilasb00}. Among all these condensations, only
BHR 138 presents a blue-asymmetric profile, while BHR 149 shows a typical
profile observed in the others condensations. The presence of a
blue-asymmetric profile in the spectrum of BHR 138 provides evidence of an
eventual gravitational collapse. No spectral line data were available
to study the other globules.

\subsection{Reduction of the Spectra}
\label{sec:redspec}

The reduction of \tco\ and \ceio\ spectra was made with the
spectral lines reduction program Drawspec\footnote{\url{http://www.cv.nrao.edu/\~{}hliszt}}.
The lines were adjusted with
gaussians. In the case of BHR 138, the fitting was made with two independent
gaussians. For more details about the reduction of the spectra and the
derivation of the line parameters, see \citet{vilasb94}.

Figure \ref{fig:spectra} shows the \tco\ and \ceio\ spectra for BHR
138 and BHR 149. The histogram corresponds to the observed spectrum and the
grey gaussians are the fitting made with Drawspec. For the \tco\ line
in BHR 138, the resulting gaussian is the sum of the two thin gaussians.
Table \ref{tab:spectra} shows the parameters extracted from the line
fitting. Column 1 is the antenna temperature corrected from atmospheric
attenuation, column 2 is the velocity of the line center, and column 3 is
the FWHM of the line.

%==============================
\begin{deluxetable}{lccc}
\tablecaption{\label{tab:spectra}Observed Parameters of Spectral Lines}
\tabletypesize{\small}
\tablewidth{0pt}
\tablecolumns{4}
\tablehead{
 & \colhead{$T_A^{\ast}$} & \colhead{$V_{LSR}$} & \colhead{$\Delta V$} \\
 & \colhead{(K)} & \colhead{(km s$^{-1}$)} & \colhead{(km s$^{-1}$)}
}

\startdata
\cutinhead{BHR 138}
\tco\  & 2.57 $\pm$ 0.11 & $-$7.62 $\pm$ 0.01 & 0.78 $\pm$ 0.03 \\
         & 1.32 $\pm$ 0.10 & $-$7.27 $\pm$ 0.04 & 2.18 $\pm$ 0.09 \\
\ceio\ & 0.46 $\pm$ 0.05 & $-$7.55 $\pm$ 0.05 & 0.82 $\pm$ 0.11 \\
\cutinhead{BHR 149}
\tco\  & 3.12 $\pm$ 0.03 & 1.11 $\pm$ 0.01 & 1.02 $\pm$ 0.01 \\
\ceio\ & 0.31 $\pm$ 0.03 & 1.08 $\pm$ 0.05 & 1.01 $\pm$ 0.12 \\
\enddata

\end{deluxetable}
%==============================

\subsection{Modelling of the \tco\ Line}
\label{sec:modline}

Comparing the spectra of \tco\ for BHR 138 and BHR 149 in Figures
\ref{fig:spectra} (a) and (b), respectively, we can see clearly the
asymmetry in the line of BHR 138, while BHR 149 shows a perfectly symmetric
line. The spectrum of BHR 138 does not presents two distinct peaks
\cite[see, e.g.,][]{evans99}, but do presents a typical profile of
collapsing clouds denominated ``red-shoulders'' by \citet{myers96}.

Given the blue-asymmetric profile shown by the \tco\ line in BHR 138,
one of the fundamental questions is to know if this is really a profile
indicative of collapse. One of the possibilities to explain that profile
will be the superposition of distinct clouds in the line-of-sight or
rotational effects of the clouds. Analysis of the field of the globule
(Figure \ref{fig:maps}) shows a completely isolated object and the antenna
beam contains only the globule. Similar situation occurs with BHR 149, which
is also isolated and the profile does not suggests any asymmetry in the
lines. Analysis of 30 globules in Scorpius also does not show profiles with
asymmetries of the kind identified in BHR 138. According to \citet{vilasb00},
only 4 condensations of their sample observed in that region present
contamination for superposition of clouds in the antenna beam. Generally,
the components that contaminate the CO emission in that region have very
wide line widths ($>$ 4 km s$^{-1}$) at different radial velocities.

Because the \tco\ and \ceio\ emissions are originated in the more
external regions of the globules, with densities lesser than 10$^5$ cm$^{-3}$
\citep[e.g.,][]{alves99}, we would also expect to identify that asymmetry in
the \ceio\ line. Nonetheless, the signal-to-noise ratio for the spectrum
of that molecule does not allow to identify a red-shifted component with
sufficient accuracy. More greater integrations will be necessaries to
improve the signal-to-noise ratio.

The first model, proposed by \citet{myers96}, is called \emph{two-layer}
and consists of two collapsing gas layers, one approaching the other with a
velocity of twice the infall velocity $v_{in}$. Each layer has a constant
excitation temperature: the front layer, moving away from the observer, has
a temperature $T_f$ equal to the background temperature ($T_f$ = 2.7 K),
while the rear layer, which is hotter, has a temperature $T_r$.
Assuming that both layers have the same velocity dispersion $\sigma$ and
optical depth $\tau_0$ at the line center, \citet{devries05} obtained the
brightness temperature emerging from the spectral line:

\begin{equation}
\label{eq:tbright}
\Delta T_B = J(T_r)(1 - e^{-\tau_r})e^{-\tau_f} + J(T_f)(e^{-\tau_f - \tau_r} - e^{-\tau_f}),
\end{equation}

\noindent
where $J(T) = T_{0}/[\mathrm{exp}(T_{0}/T)-1]$, $T_{0} = h\nu /k_{B}$,
$\nu$ is the transition frequency , and $\tau _{f}$ and $\tau _{r}$ are the
optical depth of the radiation coming from the front and rear
layers, respectively:

\begin{equation}
\tau_f(v) = \tau_0 \: \mathrm{exp} \left[ -(v - v_{LSR} - v_{in})^2 / 2\sigma^2 \right],
\end{equation}
\begin{equation}
\tau_r(v) = \tau_0 \: \mathrm{exp} \left[ -(v - v_{LSR} + v_{in})^2 / 2\sigma^2 \right],
\end{equation}

\noindent
where $v_{LSR}$ is the line-of-sight velocity, assumed to be the
same for both regions. Therefore, this model has five free parameters:

\begin{displaymath}
\tau_0, v_{in}, T_r, \sigma, v_{LSR}.
\end{displaymath}

\noindent
The second model, introduced in \citet{devries05}, called \emph{hill}
model, consists of a dense core with a peak excitation temperature
$T_p$ at the center, and a temperature $T_0$ at the core edges. The $J(T)$
profile decreases linearly from $J(T_p)$ at the center to $J(T_0)$ at the
edges. The optical depth is $\tau_C$ and the infall velocity is $v_C$.
Assuming $T_0$ = 2.7 K, this model has five free parameters:

\begin{displaymath}
\tau_C, v_C, T_p, \sigma, v_{LSR}.
\end{displaymath}

\noindent
For a graphical representation and different variants of the
models of \citet{devries05}, see their Figure 1 and their Tables 1 and 2.

\subsection{Modelling Results}
\label{sec:modres}

To perform the fitting of the parameters for each model, we used a routine
developed by \citet{devries05}, obtained from the site of
C.~H. De~Vries\footnote{\url{http://cfa-www.harvard.edu/\~{}cdevries/analytic\_infall.html}}.
Starting values must be supplied to the program to find the best fitting
parameters. Table \ref{tab:twolayer} shows the results obtained with the
\emph{two-layer} model, and Table \ref{tab:hill} shows the results
corresponding to the \emph{hill} model. The model that best fit the
observational data for BHR 138, according to the $\chi^2$ value, is the
\emph{two-layer} model. Figure \ref{fig:twolayer} shows the fitting
obtained with this model, where the grey curve represents the theoretical fit,
which suggests a velocity collapse of 0.25 km s$^{-1}$.
Considering that rotation was not taken into account, the derived collapse
velocity should be seen as upper limit.

%==============================
\begin{deluxetable}{cccccc}
\tablecaption{\label{tab:twolayer}Two-Layer Model Fitting for BHR 138}
\tabletypesize{\small}
\tablewidth{0pt}
\tablecolumns{6}
\tablehead{
\colhead{$\tau_0$} & \colhead{$v_{in}$} & \colhead{$T_r$} & \colhead{$\sigma$} &
\colhead{$v_{LSR}$} & \colhead{$\chi^2$} \\
 & \colhead{(km s$^{-1}$)} & \colhead{(K)} & \colhead{(km s$^{-1}$)} & \colhead{(km s$^{-1}$)}
}

\startdata
1.6 & 0.25 & 10.9 & 0.6 & $-$6.9 & 1.2 \\
\enddata

\end{deluxetable}
%==============================

%==============================
\begin{deluxetable}{cccccc}
\tablecaption{\label{tab:hill}Hill Model Fitting for BHR 138}
\tabletypesize{\small}
\tablewidth{0pt}
\tablecolumns{6}
\tablehead{
\colhead{$\tau_C$} & \colhead{$v_C$} & \colhead{$T_p$} & \colhead{$\sigma$} &
\colhead{$v_{LSR}$} & \colhead{$\chi^2$} \\
 & \colhead{(km s$^{-1}$)} & \colhead{(K)} & \colhead{(km s$^{-1}$)} & \colhead{(km s$^{-1}$)}
}

\startdata
4.2 & 0.49 & 7.7 & 0.3 & $-$7.1 & 3.8 \\
\enddata

\end{deluxetable}
%==============================

Possible rotational effects of the globules \citep{barranco98} would
be evaluated from maps of optically thin molecular transitions
(e.g., N$_2$H$^+$). Studies of rotation in starless globules through the transition
NH$_3$ (1,1) show maximum velocity gradients, due to rotation, of the order of
1.4 km s$^{-1}$ pc$^{-1}$ \citep{barranco98,swift05}, and a survey of 12
starless globules and 14 protostellar globules in N$_2$H$^+$ 
\citep{caselli02} suggest typical velocity gradients of 2 km s$^{-1}$ pc$^{-1}$
in both kind of globules.

Adopting a mean velocity gradient due to rotation of
1.4 km s$^{-1}$ pc$^{-1}$ as being typical for Bok globules,
it is possible that the collapse velocity derived for BHR 138
could be overestimated by 40\% due to rotational effects that were not
corrected in this work. Of the factors that would affect this conclusion,
rotation seems to be the most relevant component. Taking that gradient and
the size of BHR 138, rotation could be contaminating the spectrum with 0.14
km s$^{-1}$, assuming the angular momentum perpendicular to the
line-of-sight. In that case, the collapse velocity would be $\sim$ 0.1 km s$^{-1}$.

In a sample of 53 starless cores observed in N$_2$H$^+$, \citet{lee01} found
collapse velocities of $\sim$ 0.1 km s$^{-1}$ using the \emph{two-layer}
model. Using observations of CS, \citet{swift06} obtained a collapse
velocity of 0.15 km s$^{-1}$ for the starless core L1551-MC with the \emph{hill}
model. \citet{myers96}, utilizing the \emph{two-layer} model,
derived a collapse velocity of 0.35 km s$^{-1}$ for the dense core L1251B,
which harbors a protostar.

\section{Summary}
\label{sec:summary}

In this work, we studied a sample of 21 isolated Southern Bok globules selected among 169 globules
catalogued by BHR. In this sample, 11 are starless and 10 have associated IRAS point sources.
Employing the NICE method of \citet{lada94}, we used the 2MASS PSC to construct
visual extinction maps for the Bok globules with 20$''$ resolution. This enabled us to detect
dense cores and substructures embedded in the globules, which were modelled with Bonnor-Ebert
spheres to derive their physical parameters and investigate their stability. Two methods were used to
determine the spectral class of the IRAS sources associated to the globules. In addition, we searched
for spectral lines observations data obtained toward these globules to explore whether they have or
not evidence of collapse. We found observations of \tco\ and \ceio\ J = 1--0 lines toward two globules
of the sample showing evidences of collapse in one of them. Distances were estimated to all these
globules from plots of color excess versus distance. The main results of this paper are summarized
as follows:

\begin{enumerate}

\item Dense cores were identified embedded in the globules, with mean volumetric densities
greater than 10$^{4}$~cm$^{-3}$ and masses between $\sim $ 1-4~M$_{\odot }$.

\item We inferred distances to several globules different from that ones obtained by BHR in spite
of using the same technique. Due to the uncertain nature of this method, in order to improve
the distance determination, we used smaller areas and a catalog containing information for
approximately 10 times more stars than the catalog utilized by BHR.

 \item We found that there is a large number of cores in the stable
regime, followed by an abrupt decrease of cores in the unstable regime, but this decrease is steeper
for IRAS globules than for starless. The globules that are in the stable state have a temperature of
$T = 15 \pm 6$ K, while the globules that are in the critical and unstable states have a remarkably
different temperature of $T = 10 \pm 3$ K.

\item We found that, according to the bolometric temperature, 6 IRAS sources are classified as Class 0,
4 as Class I, and 3 as Class II. The spectral index also produced similar results. No correlation was
identified between the Bonnor-Ebert stability parameter of the condensations and the spectral class
of the associated IRAS sources.

\item By means of the \tco\ J = 1--0 molecular line, we identified a blue-assymetric line profile toward
BHR 138. Using the \emph{two-layer} collapse model, it was possible to estimate an infall speed
of 0.25 km s$^{-1}$. However, taking into account possible rotational effects, this value could be
as small as $\sim$ 0.1 km s$^{-1}$.

\end{enumerate}

\acknowledgments
We are grateful to an anonymous referee for useful comments and suggestions that
improved the manuscript. G. A. R. is grateful to a PCI-INPE/CNPq fellowship and a CAPES grant.
This publication makes use of data products from the Two Micron All Sky Survey, which
is a joint project of the University of Massachusetts and the Infrared Processing and Analysis
Center, funded by the National Aeronautics and Space Administration and the National
Science Foundation.

%--------------------------------Fig 1: BARNARD 68 MAP--------------------------------
\clearpage
%--------------------------------Fig 1: BARNARD 68 MAP--------------------------------

%--------------------------------Fig 2: EXTINCTION MAPS--------------------------------
\clearpage
\begin{figure}
\centering
\begin{minipage}[b]{0.3\textwidth}
 \centering
 \includegraphics[width=4cm]{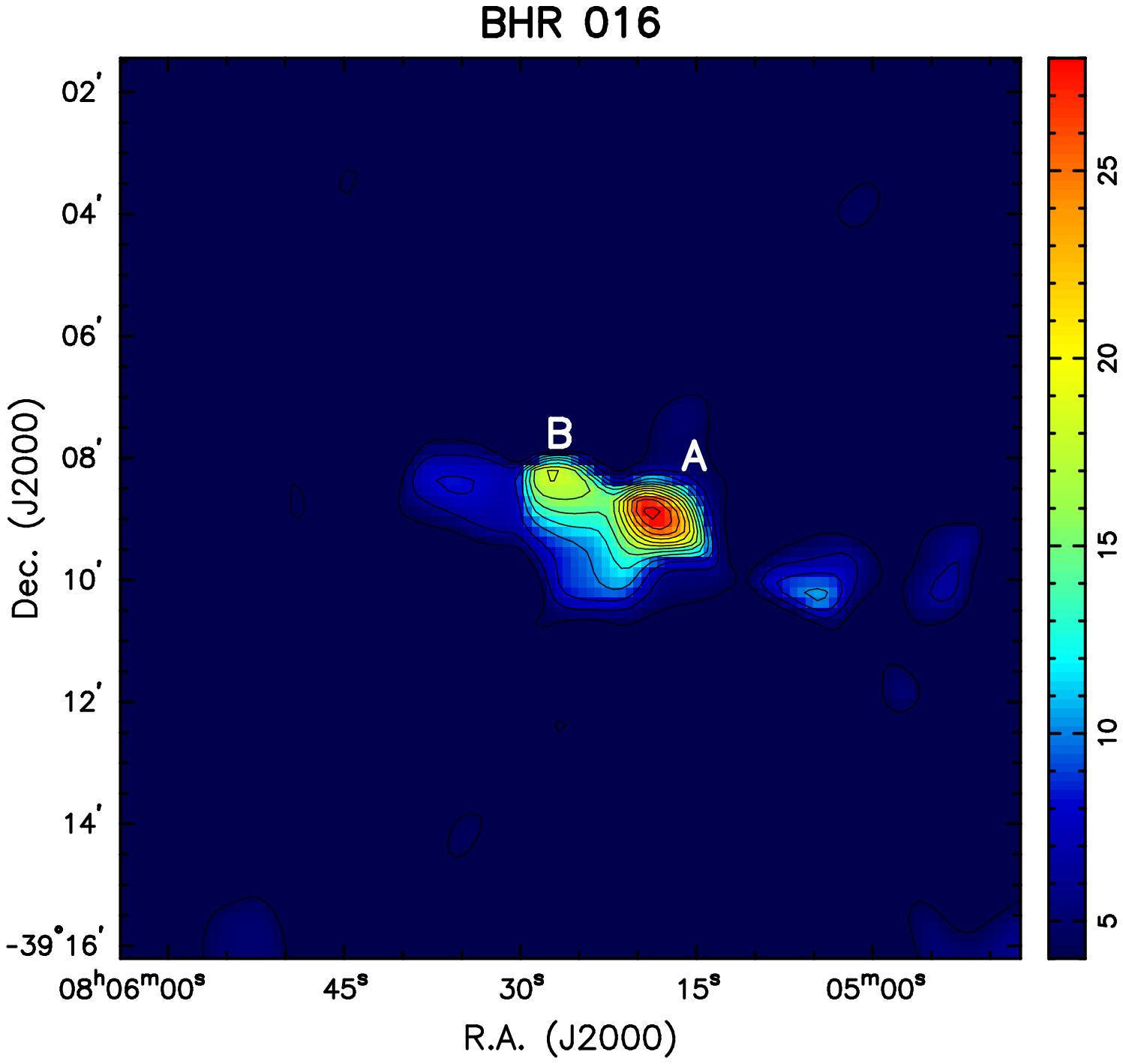}
\end{minipage}
\begin{minipage}[b]{0.3\textwidth}
 \centering
 \includegraphics[width=4cm]{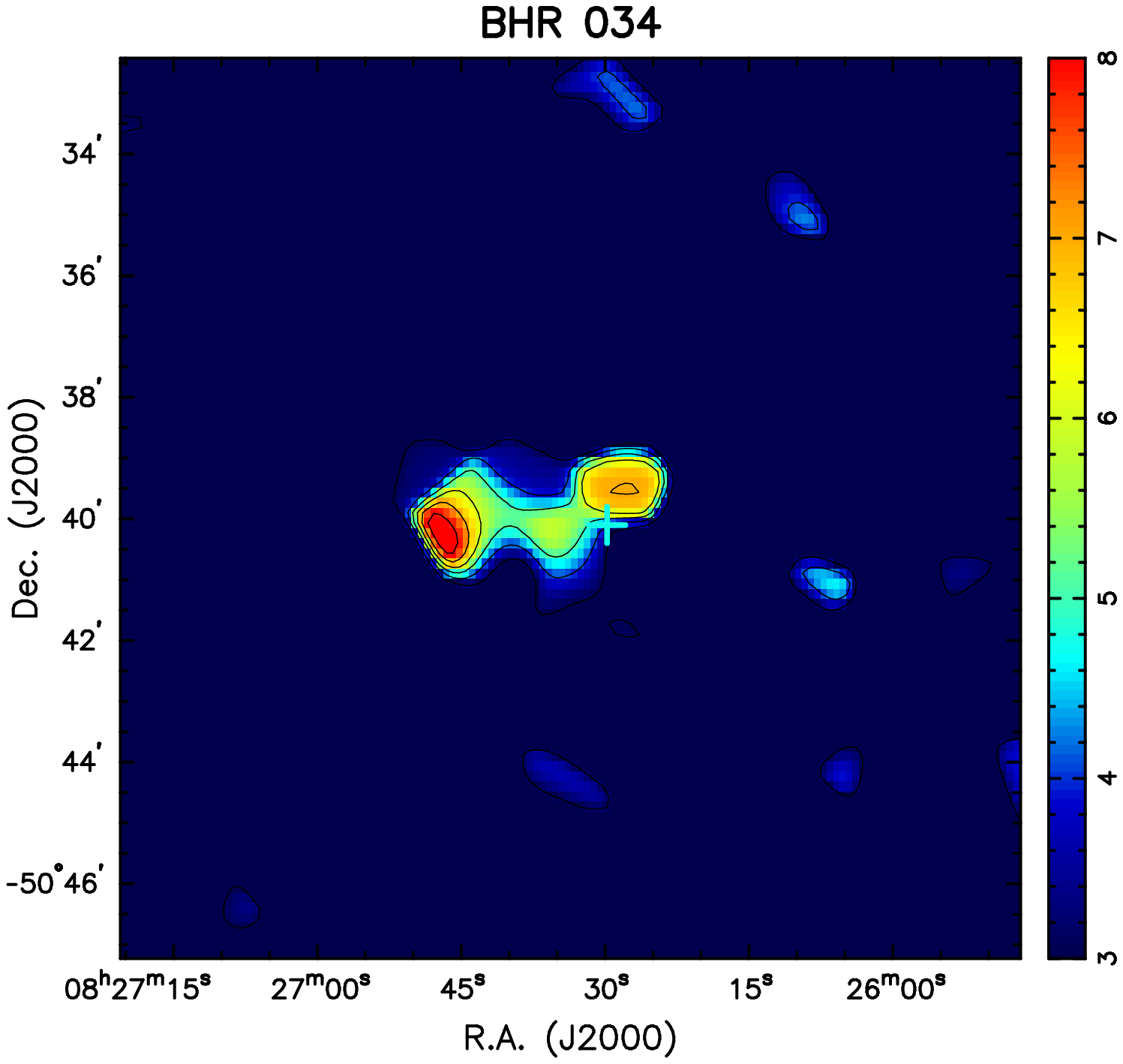}
\end{minipage}
\begin{minipage}[b]{0.3\textwidth}
 \centering
 \includegraphics[width=4cm]{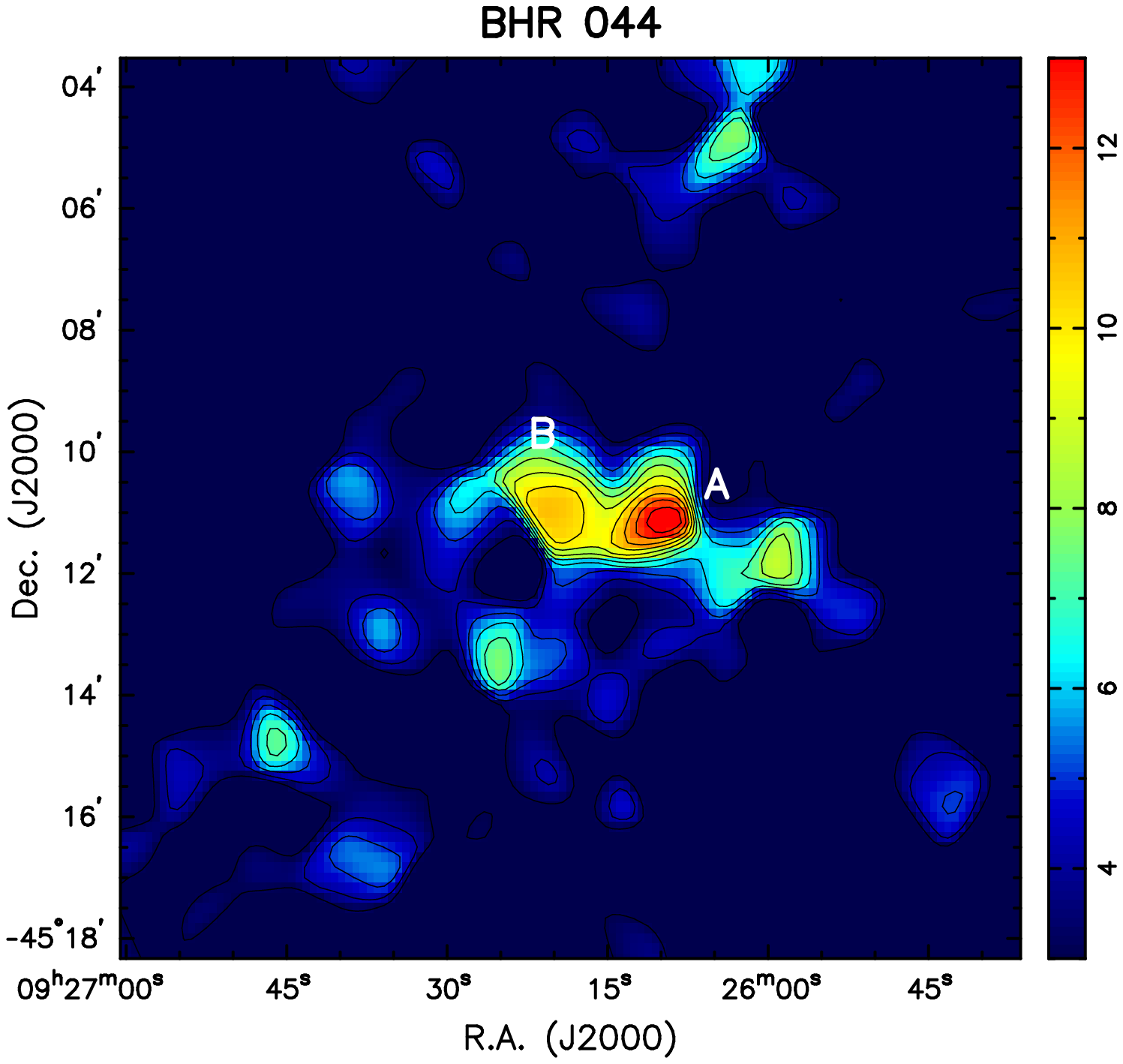}
\end{minipage}\\[0.5cm]
\begin{minipage}[b]{0.3\textwidth}
 \centering
 \includegraphics[width=4cm]{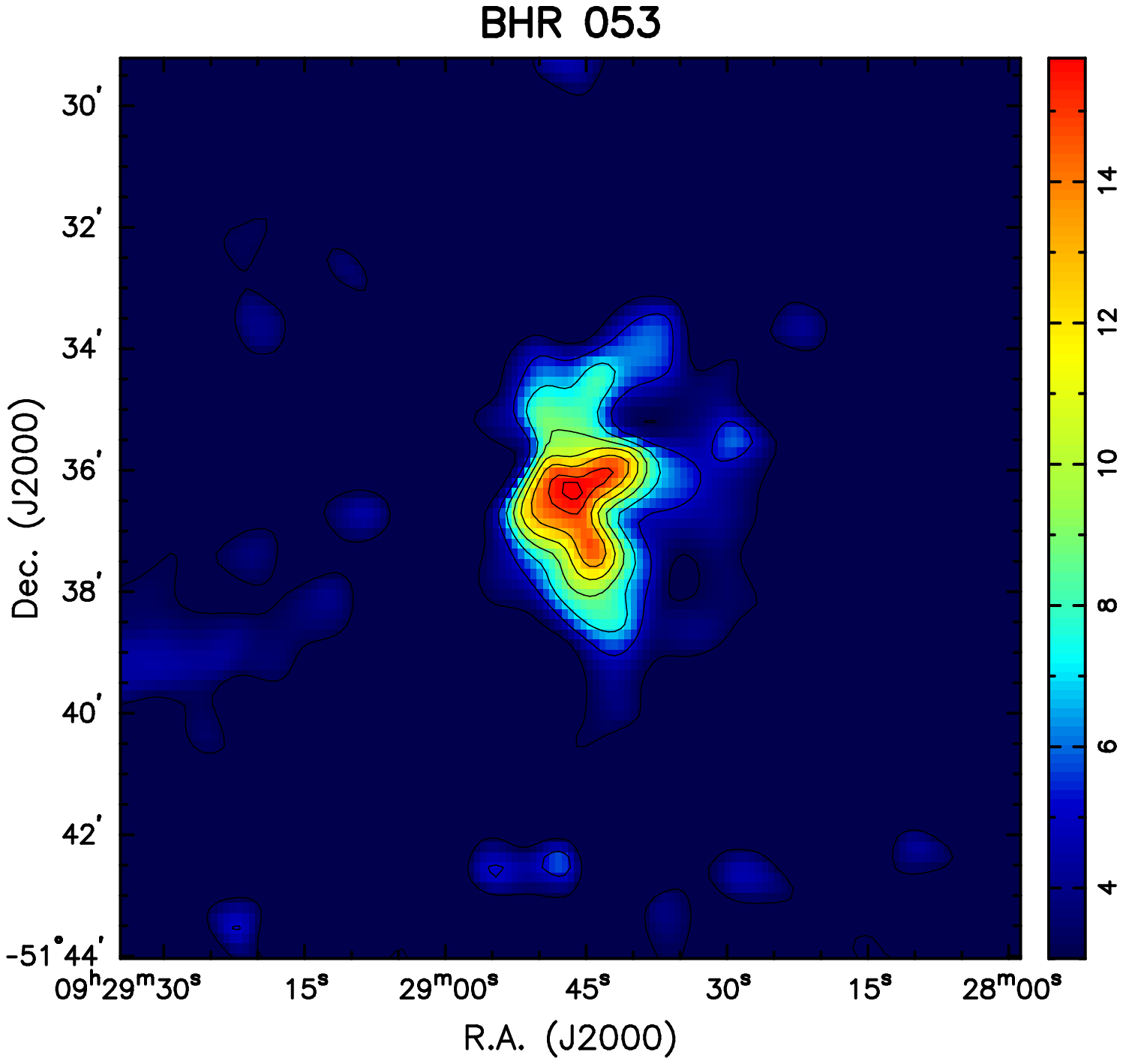}
\end{minipage}
\begin{minipage}[b]{0.3\textwidth}
 \centering
 \includegraphics[width=4cm]{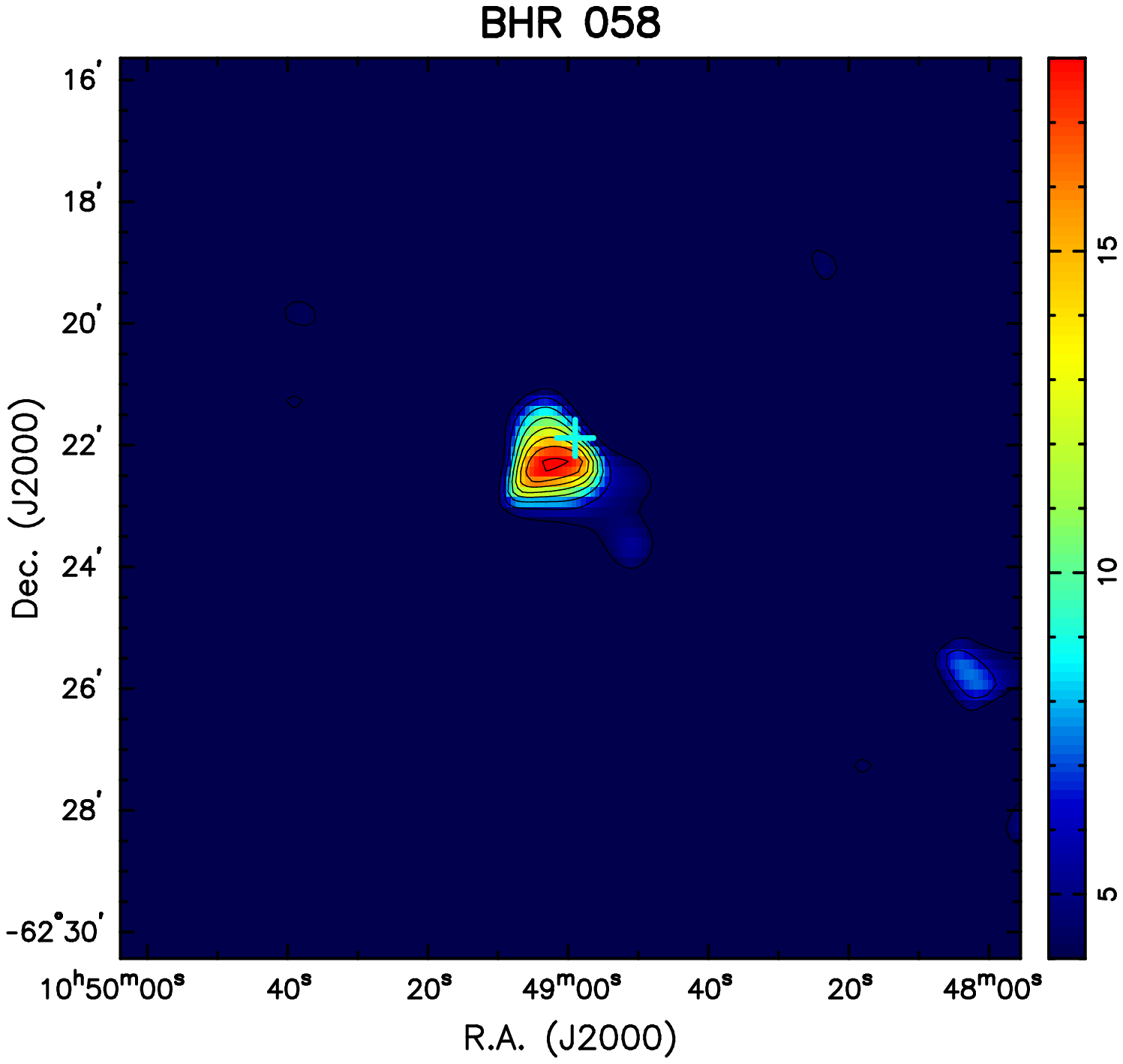}
\end{minipage}
\begin{minipage}[b]{0.3\textwidth}
 \centering
 \includegraphics[width=4cm]{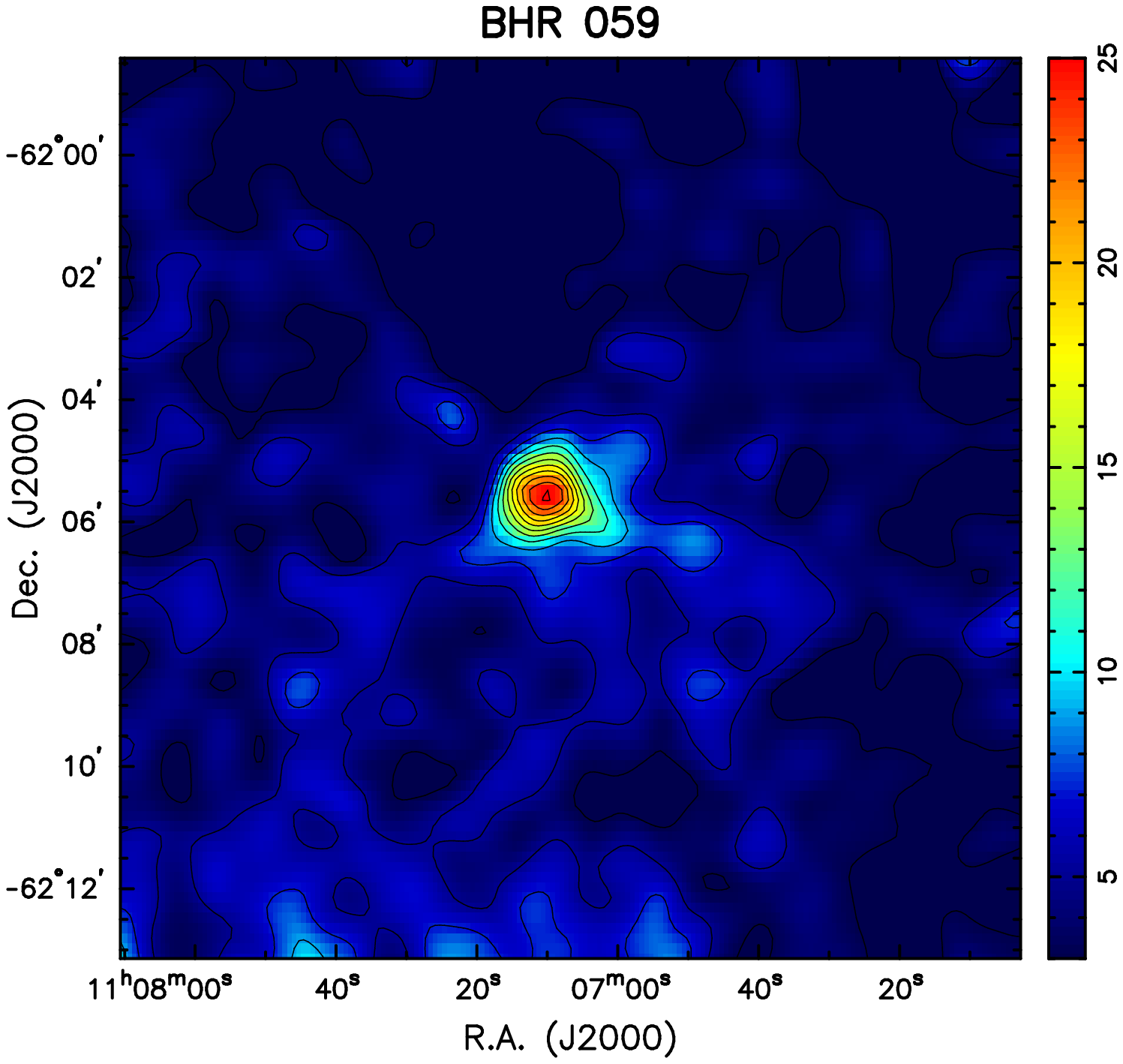}
\end{minipage}\\[0.5cm]
\begin{minipage}[b]{0.3\textwidth}
 \centering
 \includegraphics[width=4cm]{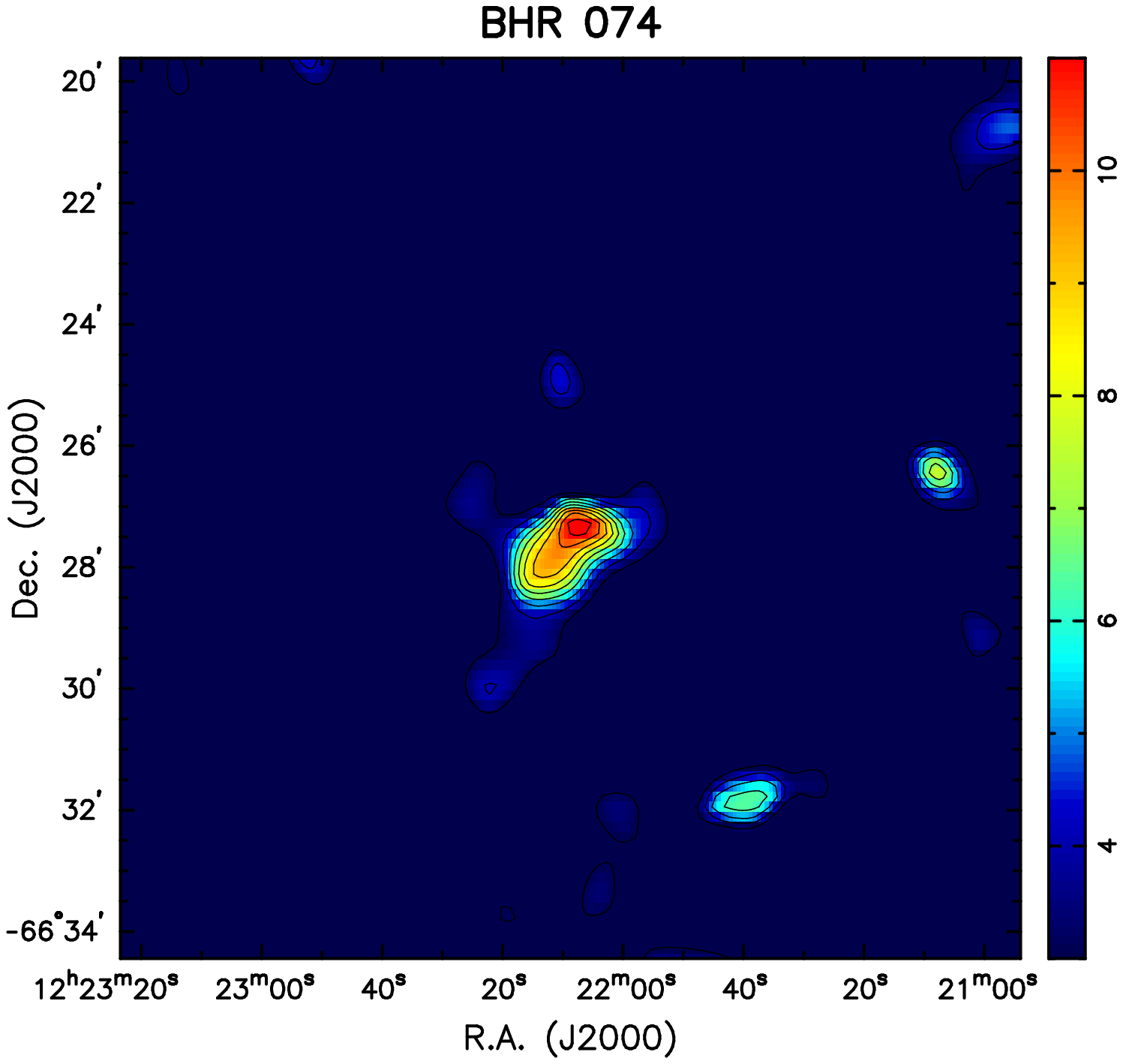}
\end{minipage}
\begin{minipage}[b]{0.3\textwidth}
 \centering
 \includegraphics[width=4cm]{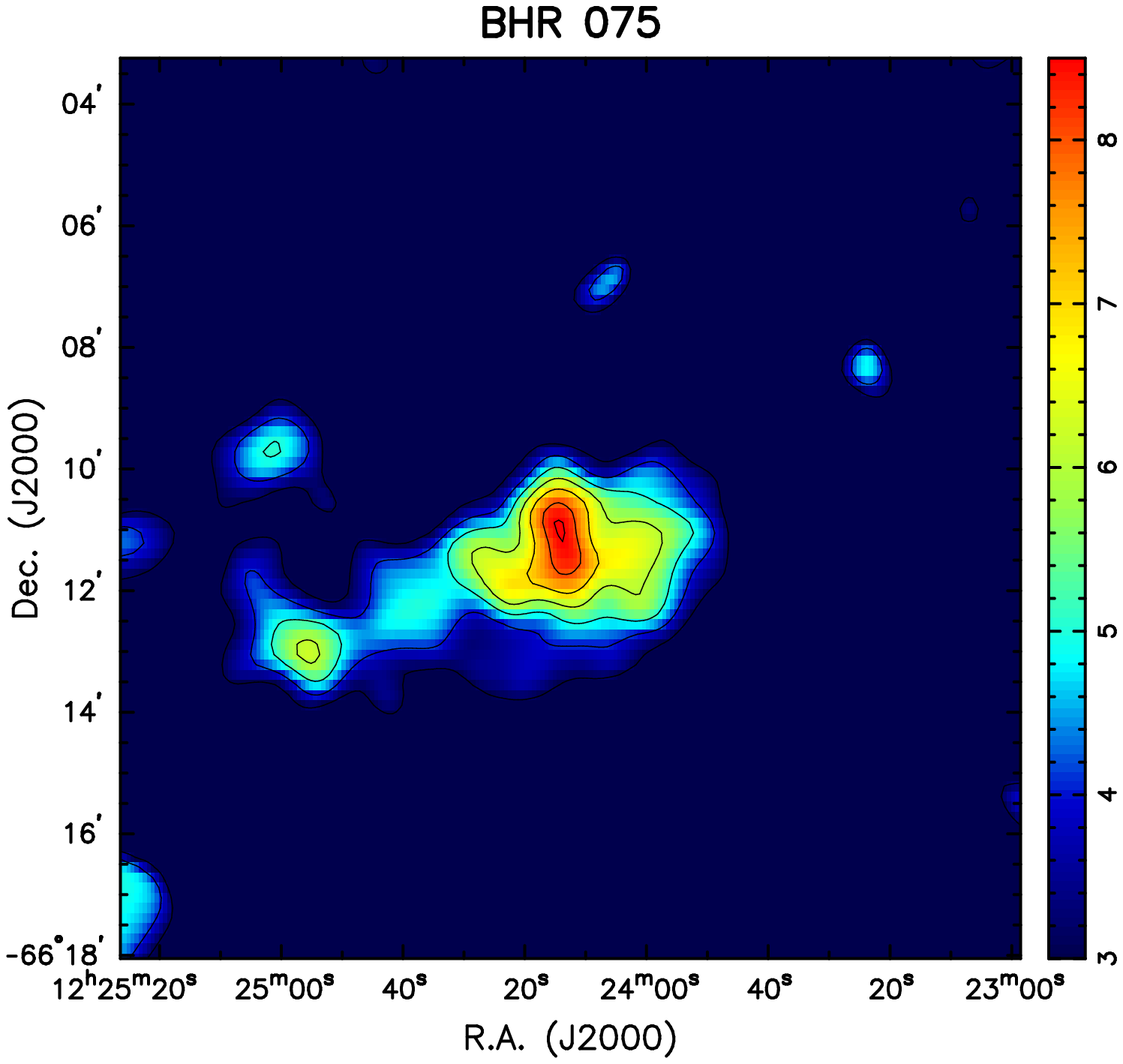}
\end{minipage}
\begin{minipage}[b]{0.3\textwidth}
 \centering
 \includegraphics[width=4cm]{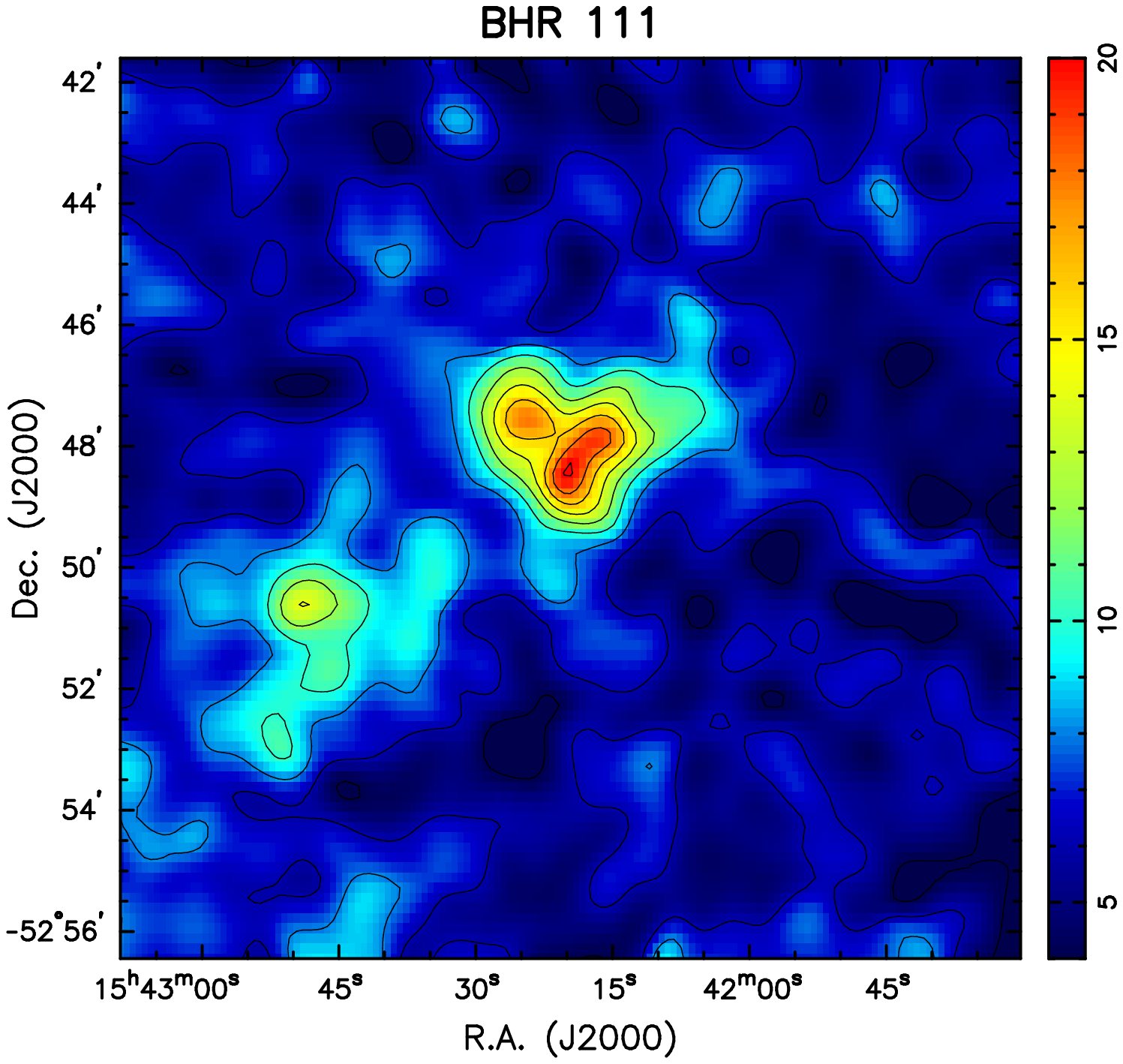}
\end{minipage}\\[0.5cm]
\begin{minipage}[b]{0.3\textwidth}
 \centering
 \includegraphics[width=4cm]{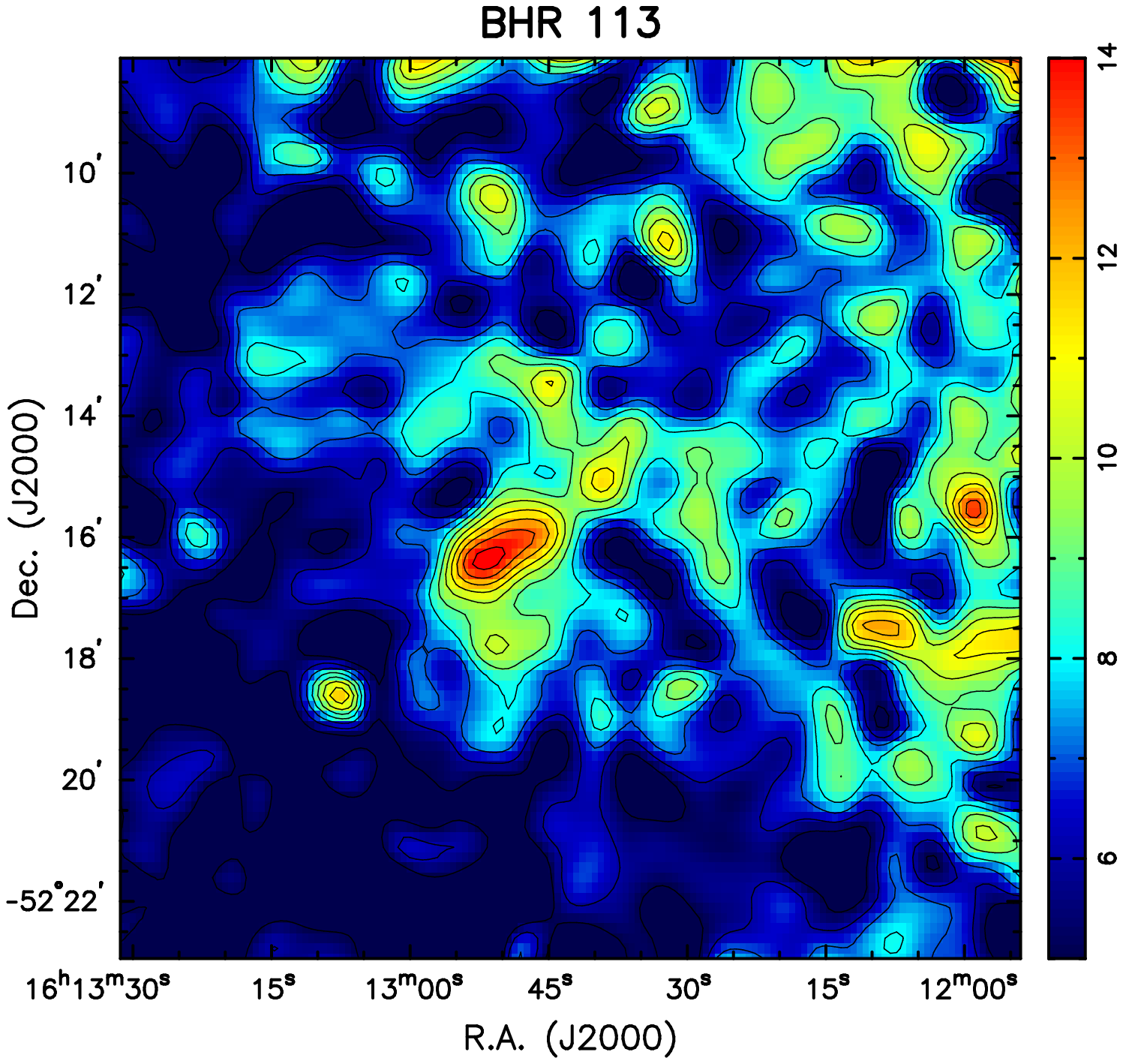}
\end{minipage}
\begin{minipage}[b]{0.3\textwidth}
 \centering
 \includegraphics[width=4cm]{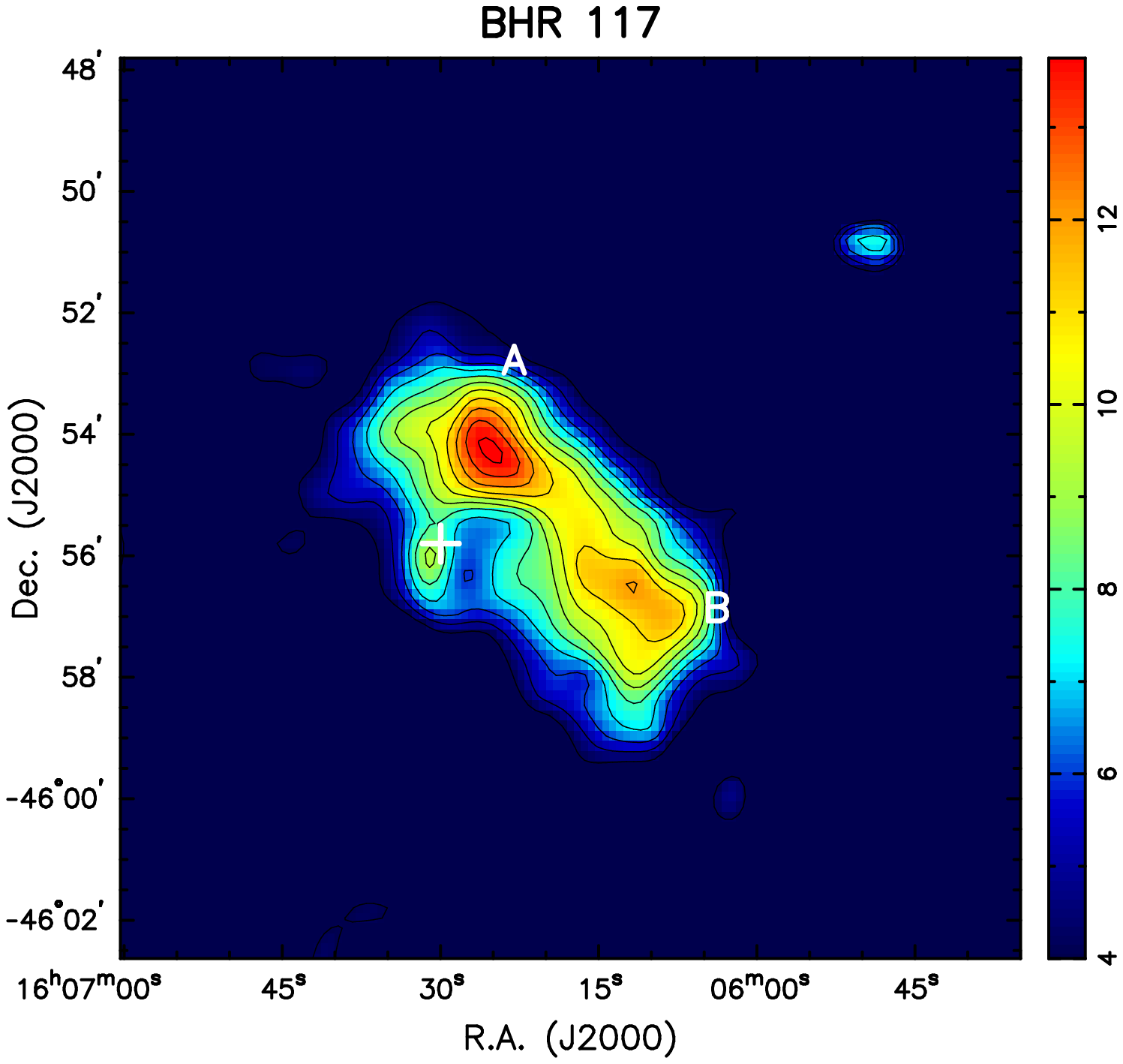}
\end{minipage}
\begin{minipage}[b]{0.3\textwidth}
 \centering
 \includegraphics[width=4cm]{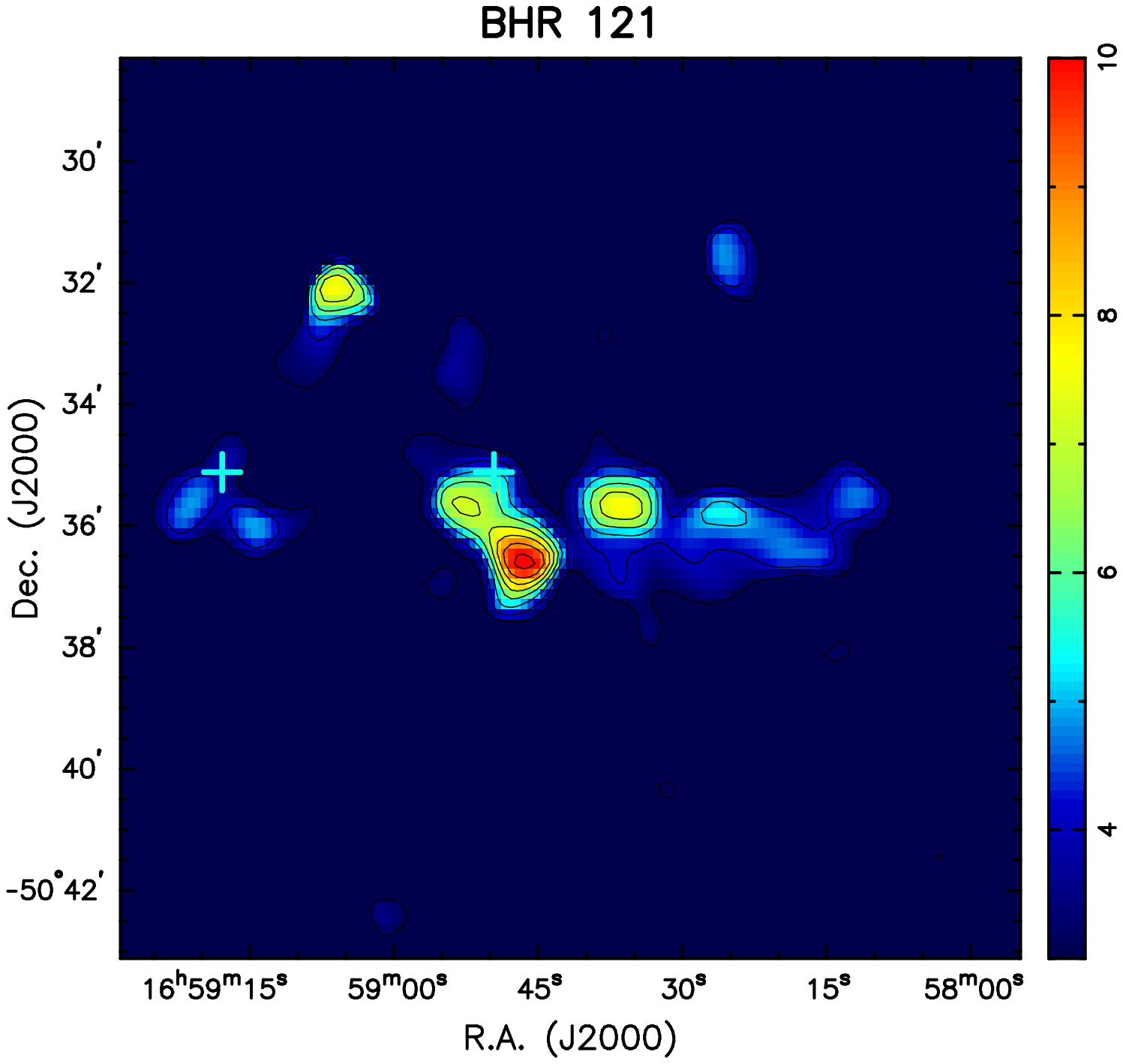}
\end{minipage}
\caption[]{Visual extinction maps constructed with 2MASS data using the NICE method.
           Each map has $\sim$~$15' \times 15'$ and resolution $h = 20''$.}
\label{fig:maps}
\end{figure}

\clearpage
\addtocounter{figure}{-1}
\begin{figure}
\centering
\begin{minipage}[b]{0.3\textwidth}
 \centering
 \includegraphics[width=4cm]{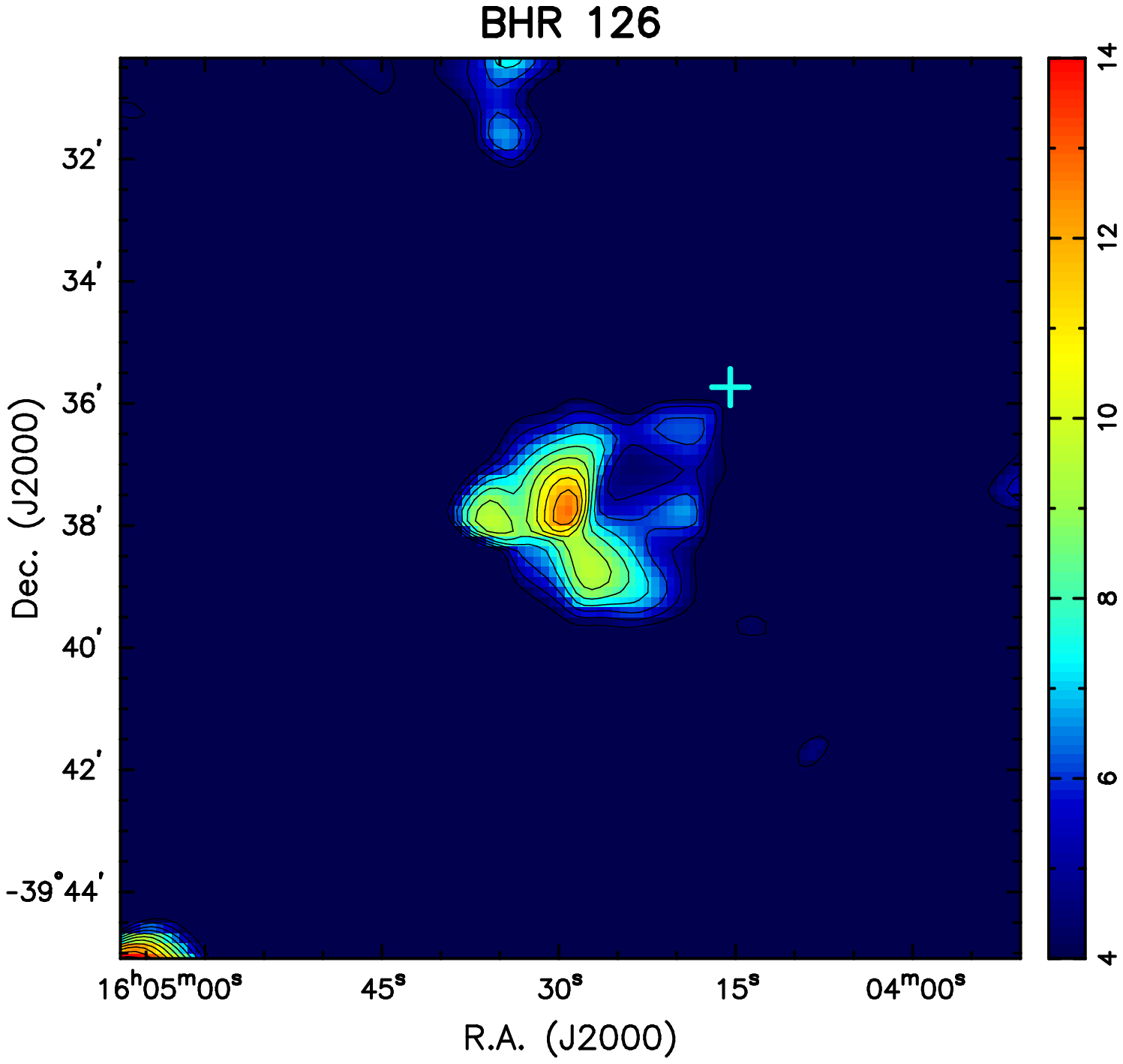}
\end{minipage}
\begin{minipage}[b]{0.3\textwidth}
 \centering
 \includegraphics[width=4cm]{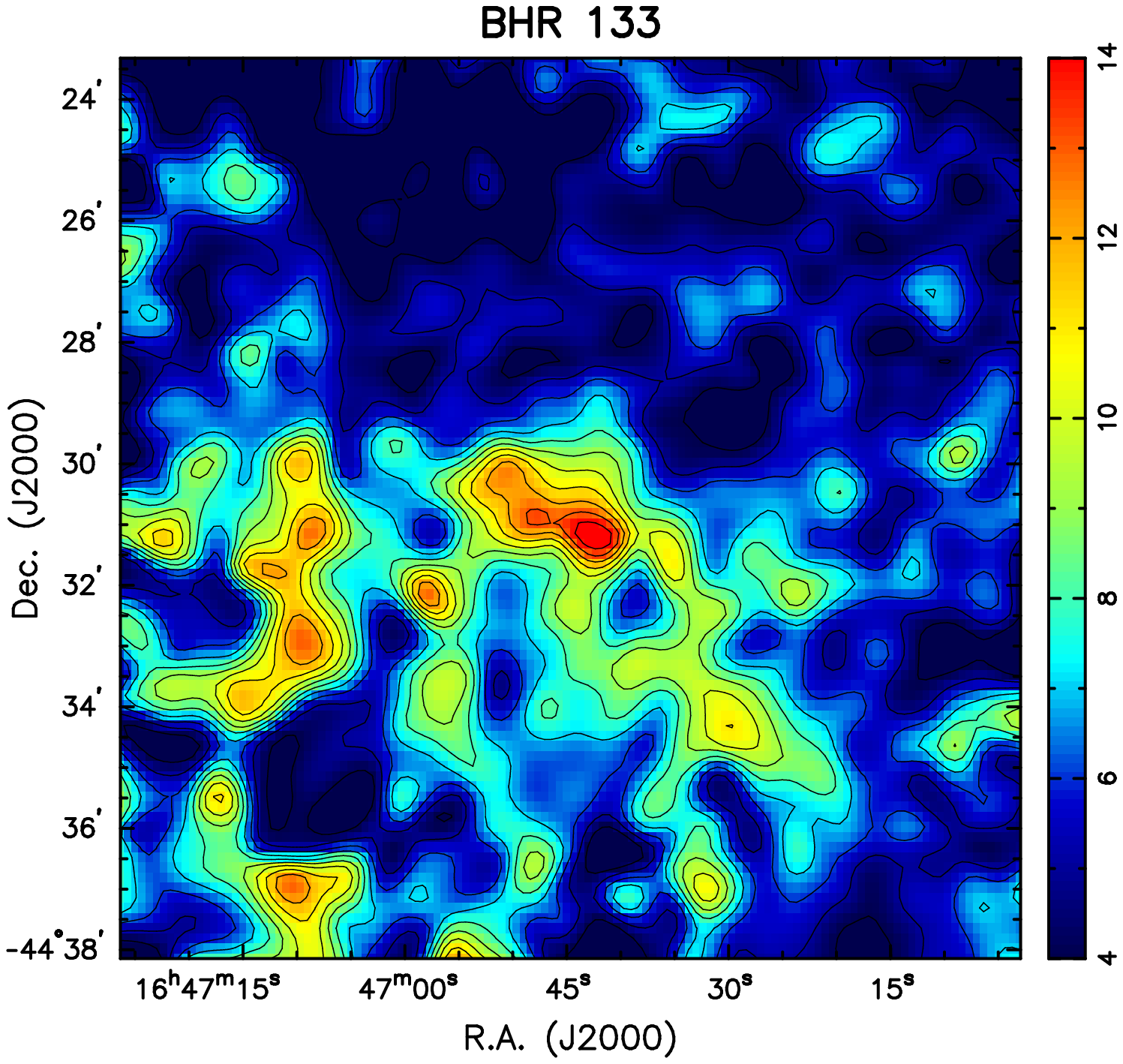}
\end{minipage}
\begin{minipage}[b]{0.3\textwidth}
 \centering
 \includegraphics[width=4cm]{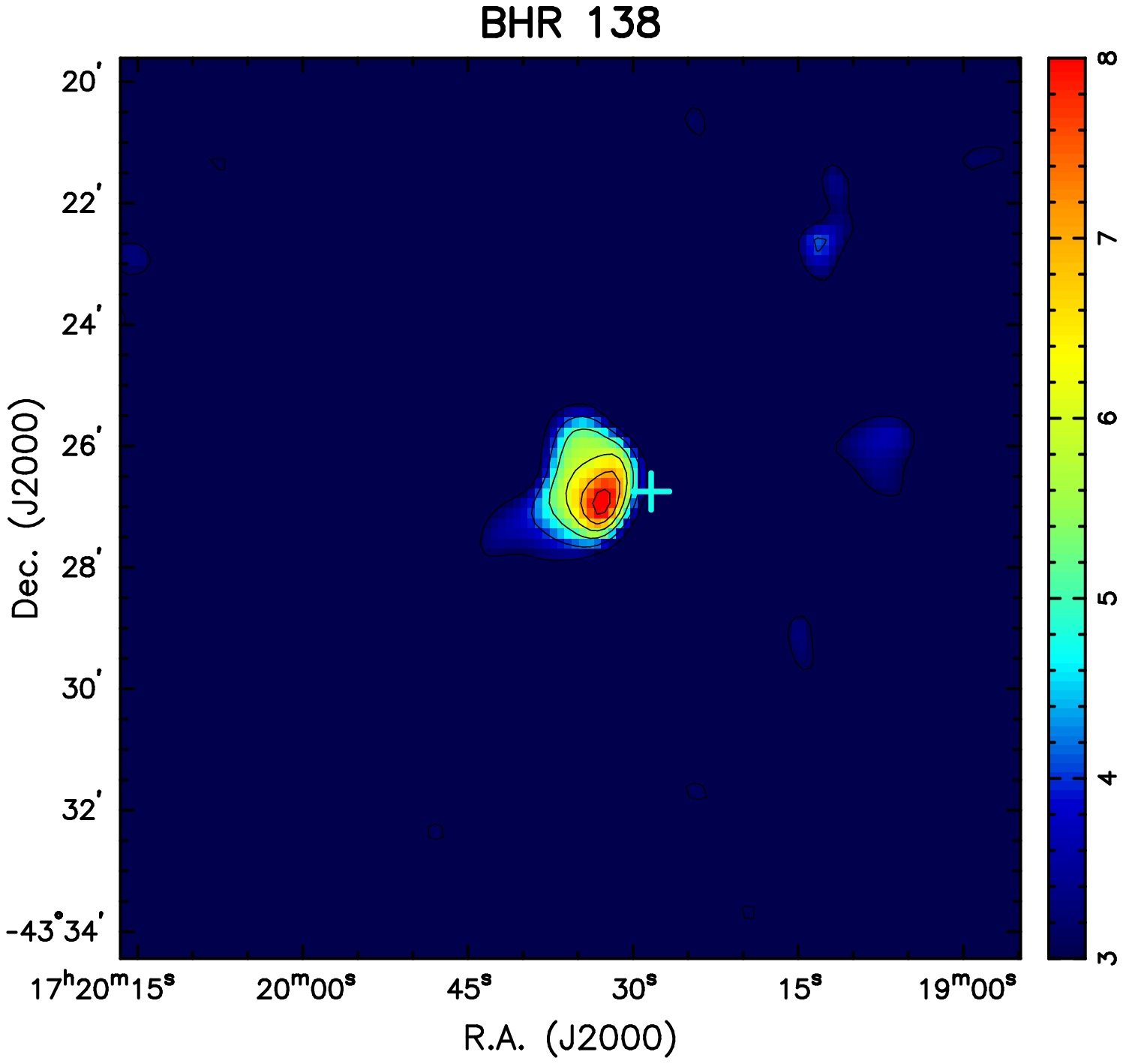}
\end{minipage}\\[0.5cm]
\begin{minipage}[b]{0.3\textwidth}
 \centering
 \includegraphics[width=4cm]{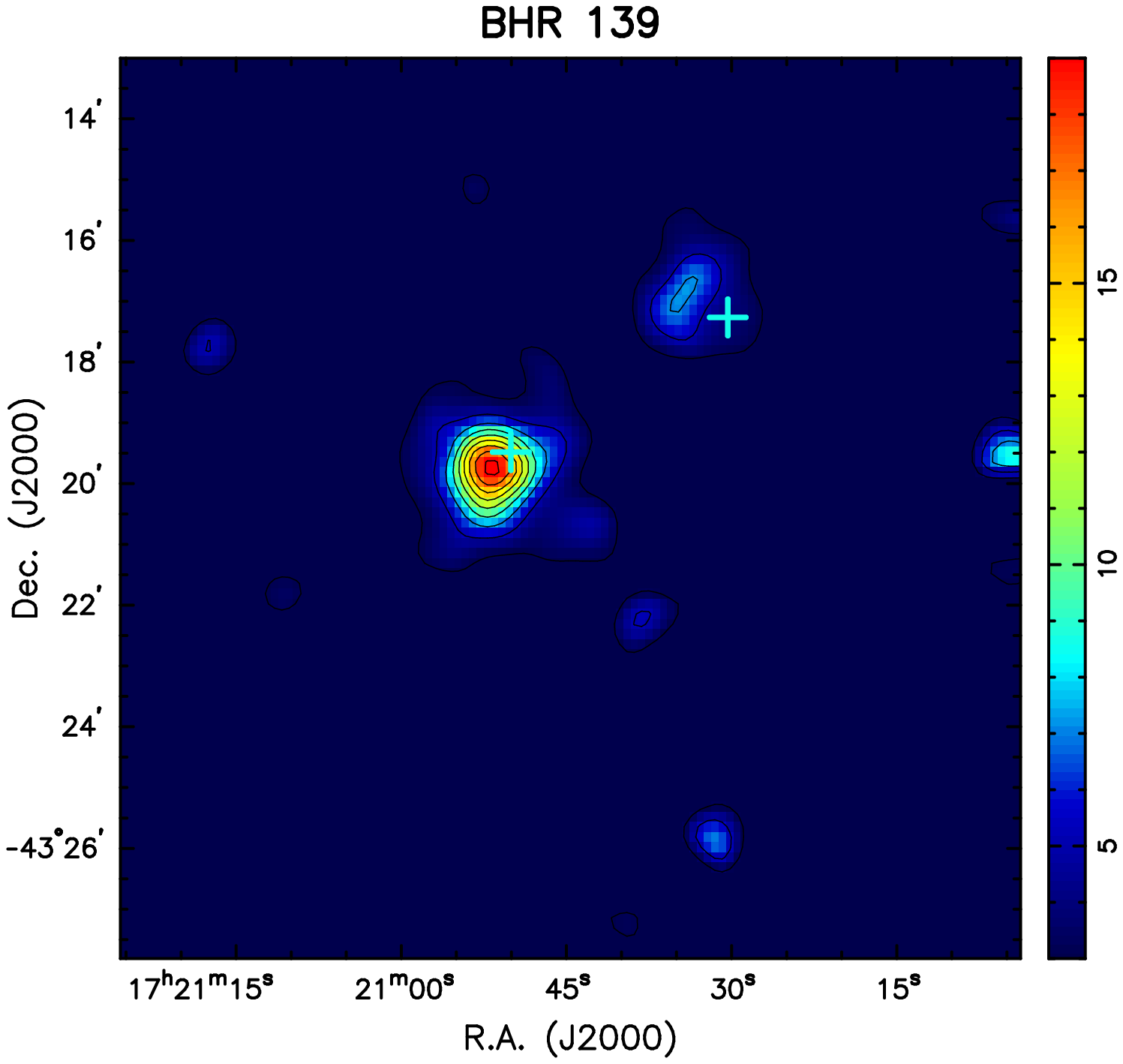}
\end{minipage}
\begin{minipage}[b]{0.3\textwidth}
 \centering
 \includegraphics[width=4cm]{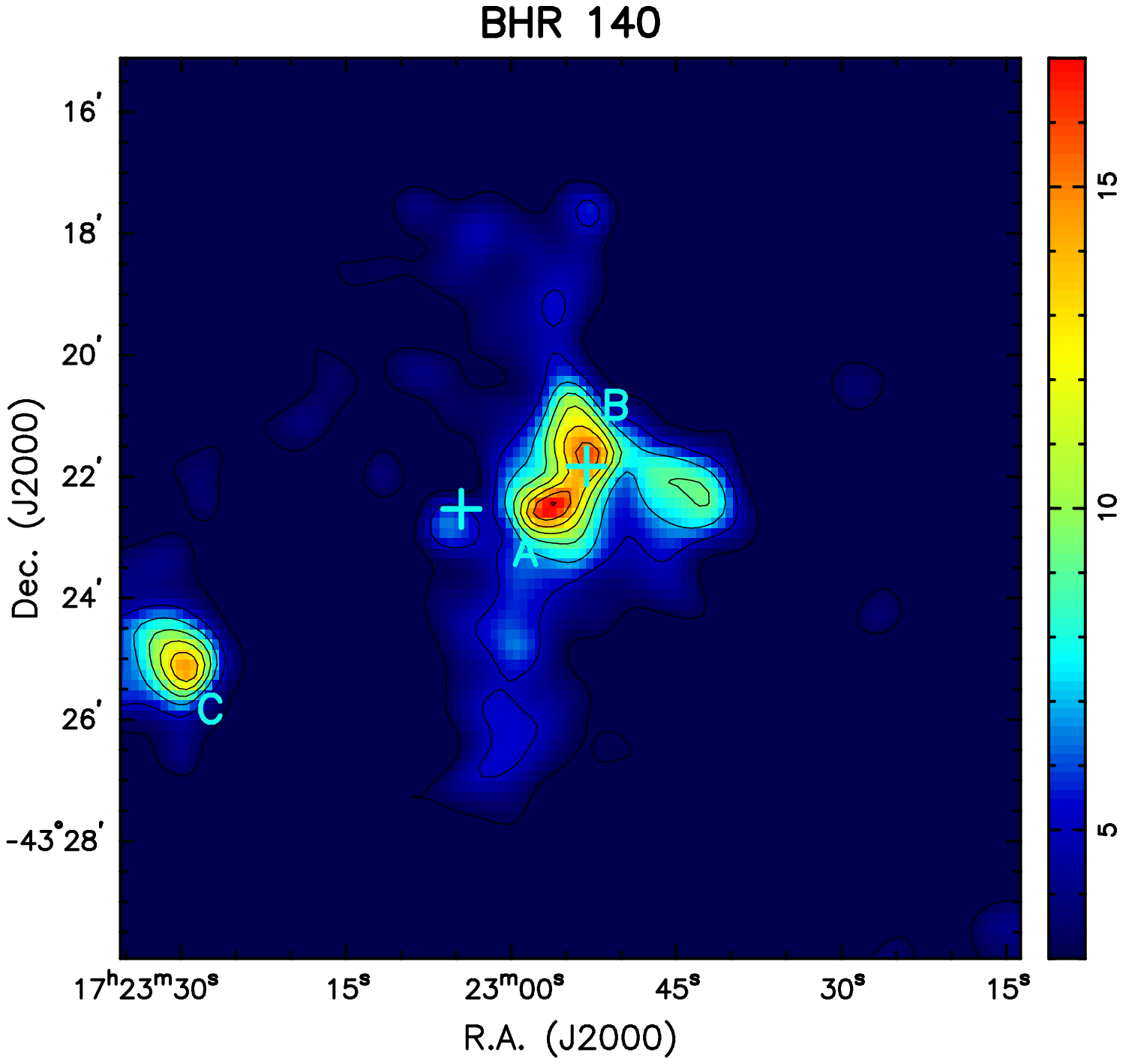}
\end{minipage}
\begin{minipage}[b]{0.3\textwidth}
 \centering
 \includegraphics[width=4cm]{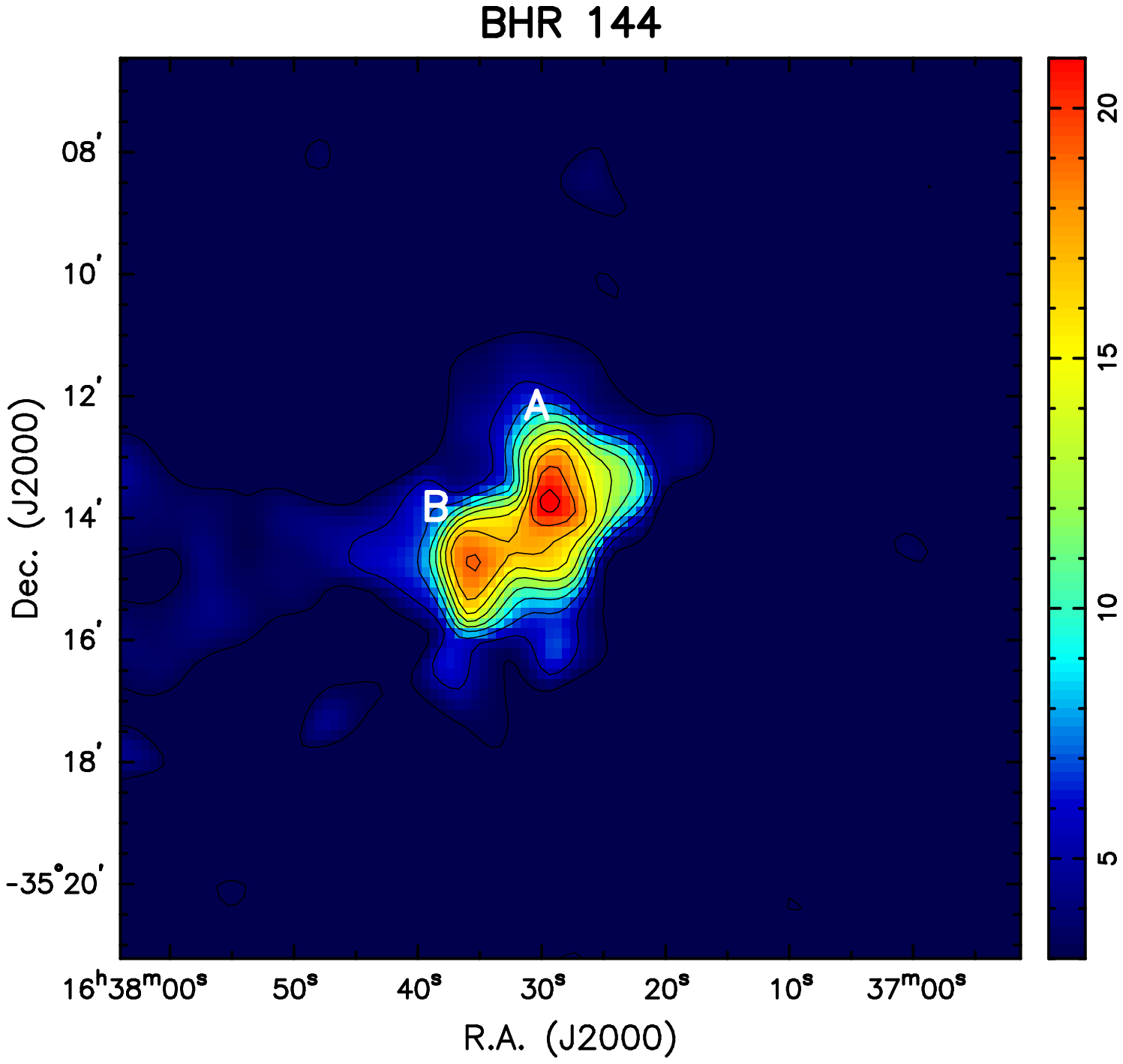}
\end{minipage}\\[0.5cm]
\begin{minipage}[b]{0.3\textwidth}
 \centering
 \includegraphics[width=4cm]{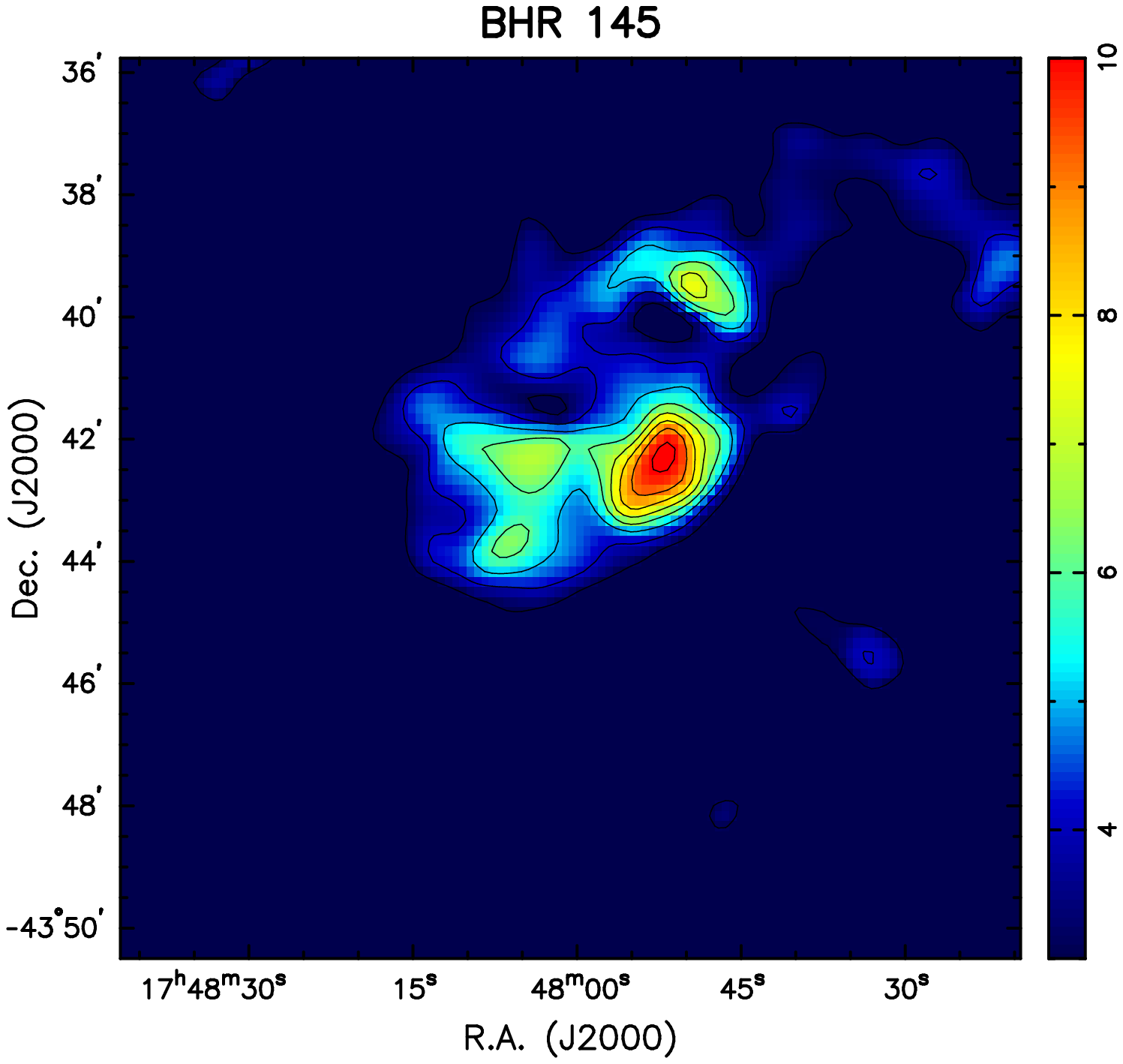}
\end{minipage}
\begin{minipage}[b]{0.3\textwidth}
 \centering
 \includegraphics[width=4cm]{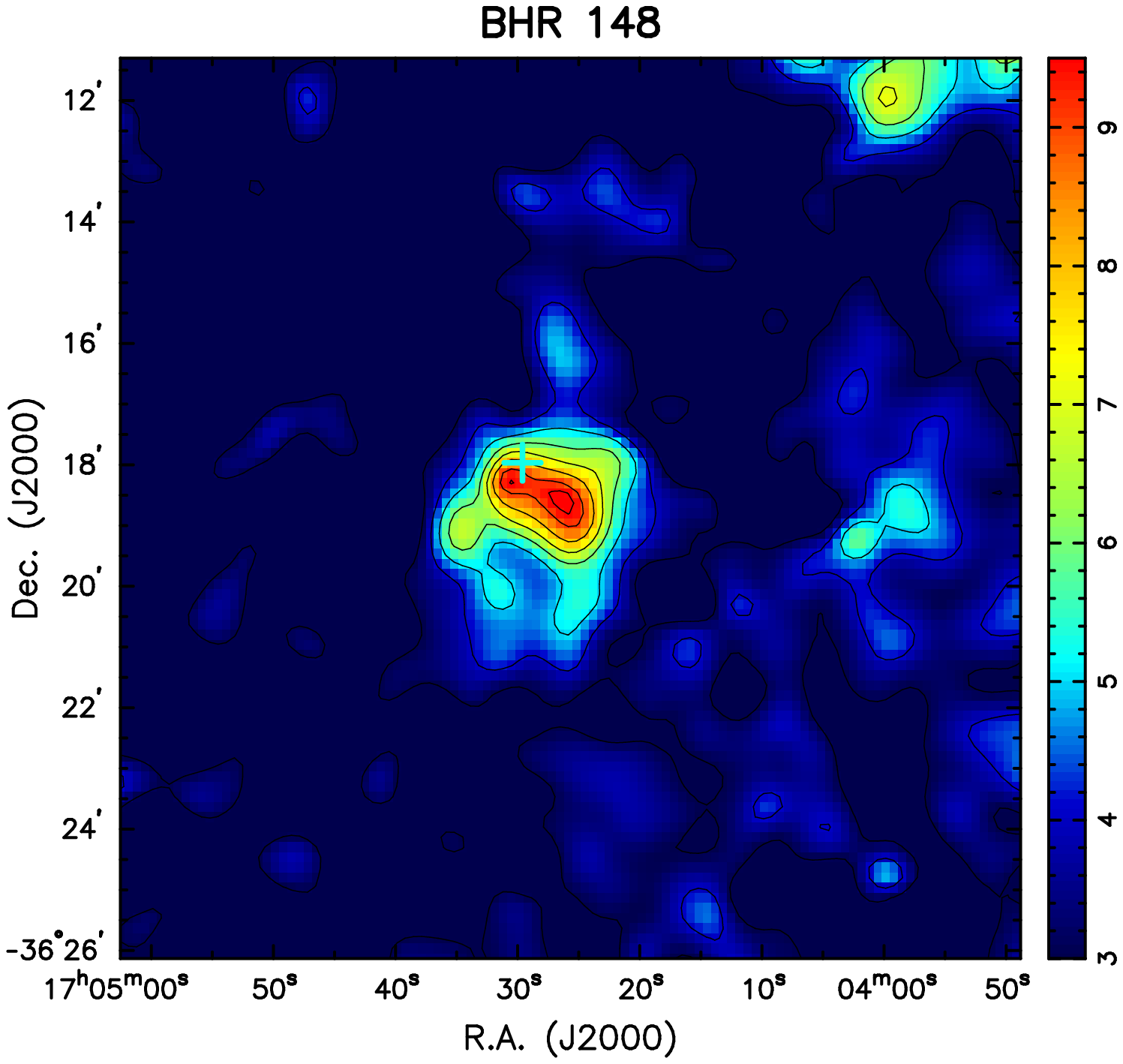}
\end{minipage}
\begin{minipage}[b]{0.3\textwidth}
 \centering
 \includegraphics[width=4cm]{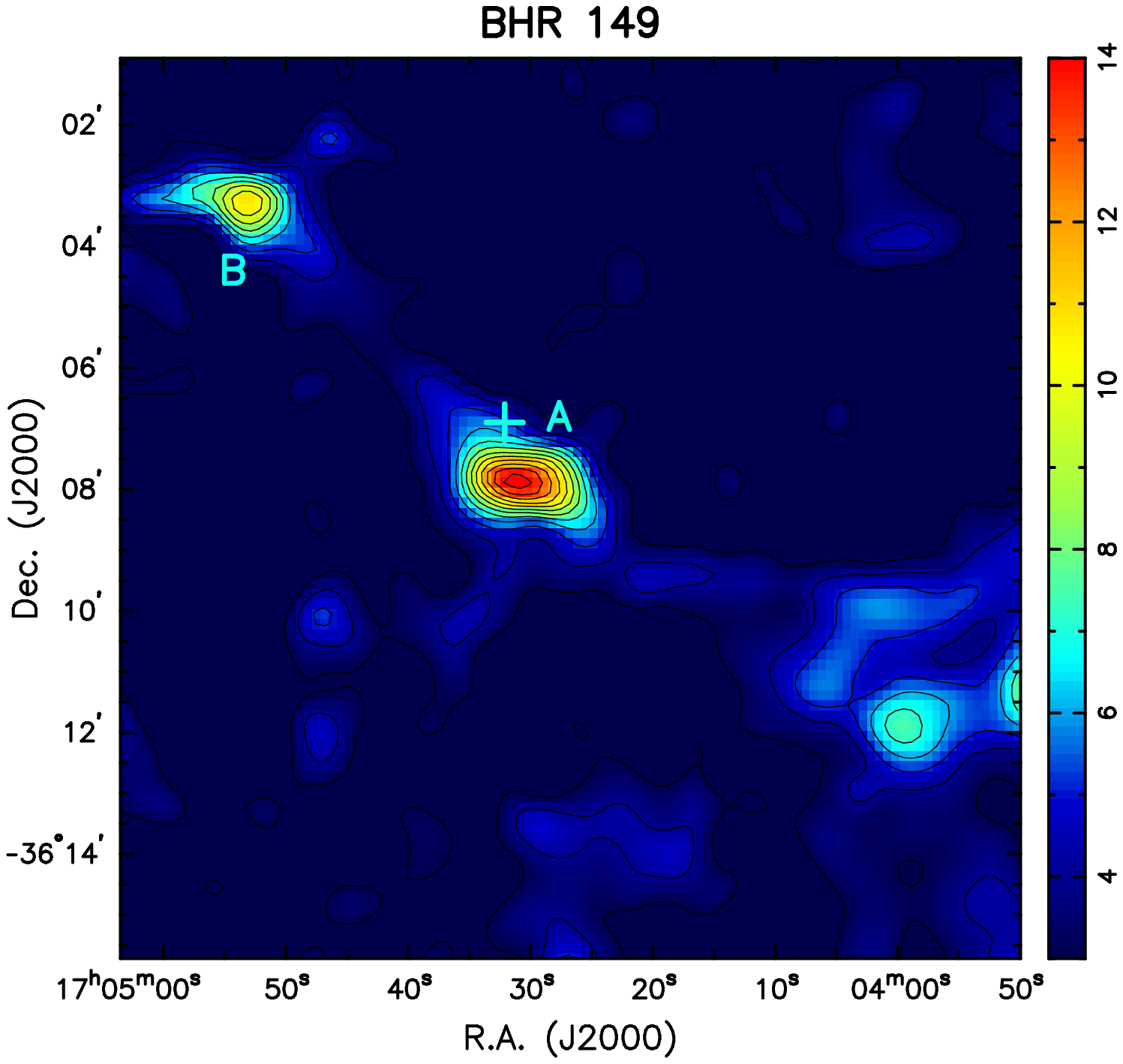}
\end{minipage}
\caption{{\it continued.}}
\end{figure}
%--------------------------------Fig 2: EXTINCTION MAPS--------------------------------

%--------------------------------Fig 3: STAR COUNTS MAPS--------------------------------
\clearpage
\begin{figure}
\centering
\begin{minipage}[b]{0.3\textwidth}
 \centering
 \includegraphics[width=4cm]{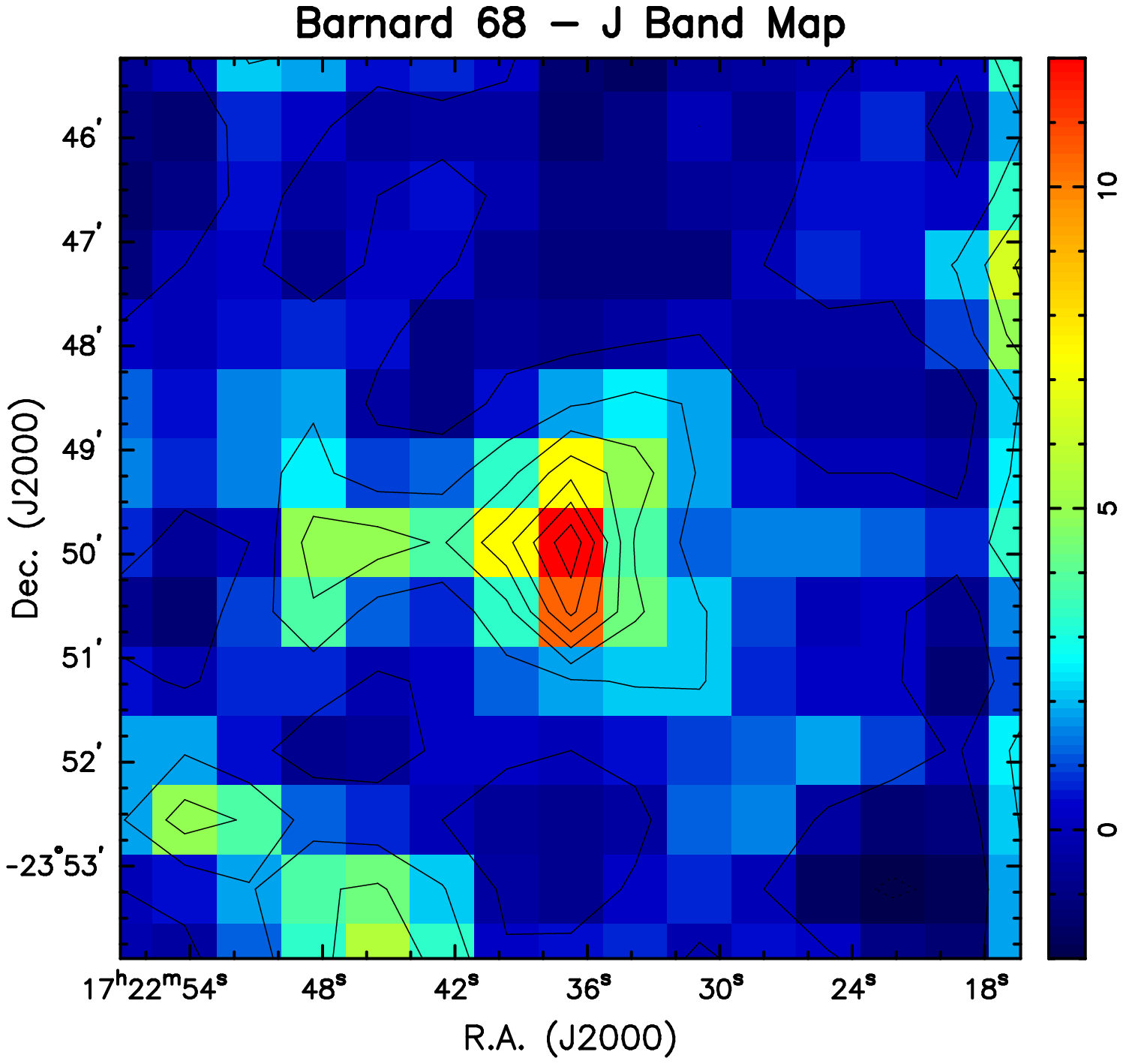}
\end{minipage}
\begin{minipage}[b]{0.3\textwidth}
 \centering
 \includegraphics[width=4cm]{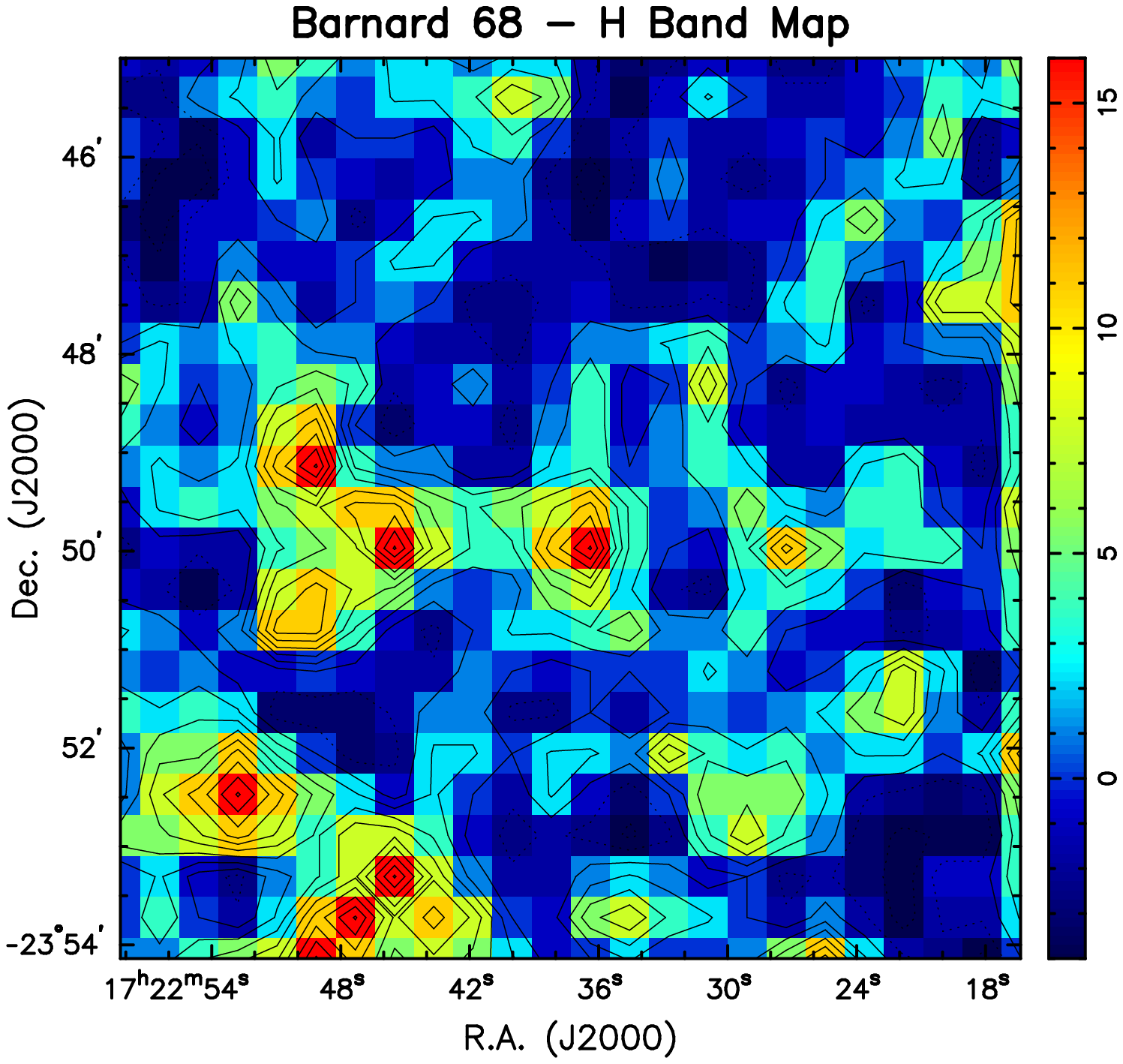}
\end{minipage}
\begin{minipage}[b]{0.3\textwidth}
 \centering
 \includegraphics[width=4cm]{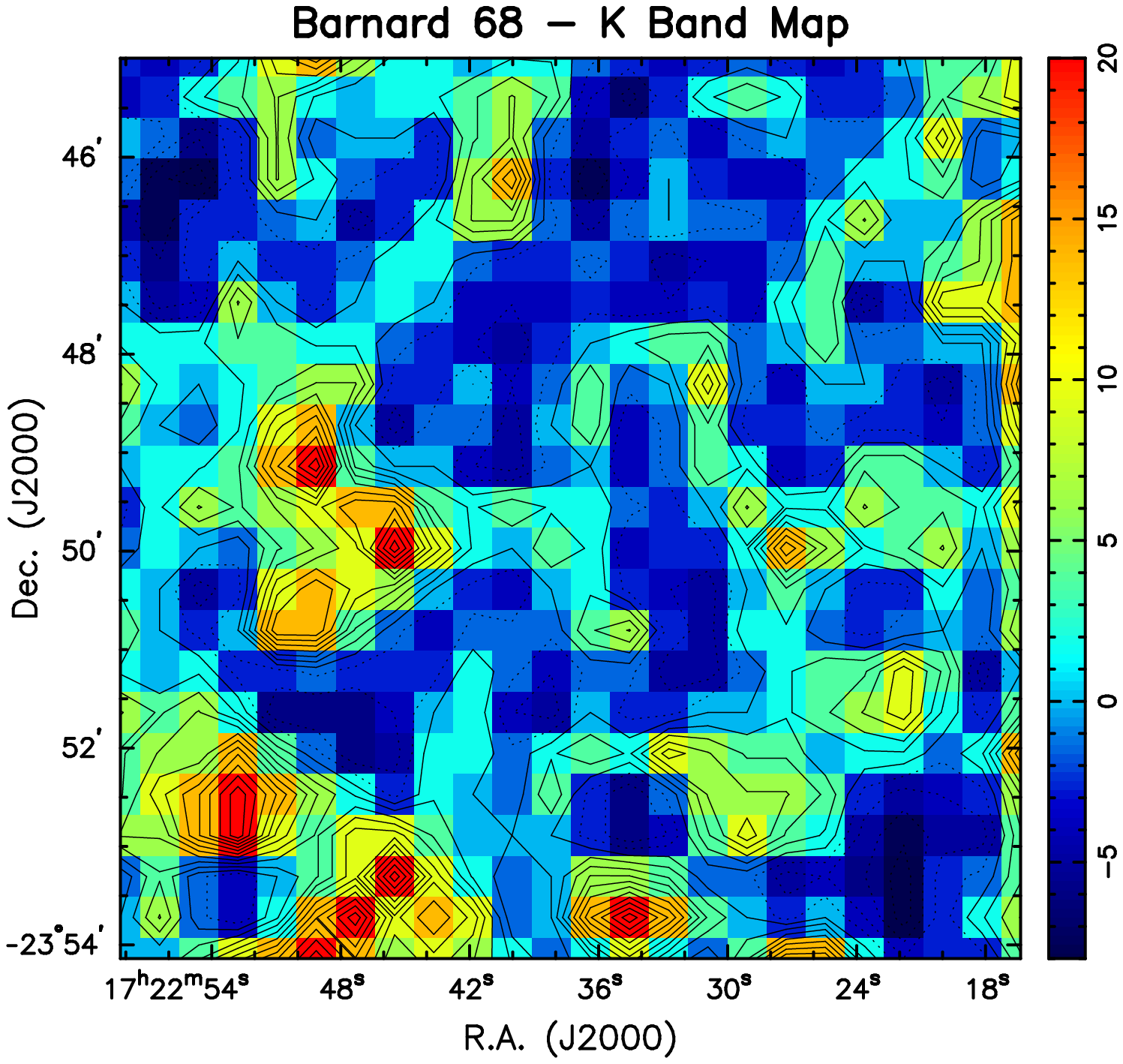}
\end{minipage}\\[0.5cm]
\begin{minipage}[b]{0.3\textwidth}
 \centering
 \includegraphics[width=4cm]{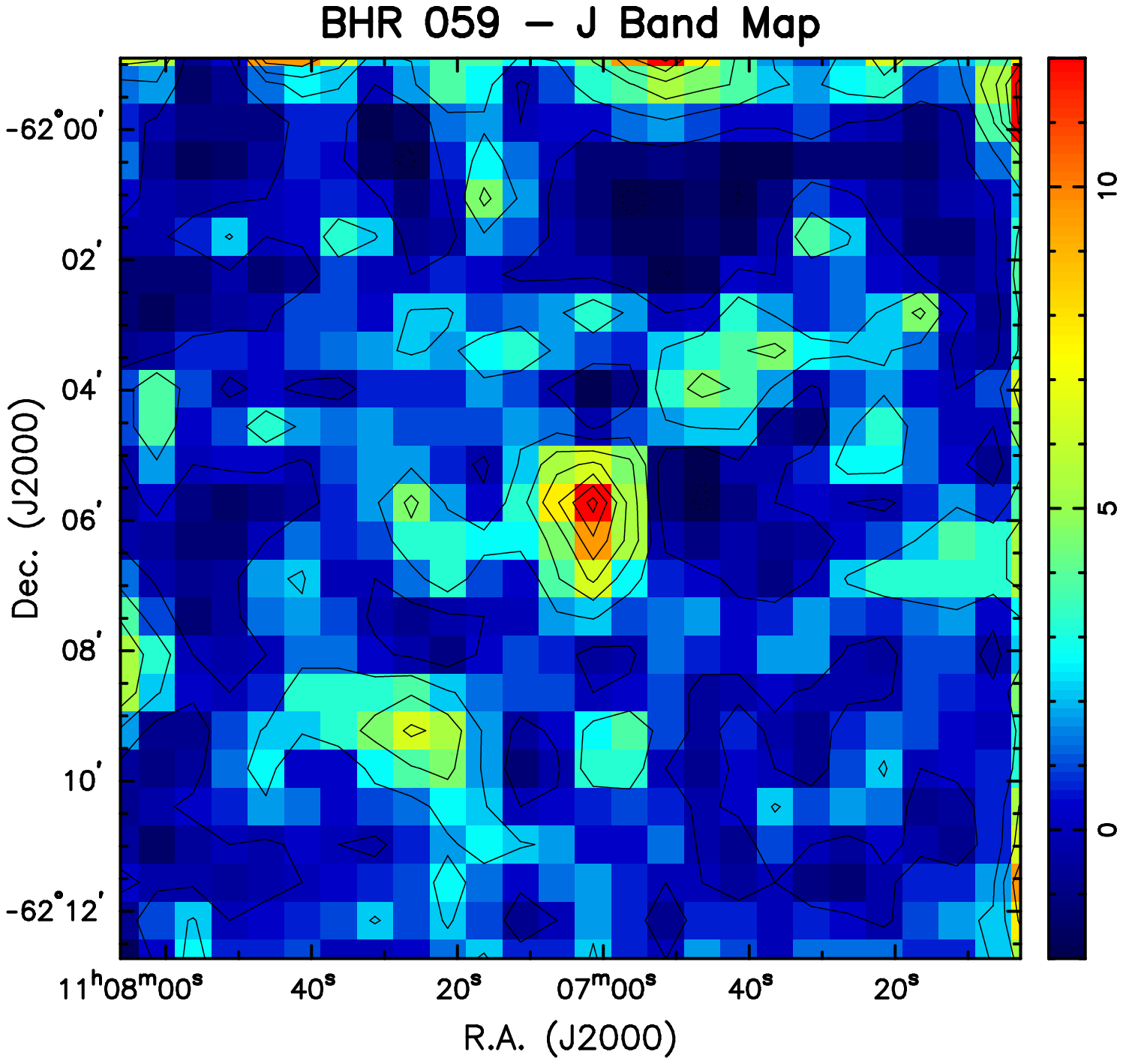}
\end{minipage}
\begin{minipage}[b]{0.3\textwidth}
 \centering
 \includegraphics[width=4cm]{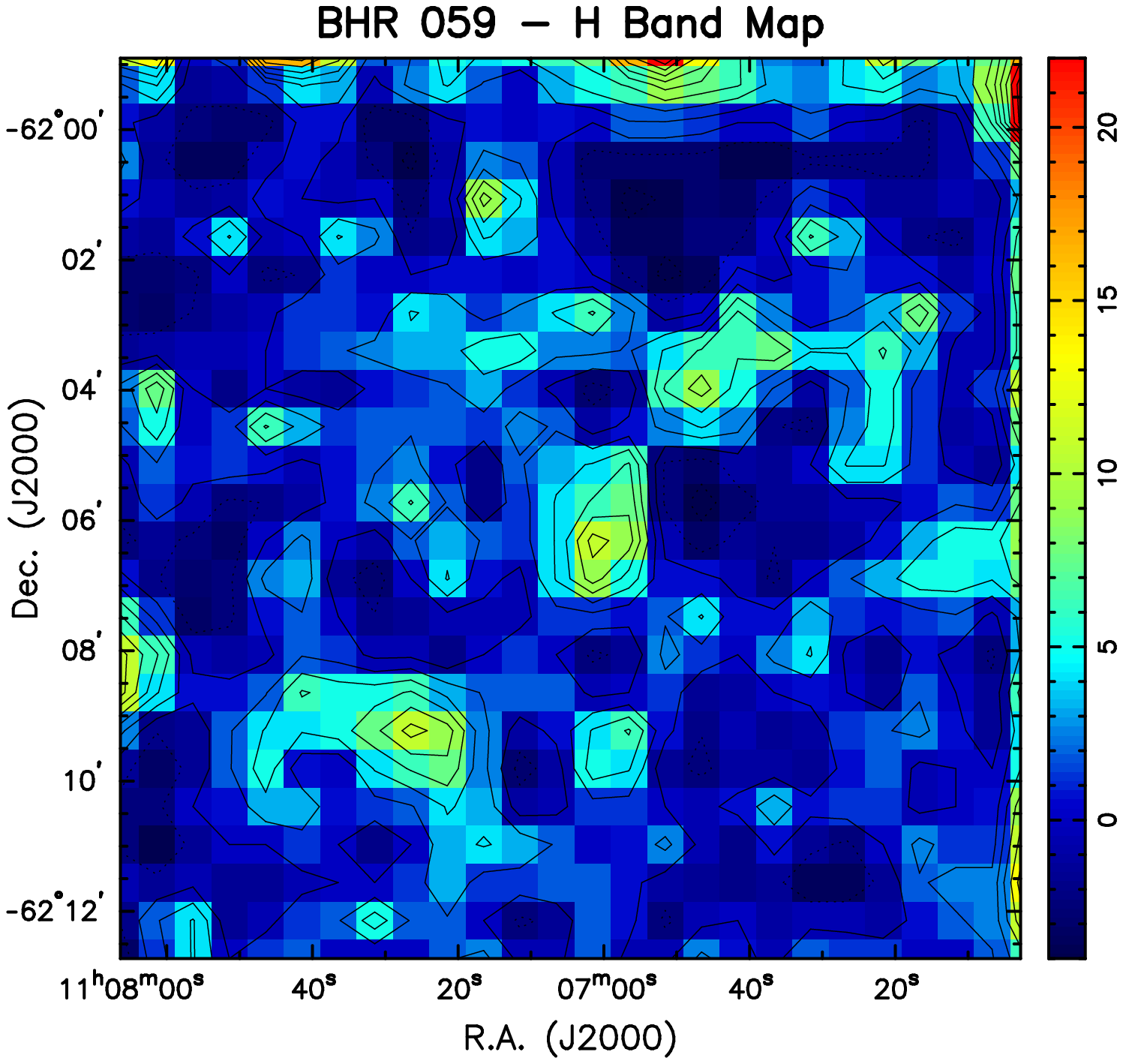}
\end{minipage}
\begin{minipage}[b]{0.3\textwidth}
 \centering
 \includegraphics[width=4cm]{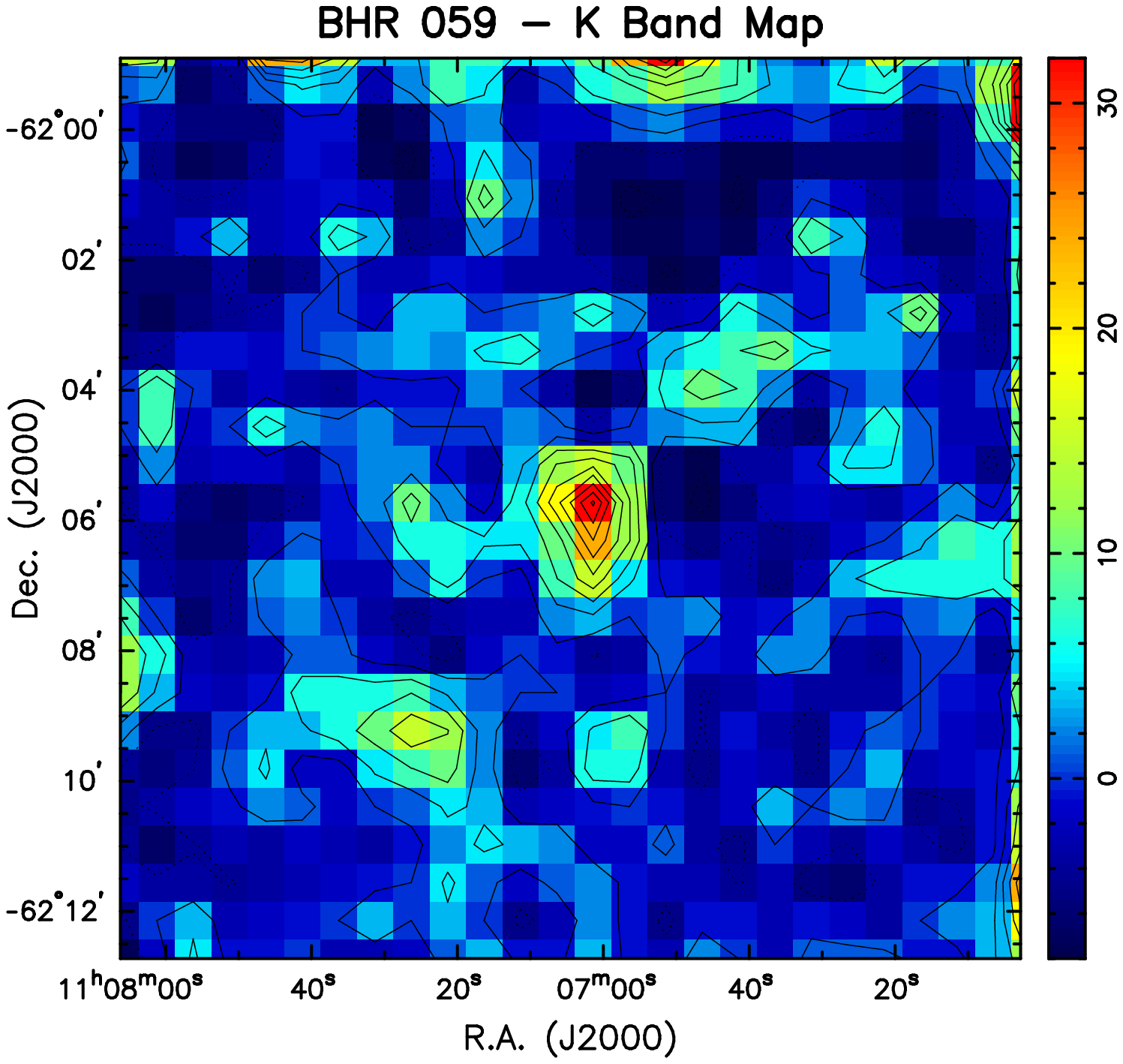}
\end{minipage}
\caption[]{Visual extinction maps constructed whith 2MASS data and
           the star counts method.}
\label{fig:countsmaps}
\end{figure}
%--------------------------------Fig 3: STAR COUNTS MAPS--------------------------------

%--------------------------------Fig 4: GALACTIC DISTRIBUTION--------------------------------
\clearpage
\begin{figure}
\centering
\includegraphics[scale=0.85]{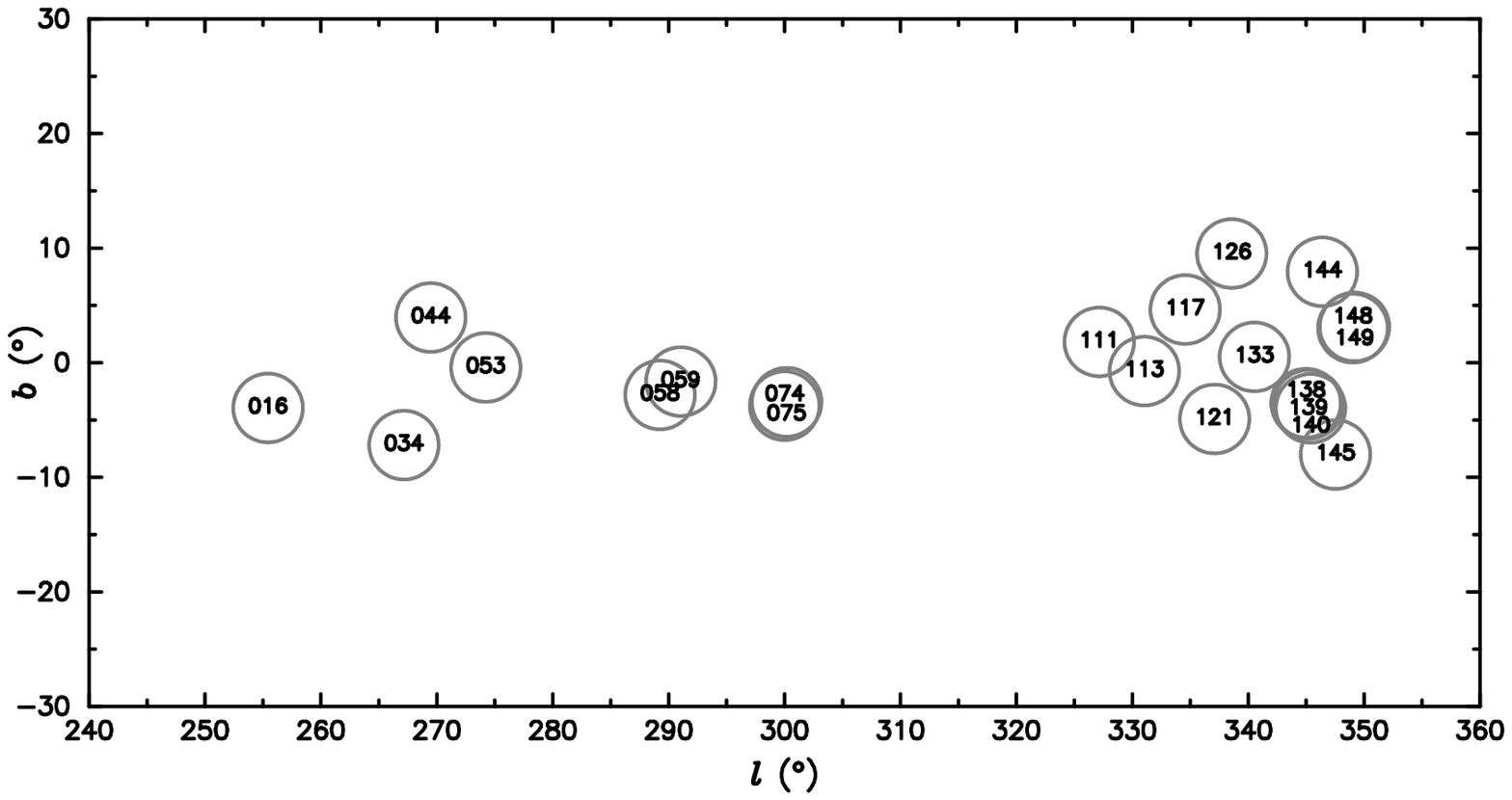}
\caption{Galactic distribution of the Bok globules of our sample. The numbres are the names
         of the globules given in Table \ref{tab:sample}, and the circles represent the
         3$^{\circ}$ regions used to search for stars in the SKY2000 catalog. In some cases
         (58-59, 74-75, 138-140, and 148-149), the same region was used.}
\label{fig:galactic}
\end{figure}
%--------------------------------Fig 4: GALACTIC DISTRIBUTION--------------------------------

%--------------------------------Fig 5: DISTANCES--------------------------------
\clearpage
\begin{figure}
\centering
\begin{minipage}[b]{0.2\textwidth}
 \centering
 \includegraphics[width=3cm]{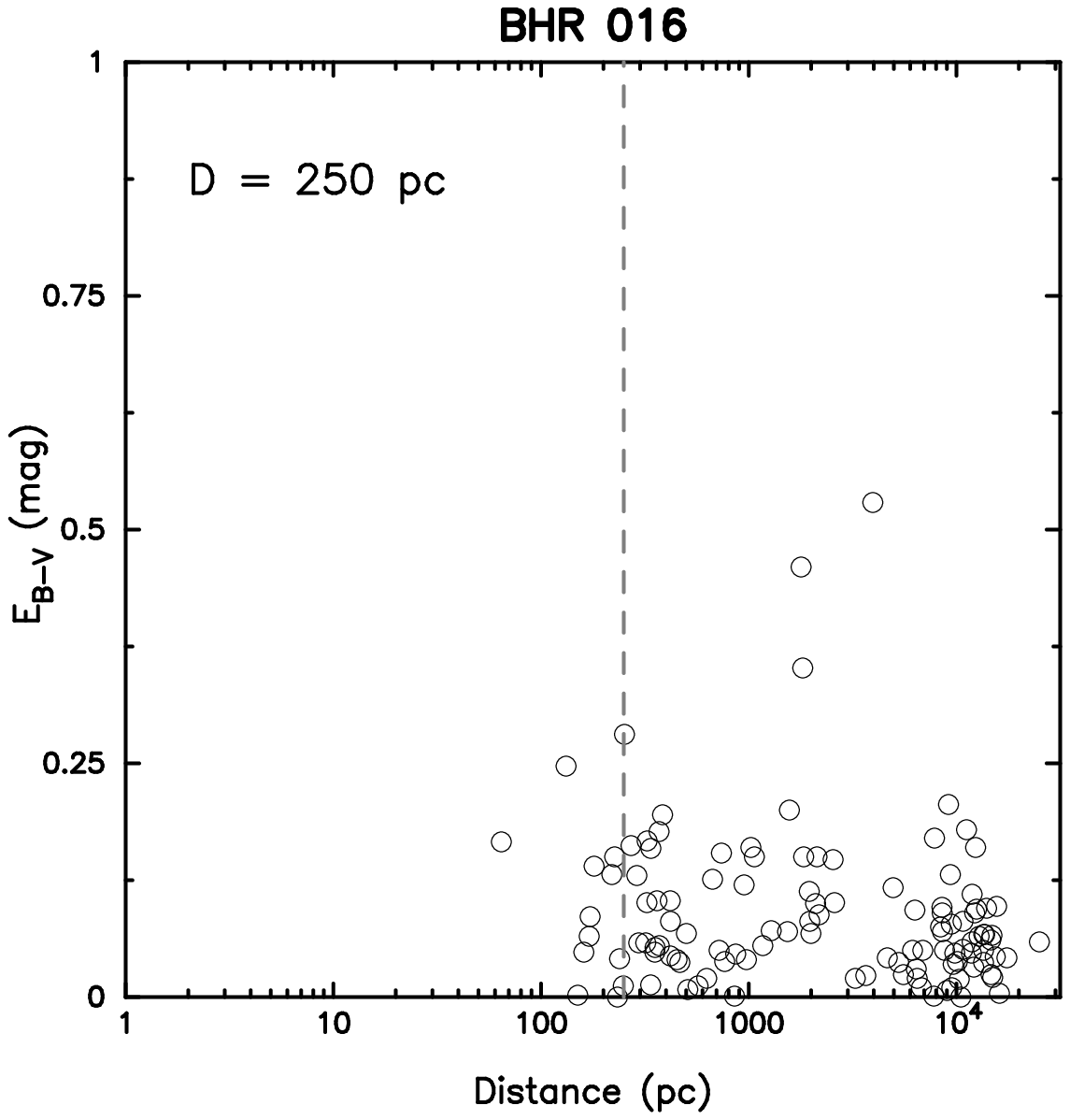}
\end{minipage}
\begin{minipage}[b]{0.2\textwidth}
 \centering
 \includegraphics[width=3cm]{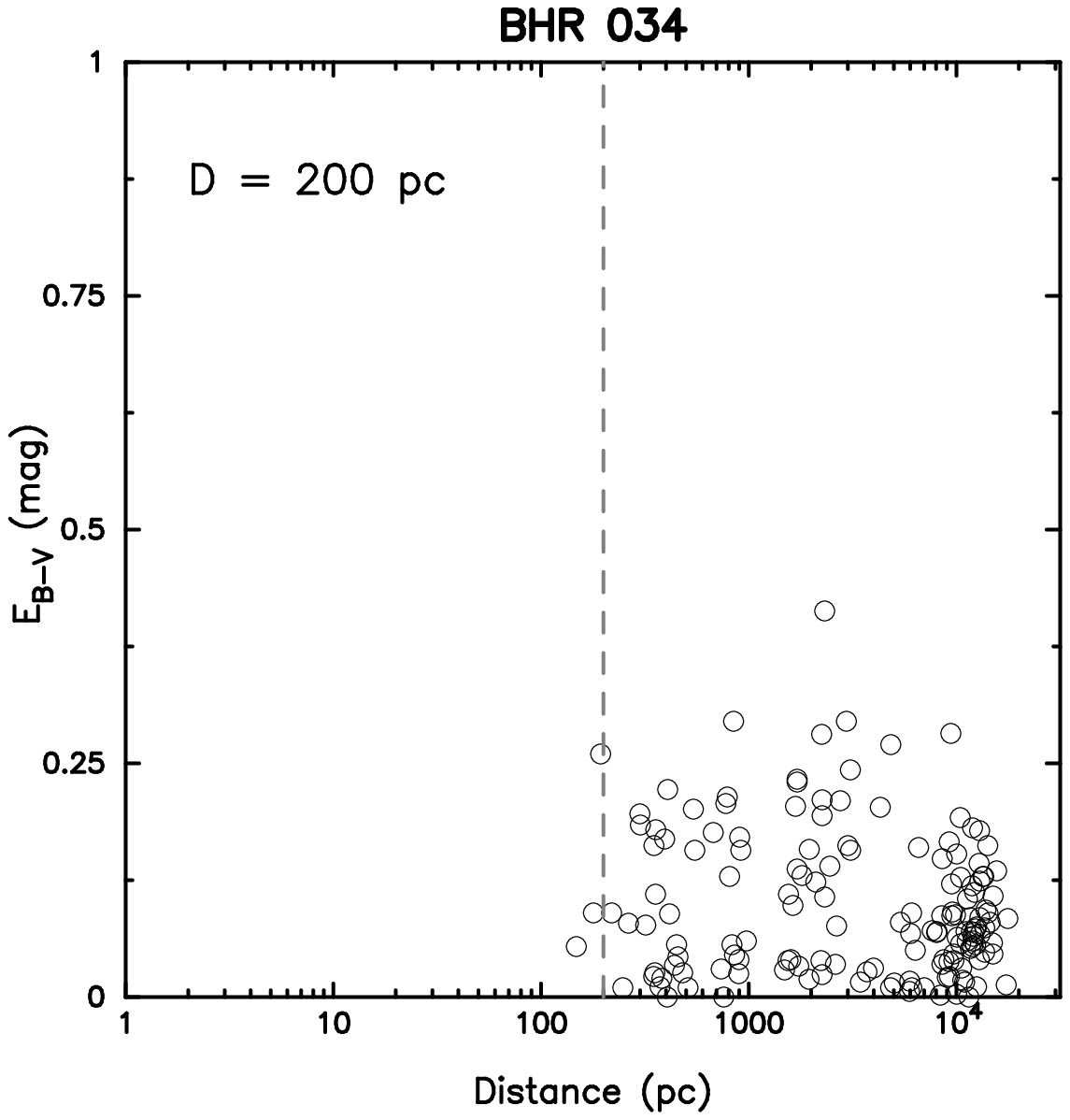}
\end{minipage}
\begin{minipage}[b]{0.2\textwidth}
 \centering
 \includegraphics[width=3cm]{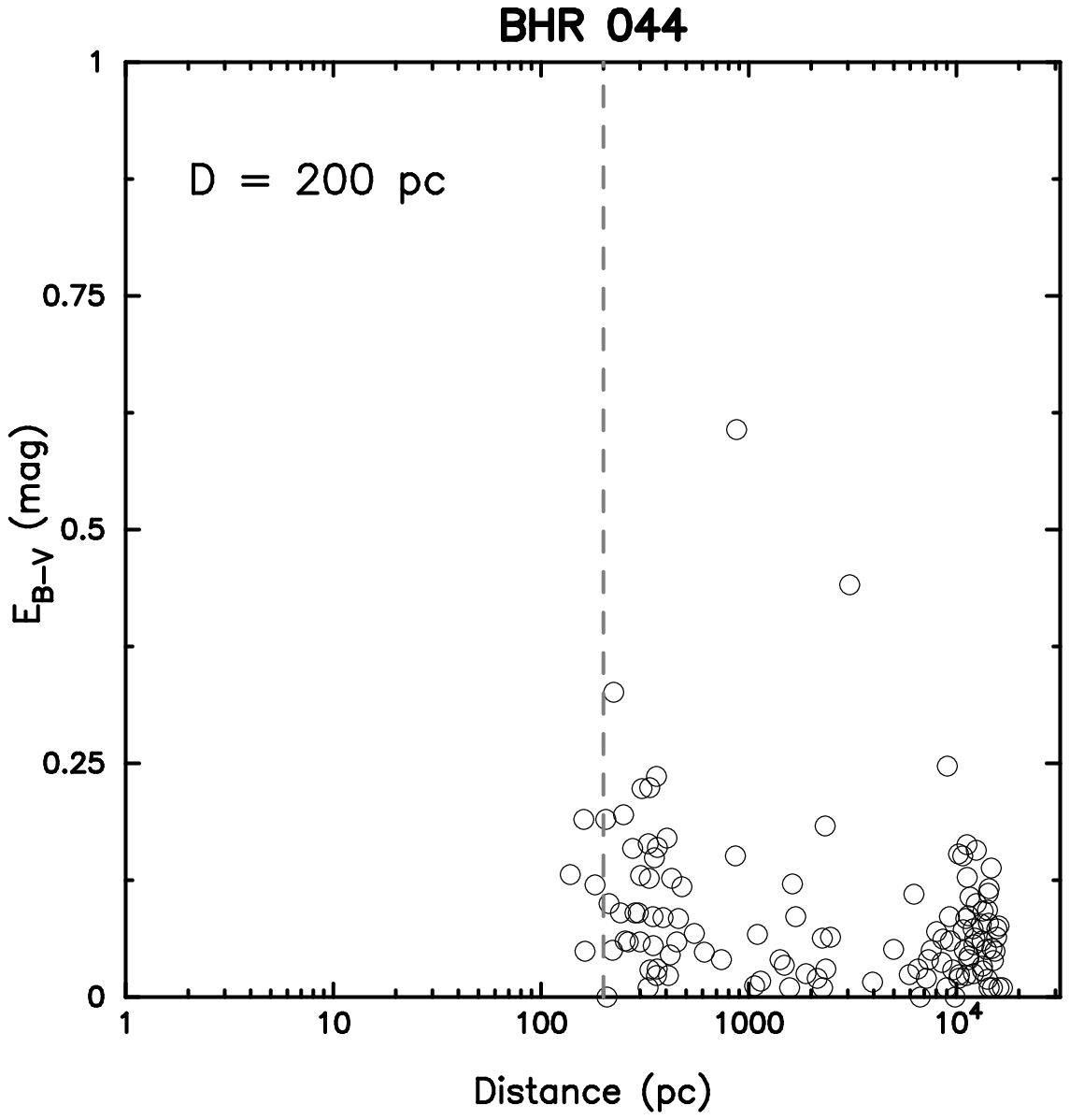}
\end{minipage}
\begin{minipage}[b]{0.2\textwidth}
 \centering
 \includegraphics[width=3cm]{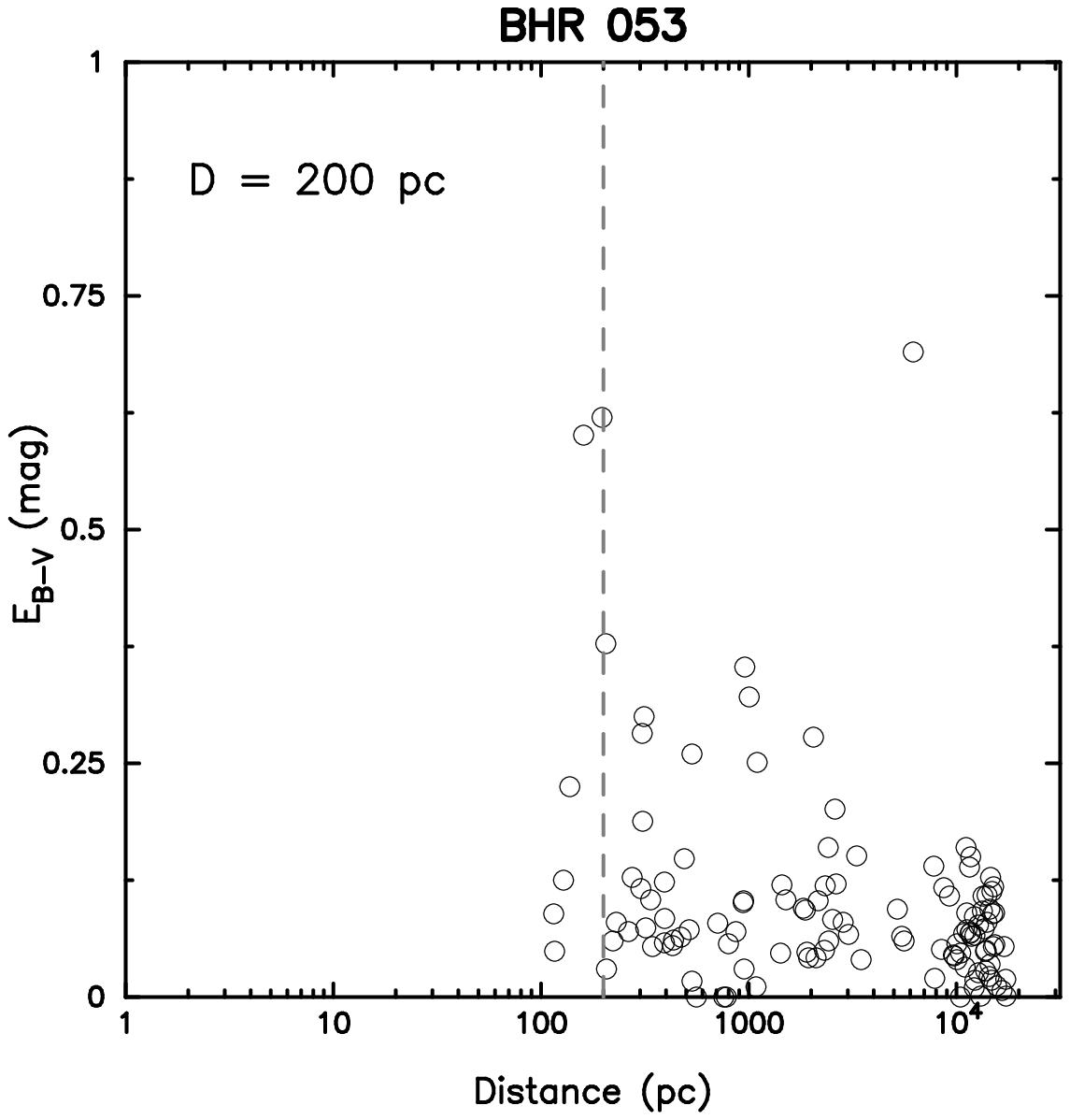}
\end{minipage}\\[0.5cm]
\begin{minipage}[b]{0.2\textwidth}
 \centering
 \includegraphics[width=3cm]{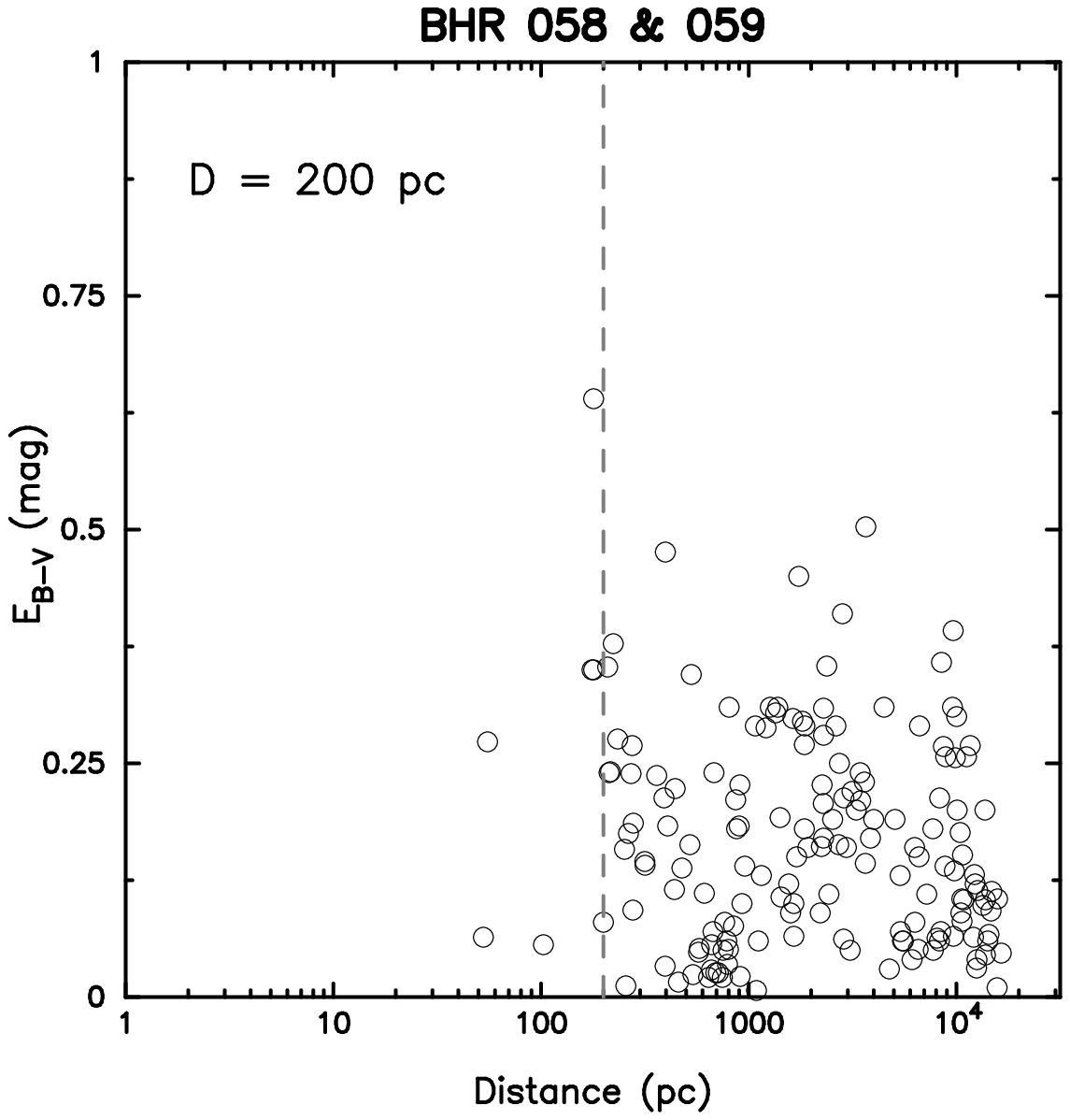}
\end{minipage}
\begin{minipage}[b]{0.2\textwidth}
 \centering
 \includegraphics[width=3cm]{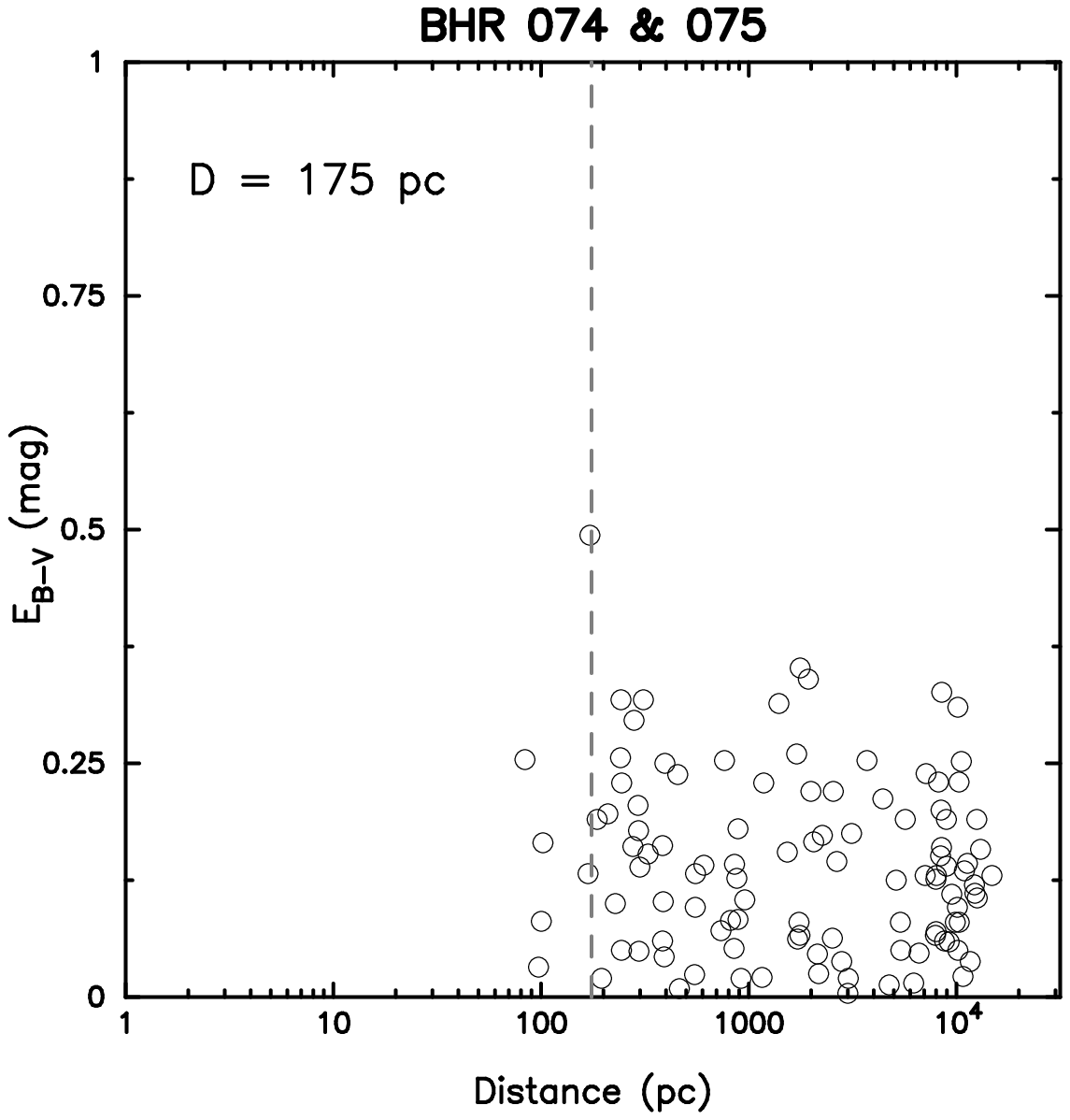}
\end{minipage}
\begin{minipage}[b]{0.2\textwidth}
 \centering
 \includegraphics[width=3cm]{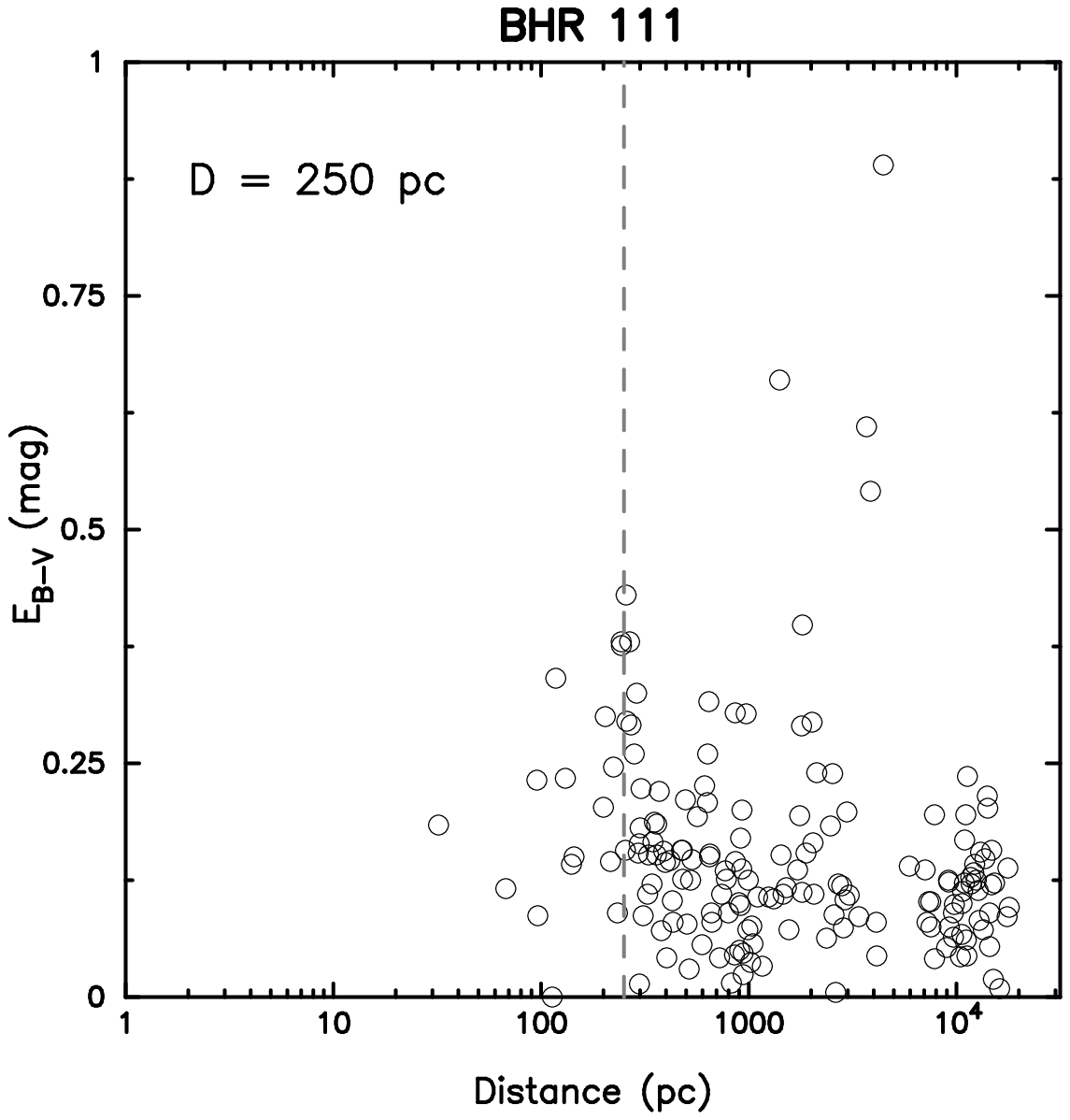}
\end{minipage}
\begin{minipage}[b]{0.2\textwidth}
 \centering
 \includegraphics[width=3cm]{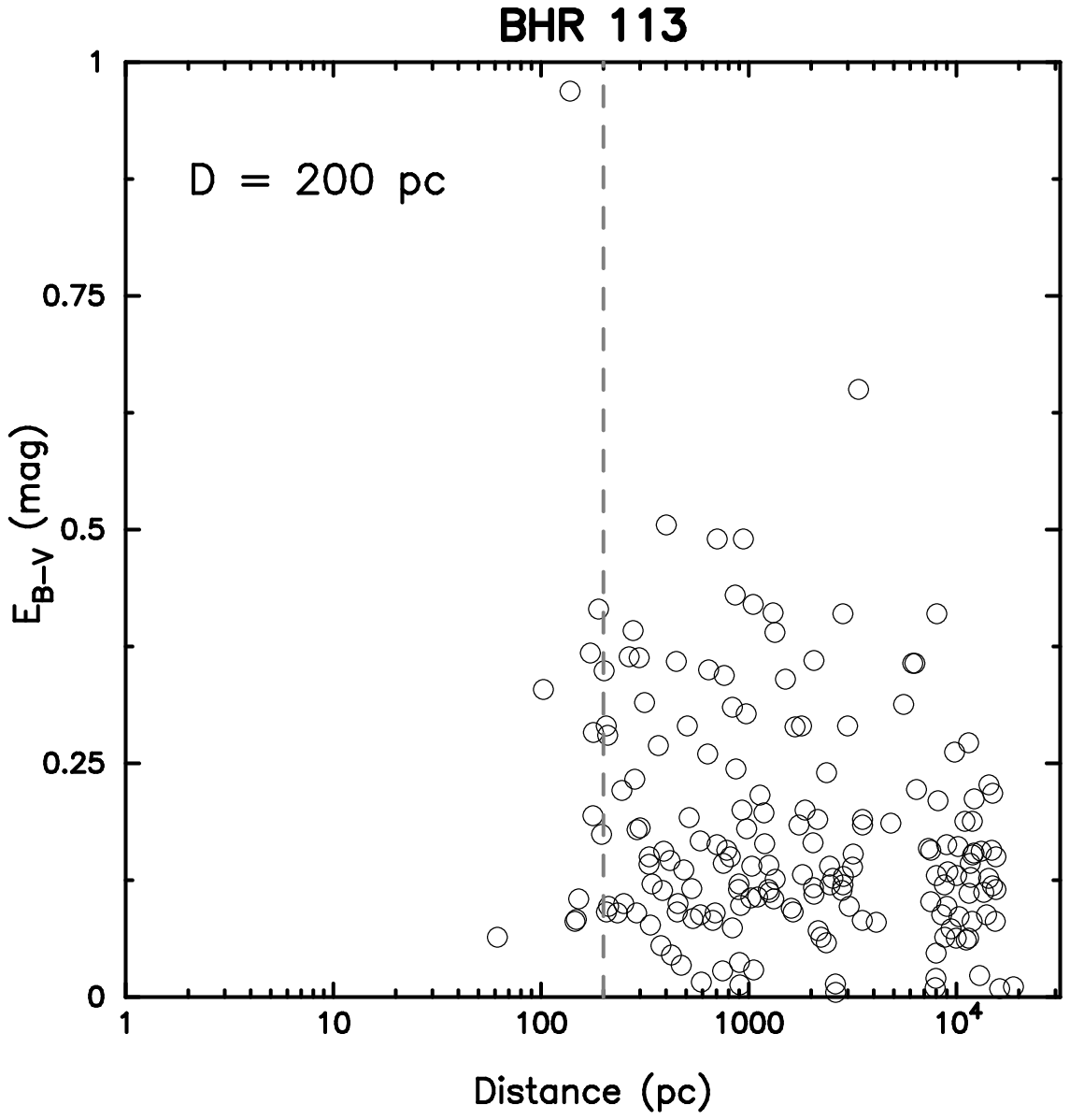}
\end{minipage}\\[0.5cm]
\begin{minipage}[b]{0.2\textwidth}
 \centering
 \includegraphics[width=3cm]{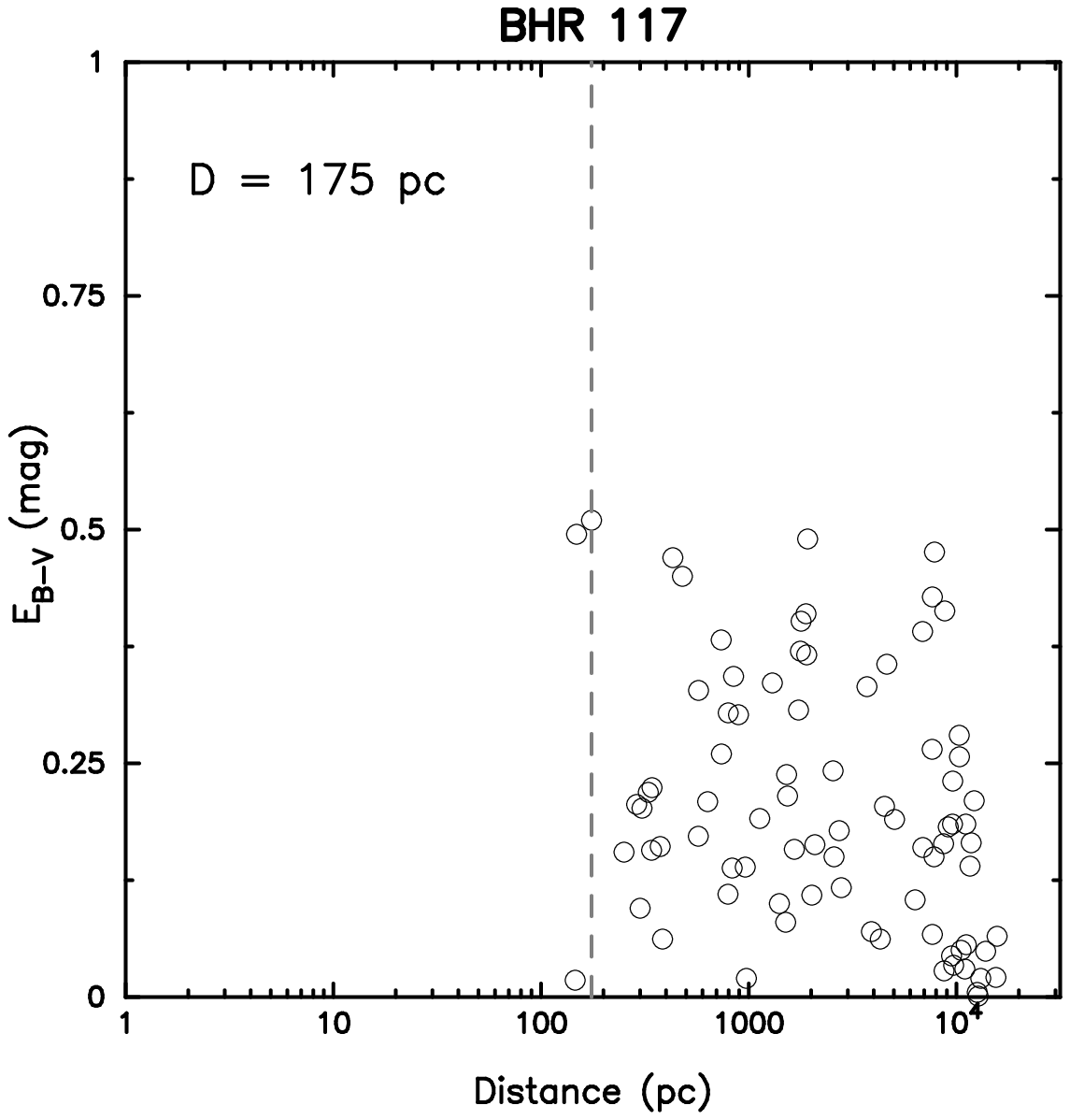}
\end{minipage}
\begin{minipage}[b]{0.2\textwidth}
 \centering
 \includegraphics[width=3cm]{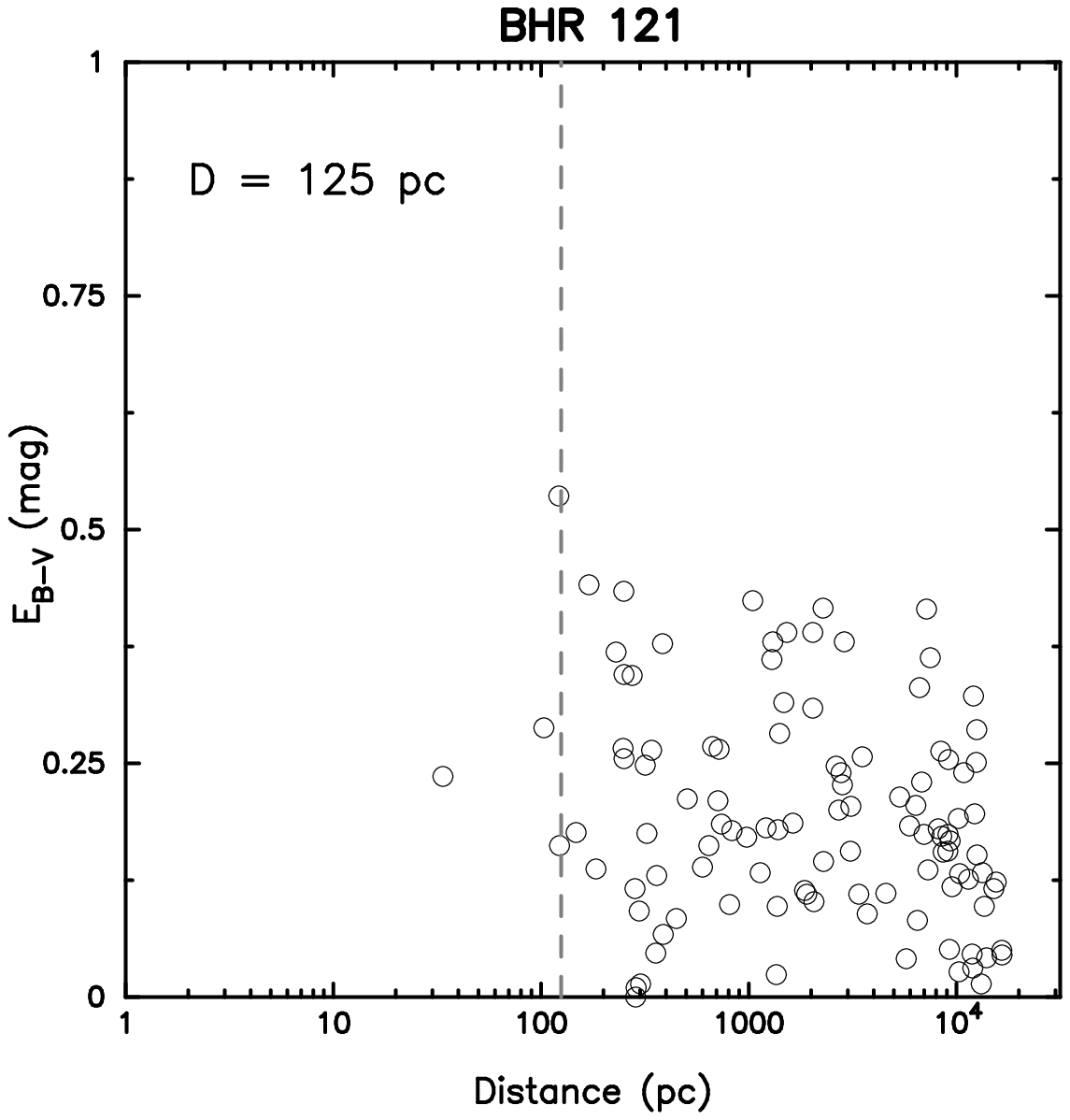}
\end{minipage}
\begin{minipage}[b]{0.2\textwidth}
 \centering
 \includegraphics[width=3cm]{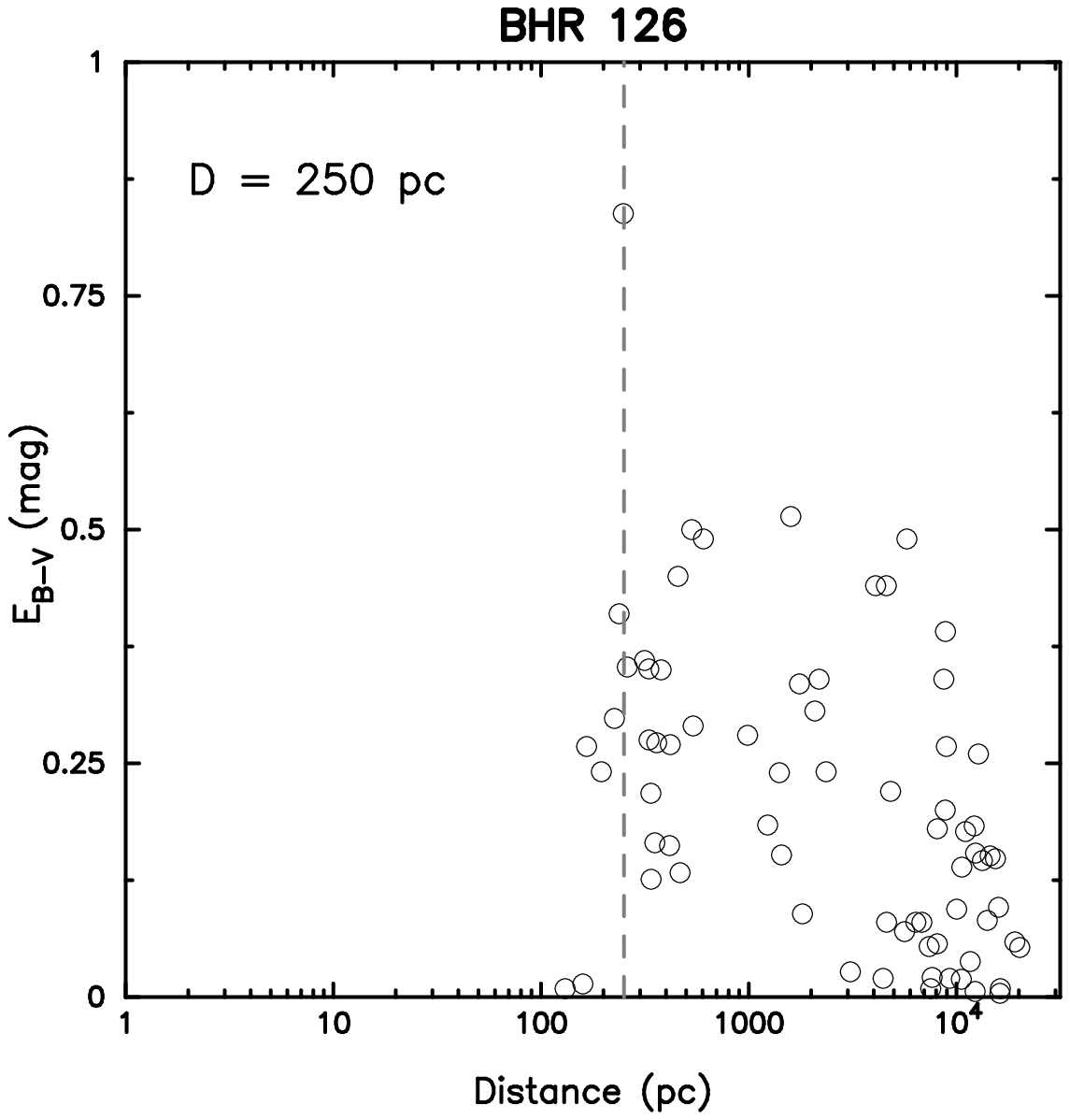}
\end{minipage}
\begin{minipage}[b]{0.2\textwidth}
 \centering
 \includegraphics[width=3cm]{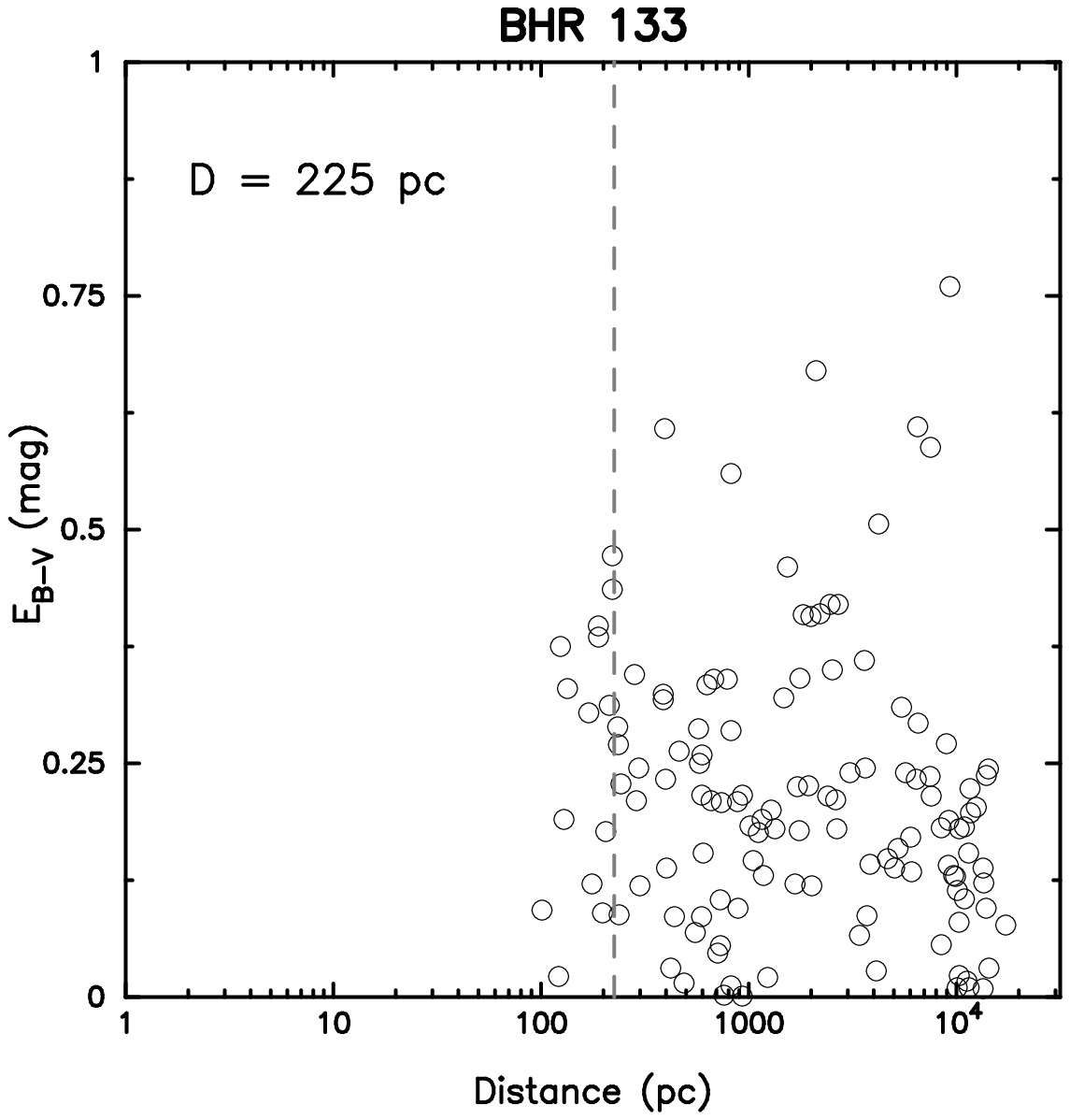}
\end{minipage}\\[0.5cm]
\begin{minipage}[b]{0.2\textwidth}
 \centering
 \includegraphics[width=3cm]{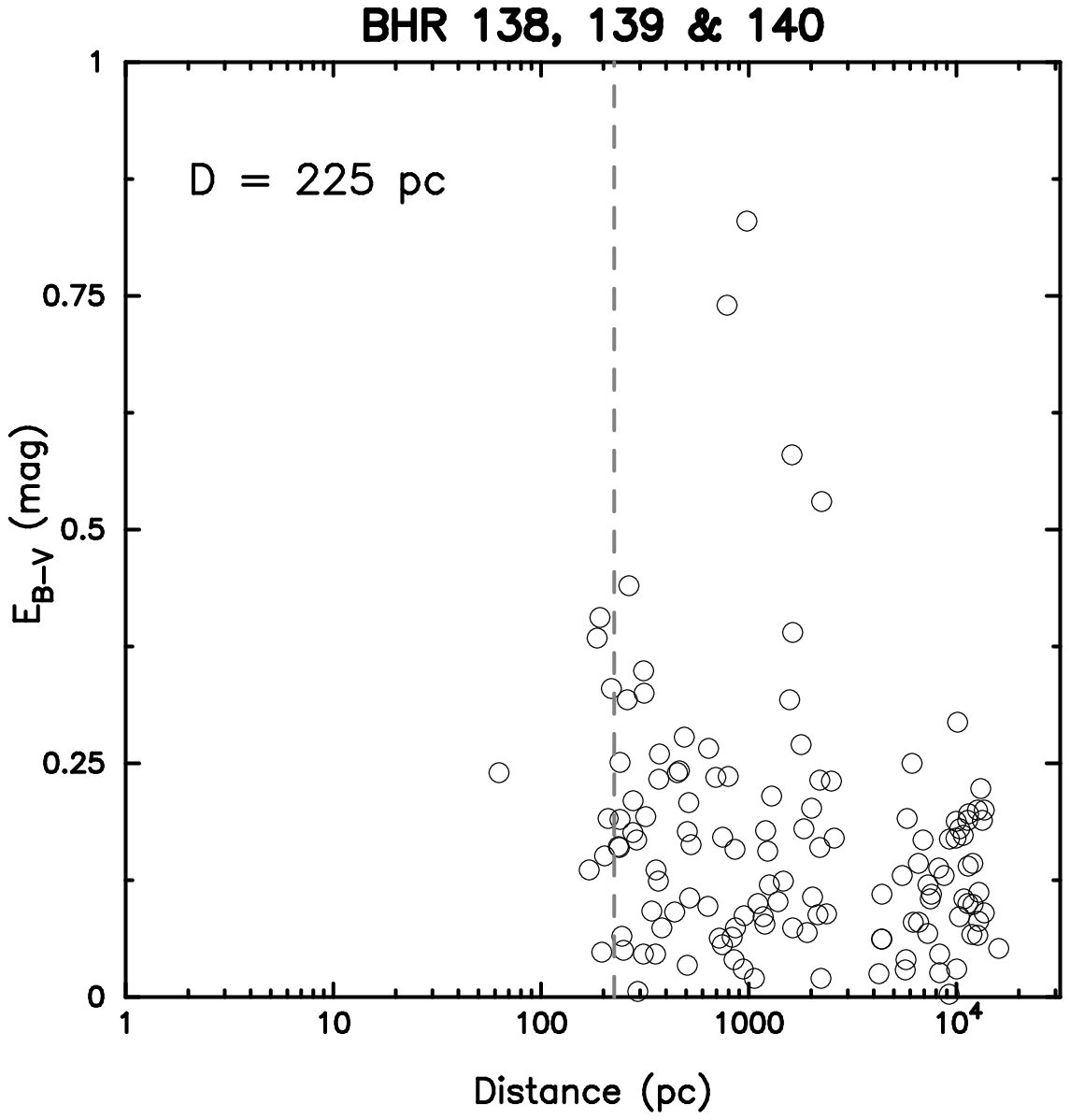}
\end{minipage}
\begin{minipage}[b]{0.2\textwidth}
 \centering
 \includegraphics[width=3cm]{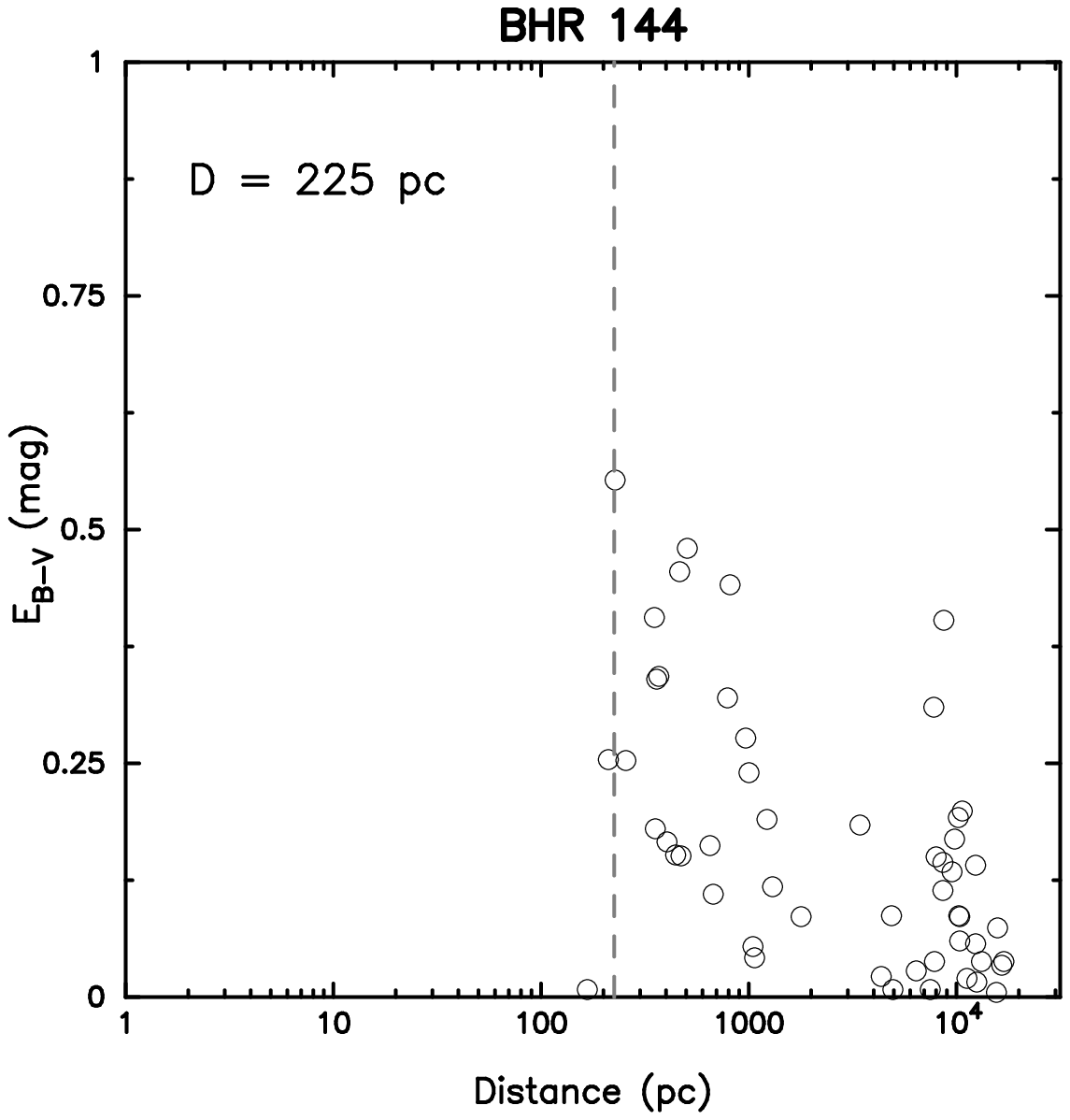}
\end{minipage}
\begin{minipage}[b]{0.2\textwidth}
 \centering
 \includegraphics[width=3cm]{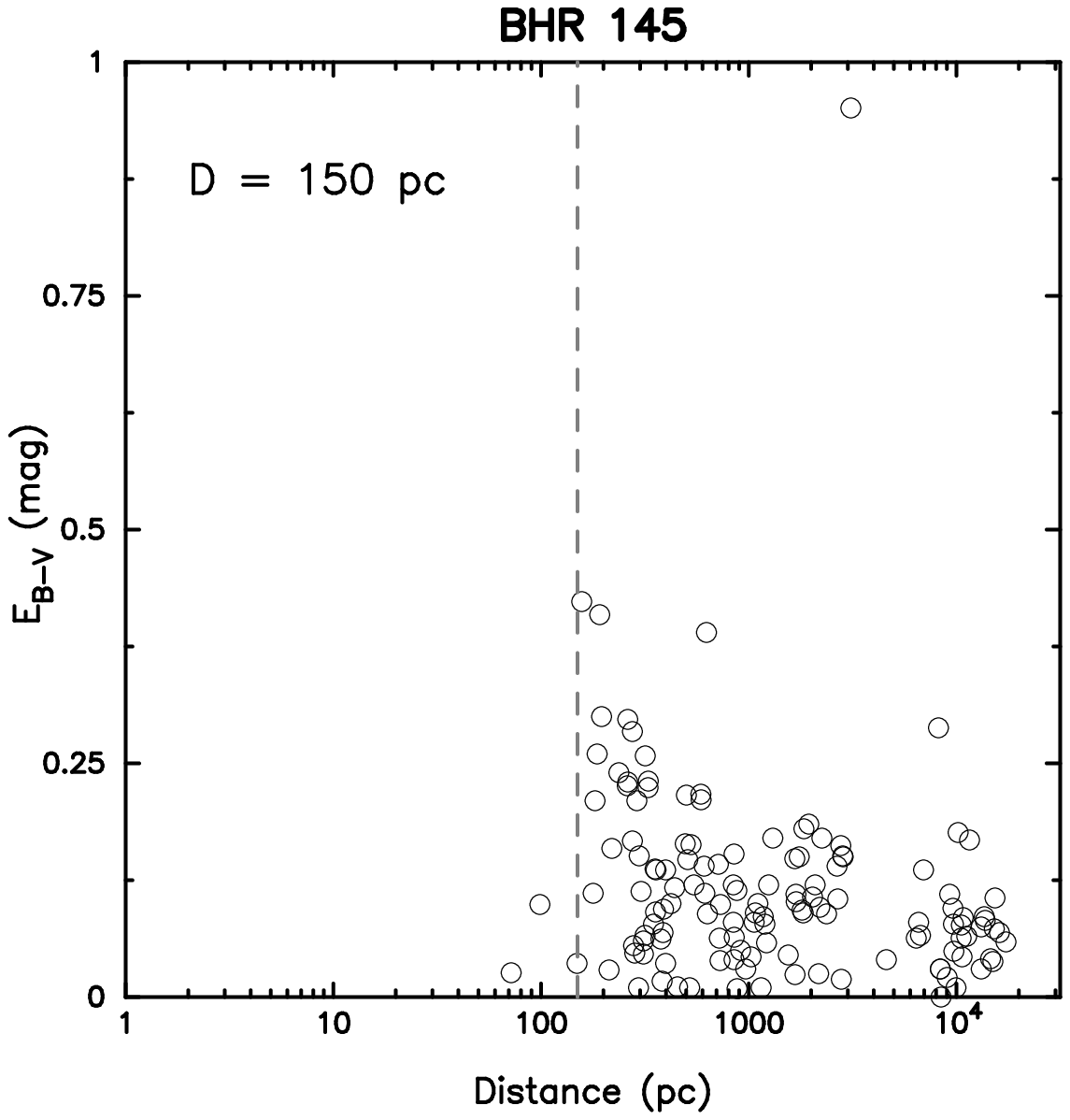}
\end{minipage}
\begin{minipage}[b]{0.2\textwidth}
 \centering
 \includegraphics[width=3cm]{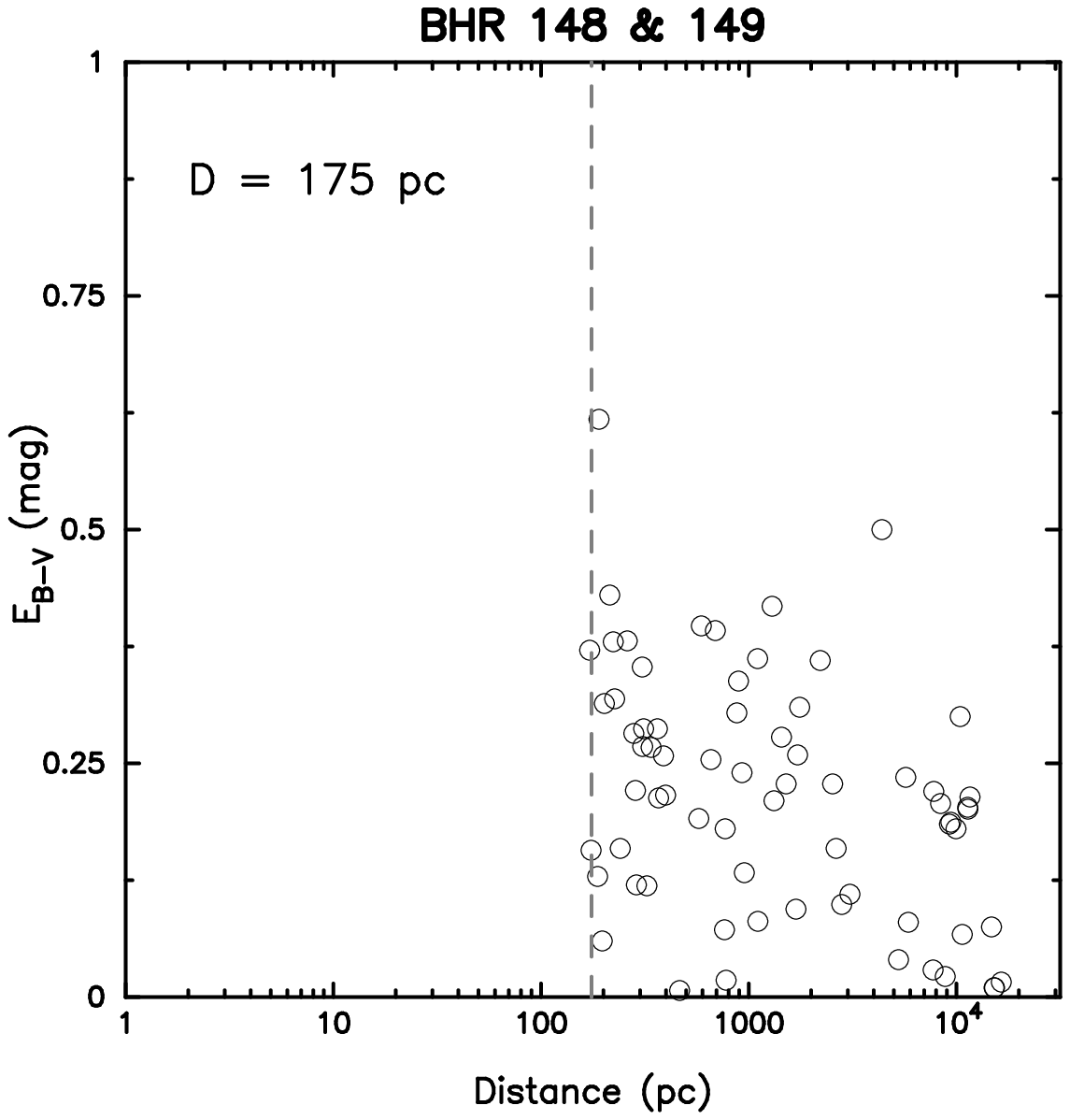}
\end{minipage}
\caption{Color excess versus distance for stars in the neighborhood of each globule.
         The dashed vertical lines indicate the distances adopted in this work. In the
         upper left corner are given the distances estimated to the globules.}
\label{fig:dist}
\end{figure}
%--------------------------------Fig 5: DISTANCES--------------------------------

%--------------------------------Fig 6: BARNARD 68 PROFILE--------------------------------
\clearpage
\begin{figure}
\centering
\includegraphics[scale=0.85]{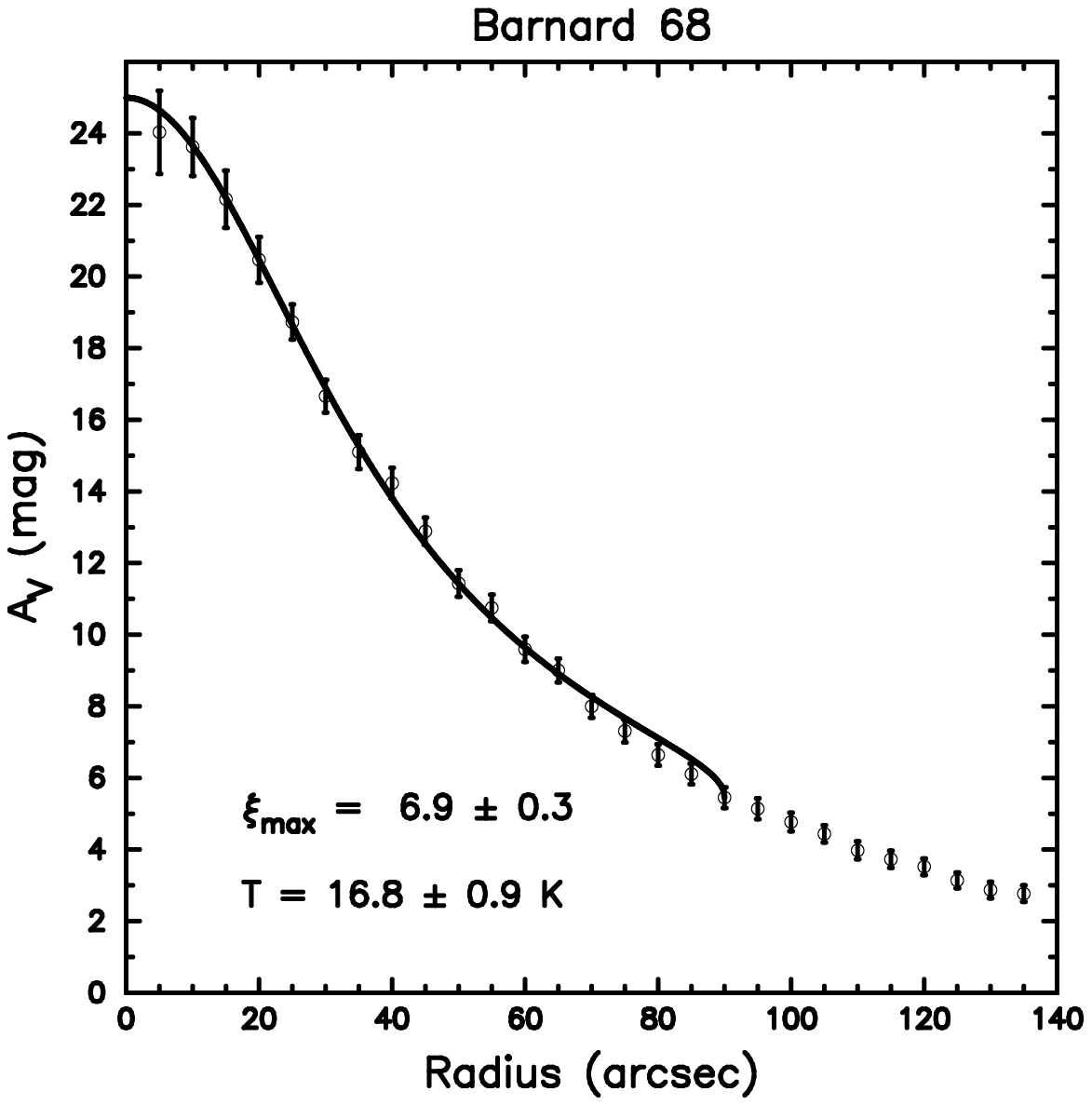}
\caption{Azimuthally-averaged radial extinction profile for Barnard 68. The solid curve
         represents the Bonnor-Ebert theoretical profile fitting.}
\label{fig:b68prof}
\end{figure}
%--------------------------------Fig 6: BARNARD 68 PROFILE--------------------------------

%--------------------------------Fig 7: EXTINCTION PROFILES--------------------------------
\clearpage
\begin{figure}
\centering
\begin{minipage}[b]{0.2\textwidth}
 \centering
 \includegraphics[width=3cm]{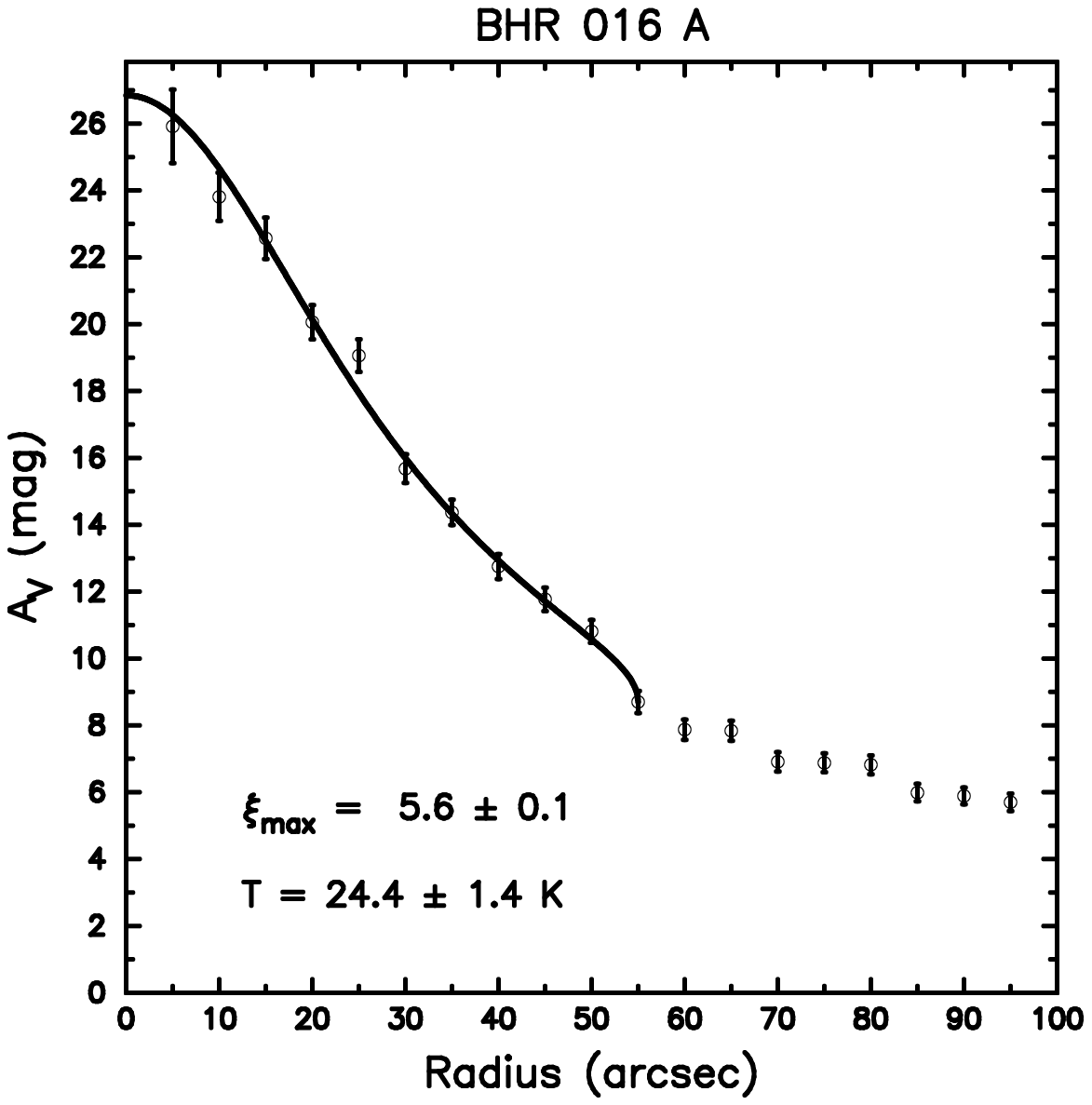}
\end{minipage}
\begin{minipage}[b]{0.2\textwidth}
 \centering
 \includegraphics[width=3cm]{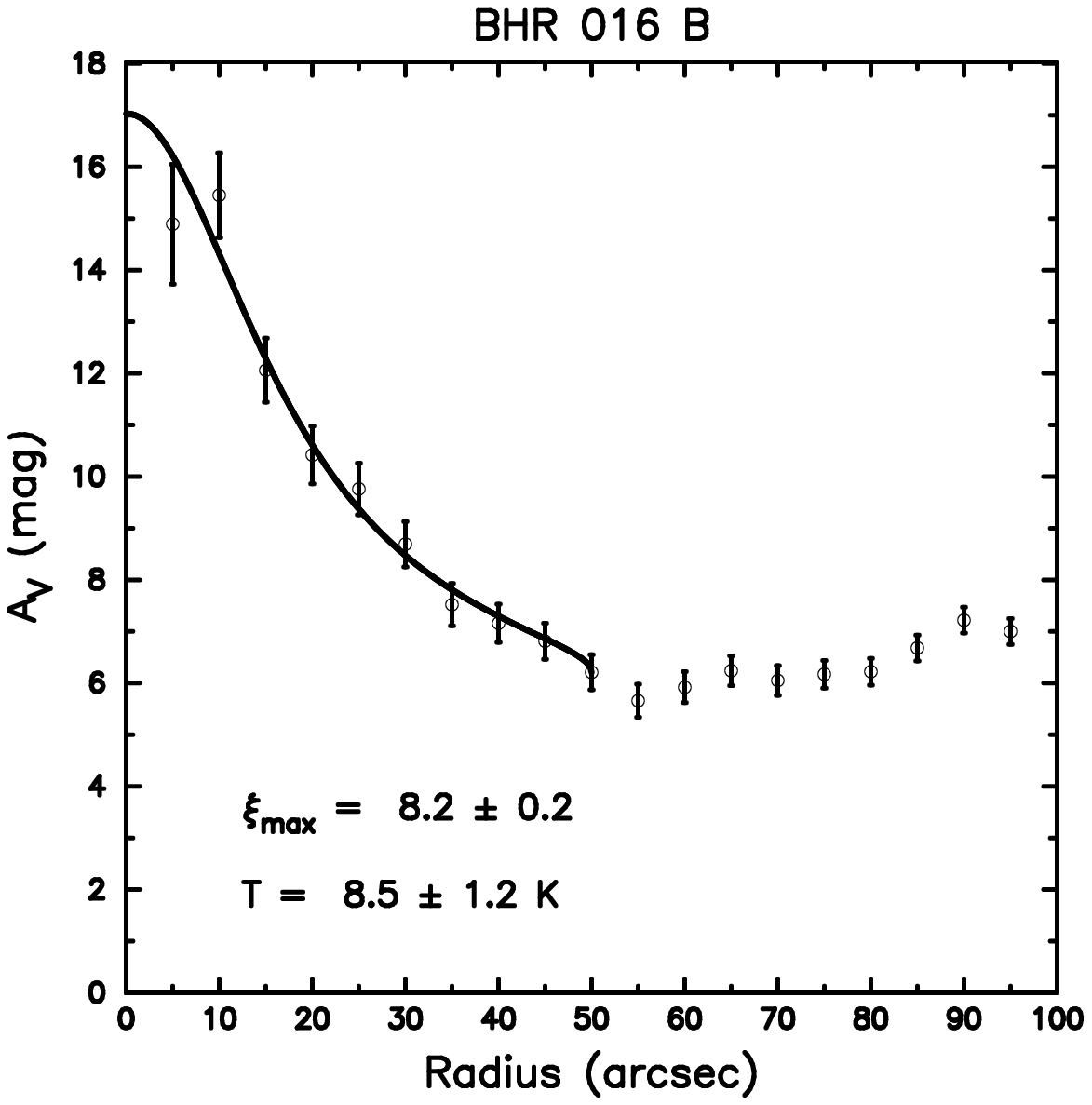}
\end{minipage}
\begin{minipage}[b]{0.2\textwidth}
 \centering
 \includegraphics[width=3cm]{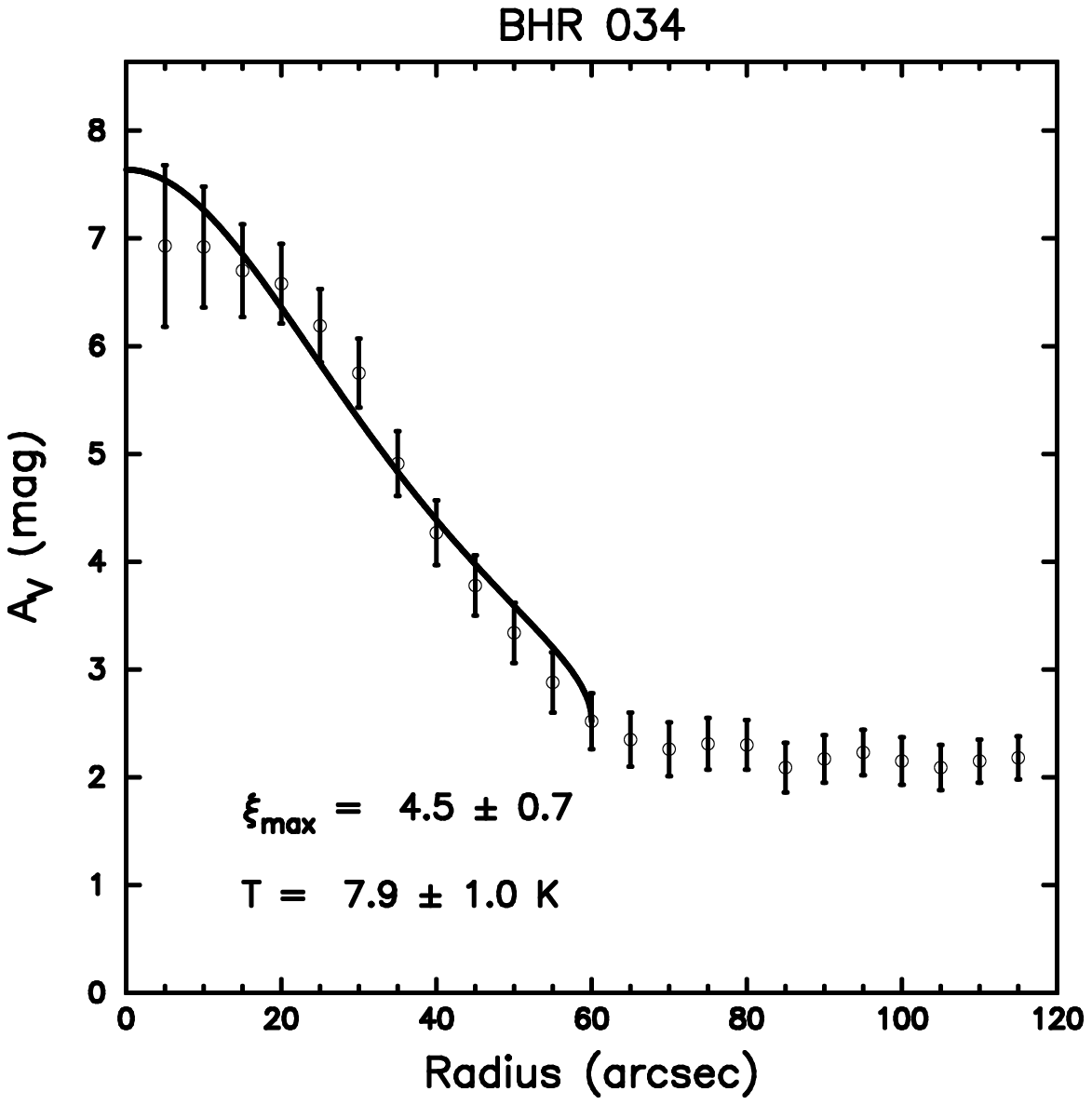}
\end{minipage}
\begin{minipage}[b]{0.2\textwidth}
 \centering
 \includegraphics[width=3cm]{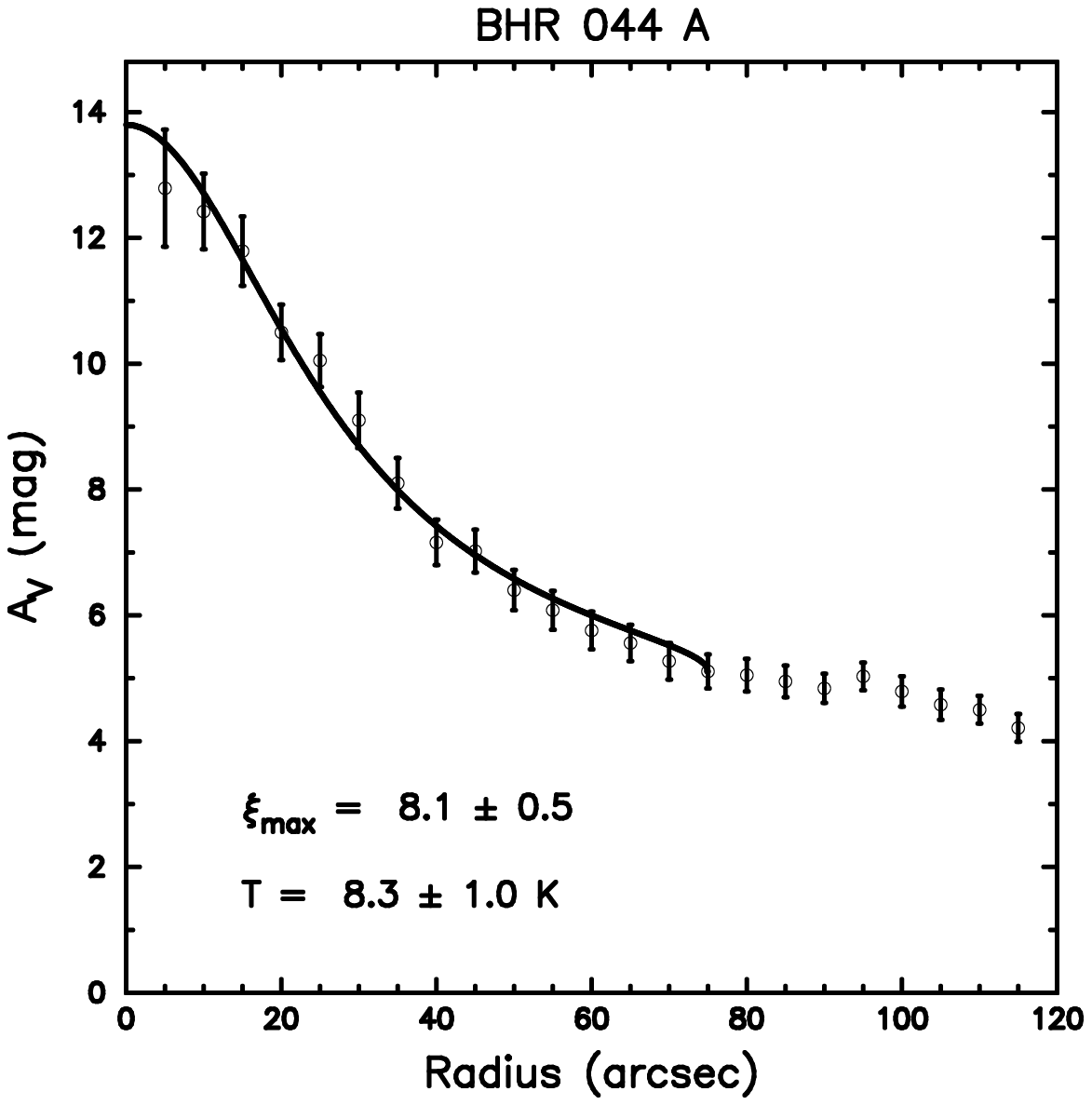}
\end{minipage}\\[0.5cm]
\begin{minipage}[b]{0.2\textwidth}
 \centering
 \includegraphics[width=3cm]{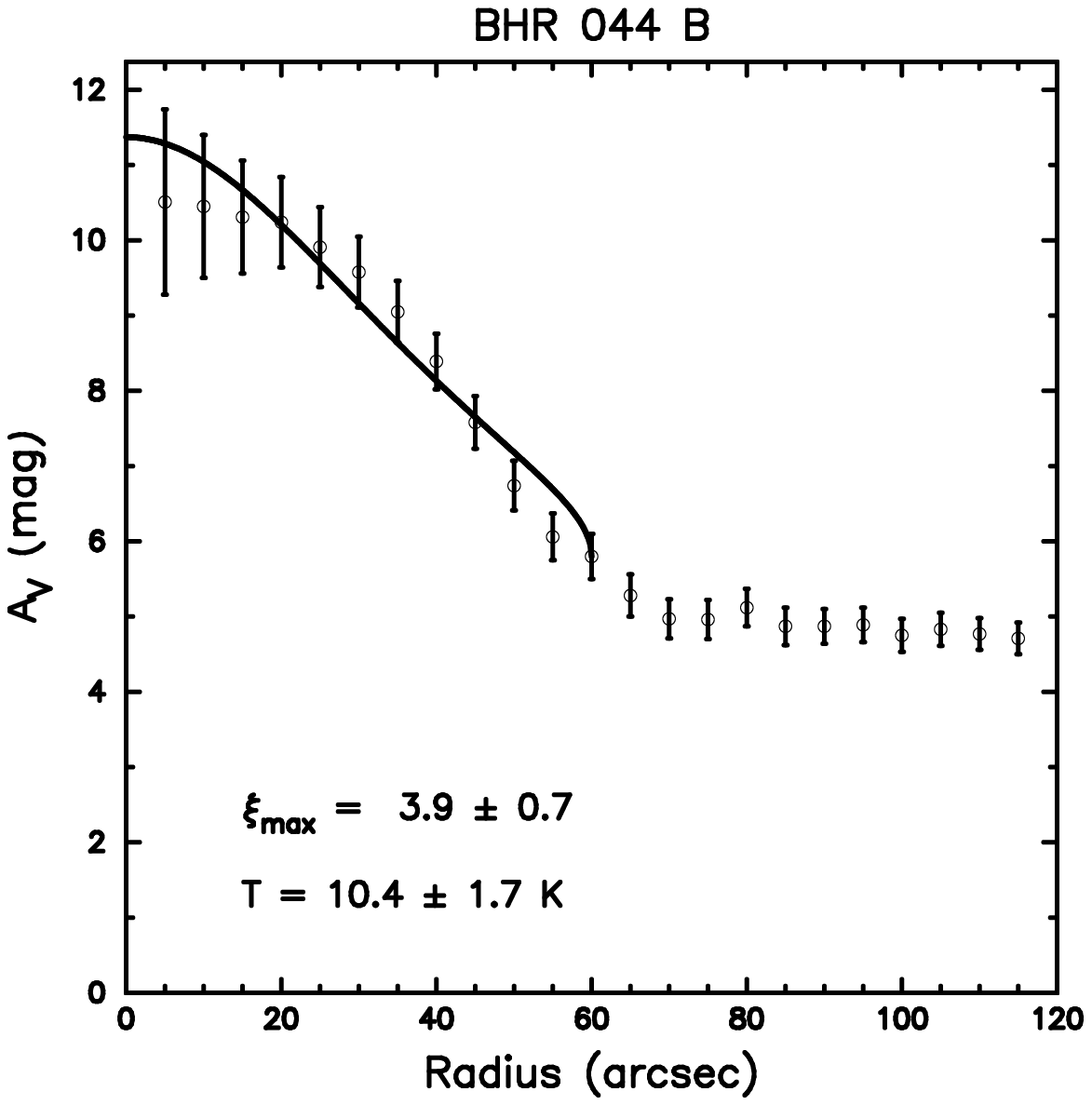}
\end{minipage}
\begin{minipage}[b]{0.2\textwidth}
 \centering
 \includegraphics[width=3cm]{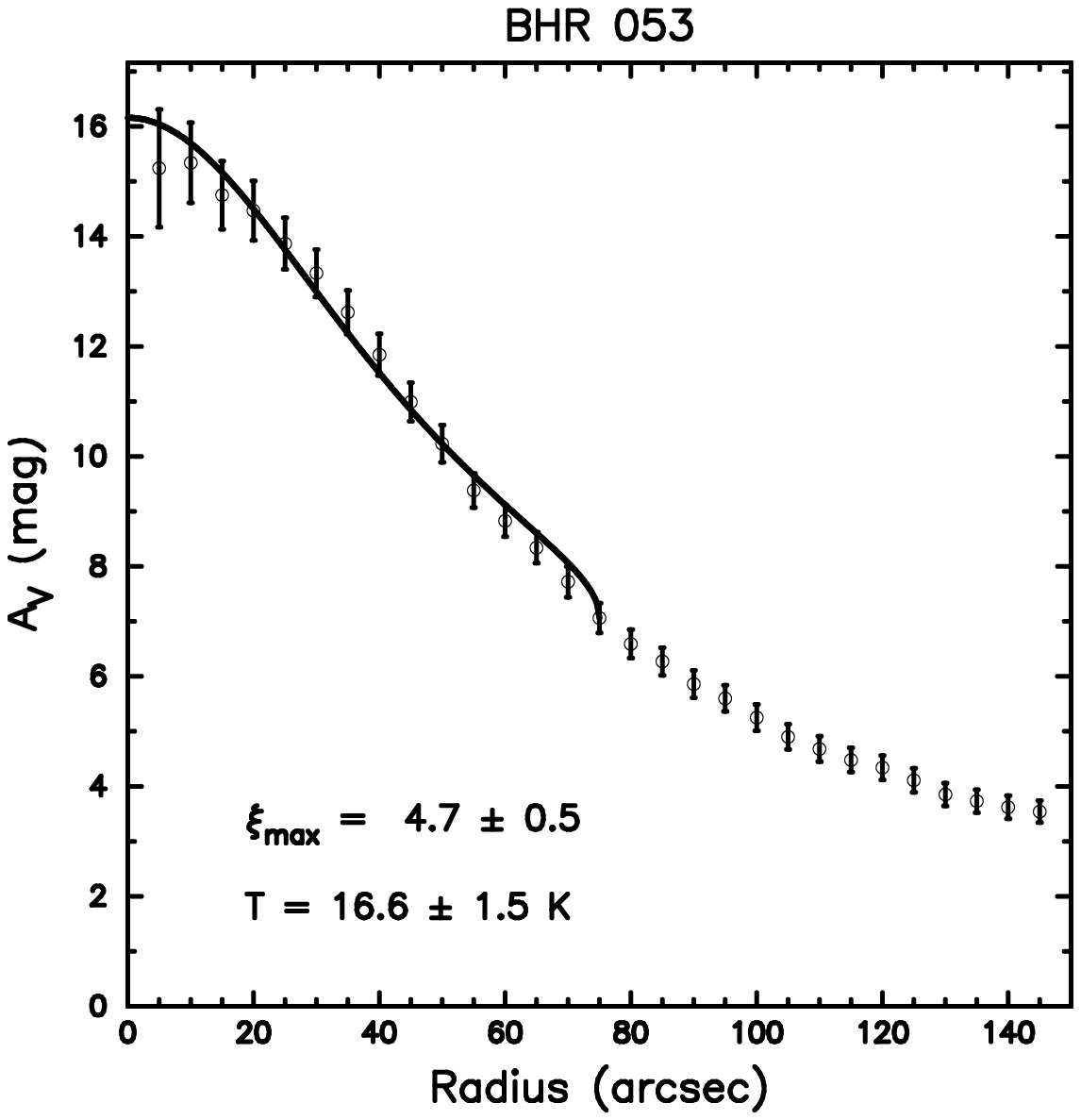}
\end{minipage}
\begin{minipage}[b]{0.2\textwidth}
 \centering
 \includegraphics[width=3cm]{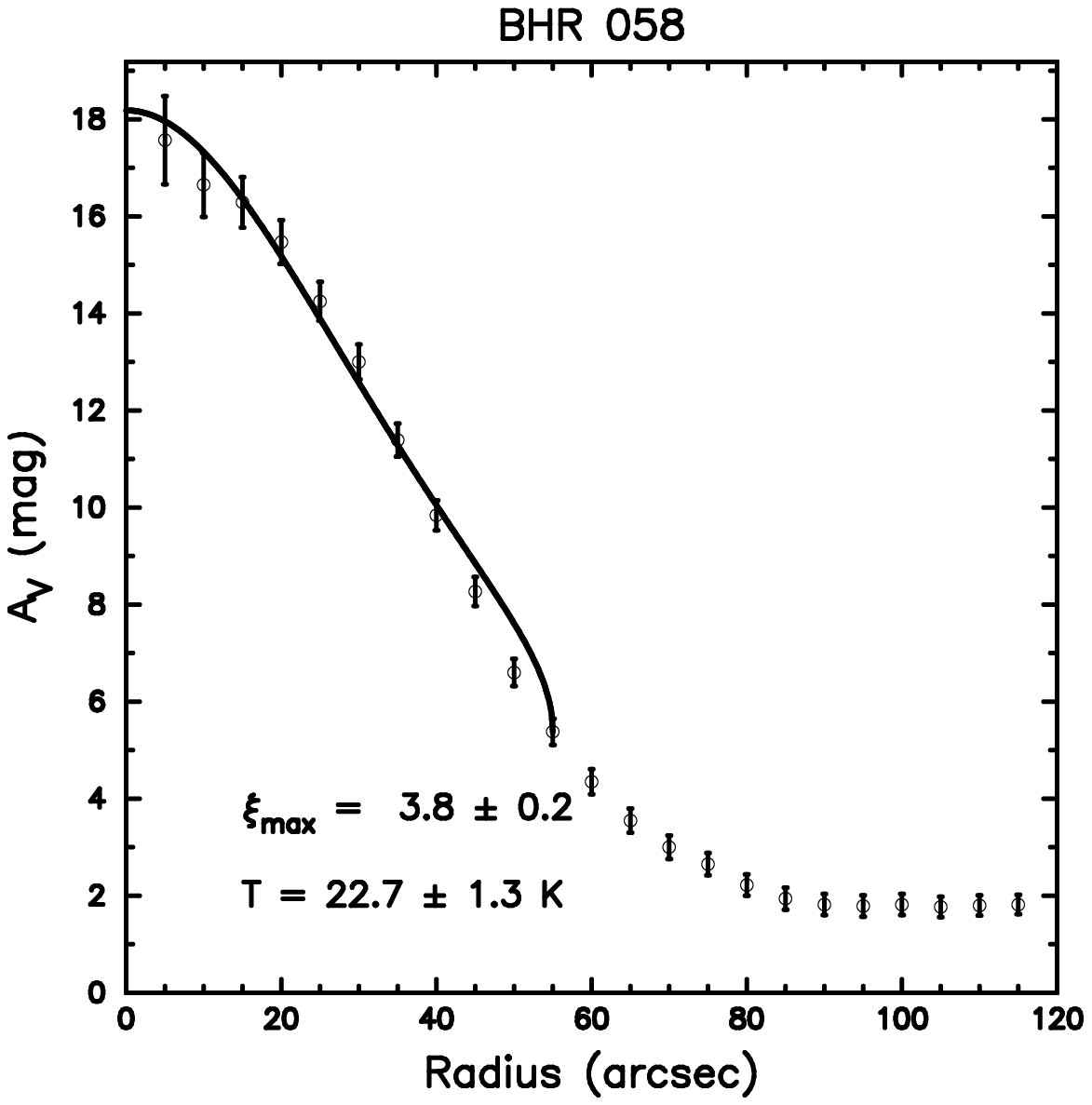}
\end{minipage}
\begin{minipage}[b]{0.2\textwidth}
 \centering
 \includegraphics[width=3cm]{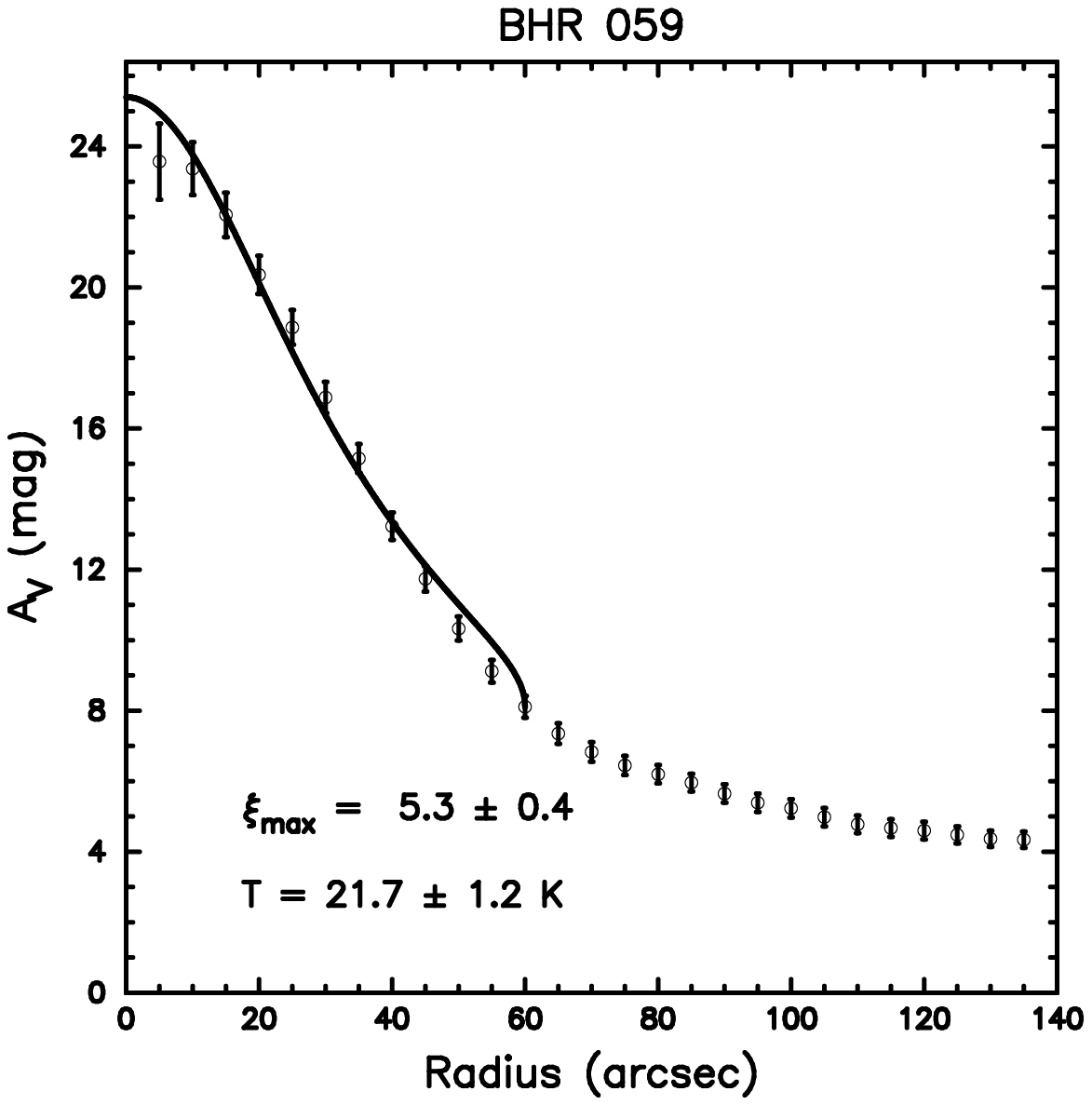}
\end{minipage}\\[0.5cm]
\begin{minipage}[b]{0.2\textwidth}
 \centering
 \includegraphics[width=3cm]{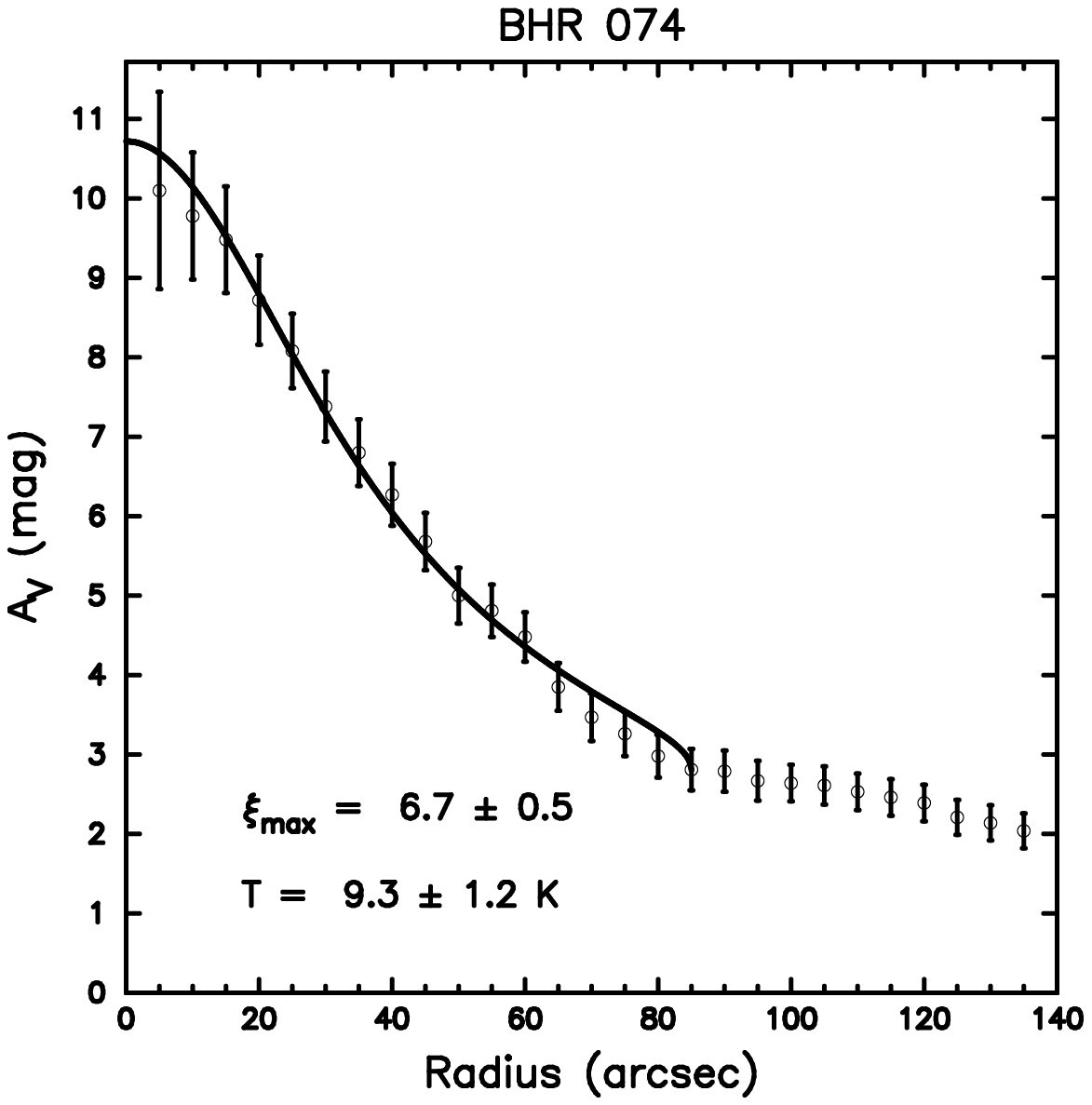}
\end{minipage}
\begin{minipage}[b]{0.2\textwidth}
 \centering
 \includegraphics[width=3cm]{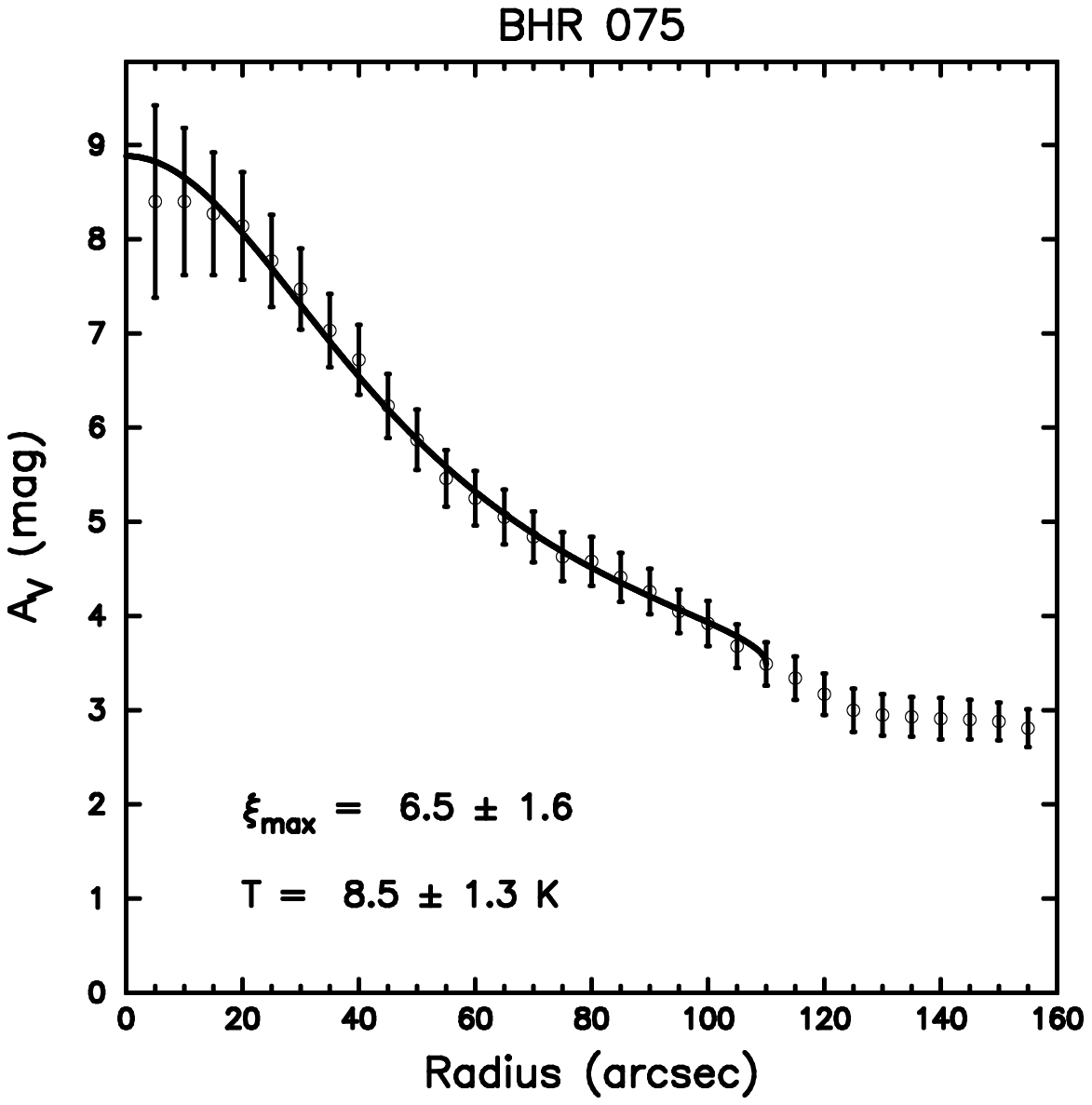}
\end{minipage}
\begin{minipage}[b]{0.2\textwidth}
 \centering
 \includegraphics[width=3cm]{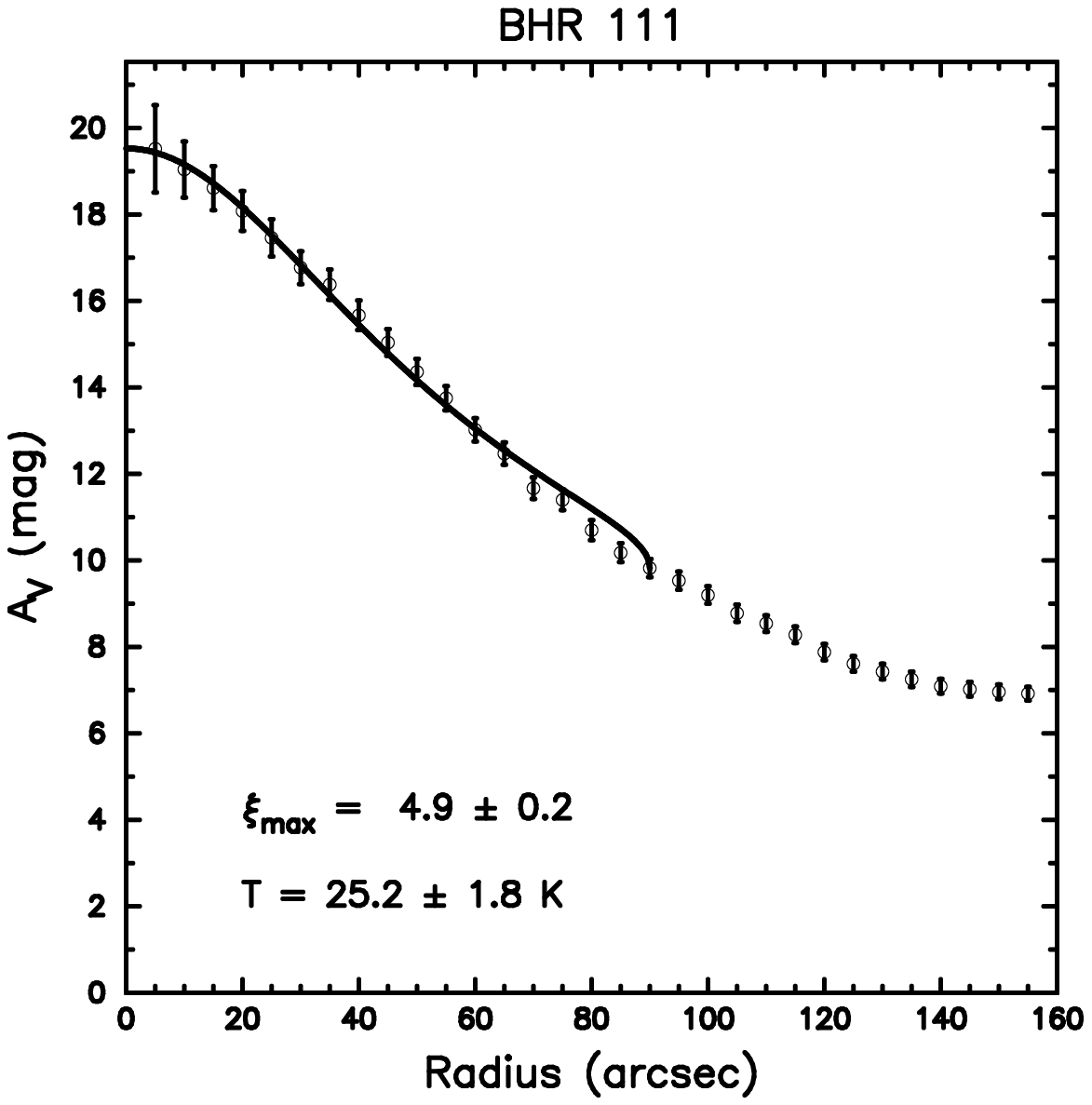}
\end{minipage}
\begin{minipage}[b]{0.2\textwidth}
 \centering
 \includegraphics[width=3cm]{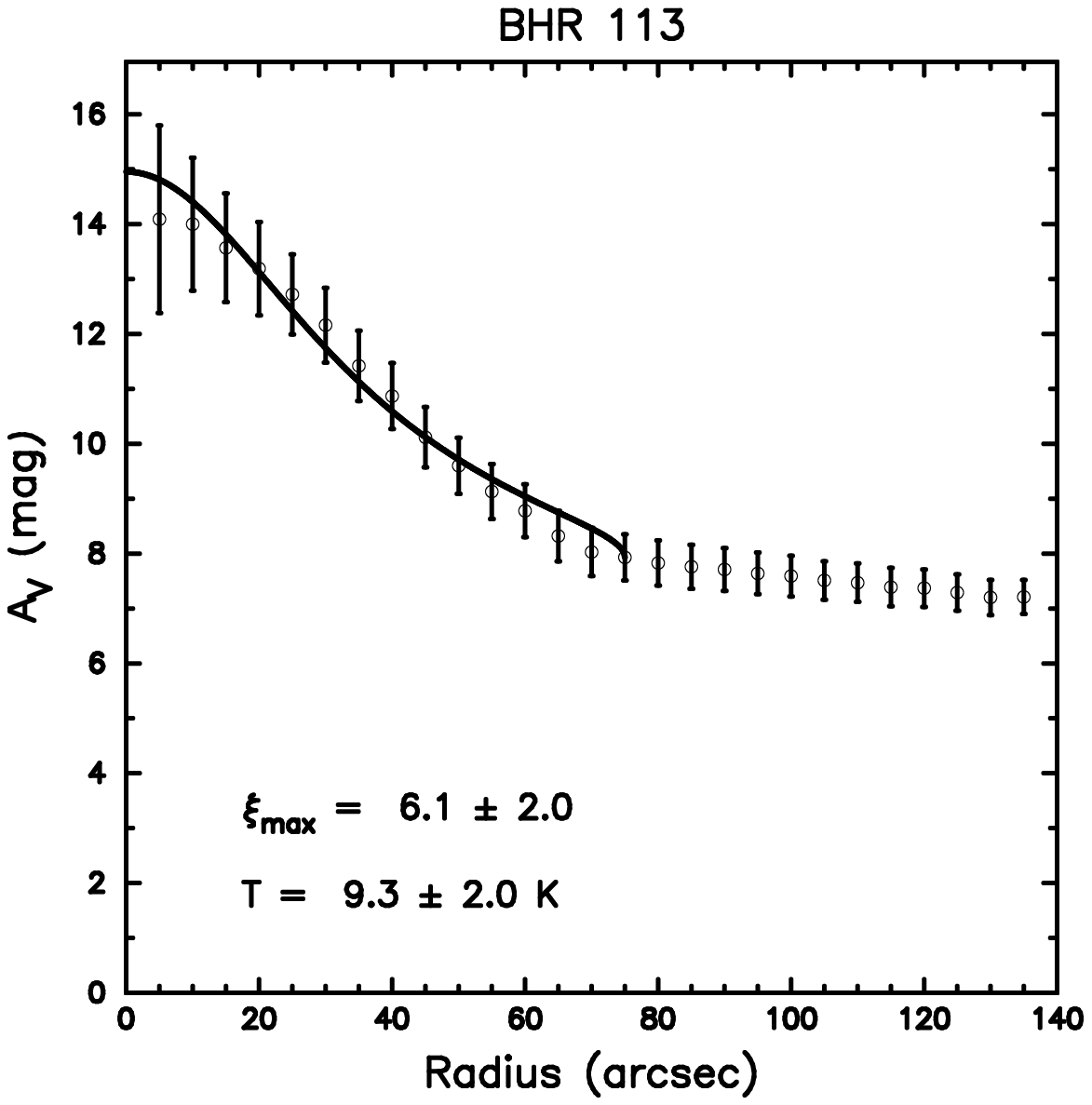}
\end{minipage}\\[0.5cm]
\begin{minipage}[b]{0.2\textwidth}
 \centering
 \includegraphics[width=3cm]{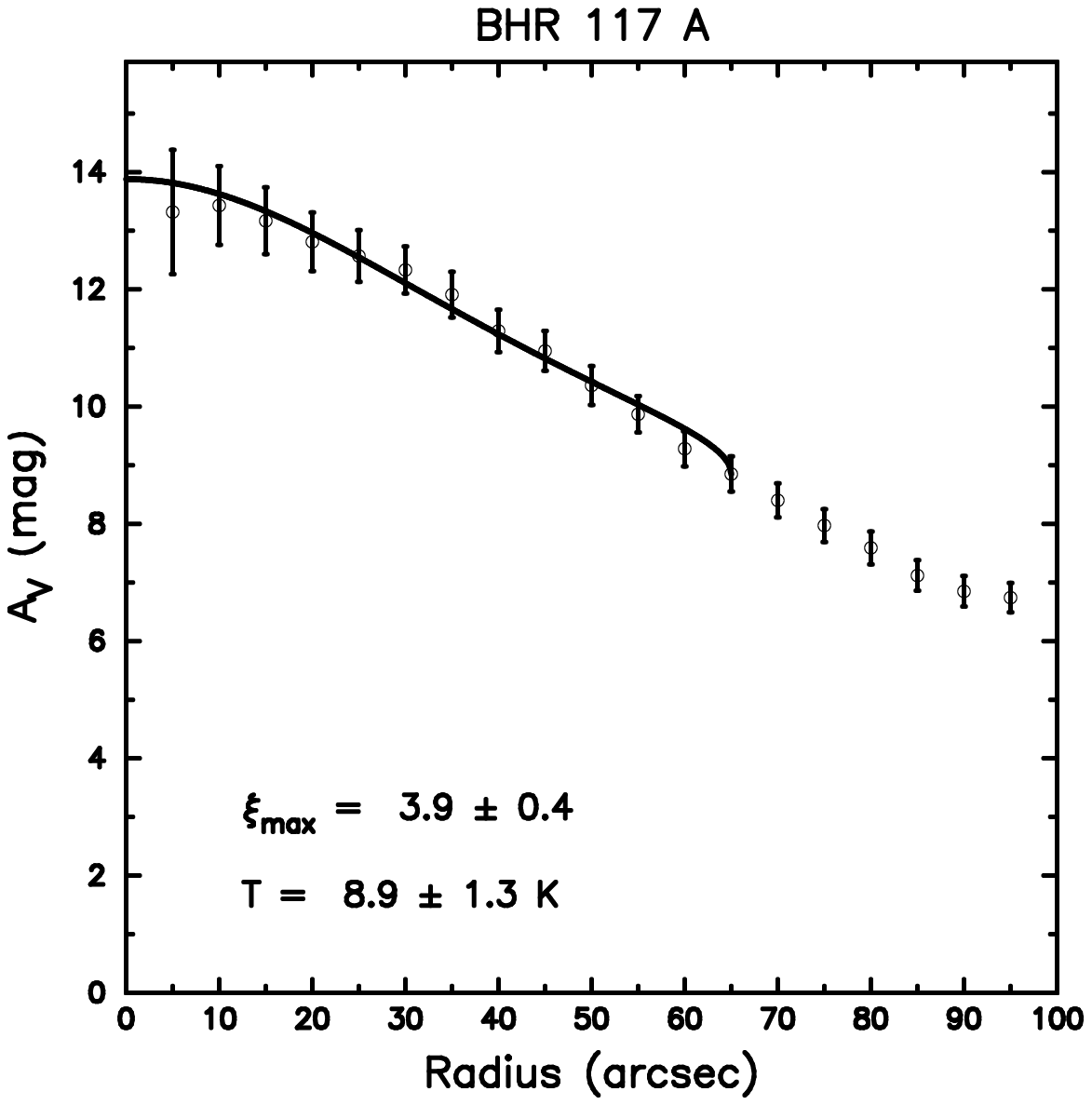}
\end{minipage}
\begin{minipage}[b]{0.2\textwidth}
 \centering
 \includegraphics[width=3cm]{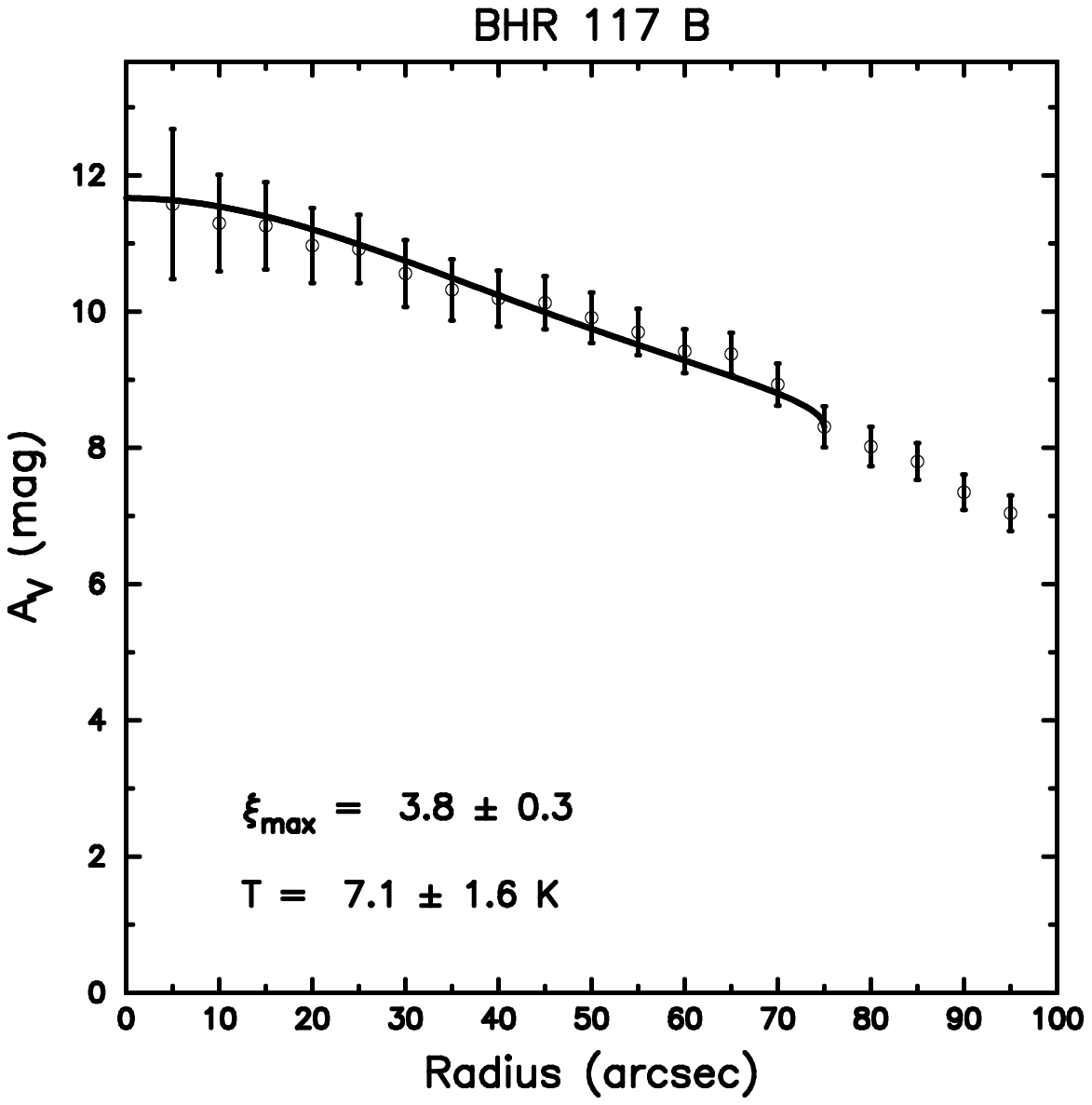}
\end{minipage}
\begin{minipage}[b]{0.2\textwidth}
 \centering
 \includegraphics[width=3cm]{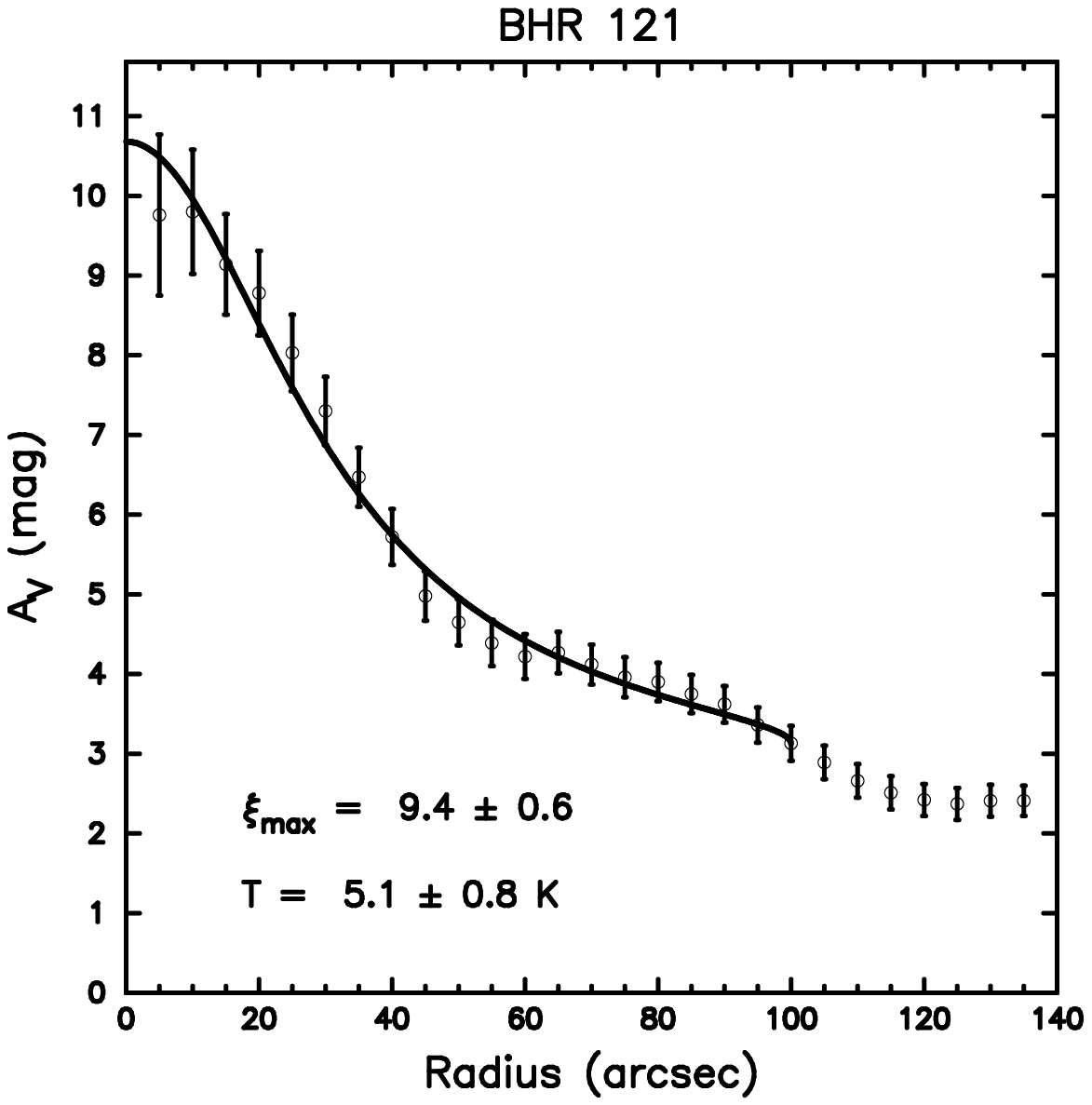}
\end{minipage}
\begin{minipage}[b]{0.2\textwidth}
 \centering
 \includegraphics[width=3cm]{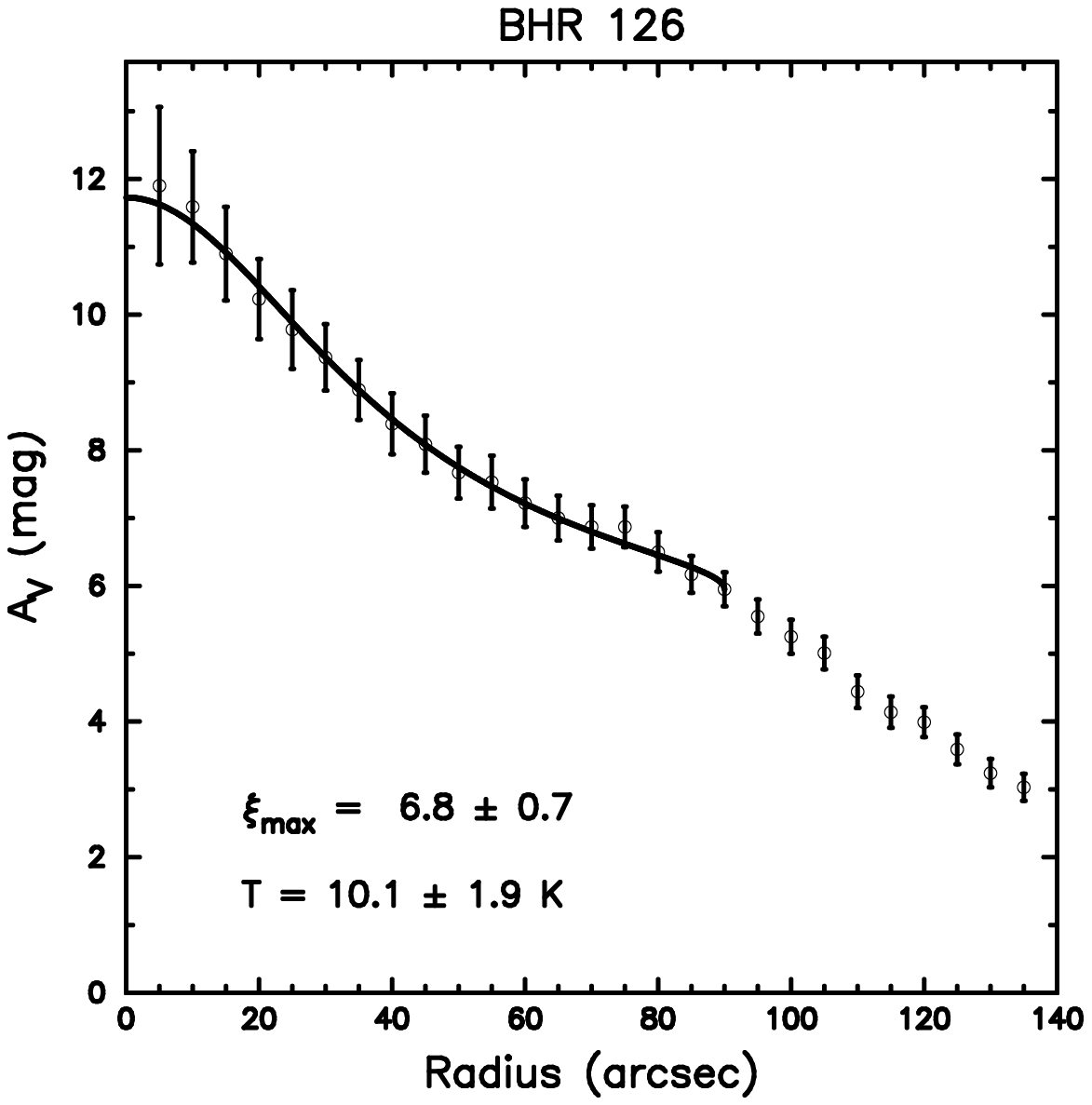}
\end{minipage}
\caption{Azimuthally-averaged radial extinction profiles for dense cores in Bok globules.
         The solid curve superimposed on each observed profile represents the theoretical
         Bonnor-Ebert profile fitting.}
\label{fig:perfs}
\end{figure}

\clearpage
\addtocounter{figure}{-1}
\begin{figure}
\centering
\begin{minipage}[b]{0.2\textwidth}
 \centering
 \includegraphics[width=3cm]{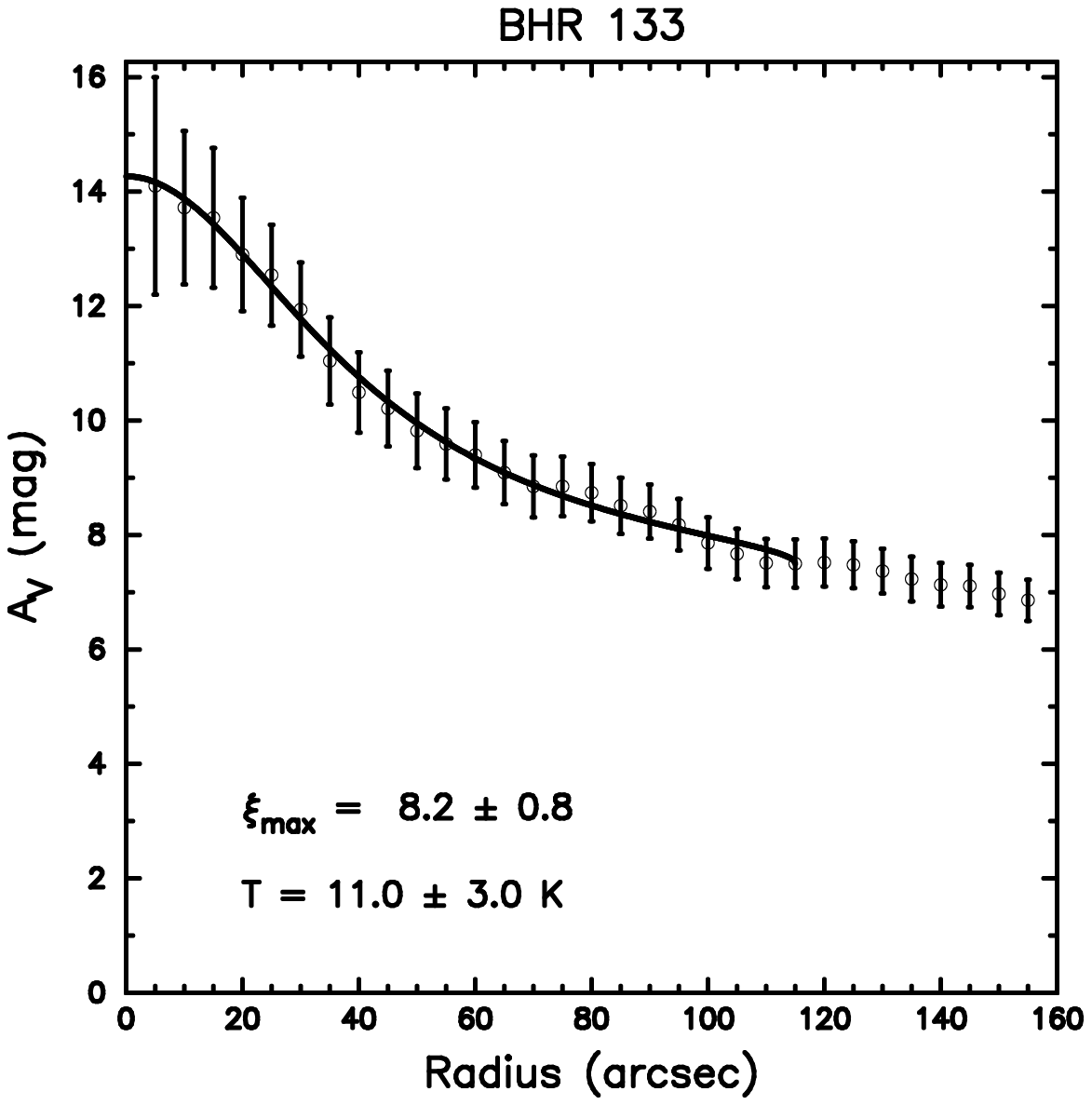}
\end{minipage}
\begin{minipage}[b]{0.2\textwidth}
 \centering
 \includegraphics[width=3cm]{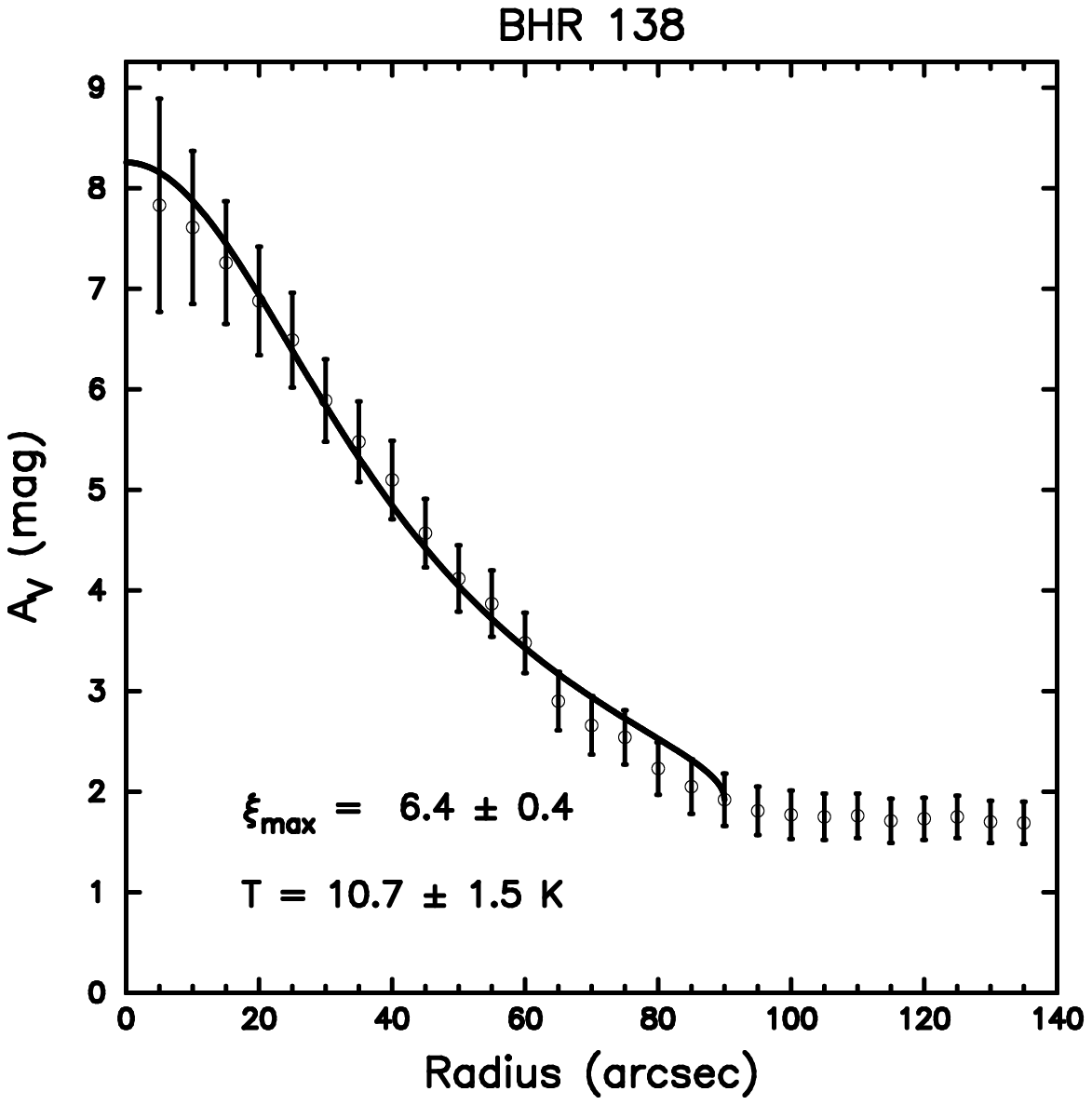}
\end{minipage}
\begin{minipage}[b]{0.2\textwidth}
 \centering
 \includegraphics[width=3cm]{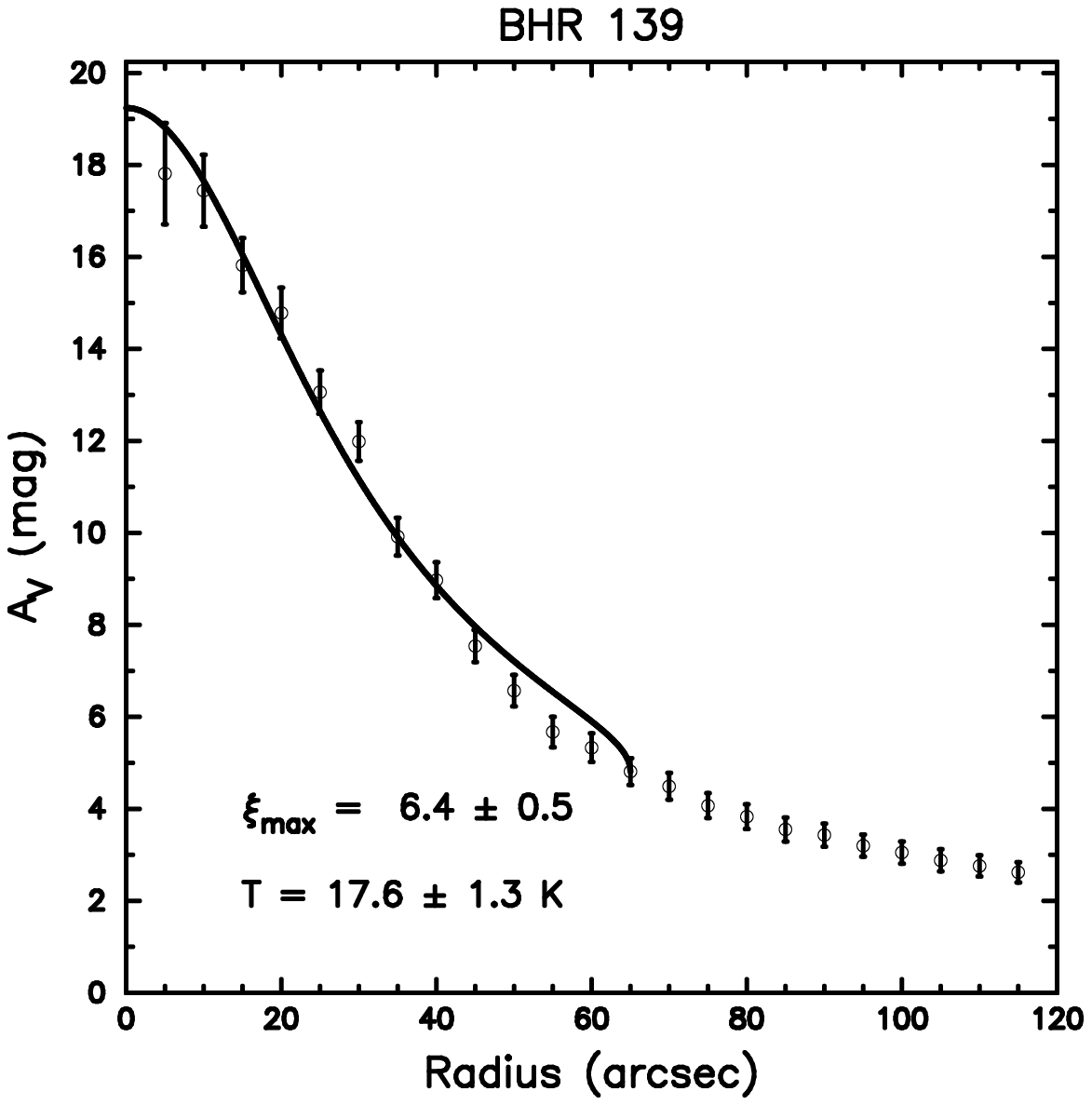}
\end{minipage}
\begin{minipage}[b]{0.2\textwidth}
 \centering
 \includegraphics[width=3cm]{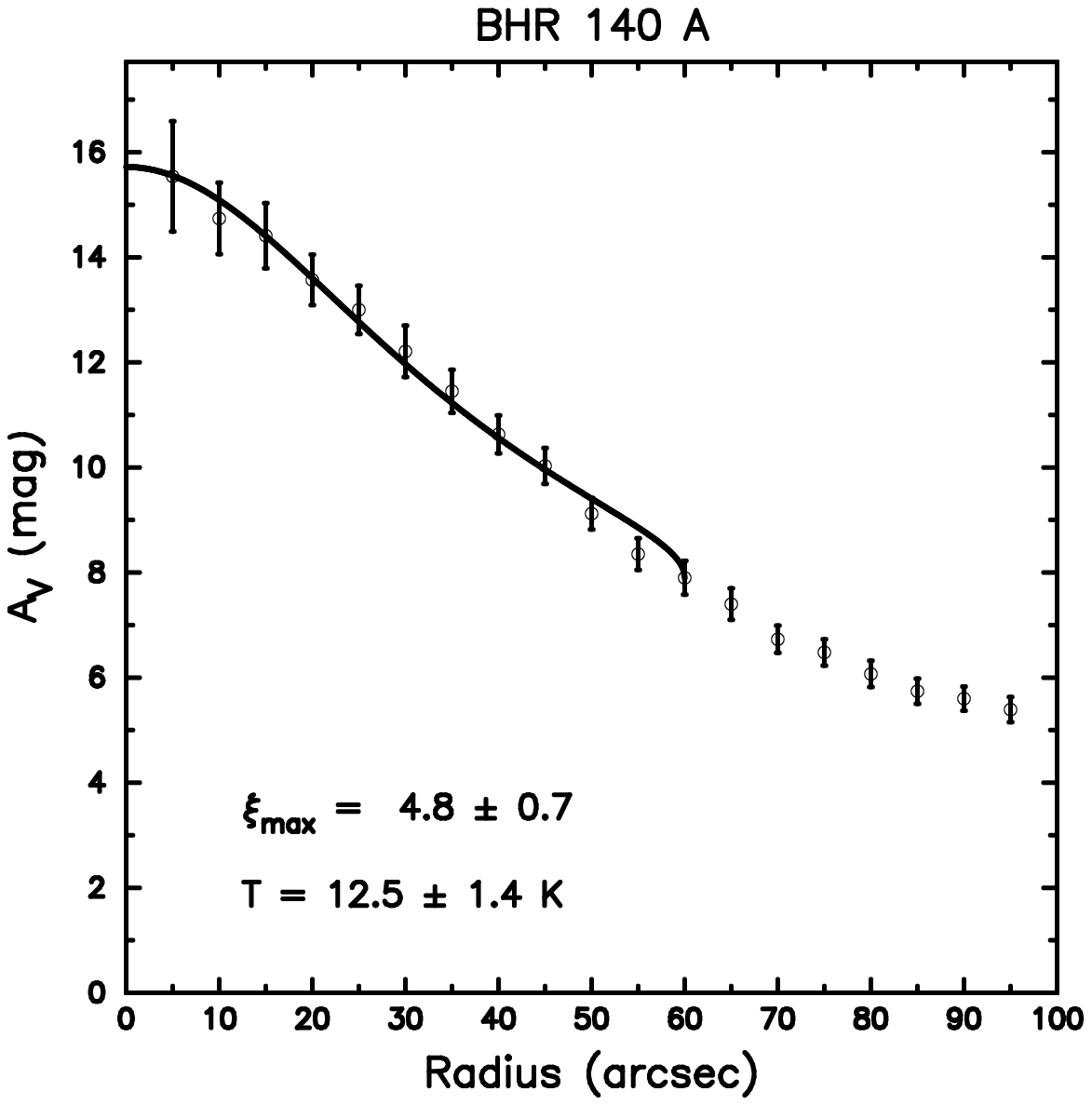}
\end{minipage}\\[0.5cm]
\begin{minipage}[b]{0.2\textwidth}
 \centering
 \includegraphics[width=3cm]{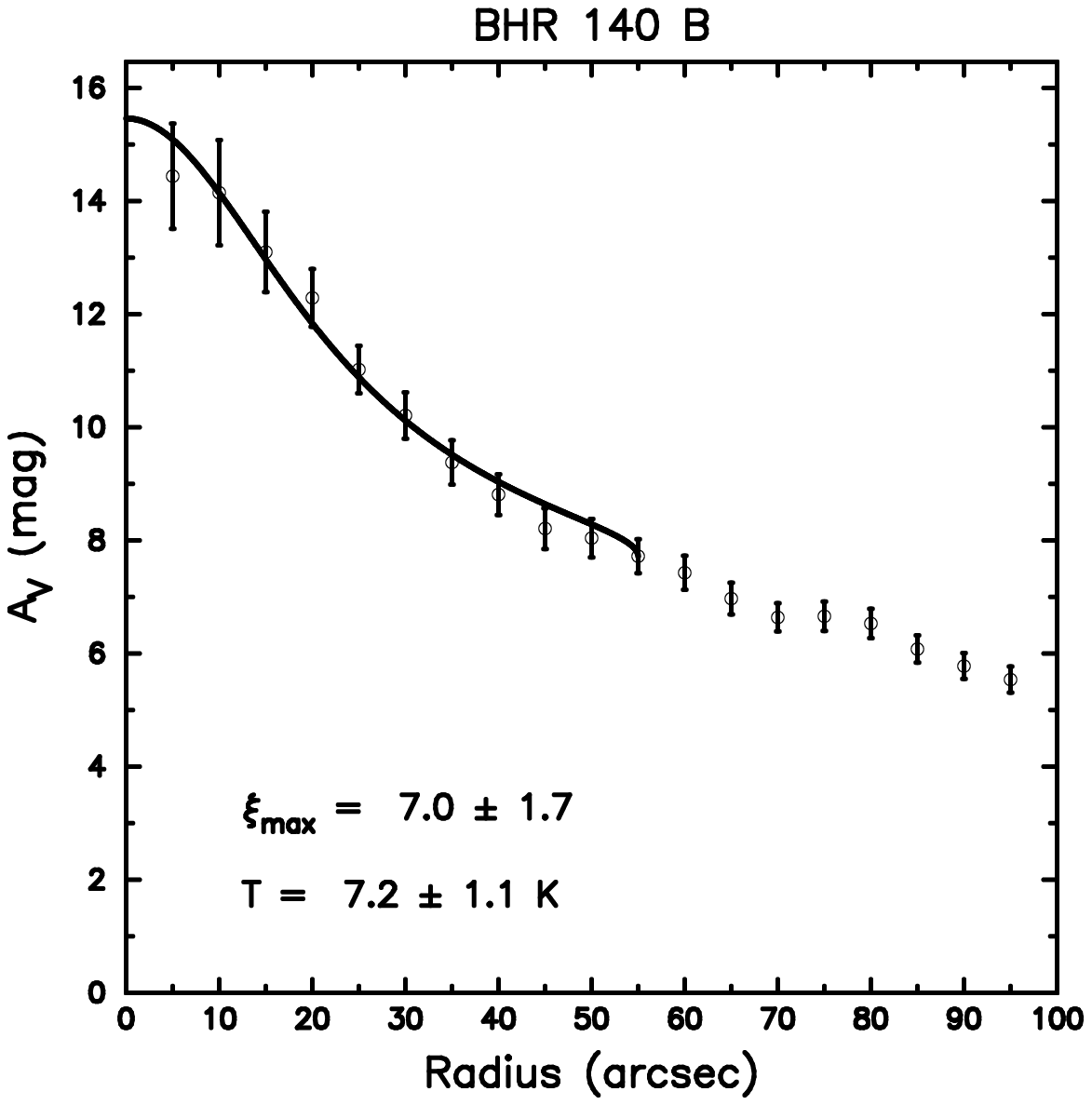}
\end{minipage}
\begin{minipage}[b]{0.2\textwidth}
 \centering
 \includegraphics[width=3cm]{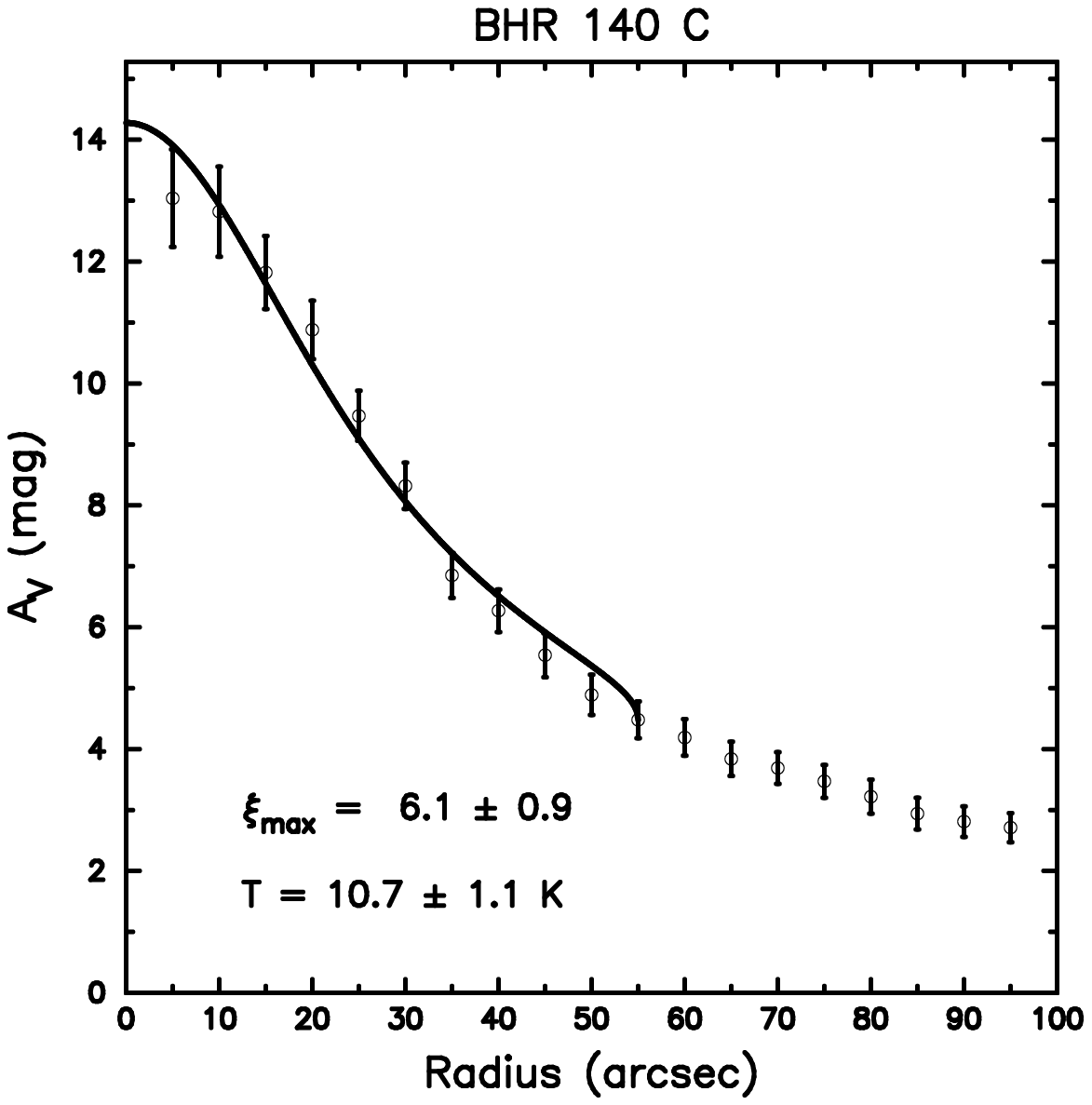}
\end{minipage}
\begin{minipage}[b]{0.2\textwidth}
 \centering
 \includegraphics[width=3cm]{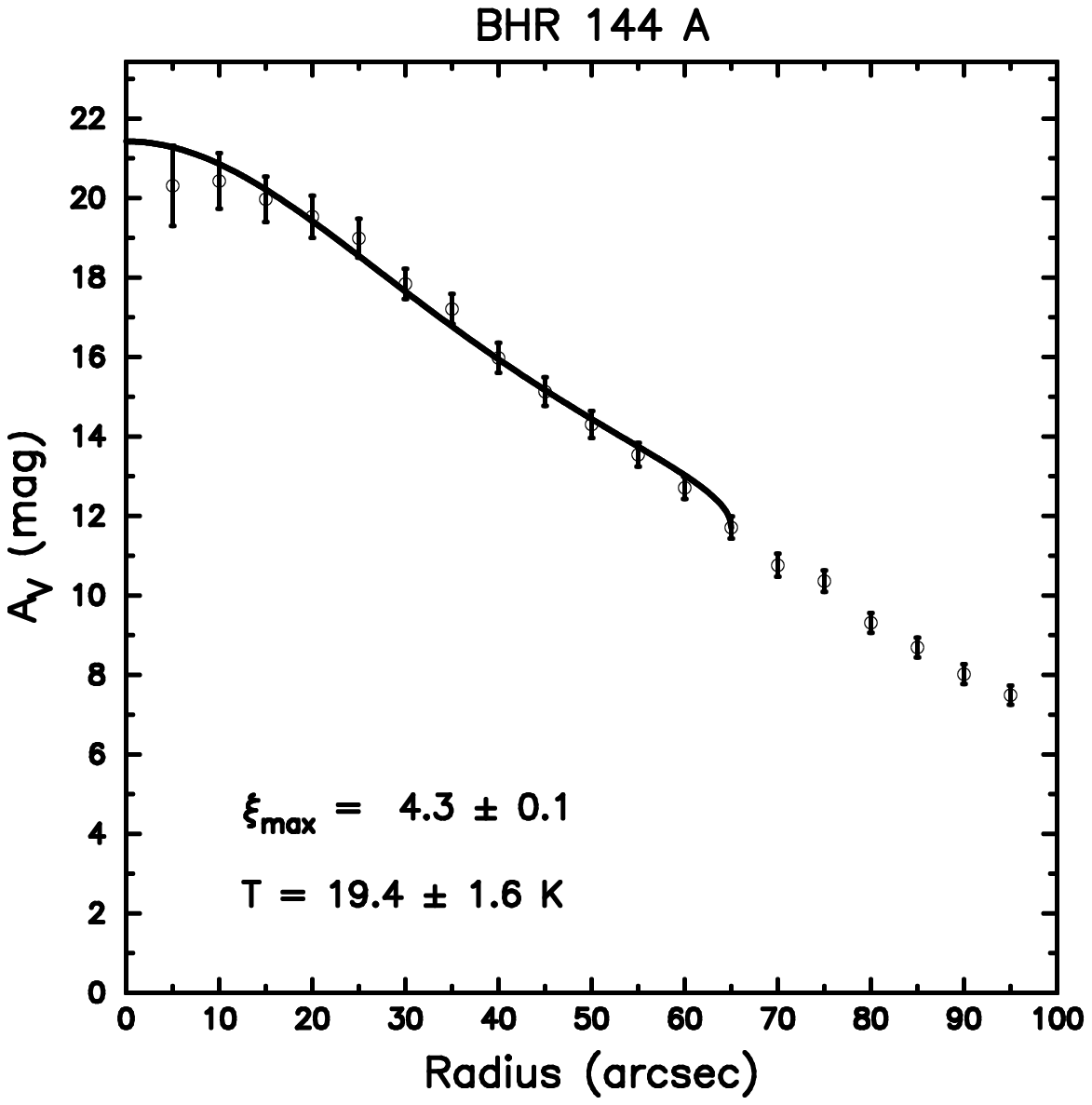}
\end{minipage}
\begin{minipage}[b]{0.2\textwidth}
 \centering
 \includegraphics[width=3cm]{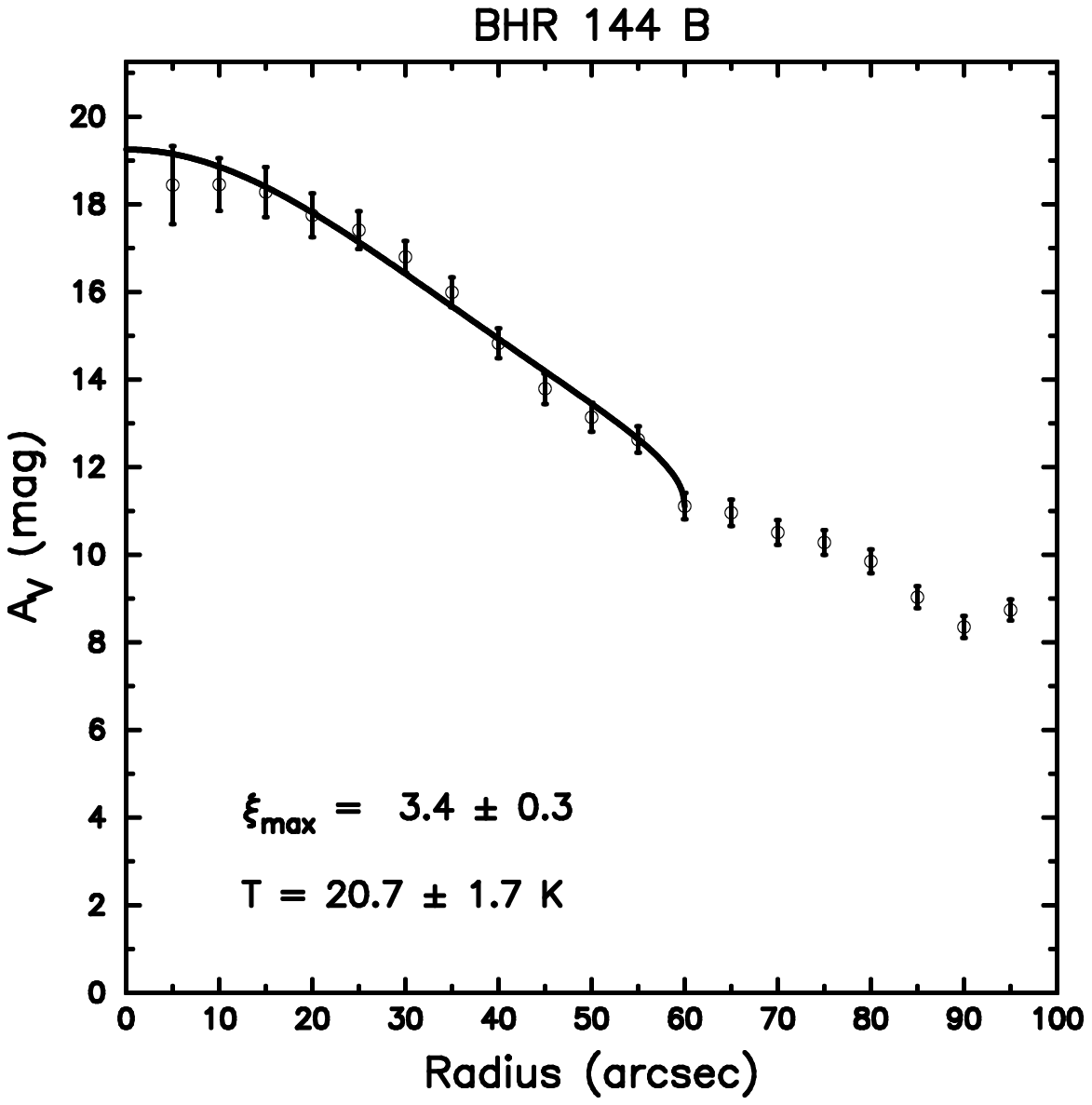}
\end{minipage}\\[0.5cm]
\begin{minipage}[b]{0.2\textwidth}
 \centering
 \includegraphics[width=3cm]{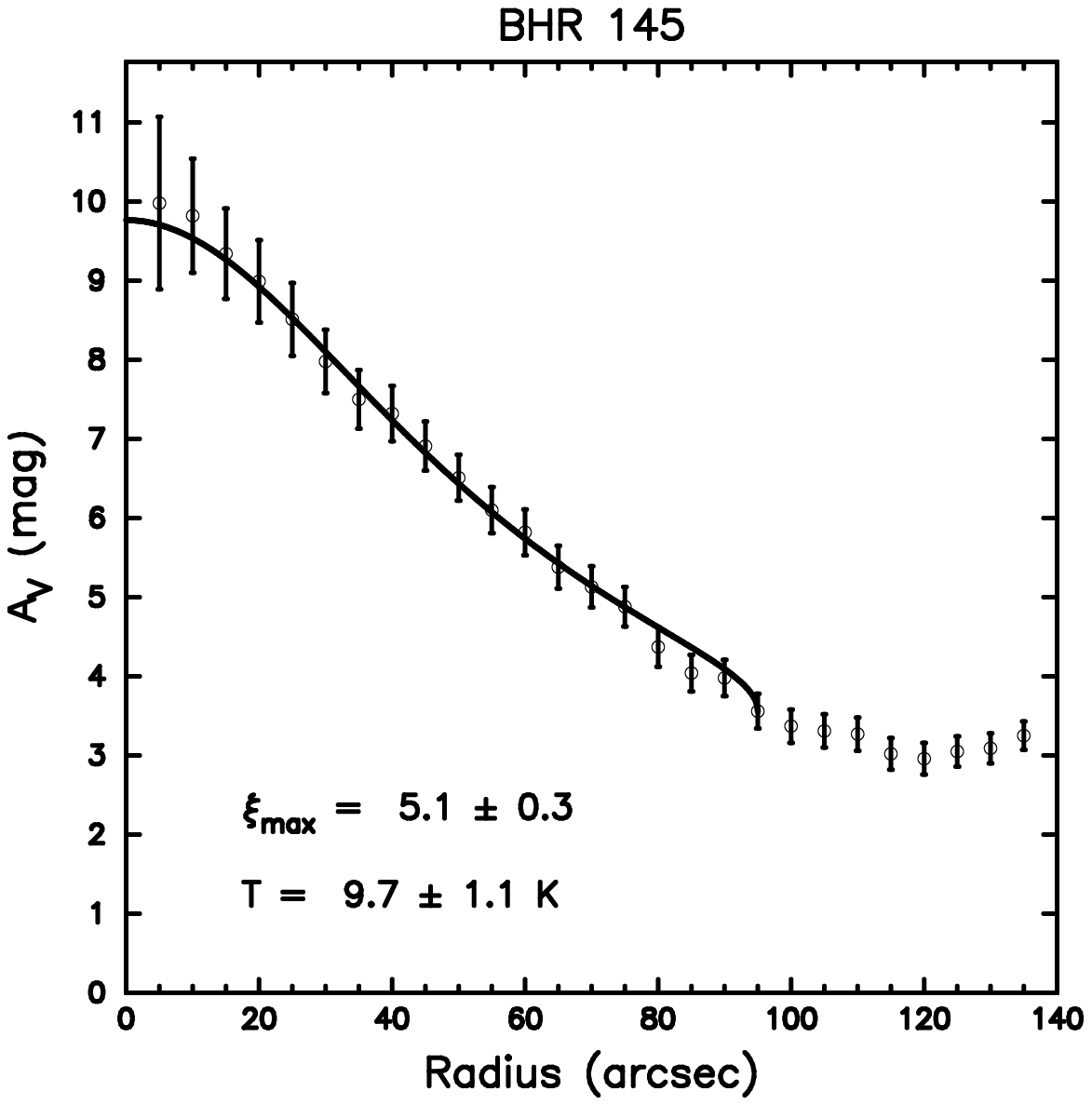}
\end{minipage}
\begin{minipage}[b]{0.2\textwidth}
 \centering
 \includegraphics[width=3cm]{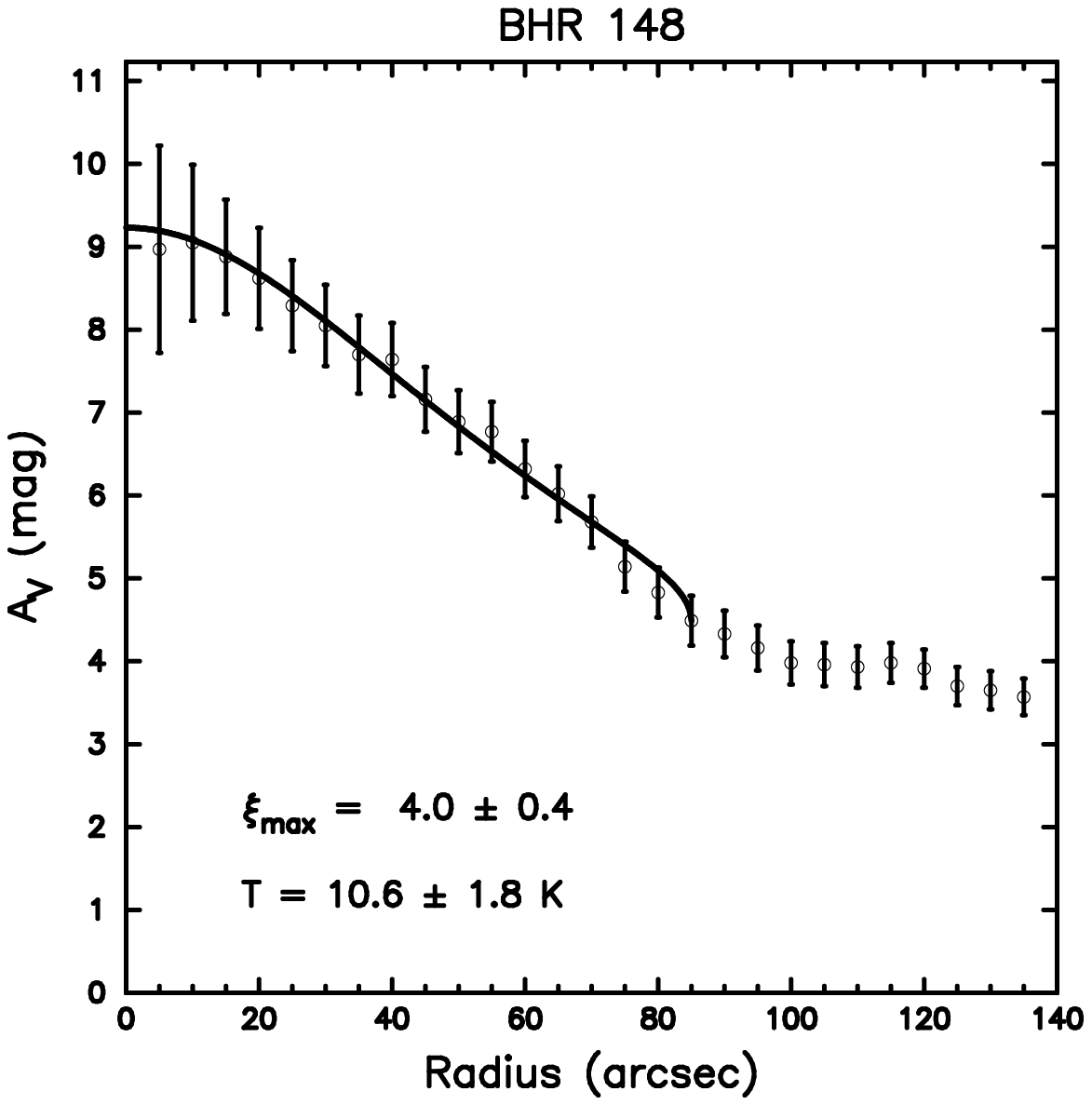}
\end{minipage}
\begin{minipage}[b]{0.2\textwidth}
 \centering
 \includegraphics[width=3cm]{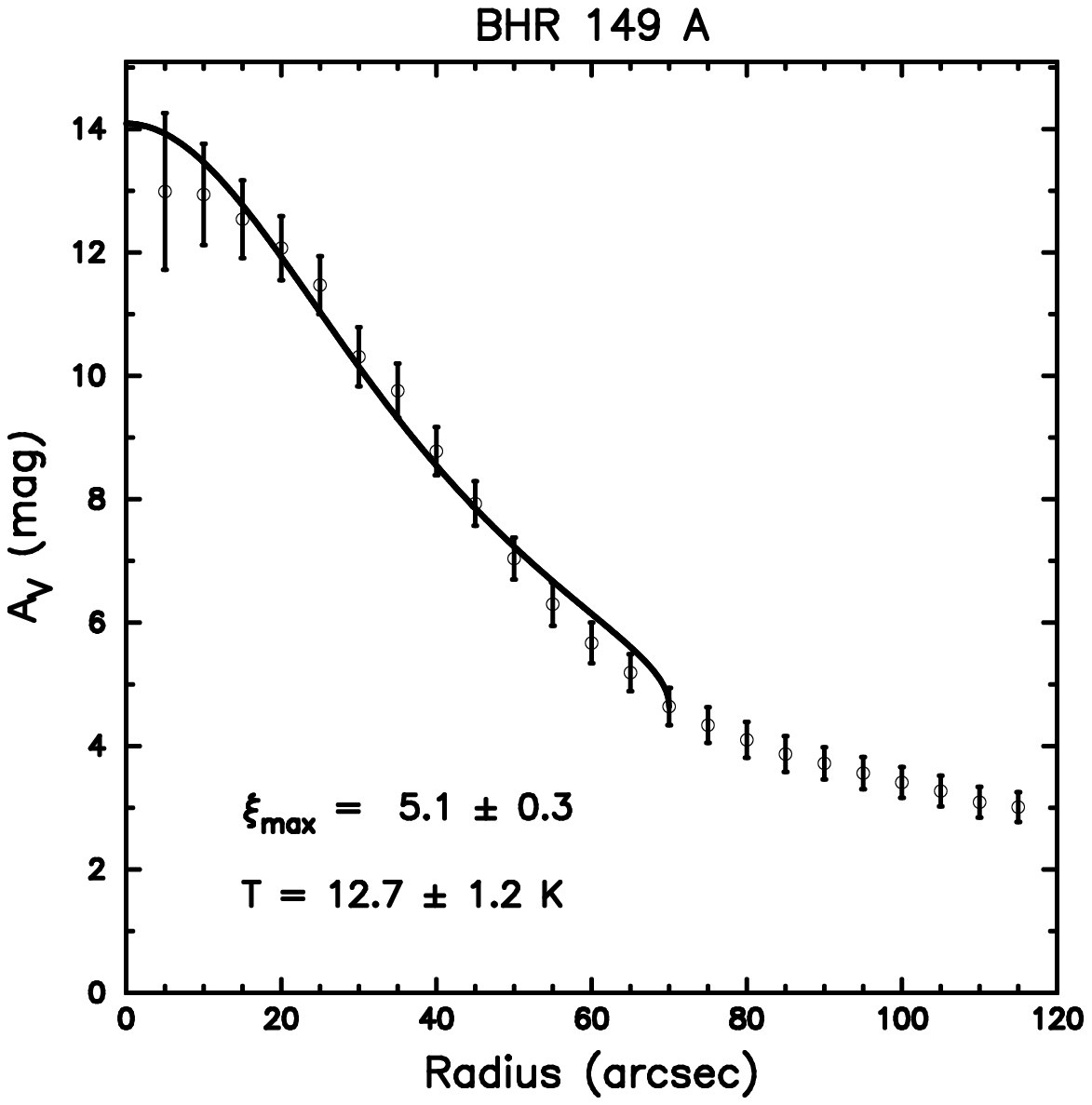}
\end{minipage}
\begin{minipage}[b]{0.2\textwidth}
 \centering
 \includegraphics[width=3cm]{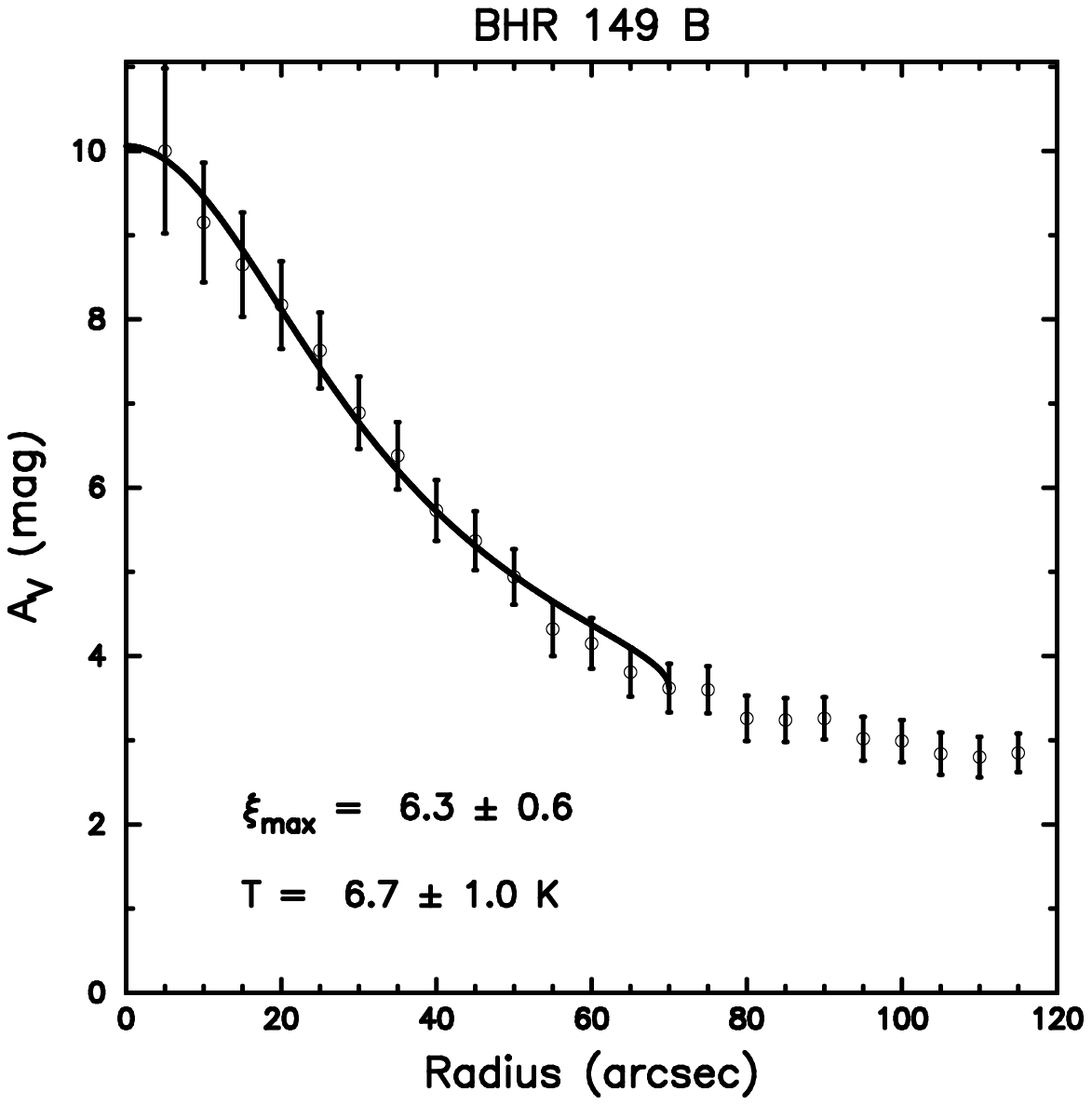}
\end{minipage}
\caption{{\it continued.}}
\end{figure}
%--------------------------------Fig 7: EXTINCTION PROFILES--------------------------------

%--------------------------------Fig 8: EVOLUTION--------------------------------
\clearpage
\begin{figure}
\centering
\includegraphics[scale=0.85]{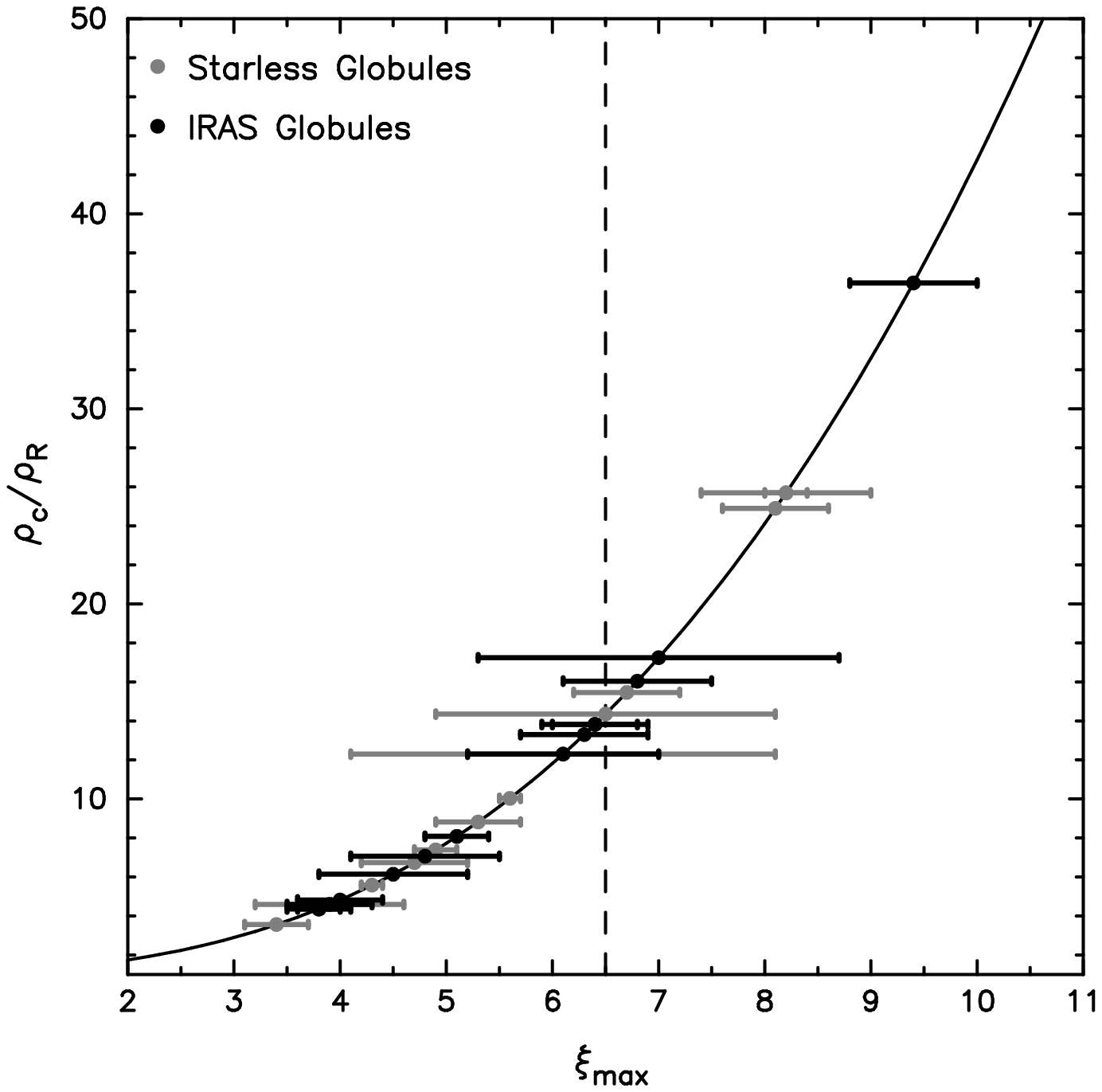}
\caption{Relation between $\xi_{max}$ and the center-to-edge density contrast for dense
         cores in Bok globules. Grey dots correspond to starless globules and black dots
         to IRAS globules. The vertical dashed line denotes the critical value
         $\xi_{max}$ = 6.5.}
\label{fig:evol}
\end{figure}
%--------------------------------Fig 8: EVOLUTION--------------------------------

%--------------------------------Fig 9: XMAX vs TEMP--------------------------------
\clearpage
\begin{figure}
\centering
\includegraphics[scale=0.85]{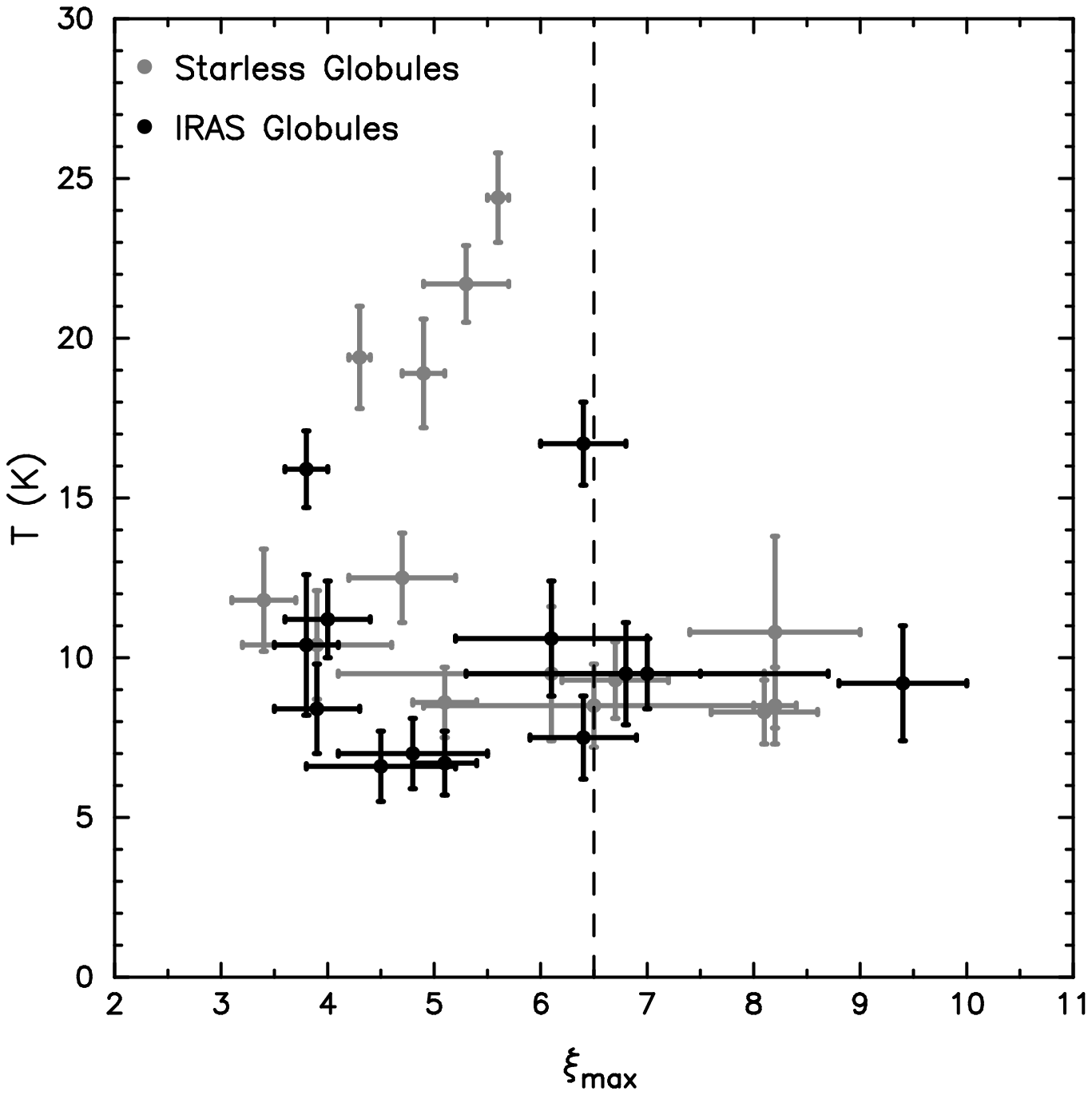}
\caption{Relation between stability parameter and temperature for dense cores in Bok
         globules. The stable globules have $T = 15 \pm 6$~K and the critical plus
         unstable ones have $T = 10 \pm 3$~K. Grey dots correspond to starless globules
         and black dots to IRAS globules. The vertical dashed line denotes the critical
         value $\xi_{max}$ = 6.5.}
\label{fig:xmax_temp}
\end{figure}
%--------------------------------Fig 9: XMAX vs TEMP--------------------------------

%--------------------------------Fig 10: HISTOGRAMS--------------------------------
\clearpage
\begin{figure}
\centering
\includegraphics[scale=0.75]{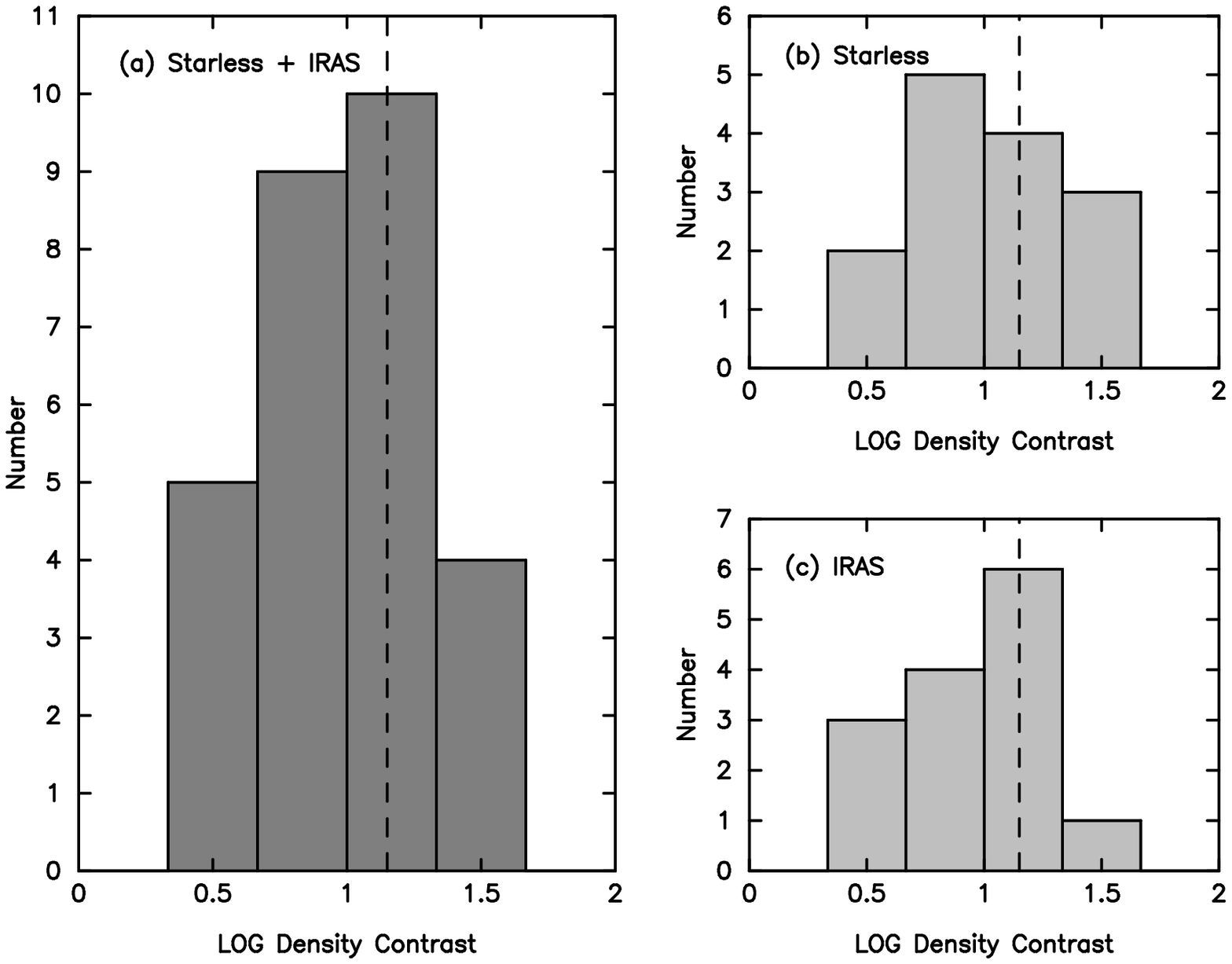}
\caption{Histograms of the logarithmic density contrast for the whole sample of Bok
         globules (a), for starless globules (b), and for IRAS globules (c). The vertical
         dashed line denotes the critical value of the density contrast,
         $\rho_c / \rho_R$ = 14.}
\label{fig:histog}
\end{figure}
%--------------------------------Fig 10: HISTOGRAMS--------------------------------

%--------------------------------Fig 11: PARAMS. vs DENS. CONT.--------------------------------
\clearpage
\begin{figure}
\centering
\begin{minipage}[b]{0.4\textwidth}
 \centering
 \includegraphics[width=5cm]{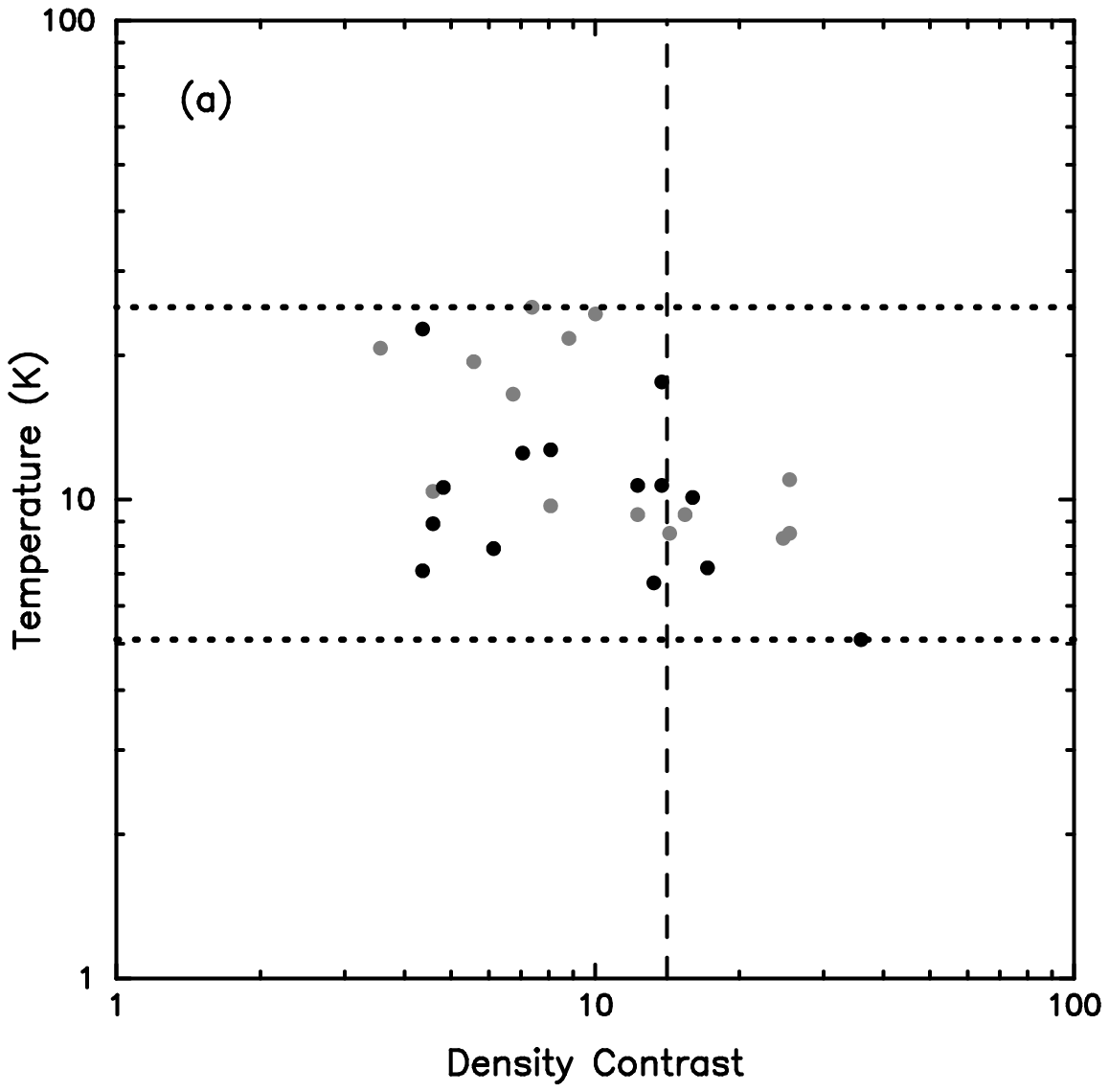}
\end{minipage}
\begin{minipage}[b]{0.4\textwidth}
 \centering
 \includegraphics[width=5cm]{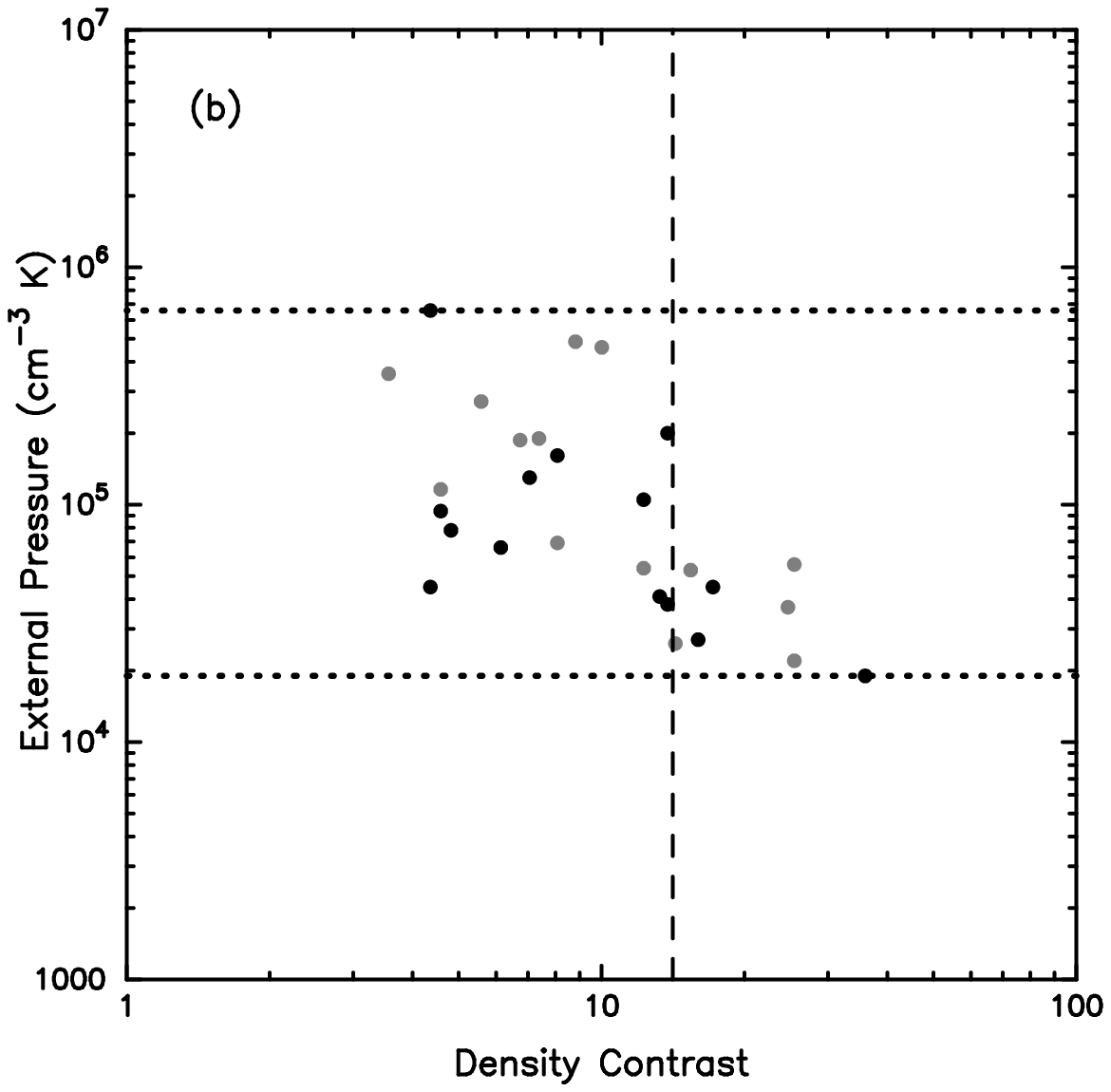}
\end{minipage}\\[1cm]
\begin{minipage}[b]{0.4\textwidth}
 \centering
 \includegraphics[width=5cm]{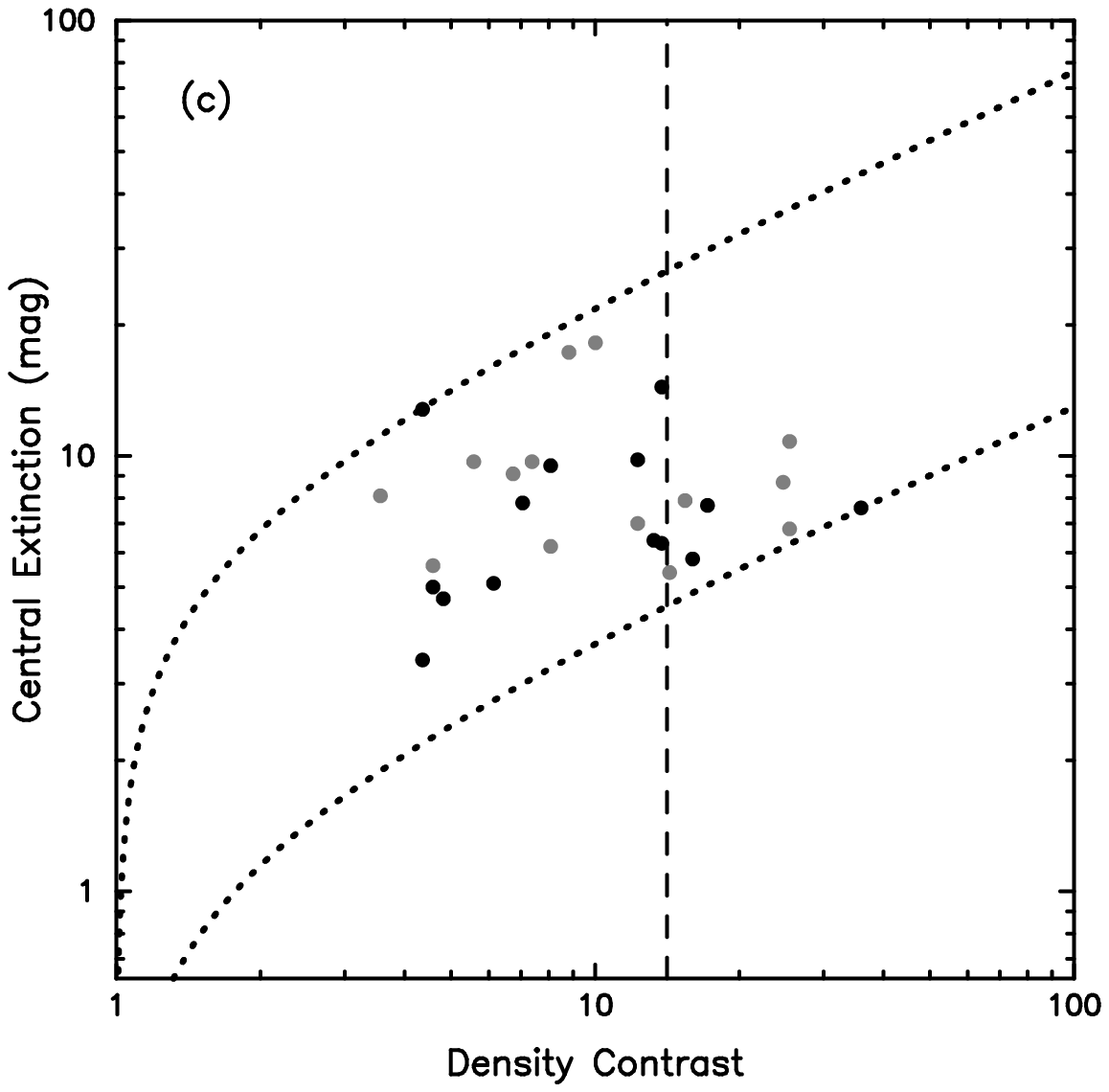}
\end{minipage}
\begin{minipage}[b]{0.4\textwidth}
 \centering
 \includegraphics[width=5cm]{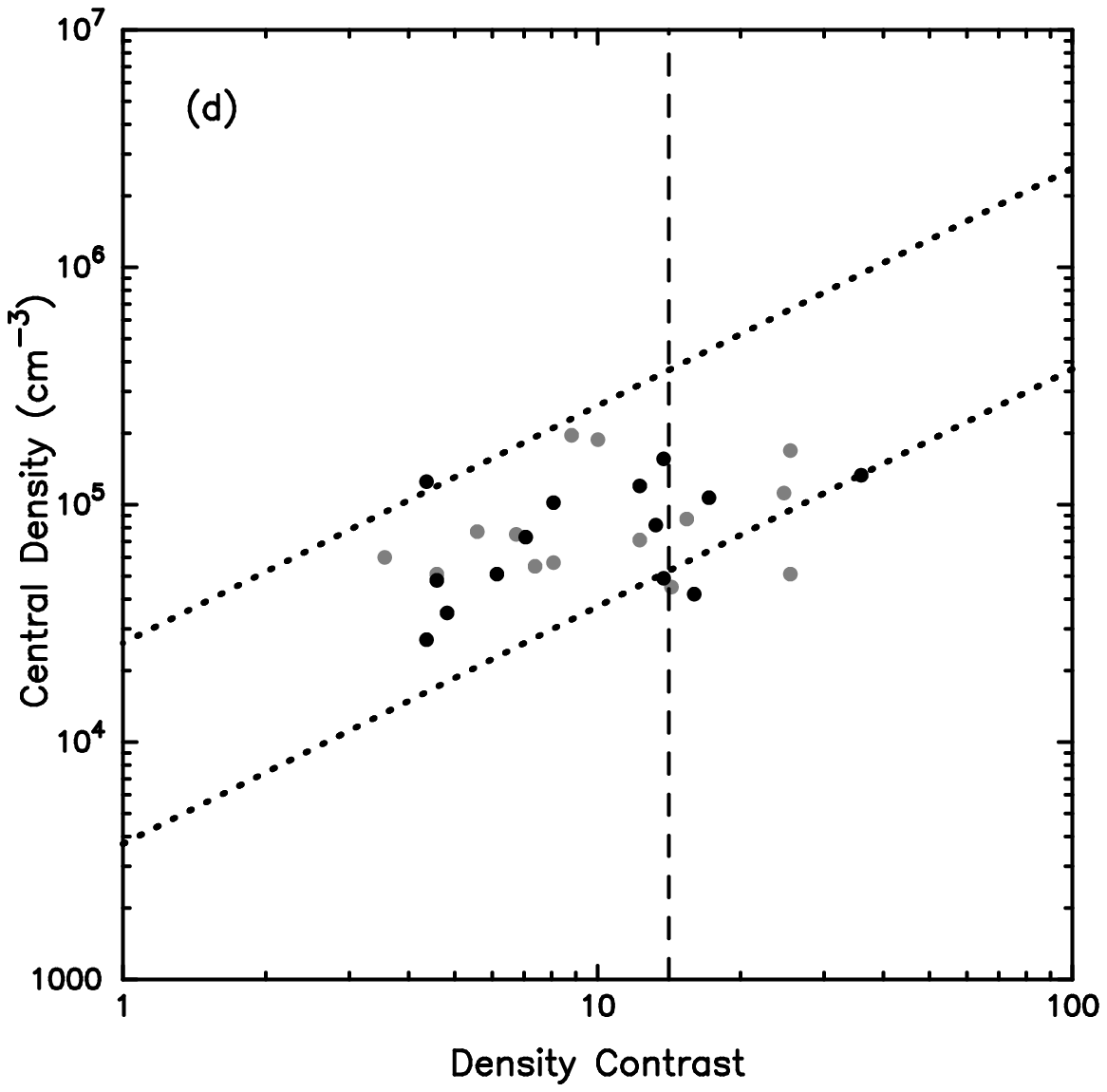}
\end{minipage}\\[1cm]
\begin{minipage}[b]{0.4\textwidth}
 \centering
 \includegraphics[width=5cm]{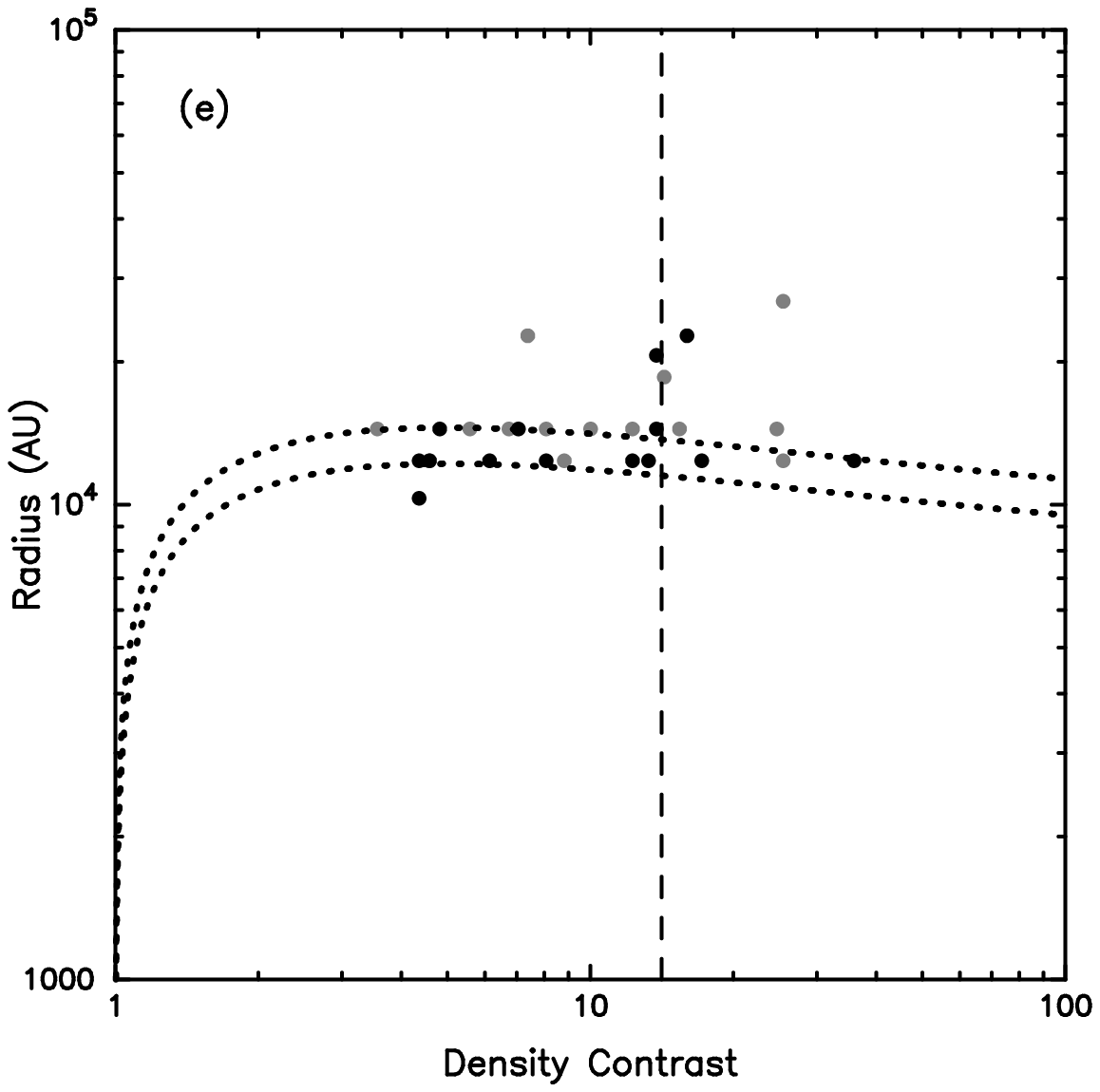}
\end{minipage}
\begin{minipage}[b]{0.4\textwidth}
 \centering
 \includegraphics[width=5cm]{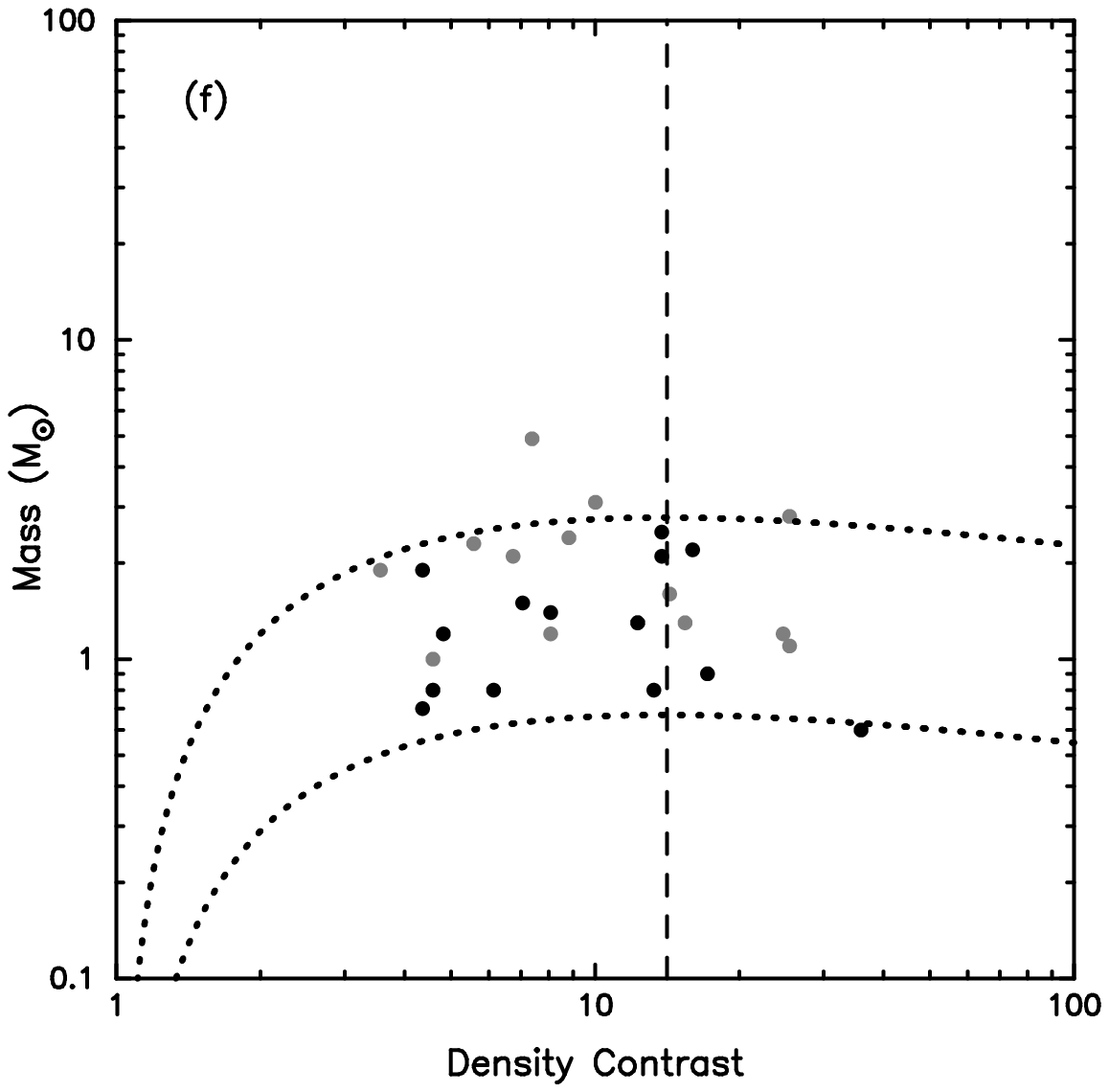}
\end{minipage}
\caption{Correlations between density contrast and the parameters determined from
         the Bonnor-Ebert fitting of dense cores in Bok globules. (a)-(f) Plots for
         temperature, external pressure, central extinction, central density, radius
         and mass, respectively.}
\label{fig:parameters}
\end{figure}
%--------------------------------Fig 11: PARAMS. vs DENS. CONT.--------------------------------

%--------------------------------Fig 12: SEDs--------------------------------
\clearpage
\begin{figure}
\centering
\begin{minipage}[b]{0.2\textwidth}
 \centering
 \includegraphics[width=3cm]{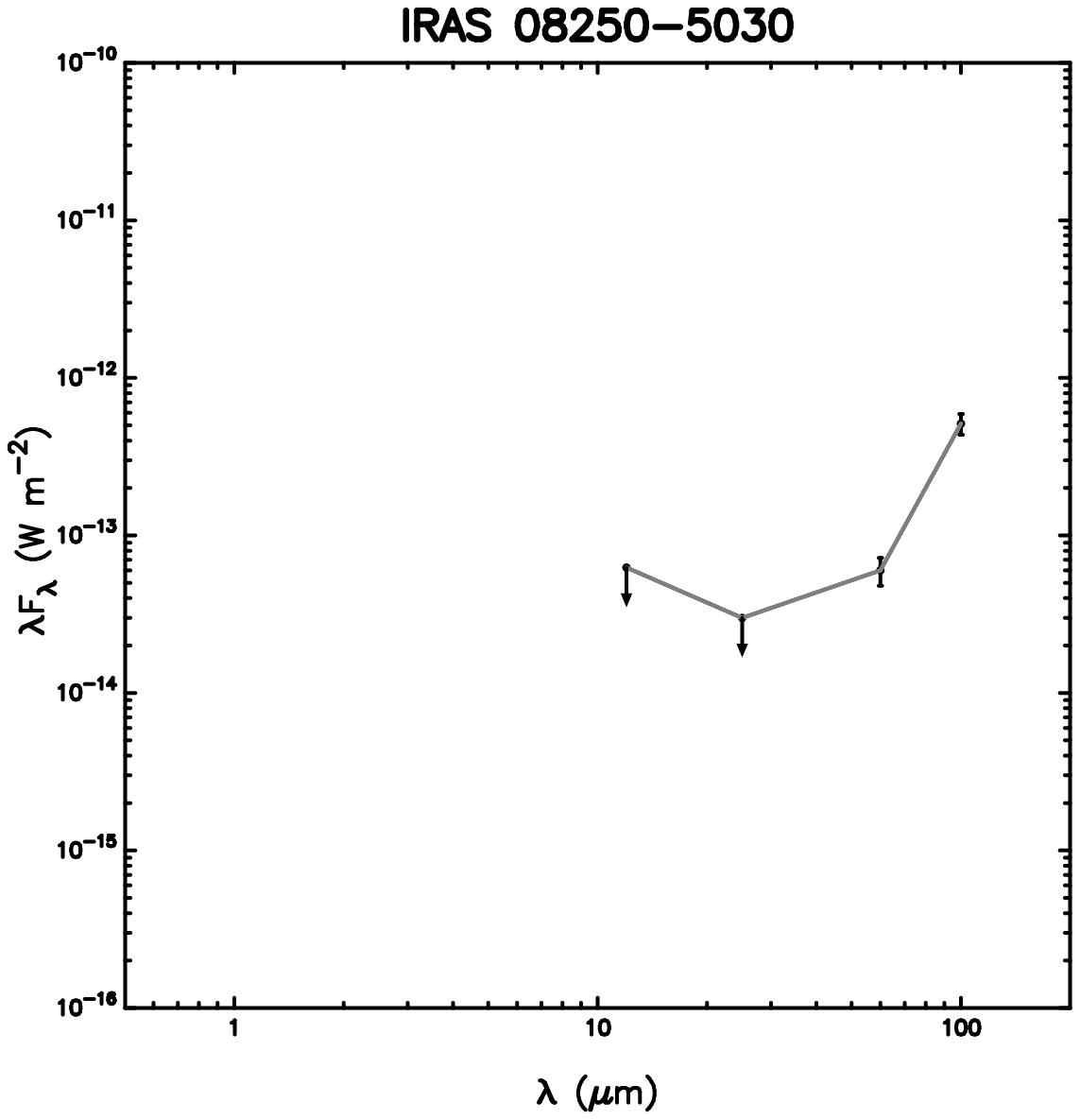}
\end{minipage}
\begin{minipage}[b]{0.2\textwidth}
 \centering
 \includegraphics[width=3cm]{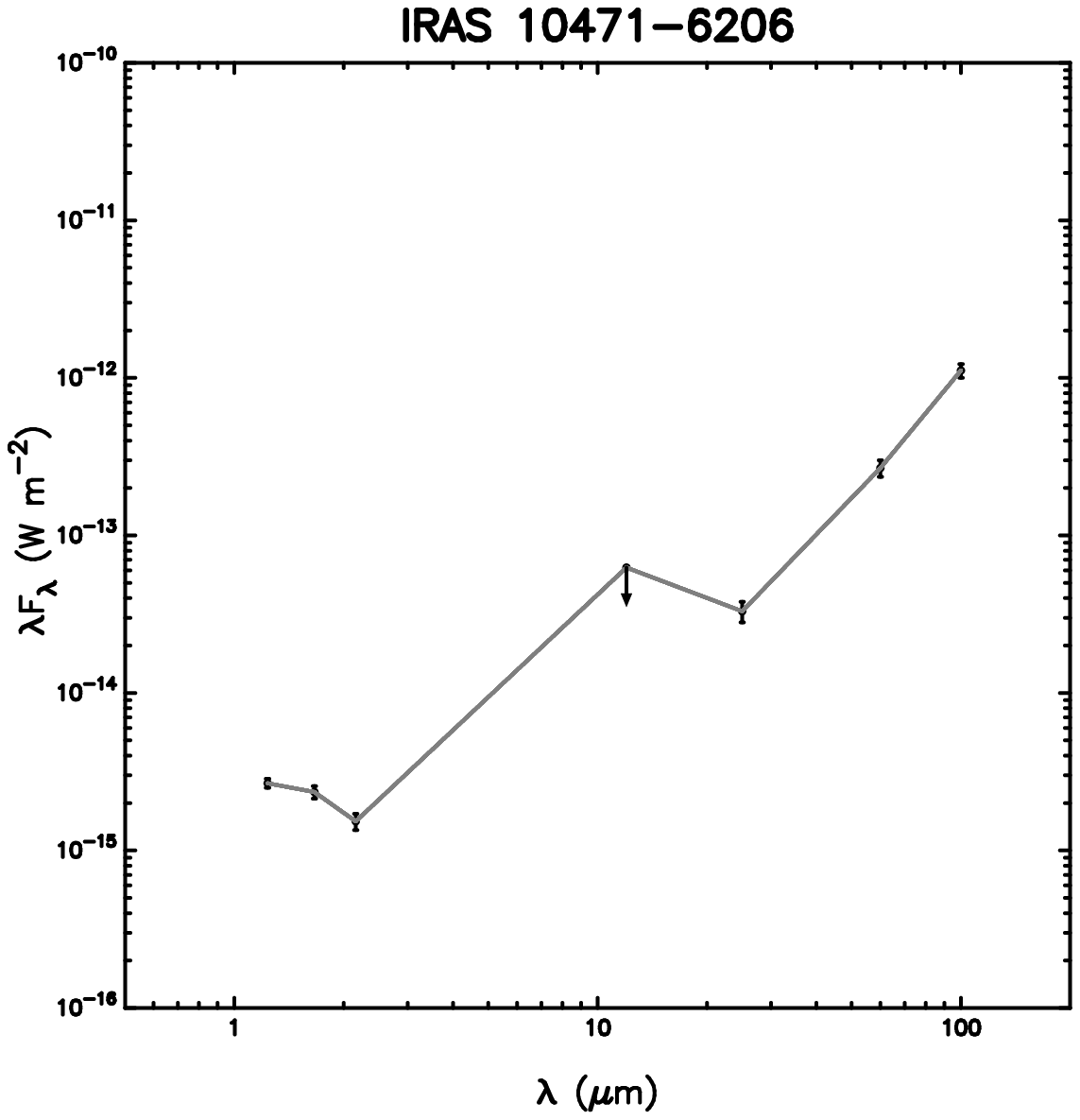}
\end{minipage}
\begin{minipage}[b]{0.2\textwidth}
 \centering
 \includegraphics[width=3cm]{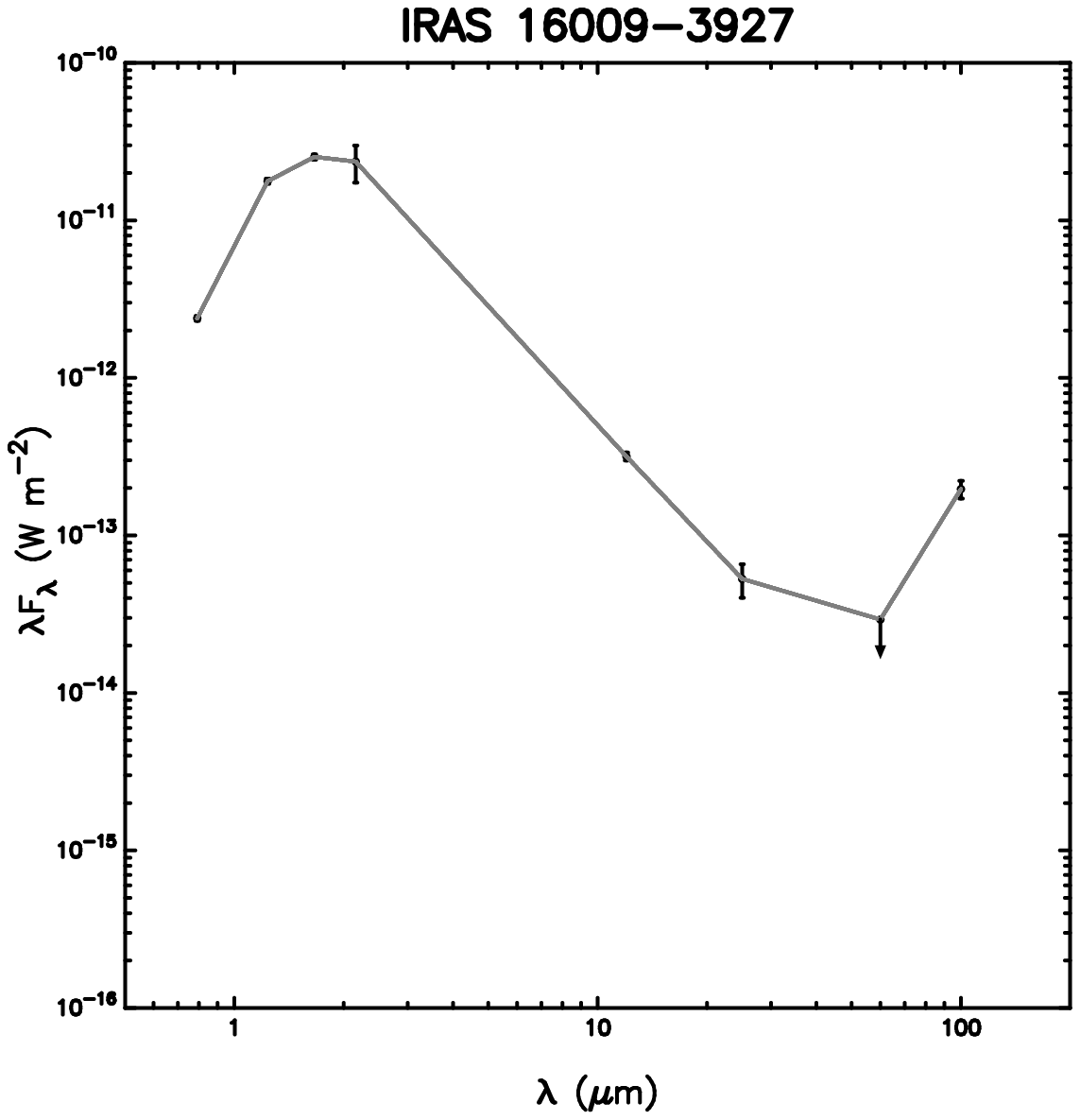}
\end{minipage}
\begin{minipage}[b]{0.2\textwidth}
 \centering
 \includegraphics[width=3cm]{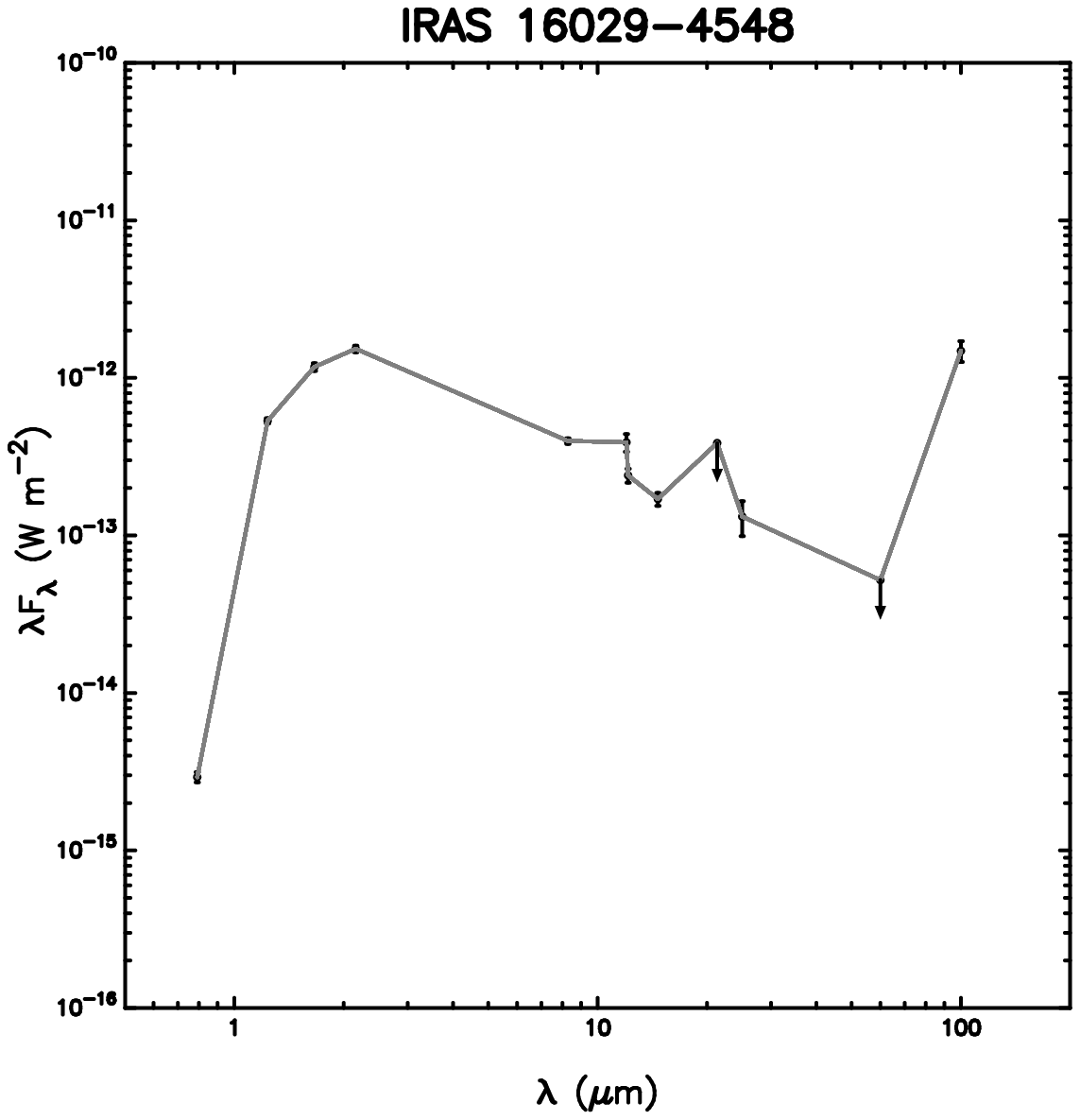}
\end{minipage}\\[0.5cm]
\begin{minipage}[b]{0.2\textwidth}
 \centering
 \includegraphics[width=3cm]{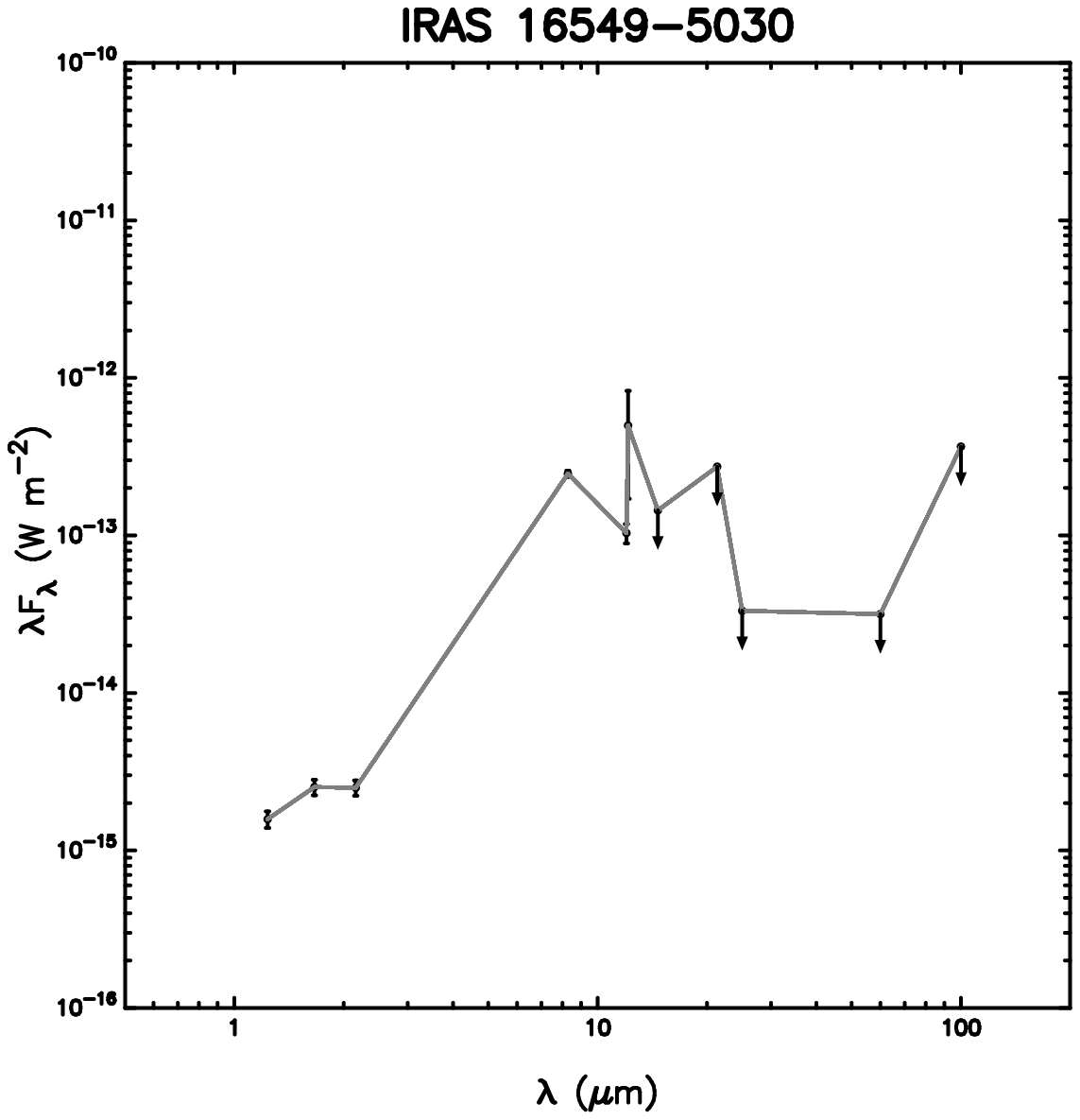}
\end{minipage}
\begin{minipage}[b]{0.2\textwidth}
 \centering
 \includegraphics[width=3cm]{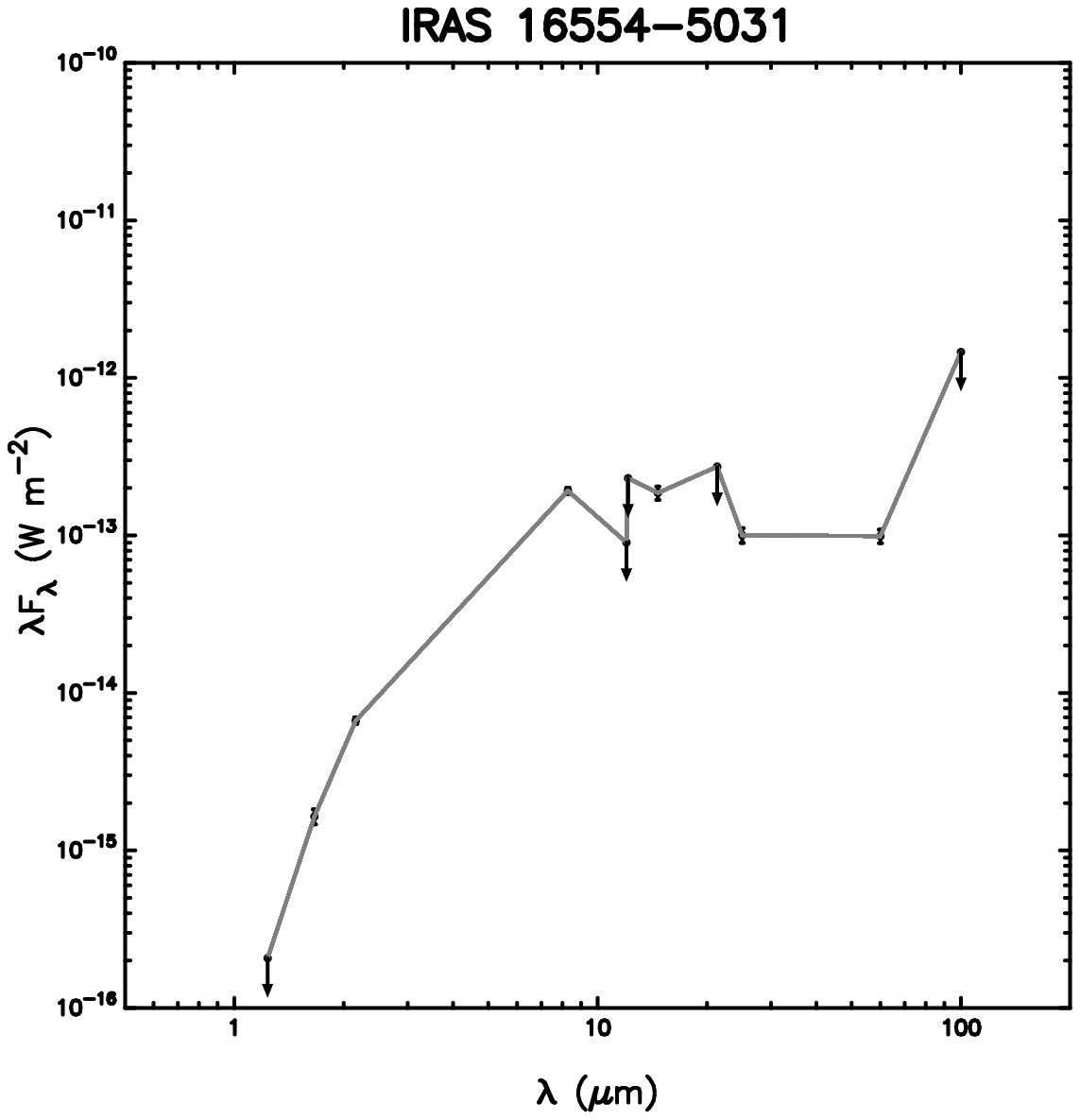}
\end{minipage}
\begin{minipage}[b]{0.2\textwidth}
 \centering
 \includegraphics[width=3cm]{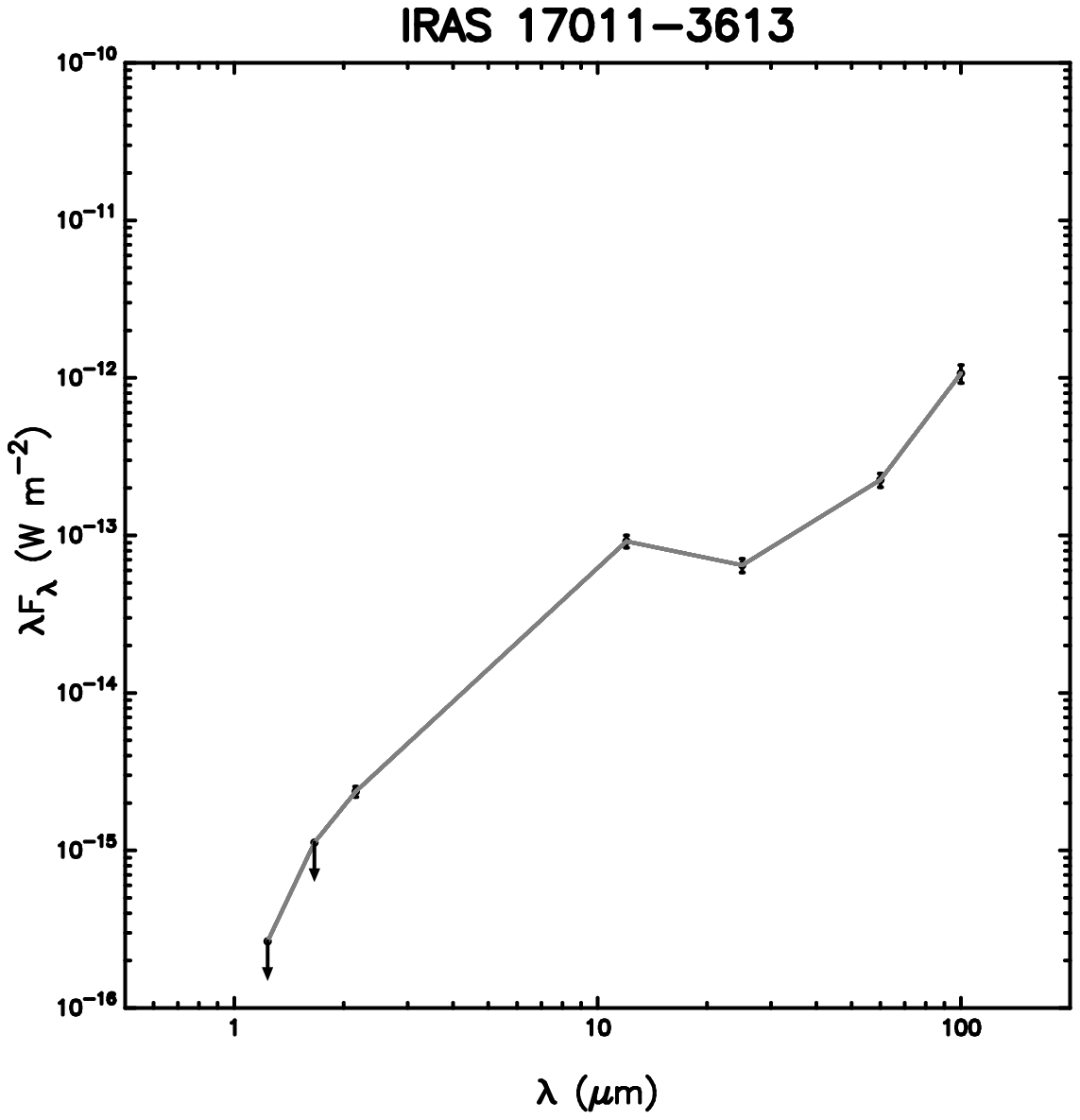}
\end{minipage}
\begin{minipage}[b]{0.2\textwidth}
 \centering
 \includegraphics[width=3cm]{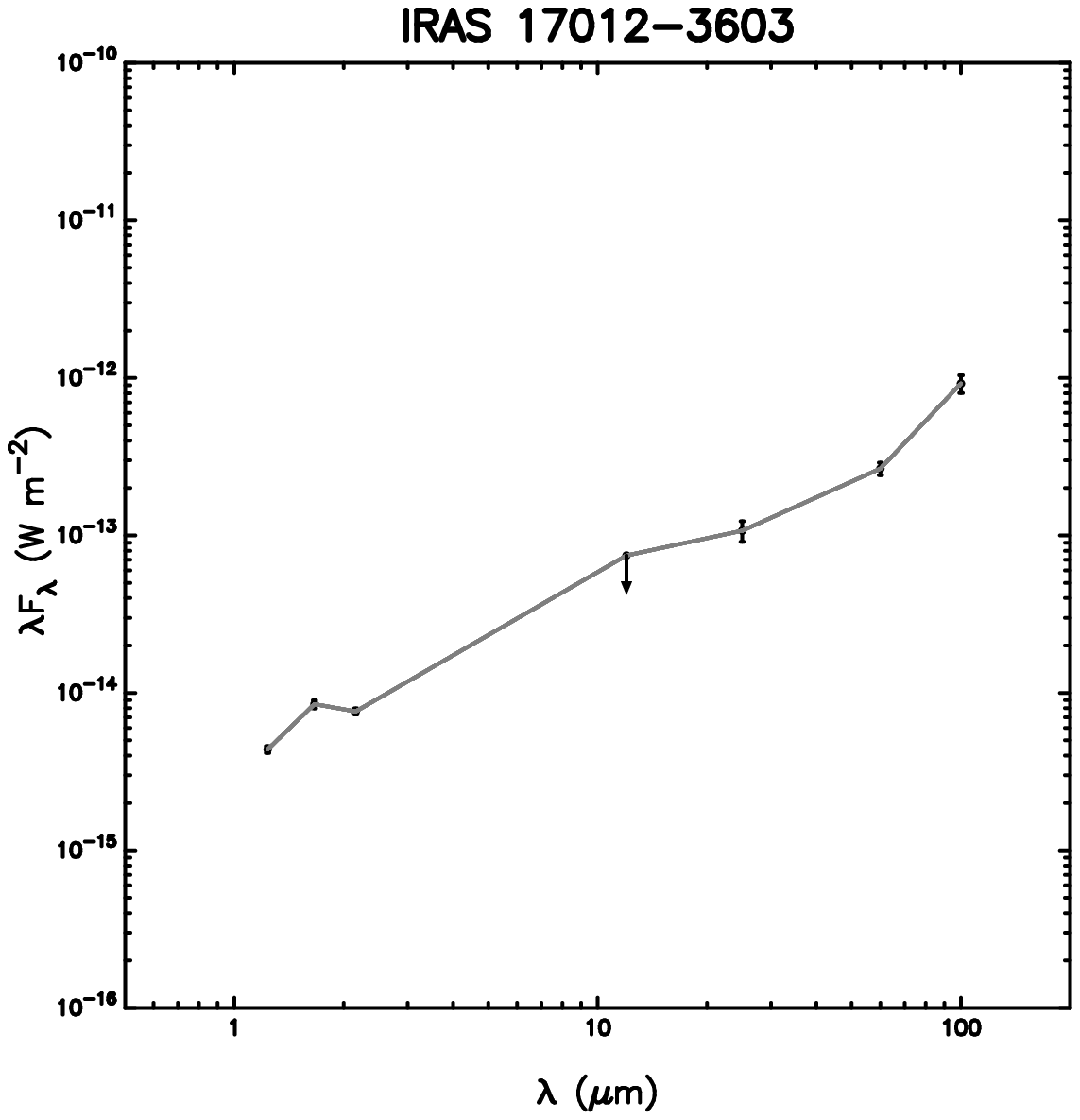}
\end{minipage}\\[0.5cm]
\begin{minipage}[b]{0.2\textwidth}
 \centering
 \includegraphics[width=3cm]{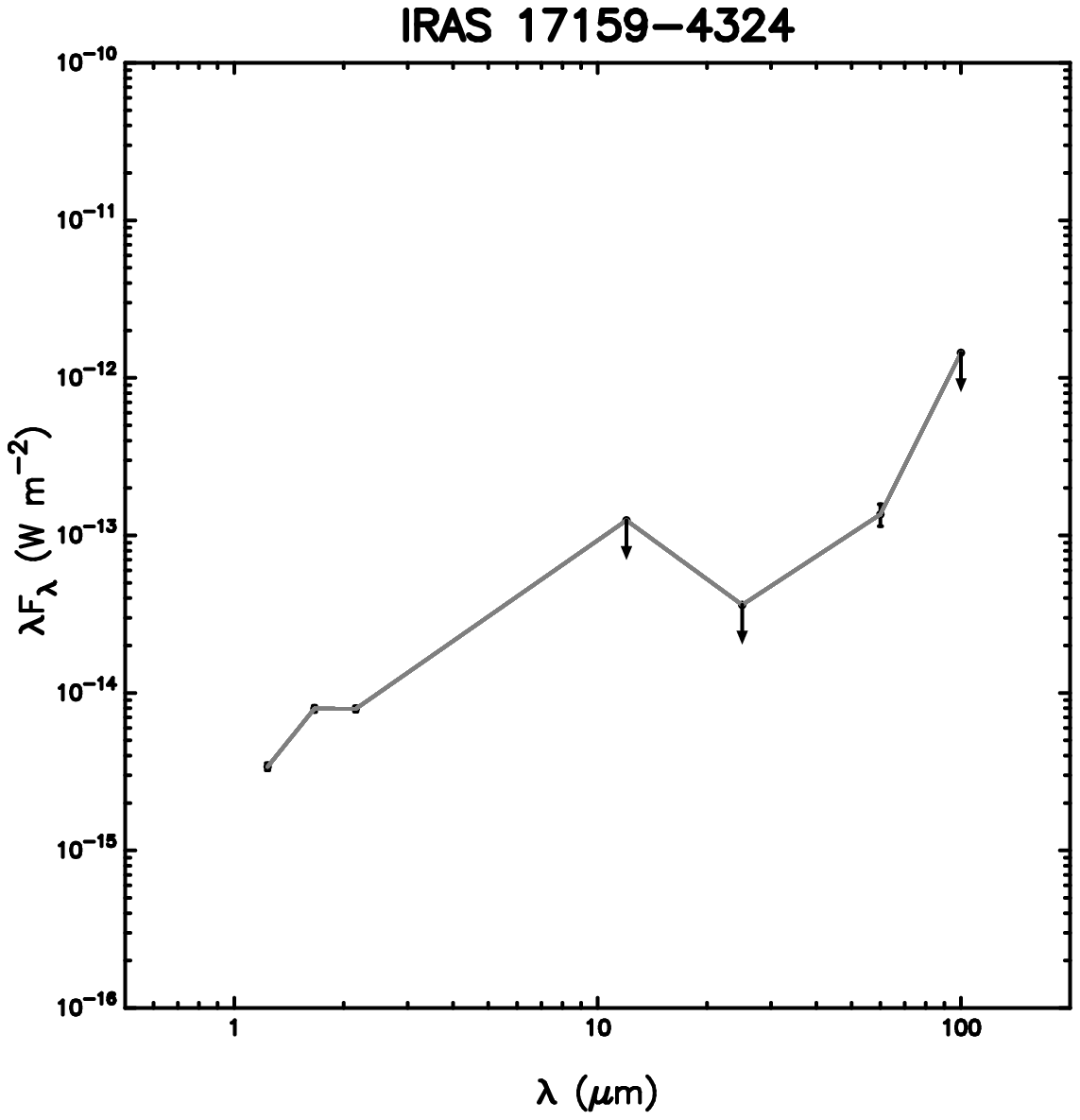}
\end{minipage}
\begin{minipage}[b]{0.2\textwidth}
 \centering
 \includegraphics[width=3cm]{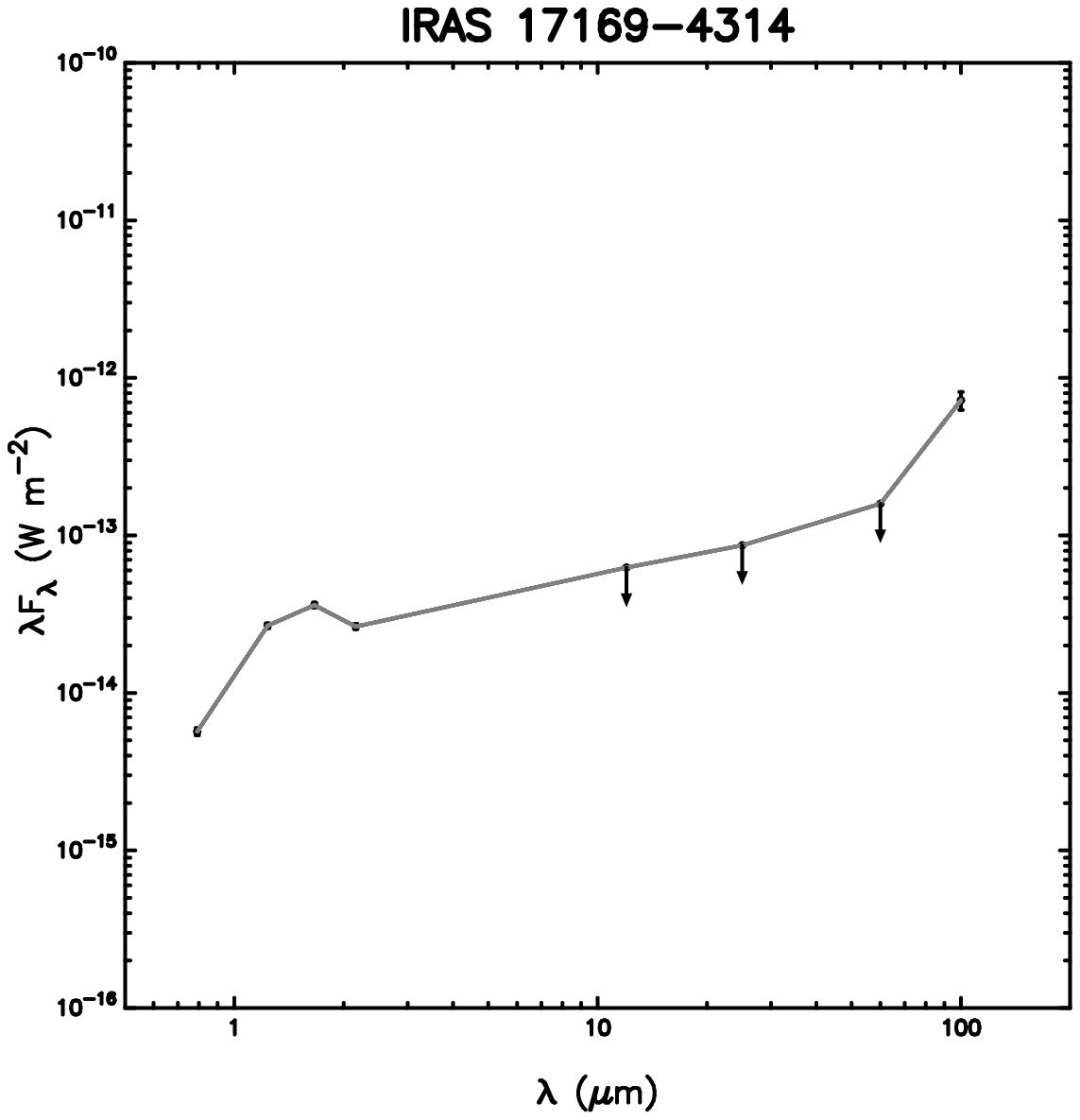}
\end{minipage}
\begin{minipage}[b]{0.2\textwidth}
 \centering
 \includegraphics[width=3cm]{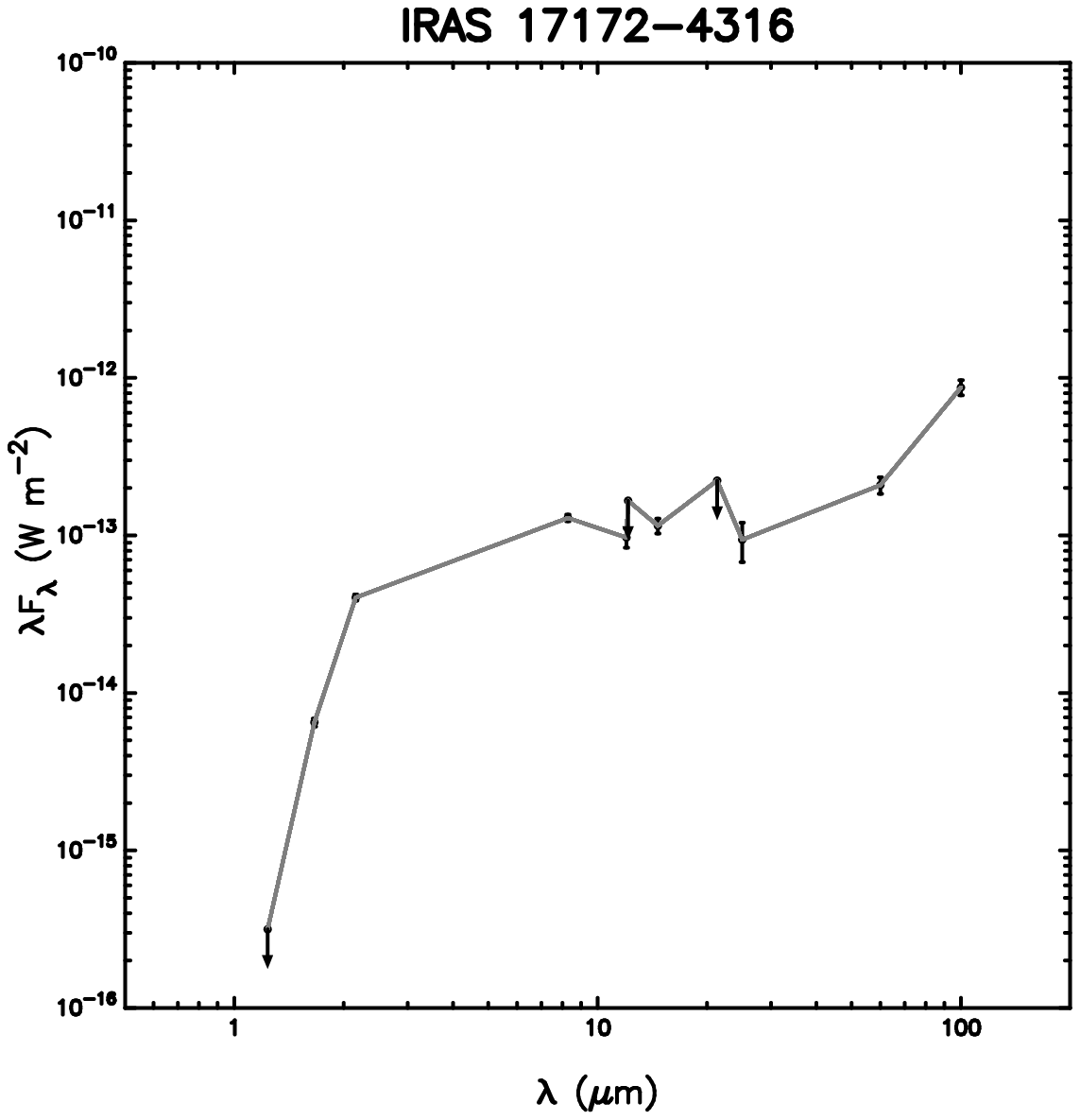}
\end{minipage}
\begin{minipage}[b]{0.2\textwidth}
 \centering
 \includegraphics[width=3cm]{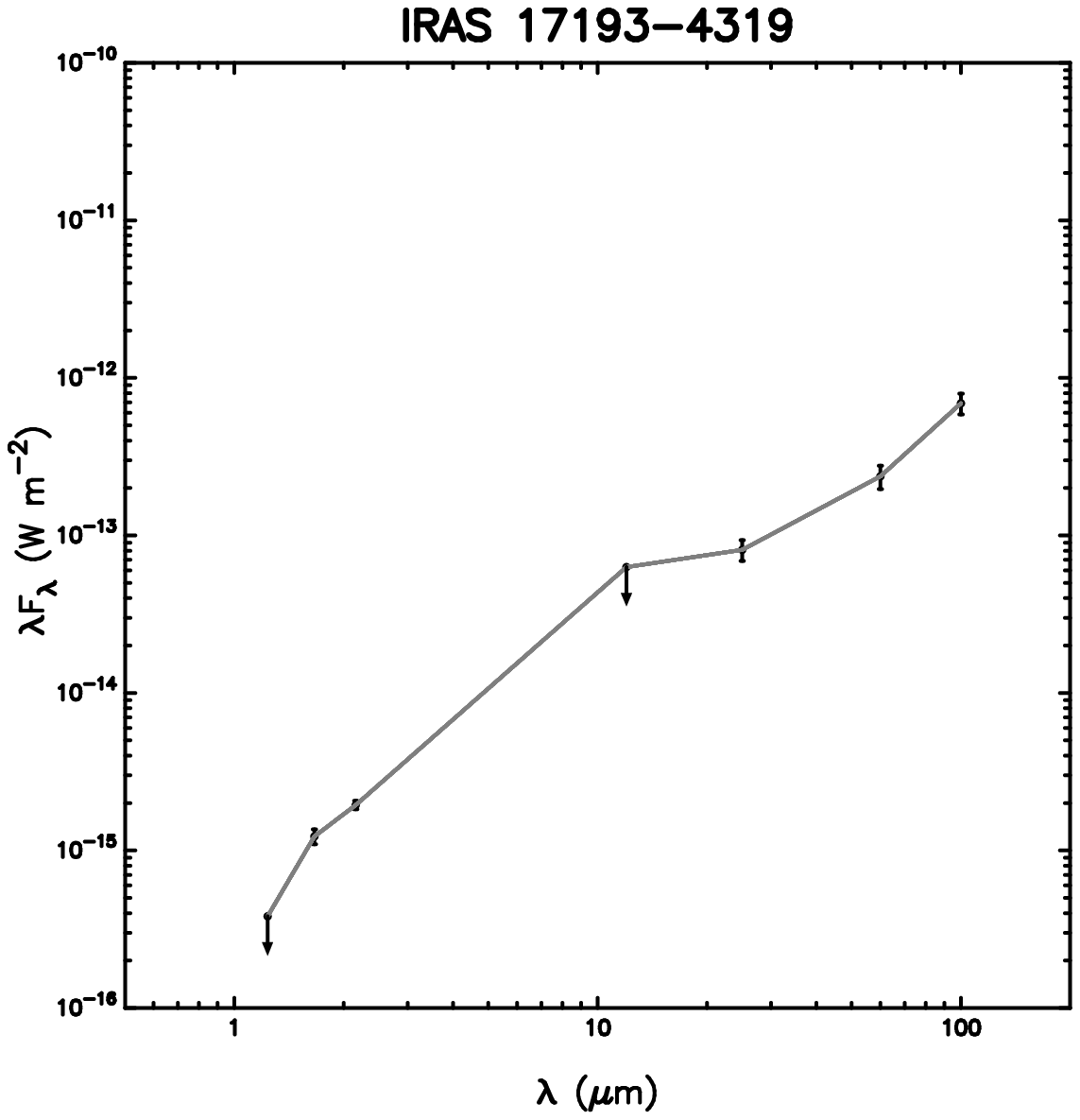}
\end{minipage}\\[0.5cm]
\begin{minipage}[b]{0.2\textwidth}
 \centering
 \includegraphics[width=3cm]{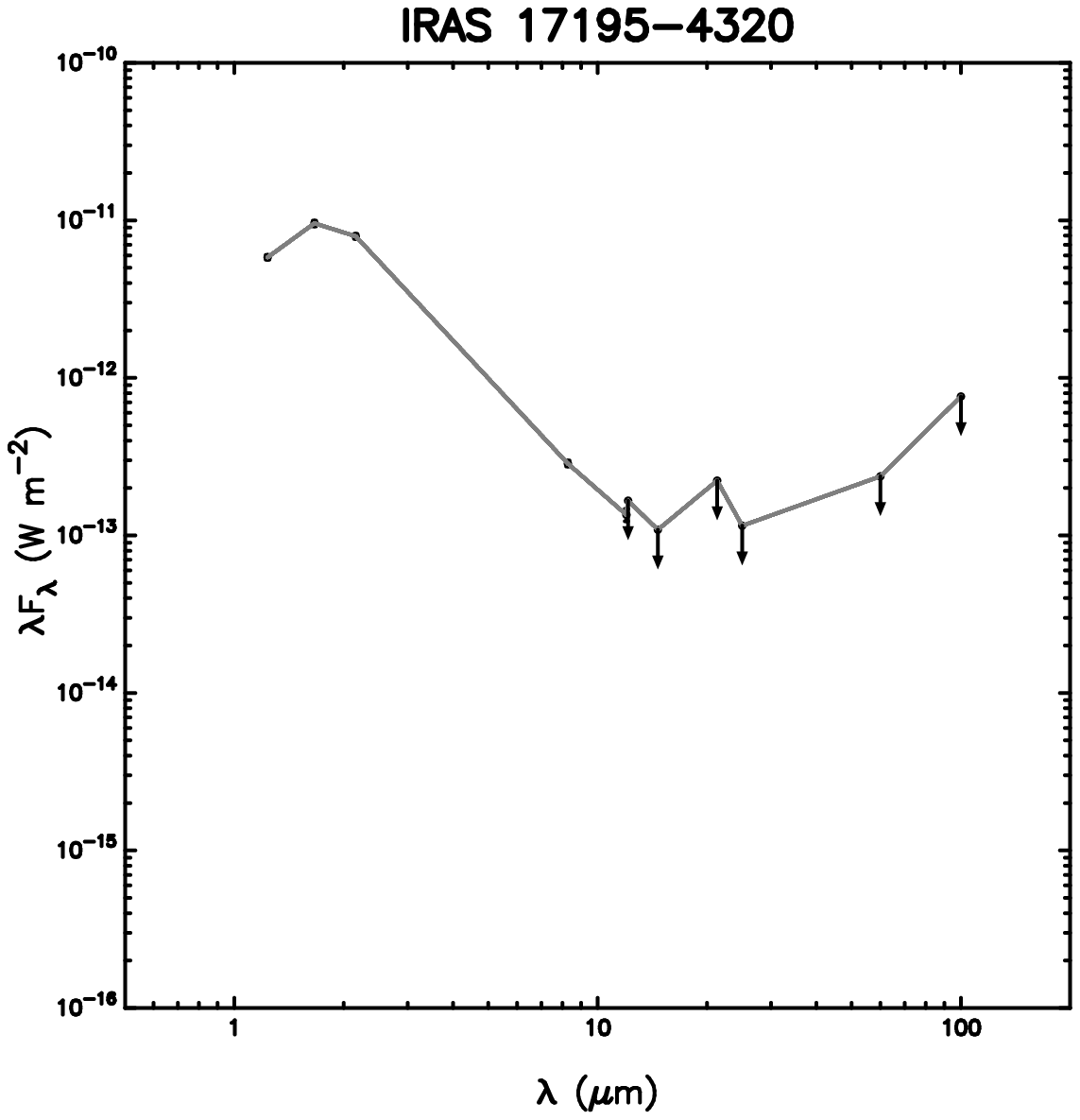}
\end{minipage}
\begin{minipage}[b]{0.2\textwidth}
 \centering
 \rule{3cm}{0pt}
\end{minipage}
\begin{minipage}[b]{0.2\textwidth}
 \centering
 \rule{3cm}{0pt}
\end{minipage}
\begin{minipage}[b]{0.2\textwidth}
 \centering
 \rule{3cm}{0pt}
\end{minipage}
\caption{Spectral energy distributions for infrared IRAS sources associated to
         Bok globules.}
\label{fig:seds}
\end{figure}
%--------------------------------Fig 12: SEDs--------------------------------

%--------------------------------Fig 13: BLT Diagram--------------------------------
\clearpage
\begin{figure}
\centering
\includegraphics[scale=0.85]{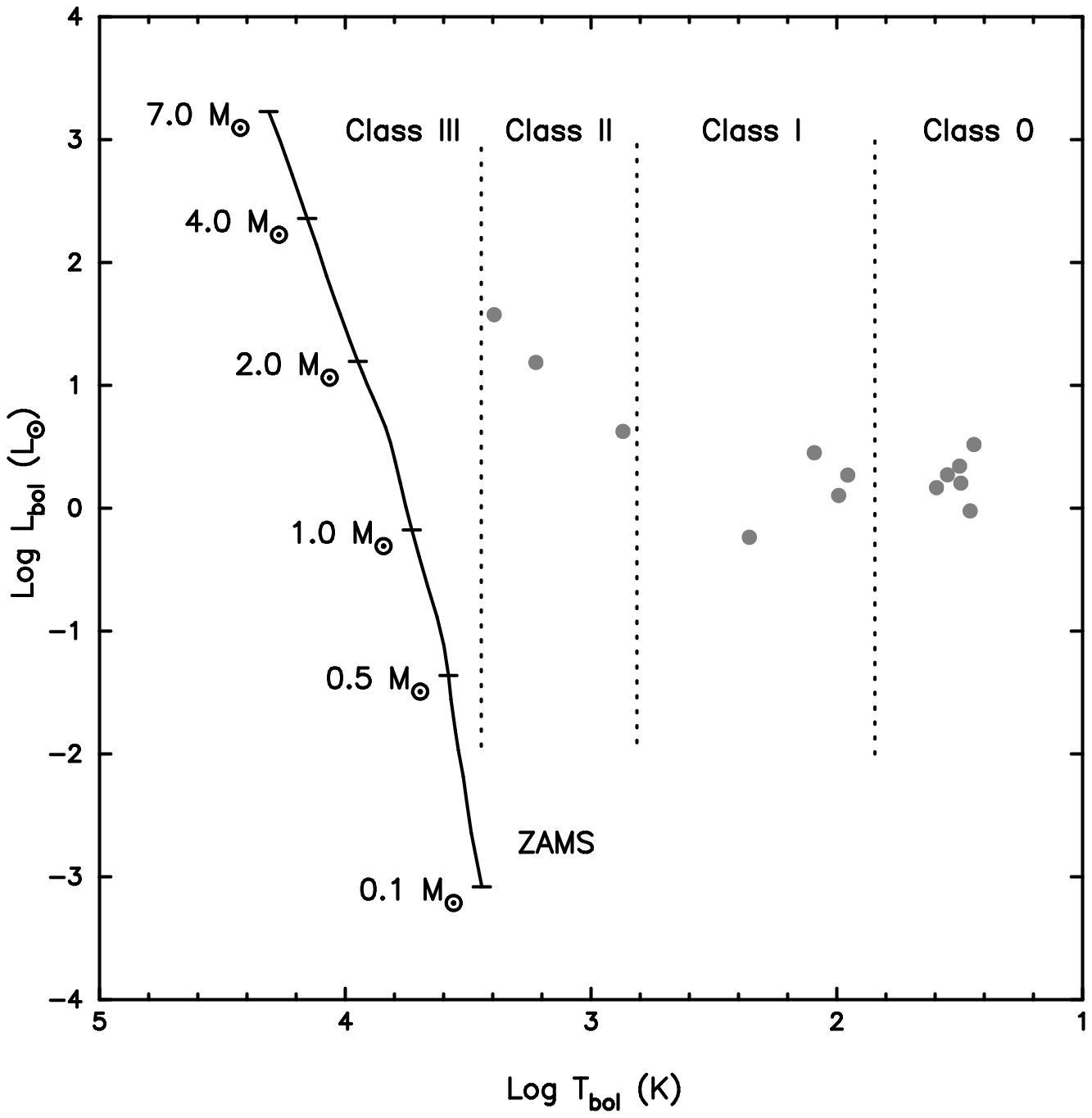}
\caption{BLT diagram of infrared IRAS sources associated to Bok globules. The solid line
         indicates the ZAMS \citep{siess00}. The vertical dashed lines determine the limits
         of the different spectral classes \citep{chen95}.}
\label{fig:blt}
\end{figure}
%--------------------------------Fig 13: BLT Diagram--------------------------------

%--------------------------------Fig 14: CO SPECTRA--------------------------------
\clearpage
\begin{figure}
\epsscale{0.85}
\plottwo{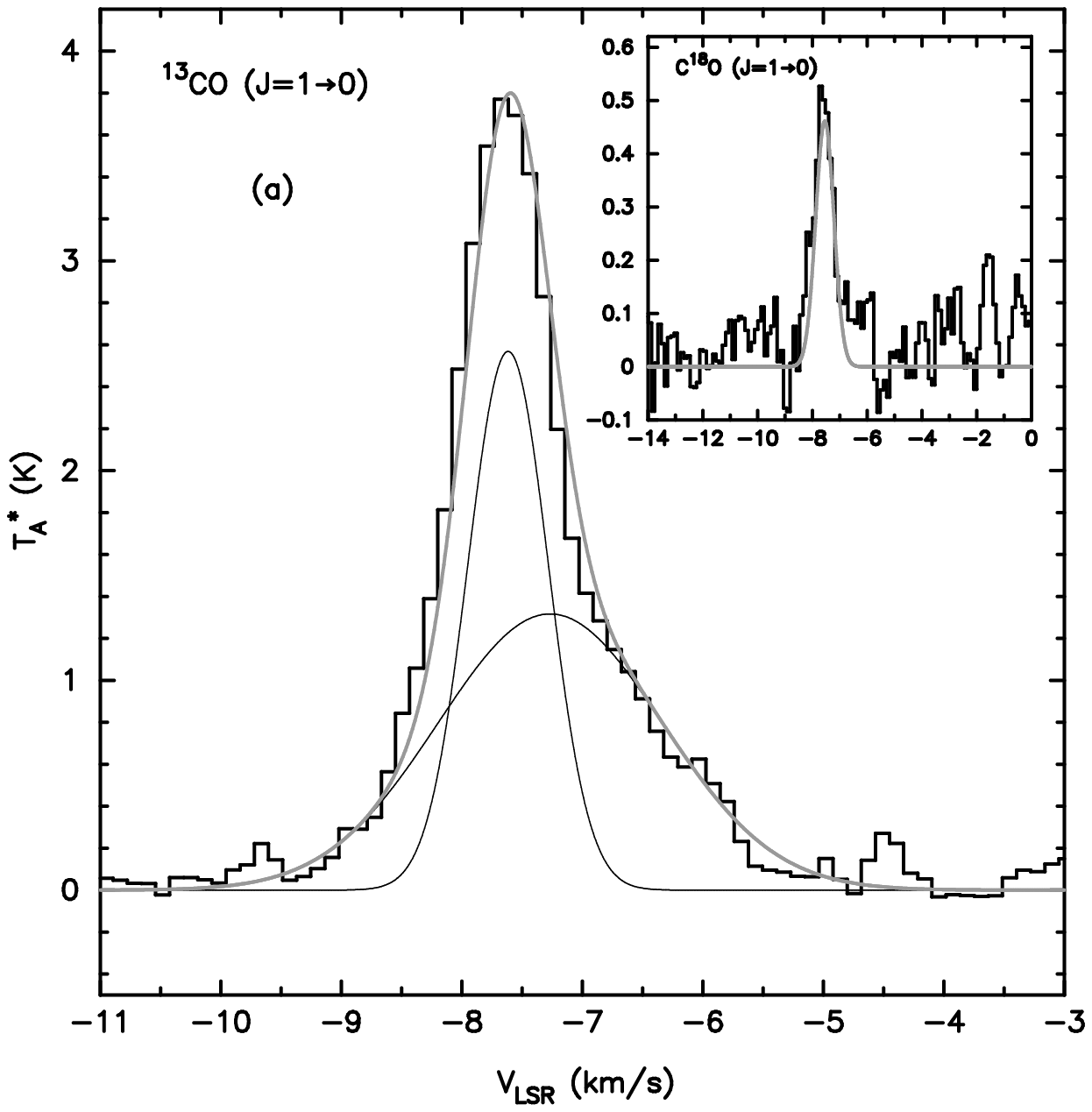}{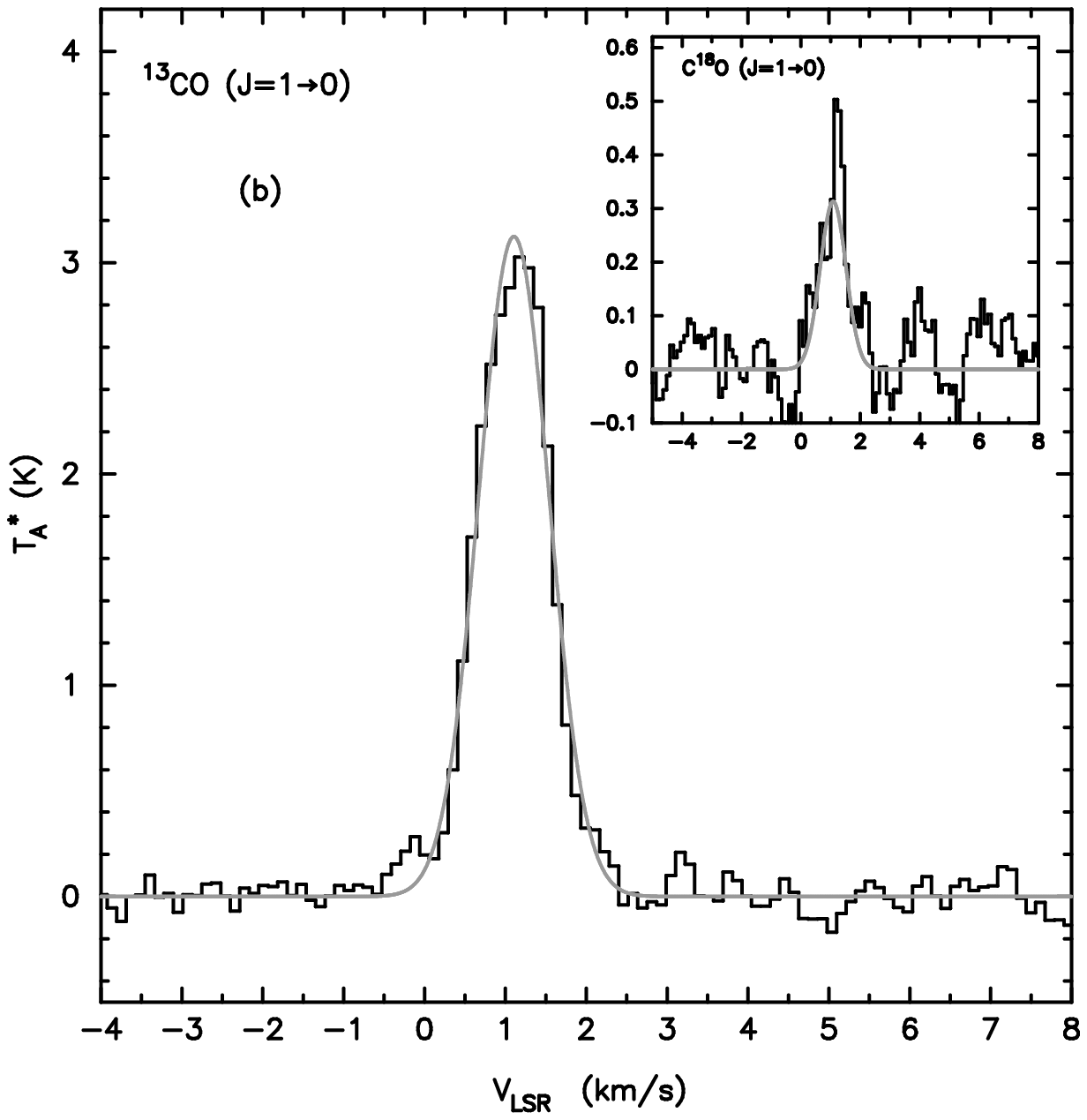}
\caption{\tco\ and \ceio\ line profiles for (a) BHR 138 and (b) BHR 149.}
\label{fig:spectra}
\end{figure}
%--------------------------------Fig 14: CO SPECTRA--------------------------------

%--------------------------------Fig 15: TWO-LAYER MODEL--------------------------------
\clearpage
\begin{figure}
\centering
\includegraphics[scale=0.85]{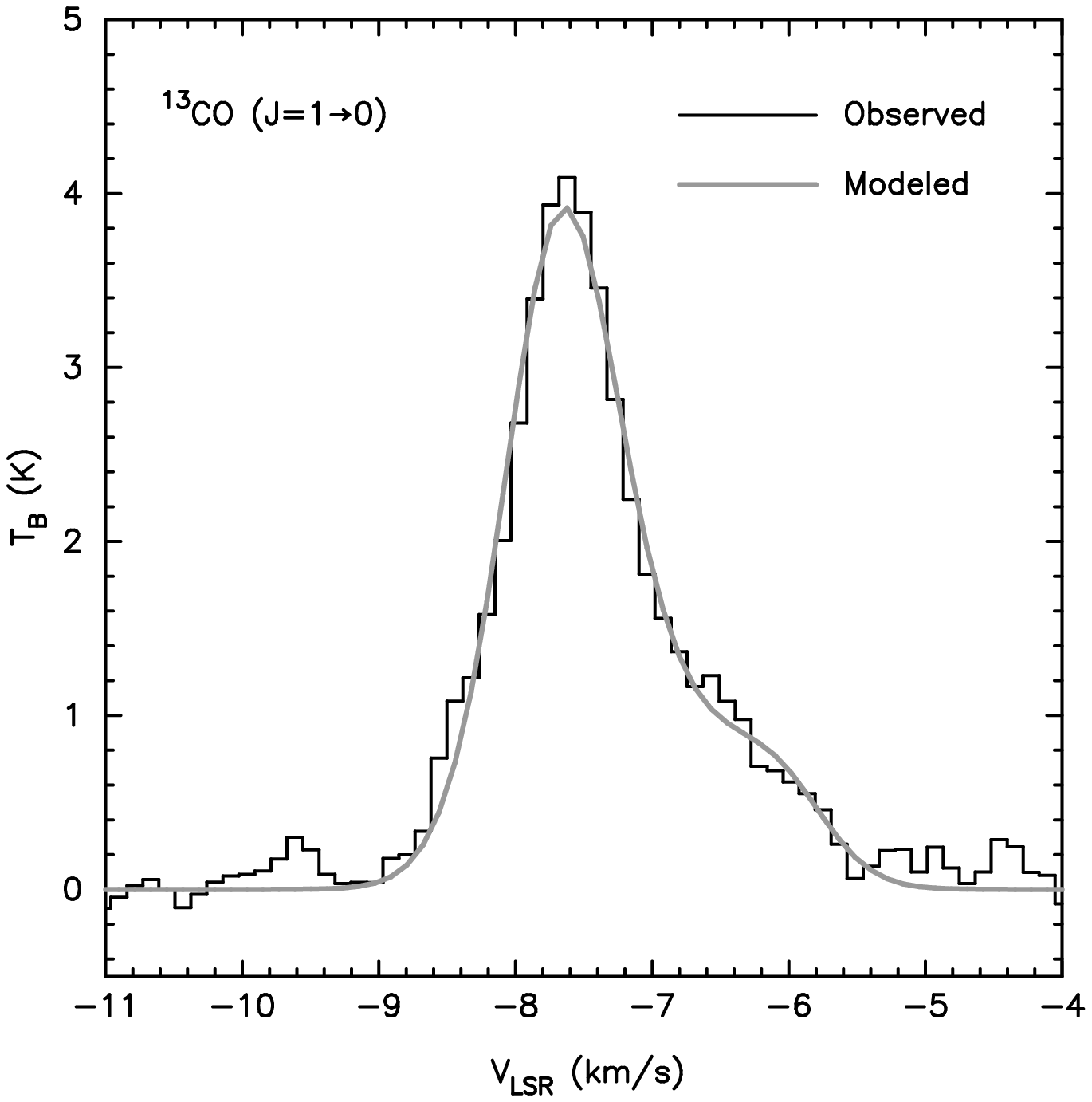}
\caption{Modelling of the \tco\ line profile for BHR 138. The histogram represents the
         observational data and the grey curve corresponds to the {\it two-layer} fit model,
         which provides an infall velocity of 0.25 km s$^{-1}$.}
\label{fig:twolayer}
\end{figure}
%--------------------------------Fig 15: TWO-LAYER MODEL--------------------------------

%--------------------------------BIBLIOGRAPHY--------------------------------

%--------------------------------BIBLIOGRAPHY--------------------------------

\end{document}